\documentclass[11pt]{article}
\newif\iffigs\figstrue

\usepackage{epsfig,latexsym}
\usepackage{amsmath}
\usepackage{verbatim}
\usepackage{mathrsfs}
\usepackage{amssymb}

\DeclareMathAlphabet{\mathpzc}{OT1}{pzc}{m}{it}

 \csname
@addtoreset\endcsname{equation}{section}

\def\gz0{\gamma^{0}}

\def\sign{\rm sign}

\def\scs#1{\section{\sc #1}}

\def\g{\gamma}

\def\vf{\varphi}

\def\be{\begin{equation}}
\def\ee{\end{equation}}
\def\bea{\begin{eqnarray}}
\def\eea{\end{eqnarray}}
\def\ba{\begin{array}}
\def\ea{\end{array}}
\def\bec{\begin{center}}
\def\ec{\end{center}}
\def\ba{\begin{align}}
\def\ena{\end{align}}

\def\12{\frac{1}{2}}

\begin{document}

\begin{flushright}
{\today}
\end{flushright}

\vspace{10pt}

\begin{center}


{\Large\sc Integrable Scalar Cosmologies}\vskip 24pt
{\sc I. Foundations and links with String Theory}\\


\vspace{25pt}
{\sc P.~Fr\'e${}^{\; a}$, A.~Sagnotti${}^{\; b}$ and A.S.~Sorin$^{\; c}$}\\[15pt]

{${}^a$\sl\small Dipartimento di Fisica, Universit\`a di Torino\\INFN -- Sezione di Torino \\
via P. Giuria 1, \ 10125 Torino \ ITALY \\}e-mail: {\small \it
fre@to.infn.it}\vspace{8pt}

{${}^b$\sl\small
Scuola Normale Superiore and INFN\\
Piazza dei Cavalieri, 7\\ 56126 Pisa \ ITALY \\
e-mail: {\small \it sagnotti@sns.it}}\vspace{10pt}

{${}^c$\sl\small Bogoliubov Laboratory of Theoretical Physics\\
Joint Institute for Nuclear Research \\ 141980 Dubna, Moscow Region, RUSSIA
\\ }e-mail: {\small \it
sorin@theor.jinr.ru}\vspace{8pt}

\vspace{24pt} {\sc\large Abstract}\end{center}
\noindent
We build a number of integrable one--scalar spatially flat cosmologies, which play a natural role in inflationary scenarios, examine their behavior in several cases and draw from them some general lessons on this type of systems, whose potentials involve combinations of exponential functions, and on similar non--integrable ones. These include the impossibility for the scalar to emerge from the initial singularity descending along asymptotically exponential potentials with logarithmic slopes exceeding a critical value (``climbing phenomenon'') and the inevitable collapse in a big Crunch whenever the scalar tries to settle at negative extrema of the potential. We also elaborate on the links between these types of potentials and ``brane supersymmetry breaking'', a mechanism that ties together string scale and scale of supersymmetry breaking in a class of orientifold models.
\setcounter{page}{1}

\pagebreak

\tableofcontents

\newpage
\section{\sc  Introduction}\label{sec:intro}
The Cosmic Microwave Background (CMB) is a mine of profound hints on the early history of our Universe \cite{XCMB}. Together with the results obtained by the Cosmological Supernova project \cite{Xsupernovaproject}, the impressive data on its temperature fluctuations recorded during the last decade or so have led in fact to a new paradigm for Cosmology reflecting three main facts:
\begin{itemize}
  \item our Universe is highly \emph{isotropic} and \emph{homogeneous} at large scales, while its current state of acceleration is well accounted for by a small positive cosmological constant;

  \item our Universe is spatially flat, which brings to the forefront metrics of the form
      \begin{equation}\label{piatttosa}
        ds^2 \ = \ e^{\,2\,{\cal B}(t)} \, \mathrm{d}t^2 \ - \ a^2(t) \ \mathrm{d}\mathbf{x}\cdot \mathrm{d}\mathbf{x} \ .
      \end{equation}
Special ``gauge functions'' ${\cal B}(t)$ can result in simpler expressions for the scale factor $a(t)$, which becomes a quantity of utmost interest for Theoretical Physics;

  \item vacuum energy accounts for about $70 \%$ of the present contents of the Universe, dark matter of unknown origin for another $24 \%$, so that only $6 \%$ is left for conventional baryonic matter in the form of luminous stars and galaxies.
\end{itemize}

The inflationary scenario is today the widely adopted paradigm to interpret these facts within a consistent framework \cite{inflation}. It ascribes the observed spatial flatness of our Universe to a primeval accelerated expansion of the scale factor $a(t)$ that was somehow injected right after the Big Bang. This epoch of acceleration was then gracefully exited leaving way to more standard epochs of decelerated expansion dominated by radiation and matter, and both the early stage of inflation and the ensuing history of the Universe can find a rationale within the framework of General Relativity. Actually, even the current acceleration could perhaps originate from a small relic of the early inflationary phase that is resuming some prominence after other forms of energy have been diluted by billions of years of cosmological expansion.

Although various attempts have been made over the years to generate inflation by different means \cite{different_inflation}, its simplest and most natural realization remains based on the coupling of Einstein gravity to scalar fields $\Phi^I $ with self interactions driven by potential functions ${V}(\Phi)$. The vacuum energy during inflation is then essentially the slowly varying value of this potential function, and the available solutions of the coupled Einstein--Klein-Gordon equations provide possible backgrounds for evolutionary histories of the Universe.  Not only does this viewpoint provide the simplest route toward an analytical formulation of the inflationary scenario, but it is also conceptually most appealing, since it opens up a direct bridge between Cosmology and the Fundamental Interactions. Scalar fields play in fact a key role in symmetry breaking mechanisms, and the recent discovery of a 126 GeV scalar boson provides a striking evidence for their role in Nature \cite{BEH}. Furthermore, they are ubiquitous and abundant in Supergravity \cite{SUGRA} in diverse dimensions, and their geometries codify the structure of supersymmetric Lagrangians. Once Supergravity is connected to Superstrings \cite{strings}, its scalar fields acquire a higher--dimensional origin, encode properties of the compactification, accompany brane and orientifold tensions and thus acquire a real status of messengers of Fundamental Physics.

Black--hole solutions of Supergravity have been widely studied and classified \cite{BH_SUGRA}, but exact time--dependent solutions of the same equations describing inflationary scenarios are comparatively scarce. Indeed, most of the work done on inflation during the last two decades and its testable implications for the CMB power spectrum were derived within the \emph{slow--roll} approximation \cite{slow_roll}, which imposes rather mild conditions on the potential function $V(\Phi)$. As a result, most models rest on a variety of \emph{ad hoc} potentials, typically of polynomial form, for a special scalar field usually termed the \emph{inflaton} that is supposed to have driven inflation and part of the subsequent history of the Universe.
This state of affairs might seem rather disappointing, but actually it has long been regarded as a sign of robustness, and indeed it is usually stressed how key consequences of inflation are rather insensitive to the detailed form of the potential. Yet, the history of Mathematical Physics shows that exact analytic solutions of simplified and idealized models can often provide deeper insights, since after all a real understanding of fundamental physical processes is only attained when one controls entire moduli spaces, whose corners can hide much significant information.

Exact solutions of one--field cosmological models are now acquiring more interest in view of the comparison with CMB data, since the precision reached by the PLANCK and WMAP experiments \cite{XCMB} allows more detailed tests of the Power Spectrum of primeval quantum fluctuations. Moreover, standard treatments do not address the key issue of the onset of the inflationary phase, let alone the possibility that it be an inevitable fate for our Universe, while exact solutions for a scalar coupled to gravity may have some bearing on the low--$\ell$ end of the CMB fluctuations and their possible non--gaussian features. On a more mathematical note, exact solutions of the Einstein--Klein-Gordon equations could in principle be accompanied by exact solutions for the fluctuation equations of various types of fields, which could bring along further insights into their behavior.

With these motivations in mind, in this paper we have performed a wide search for systems involving a single scalar field $\phi$ minimally coupled to gravity whose potentials $V(\phi)$ result in complete classical integrability.  Our work rests, to a large extent, on important results obtained in Mathematical Physics on integrable dynamical systems with two degrees of freedom \cite{hietarinta} -- \cite{inozem}. Families of integrable dynamical systems depending on one or more parameters, as well as a number of sporadic examples, can indeed be connected, within suitable ranges for the independent variable $t$, to Einstein--Klein-Gordon systems whose potentials $V(\phi)$ involve rational or irrational combinations of exponentials of $\phi$ in the metrics of eq.~\eqref{piatttosa}. The detailed analysis of critical points and the explicit solutions of the resulting cosmological models reveal clearly two key features of these systems:
\begin{itemize}
\item the \emph{climbing phenomenon}, whereby the scalar field \emph{cannot emerge} from the initial singularity climbing down potentials that are asymptotically exponential with logarithmic slopes exceeding a critical value. Or, if you will, the impossibility for scalar fields to overcome, in a contracting phase, the attractive force of such potential ends. The physical meaning of this phenomenon was first elucidated in \cite{dks} in the simple exponential potential, although the corresponding solutions have a long history \cite{lm,exp_sol}. Possible imprints on the low--$\ell$ tail of the CMB power spectrum were then discussed in \cite{dkps}, while an analysis of the mechanism near the initial singularity was recently presented in \cite{cd};
\item the eventual collapse in a Big Crunch of systems of this type whenever the scalar tends to settle at a negative extremum of the potential $V(\phi)$. This was expected: it reflects the fact that AdS has no spatially flat metrics, or that negative extrema are non--admissible fixed points for the corresponding dynamical systems.
\end{itemize}
This insight made it possible to also understand the gross features of a number of complicated sporadic systems whose Liouville integrability does not translate into handy ways of solving their field equations.
Moreover, general features of two-dimensional dynamical systems suggest that integrable models capture at least the gross features of these systems, where chaotic behavior can be generically excluded.

Let us stress that potentials involving combinations of exponentials play an important role in Supergravity and in String Theory. In particular, potentials involving a single exponential emerge at tree level in brane--antibrane systems, where, however, they are generally accompanied by tachyonic excitations, but also in classically stable ``brane supersymmetry breaking'' \cite{bsb} (BSB) vacuum configurations of (anti)branes and orientifolds \cite{orientifolds}. In these models SUSY is broken at the string scale and is non--linearly realized in the low--energy Supergravity \cite{dmpr}. Moreover, in general gauged supergravity models based on non--compact coset manifolds $\mathrm{G/H}$ one can always resort to a \textit{solvable} parametrization, and the scalar fields then fall in two classes \cite{SUGRA_solvable}:
 \begin{itemize}
  \item the fields $\mathfrak{h}^i$ associated with the Cartan generators of the Lie algebra of $G$, whose number equals the rank $r$ of the coset and whose kinetic terms, determined by the invariant metric of $\mathrm{G/H}$, are canonical up to an overall constant;
  \item the axions $b^I$ associated with the roots of the Lie algebra of $G$, whose kinetic terms depend instead on both the Cartan fields $\mathfrak{h}^i$ and the $b^I$.
\end{itemize}
The scalar potentials of gauged Supergravity are in general polynomial functions of coset representatives, so that once the axions are set to constant values, which solves their field equations, one is left with combinations of exponentials of Cartan fields. Moreover, as shown in \cite{dks}, the non--minimal axion couplings can have the effect of ``freezing'' these fields close to the initial singularity. All in all, a final consistent truncation to a single Cartan field $\phi$, with all others stabilized at extremal values, leads generically to potentials involving combinations of exponentials of $\phi$.

Having ascertained that exponential functions play a role in Supergravity and in String Theory, three intriguing questions emerge:
\begin{enumerate}
  \item Can the integrable models that we have identified be realized within conventional gauged Supergravity, and for what choices of fluxes? This proviso is important, since some of the simplest potentials in our list do appear, albeit in versions where SUSY is non--linearly realized.
  \item Can integrable potentials provide interesting insights on inflationary scenarios behind the slow--roll regime, in addition to those encoded by the single--exponential potential, the simplest member of the set, that already revealed the existence of the climbing phenomenon?
  \item How much can one learn from integrable potentials about Cosmology with similar non--integrable potentials?
\end{enumerate}

The first question is perhaps the most difficult one, but it is also particularly interesting since a proper understanding of the issue will encode low--energy manifestations of non--perturbative string effects present in these contexts even with supersymmetry broken at high scales. It will be dealt with in detail elsewhere \cite{prep}.

The second question has encouraging answers.  There are indeed two classes of handily integrable models where an early climbing phase leaves way to inflation during the ensuing descent (models (2) and (9) in Table 1). This setting can leave interesting imprints on the low--$\ell$ portion of the CMB power spectrum \cite{dkps} that are qualitatively along the lines of WMAP and PLANCK data and is close to BSB orientifold models, although not quite identical to them. Model (6) in Table 1 is perhaps the most interesting of all the examples that we are presenting, since it can even combine, in a rather elegant and relatively handy fashion, an early climbing phase with tens of $e$--folds of slow--roll inflation and with a graceful exit to an eventual phase of decelerated expansion.

Finally, the extensive literature on two--dimensional dynamical systems implies a positive answer to the third question. It turns out, in fact, that the dynamical system counterparts of our cosmological equations experience behaviors that are largely determined by the nature of their fixed points, and more specifically by the eigenvalues of their linear approximations in the vicinity of them. As a result, when an integrable potential has the same type of fixed points as a physically interesting non--integrable one, its exact solutions are expected to provide trustable clues on the actual physical system. This result is very appealing, despite the absence of general estimates of the error, and will be illustrated further in \cite{prep} comparing analytical and numerical solutions for interesting families of potential wells that include the physically relevant case of the $STU$ model \cite{mapietoine}.

Summarizing, we have constructed a wide list of one--field integrable cosmologies and we have examined in detail the properties of their most significant solutions, arriving in this fashion at a qualitative grasp of the general case. We have also addressed the question of whether the integrable models provide valuable approximations of similar non--integrable models, and in this respect we have obtained encouraging results that find a rationale in the ascertained behavior of corresponding two--dimensional dynamical systems.

The structure of the paper is as follows. In section \ref{sec:integpotenti} we derive an effective dynamical model that encompasses the possible
$d$--dimensional Friedman--Lemaitre--Robertson--Walker (FLRW) spatially flat cosmologies driven by a scalar field
$\phi$ with canonical kinetic term and self interaction produced by a potential function $V(\phi)$.
In Section \ref{sec:integrable} we describe the methods used to build integrable dynamical systems
and identify nine different families of one--scalar cosmologies that are integrable for suitable choices of the gauge function ${\cal B}(t)$ of eq.~\eqref{piatttosa}. In Section \ref{sec:properties_exact} we analyze the generic properties of dynamical systems in two variables, we describe the general classification of their fixed points and we illustrate the corresponding behavior of the solutions of Section \ref{sec:integrable}. We then discuss in detail the exact solutions of several particularly significant systems identified in Section \ref{sec:integrable} and illustrate a number of instructive lessons that can be drawn from them. In Section \ref{sec:sporadic_pot} we describe the gross features of 26 additional sporadic potentials and elaborate on the qualitative behavior of their solutions, on the basis of the key lessons drawn from the simpler examples of Section \ref{sec:properties_exact}. We also elaborate briefly on the links with other integrable systems. In Section \ref{sec:orientifolds} we illustrate how exponential potentials accompany
in String Theory a mechanism for supersymmetry breaking brought about by classically stable vacuum configurations of $D$-branes and orientifolds with broken supersymmetry and discuss their behavior in lower dimensions. Under some assumptions that are spelled out in Section 6, we also describe the  types of exponential potentials that can emerge, in four dimensions, from various types of branes present in String Theory. Insofar as possible, we work in a generic number of dimensions, but with critical superstrings in our mind, so that in most of the paper $4 \leq d \leq 10$. Finally section \ref{sec:conclusion} contains our conclusions, an assessment of our current views on the role of integrability in cosmological models emerging from a Fundamental Theory and some anticipations of results that are going to appear elsewhere \cite{prep,FayetIlio}.

\section{\sc  The Role of Integrable One-Field Models}\label{sec:integpotenti}


In this Section we set up our notation before turning to a systematic search for integrable single--scalar
cosmologies. We begin by reviewing some standard facts about FLRW
cosmologies, a useful generalization of this classic setup and some basic aspects of
Supergravity and String Theory connected to the role of exponential potentials.


\subsection{\sc  The effective dynamical system for scalar cosmologies}
\label{sec:effelag}


Let us begin with a derivation from first principles of the effective dynamical model whose integrability will be our main concern in subsequent sections, in order to make contact with the notation commonly used in Cosmology. Our starting point is provided by the action principles for Einstein
gravity minimally coupled to $n$ scalars in a ``mostly negative'' signature,
\be
  \mathcal{S} \,= \, \int \, d^{\,d} \,x \, \sqrt{- \, \mbox{det} \,g} \left[ \frac{1}{2\, k_d^2} \ R \, + \ \frac{1}{2}\ g^{\mu\nu} \, \partial_\mu \,\Phi^A \ \partial_\nu \, \Phi^B \,
  {g}_{AB}(\Phi) \, - \, V(\Phi) \right] \ ,
  \label{scalanorma}
\ee
where
\be
{R^\mu}_{\nu\rho\sigma} \ = \ \partial_\sigma \Gamma^{\mu}_{\nu\rho} \ - \ \partial_\rho \Gamma^{\mu}_{\nu\sigma} \ + \ \Gamma^{\mu}_{\sigma\tau} \, \Gamma^{\tau}_{\nu\rho} \ - \ \Gamma^{\mu}_{\rho\tau} \, \Gamma^{\tau}_{\nu\sigma}
\ee
and $R={\delta_\mu}^\rho \, g^{\,\nu\sigma}\, {R^\mu}_{\nu\rho\sigma}$.
Actions of this type, where the scalars describe $\sigma$--models with target--space metrics ${g}_{AB}$, emerge generically from Supergravity in various dimensions. In most of our examples, however, we shall focus on systems where a single scalar field $\phi$ is present to search for potential functions $V(\phi)$ that results in integrable cosmologies.

We shall focus on spatially flat cosmologies, which are of special relevance in the inflationary scenario, but we shall generalize slightly the standard Freedman--Lemaitre--Robertson--Walker (FLRW) setup
\begin{equation}\label{FLRWmet}
    ds^2 \, = \, dt^2 \, - \, e^{\,2\,A(t)} \, \mathrm{d}\mathbf{x}\cdot \mathrm{d}\mathbf{x}
\end{equation}
allowing for a wider class of metrics involving ``gauge functions'' ${\cal B}(t)$,
\begin{equation}\label{FLRWgen}
    ds^2 \, = \, e^{\,2\,{\cal B}(t)} dt^2 \, - \, e^{\,2\,A(t)} \, \mathrm{d}\mathbf{x}\cdot \mathrm{d}\mathbf{x} \ ,
\end{equation}
so that in the models that we shall examine the ``parametric'' time $t$ and the actual cosmic time $t_c$ measured by comoving observers will be related, in general, according to
\be
dt_c \ = \ e^{\,{\cal B}(t)} dt \ , \label{cosmic_parametric}
\ee
up to a ``minus'' sign that we shall introduce in some cases.
This slight departure from the standard setting will prove essential to arrive at most of our results, since the actual solutions of eq.~\eqref{cosmic_parametric} will be rather complicated in general.

As the wider class of metrics in eq.~\eqref{FLRWgen} involves a non--trivial $g_{00}$, the corresponding dynamical system for the cosmological equations
\be
  \mathcal{S} \,= \, \int \, d t \, e^{\,(d-1)\,A \ - \ {\cal B} } \, \left[ \ - \ \frac{(d-1)(d-2)}{2\,k_d^2}\ \dot{A}^{\,2} + \ \frac{1}{2}\ g^{\mu\nu} \, \dot{\Phi}^A \, \dot{\Phi}^B \,
  {g}_{AB}(\Phi) \, - \, e^{\,2\,{\cal B}} \ V(\Phi) \right] \ .
  \label{scalanorma2}
\ee
follows directly once eq.~\eqref{scalanorma} is specialized to the case of mere time dependence for the scalar fields in the background \eqref{FLRWgen}. Moreover, after the redefinitions
\bea
{\cal A} &=& (d-1)\, A \ , \nonumber \\
\varphi^A &=& k_d\ \sqrt{\frac{d-1}{d-2}} \ \Phi^A \ , \nonumber \\
\mathcal{V}(\varphi) &=& k_d^2\ \frac{d-1}{d-2} \ V(\Phi) \ , \label{redefin}
\eea
the independent equations of motion that follow from eq.~\eqref{scalanorma2} take the simple universal form
\bea
&& \dot{\cal A}^{\,2} \ - \ g^{\mu\nu} \, \dot{\varphi}^A \, \dot{\varphi}^B \, {g}_{AB} \ = \ 2\ e^{\,2\,{\cal B}} \ \mathcal{V}(\varphi) \ , \nonumber \\
&& \frac{d}{dt} \left( {g}_{AB} \ \dot{\varphi}^B \right) \ +  \ \left( \dot{\cal A} \ - \ \dot{\cal B}\right)\, {g}_{AB} \ \dot{\varphi}^B
\ + \ e^{\,2\,{\cal B}} \ \frac{\partial \mathcal{V}(\varphi)}{\partial \varphi^A} \ = \ 0 \ , \label{scalanorma_dindep}
\eea
where the dependence on the space--time dimension $d$ has disappeared.

In this case the Freedman equation or Hamiltonian constraint, the first of eqs.~\eqref{scalanorma_dindep}, follows varying the reduced action \eqref{scalanorma2} with respect to ${\cal B}$, while a second--order equation for ${\cal A}$ could be derived combining eqs~\eqref{scalanorma_dindep}, as in the standard FLRW setting. Special choices for ${\cal B}$ lead
to different classes of exact solutions for this type of systems, as we shall see in Section \ref{sec:integrable}. However, the physical interpretation of the results will often require some care, since the relation between $t$ and the actual cosmic time $t_c$, which is determined in principle by eq.~\eqref{cosmic_parametric}, will be rather complicated in general, as we have already stressed, and actually in some examples $t$ and $t_c$ will even run in opposite directions.

Starting from the general Lagrangians of eq.~\eqref{scalanorma}, one is interested in principle in the details of the consistent truncations to effective Lagrangians for a single minimally coupled scalar field. The truncation to a single scalar mode is a common simplification in Cosmology, and several of our exact solutions can prove potentially instructive in this context. Moreover, our techniques to identify integrable potentials will prove particularly powerful in this case and in a
handful of others involving only a few scalars, with potentials $\mathcal{V}$ that are combinations of exponential functions. The key consistency requirement for the truncation brings into the game $n-1$ constants $\Phi^a_0$ such that
\begin{eqnarray}
     \mathfrak{g}^{AB}(\Phi) \, \frac{\partial}{\partial \Phi^B} \, \mathcal{V}(\Phi) \, |_{\Phi^A = \Phi^A_0} &=&0 \quad \quad \quad \quad\quad \quad \quad (a=1,\dots, n-1) \ , \nonumber\\
    g^{\mu\nu} \, \partial_\mu \Phi^A \, \partial_\nu \Phi^B \, \mathfrak{g}_{AB}(\Phi) \, |_{\Phi^a = \Phi^a_0} &=& \alpha^2 \ g^{\mu\nu} \, \partial_\mu \phi \, \partial_\nu \phi \qquad \quad (\phi \, \equiv
    \, \Phi^{n}) \ , \label{rulando}
\end{eqnarray}
where $\phi$ can be identified with the $n$-th original field, up to a positive proportionality constant $\alpha^2$. The additional parameter is useful since this type of unconventional normalization presents itself naturally in Supergravity, as we shall see in detail in \cite{prep,FayetIlio}, so that it is worth emphasizing that the actual link between $(A,\phi,V)$ and $({\cal A},\varphi,{\cal V})$ in Supergravity will be in general
\bea
{\cal A} &=& (d-1)\, A \ , \nonumber \\
\varphi &=& \alpha \ k_d\ \sqrt{\frac{d-1}{d-2}} \ \phi \ , \nonumber \\
\mathcal{V}(\varphi) &=& k_d^2\ \frac{d-1}{d-2} \ V(\phi) \ \equiv \ k_d^2\ \frac{d-1}{d-2} \,
V \left( \frac{\varphi}{\alpha \, k_d} \, \sqrt{\frac{d-2}{d-1}} \right) \ . \label{redefin2}
\eea
Aside from this subtlety, the first of eqs.~\eqref{rulando} guarantees that setting $n-1$ scalar fields to constant values solves their equations of motion, which is clearly a necessary condition for a consistent truncation, while the second states instead that the kinetic term of the leftover field becomes canonical on the (constant) solutions for the others. As we shall see in \cite{prep}, in supergravity models where the scalar manifold is a symmetric coset space these conditions can naturally be satisfied, and one is led to associate the leftover field with a Cartan generator of the isometry Lie algebra.

Before moving further, let us also anticipate that exponential potentials emerge in orientifold vacua where supersymmetry is broken at the string scale due to the simultaneous presence of special collections of branes and orientifolds dictated by consistency conditions. Although we shall return in Section \ref{sec:orientifolds} to this phenomenon, usually referred to as ``brane supersymmetry breaking'' \cite{bsb}, we ought to stress right away that the resulting dynamics is in general very complicated, so that one can only arrive at eq.~\eqref{onefield} under two assumptions. Namely, that the stabilization of the additional moduli does somehow take place as a result of string corrections, and moreover that the ubiquitous higher--derivative corrections to the low--energy effective string Lagrangians play a subdominant role. Rigorous and fully convincing arguments to this effect are unfortunately not available at this time, so that a systematic investigation of these phenomena from the vantage point of Supergravity appears timely and can be potentially very instructive. We shall return to this point in the near future, starting from the more familiar case of linear supersymmetry and gauged Supergravity \cite{prep}.
\par
All in all, in the next sections we shall analyze in detail mechanical systems for the $({\cal A}, \varphi)$ variables whose equations of motion follow from the class of Lagrangians
\be
{\cal L} \ = \ e^{\,{\cal A} \ - \ {\cal B} } \left[ \ - \ \frac{1}{2} \ \dot{\cal A}^{\,2} \ + \ \frac{1}{2} \ \dot{\varphi}^2 \ - \ e^{\,2\,{\cal B}} \ \mathcal{V}(\varphi) \right] \ , \label{onefield}
\ee
where the potential function $\mathcal{V}$ involves combinations of exponentials. We have just stressed that this class of mechanical models, whose equations of motion can be cast in the form
\bea
&& \ddot{\varphi} \ + \  \left( \dot{\cal A} \ - \ \dot{\cal B} \right) \dot{\varphi} \ + \ e^{\,2\,{\cal B}} \ \mathcal{V}^{\,\prime}(\varphi) \ = \ 0 \ , \nonumber \\
&& \ddot{\cal A} \ = \ \dot{\cal A} \, \dot{\cal B} \ - \ \dot{\varphi}^{\, 2} \  , \nonumber \\
&& \dot{\cal A}^{\,2} \ - \ \dot{\varphi}^{\,2} \ = \ 2\, e^{\,2\,{\cal B}} \ \mathcal{V}(\varphi) \label{onefield_eqs}\ ,
\eea
the last of which is usually called the Friedman or Hamiltonian constraint, reflects the cosmological behavior of truncated $d$--dimensional Supergravity in space--time metrics of the form \eqref{FLRWgen}. Moreover, these equations have an interesting general feature: if the (parametric--)time variable $t$ is continued to imaginary values, their form is preserved while the class of metrics \eqref{FLRWgen} is turned into another of Euclidean signature, provided one also flips the sign of $\mathcal{V}$. In other words, Minkowski solutions in a given potential $\mathcal{V}$ afford an interesting alternative interpretation as Euclidean solutions in the inverted potential $-\mathcal{V}$, whenever the analytically continued functions remain real.

The standard four--dimensional FLRW setting for spatially flat cosmologies can be recovered inserting the definitions of scale factor and Hubble function,
\be
a(t) \ = \ e^{\, A(t)} \ , \qquad H(t) \, \equiv \, \frac{\dot{a}(t)}{a(t)} \ = \ \dot{A}(t) \ ,
\ee
in eqs.~\eqref{onefield_eqs} with ${\cal B}=0$. In particular, specializing eqs.~\eqref{scalanorma_dindep} to four dimensions and letting $k_4=\sqrt{2}$ one can recover the familiar expressions \cite{pielibro}
\begin{eqnarray}
 && H^2 \ = \  \frac{1}{3} \ \dot{\phi}^2 \, + \, \frac{2}{3} \ V(\phi)  \ , \nonumber\\
 &&  \dot{H}  \ = \ - \, \dot{\phi}^2   \ , \nonumber\\
 && \ddot{\phi} \,+ \,  3 \, H \, \dot{\phi} \, + \, V^{\,\prime}  \ = \ 0 \ , \label{fridmano}
\end{eqnarray}
the second of which can be deduced from the others.

In the instructive hydrodynamical picture, the energy density and the pressure of the fluid described by the scalar matter can be identified with the two combinations
 \begin{eqnarray}
&& \rho  \ = \  \frac{1}{4} \ \dot{\phi}^2 \, + \, \frac{1}{2}\ V(\phi)\ , \nonumber\\
&&   p  \ = \ \frac{1}{4} \ \dot{\phi}^2 \, - \, \frac{1}{2} \ V(\phi)\ ,\label{patatefritte}
 \end{eqnarray}
since in this fashion the first of eqs.~\eqref{fridmano} translates into the familiar link between the Hubble constant and the energy density of the Universe,
\begin{equation}\label{gordilatinus}
    H^2 \, = \, \frac{4}{3} \ \rho \ .
\end{equation}
A standard result in General Relativity (see for instance \cite{pielibro}) is that for a fluid whose equation of state is
\begin{equation}\label{equatastata}
    p\, = \, w \, \rho \qquad  \quad  w\, \in \,\mathbb{R}
\end{equation}
the relation between energy density and scale factor takes the form
\begin{equation}\label{forense2}
    \frac{\rho}{\rho_0} \, = \, \left(\frac{a_0}{a} \right)^{3(1+w)} \ ,
\end{equation}
where $\rho_0$ and $a_0$ are their values at some reference time $t_0$.
Combining eq.~(\ref{equatastata}) with the first of eqs.~(\ref{fridmano}) one can then deduce that
\begin{equation}\label{andamentus}
    a(t) \, \sim \, \left(t-t_i\right)^{\frac{2}{3 (w+1)}} \ ,
\end{equation}
where $t_i$ is an initial cosmic time. All values $-1\leq w \leq 1$ can be encompassed by eqs.~\eqref{patatefritte}, including the two particularly important cases of a dust--filled Universe, for which $w=0$ and $a(t) \, \sim \, \left(t-t_i\right)^{\frac{2}{3}}$, and of a radiation--filled Universe, for which $w=\frac{1}{3}$ and $a(t) \, \sim \, \left(t-t_i\right)^{\frac{1}{2}}$. Moreover, when the potential energy $V(\phi)$ becomes negligible with respect to the kinetic energy in eqs.~\eqref{patatefritte}, $w \approx 1$. On the other hand, when the potential energy $V(\phi)$ dominates $w\approx-1$, and eq.~(\ref{forense2}) implies that the energy density is an approximately constant vacuum energy, $\rho \, \approx \, \rho_0$. The behavior of the scale factor is then exponential, since the Hubble function is also a constant $H_0$ on account of eq.~\eqref{gordilatinus}, and therefore
      \be a(t) \ \approx \ e^{\,H_0 \, t} \ ; \qquad H_0 \ = \ \sqrt{\frac{4}{3}\ \rho_0} \ . \ee

The actual solutions of the non--linear Friedman equations that we shall come to with typically originate from potentials involving combinations of exponential functions. Hence, they will correspond to complicated equations of state whose index $w$ will vary in time, but nonetheless they will be qualitatively akin, at different epochs, to these simple types of behavior. In most of the exact solutions that we shall describe, a Universe undergoing initially a decelerated expansion will enter an eventual de Sitter phase, so that the integrable models at stake will typically address the onset of inflation rather than its end, but we shall also come to an amusing example of graceful exit.

\scs{Integrable Families of Scalar Cosmologies}\label{sec:integrable}


In this section we describe our systematic search for integrable families of scalar cosmologies, explaining in detail the methods that we relied upon. We begin by reviewing, as an illustration, the case of a single minimally--coupled scalar with an exponential potential. This affords relatively simple and yet very instructive exact solutions in terms of a suitable parametric time $t$, which exhibit a sharp transition in their behavior when the logarithmic slope of the exponential potential reaches a ``critical'' value. A similar setup will guide our subsequent search, in the remainder of this section, for integrable families of single--scalar cosmologies. Although the actual behavior in terms of the cosmological time $t_c$ will generally not be available in closed form, the key features of all these solutions will also surface clearly, in Section \ref{sec:properties_exact}, from their dependence on the parametric time $t$, as for a single exponential. The results of this section are summarized in Table \ref{tab:families}.


\subsection{\sc``Climbing'' scalars in an exponential potential}\label{sec:climbing}


Before proceeding to discuss more general exactly solvable exponential potentials, it is instructive to review how the special gauge choice
\be
{\cal V} \ e^{\, 2\, {\cal B}} \ = \ \frac{1}{2} \  \label{int8}
\ee
leads to a very interesting class of exact solutions with the exponential potentials
\be
{\cal V} \ = \ {\cal V}_0 \ e^{\,2\,\gamma\,\vf} \ . \label{int1001}
\ee
In doing so, we shall pay due attention to some peculiar features of these solutions that will then surface again in more complicated examples. This discussion will also illustrate the power and the limitations of our approach in a relatively simple context.

The key consequence of the gauge \eqref{int8} is that the field equations following from the Lagrangian \eqref{onefield} reduce to the autonomous system
\bea
&& \dot{\cal A}^2 \ -  \ \dot{\vf}^2 \ = \ 1  \ , \nonumber \\
&& \ddot{\vf} \ + \dot{\vf}\, \sqrt{1 \, + \, \dot{\vf}^2} \ + \ \gamma \left( 1 \, + \, \dot{\vf}^2 \right) \ = \ 0 \ ,
\label{int9}
\eea
where the equation for $\varphi$ is effectively of first order,
since the logarithmic derivative of the potential function ${\cal V}$  is simply a constant, $\gamma$, for the exponential potentials in eq.~\eqref{int1001}. Let us also note that the substitutions
\be
\dot{\cal A} \ = \ \cosh v \ , \qquad \dot{\vf} \ = \ \sinh v  \label{int11}
\ee
solve identically the first of eq.~\eqref{int9}, which is the form taken by the Hamiltonian constraint in this gauge. Moreover, up to the field redefinition $\vf \to \ - \ \vf$, one can confine the attention to positive values of $\gamma$.

All in all, for $0<\gamma<1$ the resulting first--order equation for $v$ leads to the \emph{two} classes of solutions
\be
\dot{\vf} \, = \, \frac{1}{2} \left[ \sqrt{\frac{1\,-\, \g}{1\,+\, \g}}\, \coth \left( \frac{t}{2}\ \sqrt{1\,-\, \g^{\,2}}\, \right) \ - \ \sqrt{\frac{1\,+\, \g}{1\,-\, \g}}\, \tanh
\left( \frac{t}{2}\ \sqrt{1\,-\, \g^{\,2}}\, \right)\right]\ . \label{int12}
\ee
and
\be
\dot{\vf} \, = \, \frac{1}{2} \left[ \sqrt{\frac{1\,-\, \g}{1\,+\, \g}}\, \tanh \left( \frac{t}{2}\ \sqrt{1\,-\, \g^{\,2}} \,\right) \ - \ \sqrt{\frac{1\,+\, \g}{1\,-\, \g}}\,
\coth \left( \frac{t}{2}\ \sqrt{1\,-\, \g^{\,2}}\, \right)\right] \label{int13} \ .
\ee
Notice that close to the initial singularity these two classes solutions behave as
\be
\dot{\varphi} \ \sim \ \pm \ \frac{1}{(1 \,\pm \, \gamma)\, t} \ , \label{int1352}
\ee
but the dependence on $\gamma$ disappears when working in terms of the cosmological time $t_c$, and
\be
\frac{d \varphi}{d t_c} \ \sim \ \pm \ \frac{1}{t_c} \ . \label{int13522}
\ee
since in this coordinate system the dynamics is initially dominated by the kinetic terms.

Eq.~\eqref{int12} describes a scalar field that emerges from the initial singularity \emph{climbing up} the exponential potential (a climbing scalar), while \eqref{int13} describes a scalar field that
emerges \emph{climbing down} (a descending scalar). Notice that eqs.~\eqref{int12} and \eqref{int13} are mapped into each other by the combined ${\mathbb{Z}_2}$ redefinitions $\varphi \to - \, \varphi$
and $\gamma \to - \, \gamma$, which are a manifest symmetry of the action \eqref{onefield} with the potential \eqref{int1001}. At the same time for $0<\gamma < 1$ the system possesses a special
\emph{exact} solution, the Lucchin--Matarrese (LM) attractor \cite{lm}, which in this gauge takes the particularly simple and suggestive form
\be
\dot{\varphi} \ = \ - \ \frac{\gamma}{\sqrt{1-\gamma^2}} \ \label{int135}
\ee
and captures the \emph{late--time behavior} of generic solutions. In cosmic time the LM attractor takes the form
\be
\varphi \ = \ - \ \frac{1}{\gamma} \ \log \left[ \, \gamma^{\,2}\, t_c\, \sqrt{\frac{2\, {\cal V}_0}{1 \, - \, \gamma^{\,2}}}\ \right] \ .
\ee
Notice that the speed of $\varphi$ apparently increases as $\gamma$ decreases. However, if the speeds are compared at points where the corresponding potentials assume identical values one is led to consider
\be
\frac{d{\varphi}}{dt_c} \ = \ - \ \frac{1}{\gamma\, t_c} \ = \ \frac{\gamma}{\sqrt{1 \, - \, \gamma^{\,2}}} \ \sqrt{2\, {\cal V}(\varphi)} \ ,
\ee
so that the resulting dependence on $\gamma$ follows eq.~\eqref{int135}.

At ny rate, as $\gamma$ approaches one from below something dramatic happens: both the descending solution and the LM attractor disappear, and one is left with a single type of solution,
\be
\dot{\vf} \, = \, \frac{1}{2t} \ - \ \frac{t}{2} \ , \label{int14}
\ee
which is a limiting case of those in eq.~\eqref{int12} so that it describes a ``critical'' scalar whose initial climbing phase soon leaves way, in this gauge, to a descent that is essentially driven by a uniform acceleration. In cosmic time, however, for large $t_c$
\be
\varphi \ \sim \ - \ \log t_c \ ,
\ee
a limiting behavior that is actually attained for all values of $\gamma$. Finally, only climbing solutions continue to be available for $\gamma >1$, with
\be \dot{\vf}\emph{} \, = \, \frac{1}{2} \left[ \sqrt{\frac{\g \,-\, 1}{\g \,+\, 1}}\, \cot \left( \frac{t}{2}\ \sqrt{\g^{\,2} \,-\, 1}\, \right) \ - \
\sqrt{\frac{\g\,+\, 1}{\g \,-\, 1}}\, \tan \left( \frac{t}{2}\ \sqrt{\g^{\,2} \,-\, 1}\, \right)\right]\ , \label{int15}
\ee
For this ``overcritical'' climbing solution near the initial singularity
\be
\dot{\varphi} \ \sim \ \frac{1}{(1 \, + \, \gamma)\, t} \ , \label{int1352_o}
\ee
and again
\be
\frac{d \varphi}{d t_c} \ \sim \ \frac{1}{t_c} \ , \label{int13522_0}
\ee
but the whole cosmological history now takes place within the finite interval
$$0 \ < \ t \  < \ \frac{\pi}{\sqrt{\g^{\,2} \,-\, 1}} $$
of parametric time. Notice that in this region the ${\mathbb{Z}_2}$ redefinitions $\varphi \to - \, \varphi$ and $\gamma \to - \, \gamma$ map the solutions into themselves, once they are combined with finite translations of $t$.
The sharp change in the nature of the classical solutions thus presents some formal analogies with phase transitions, which finds a rationale in the combined ${\mathbb{Z}_2}$ transformations that for $0<\gamma<1$ map climbing
and descending solutions into one another, and insisting on this analogy one would thus conclude that the symmetry is somehow recovered for $\gamma >1$, where a single solution exists. Amusingly, the similarities go even further, since at the transition point $\gamma=1$ the Lucchin--Matarrese attractor \cite{lm} follows a fate similar to the Euclidean instanton, in that it also disappears. Notice that the integration constant expected to be present in the velocities of eqs.~\eqref{int12}--\eqref{int15} is merely the time of the initial singularity, which we have set to zero in all cases for brevity. This effective transmutation underlies the peculiar behavior of the system for $\gamma \geq1$. Although it emerged in this relatively simple setting, as stressed in \cite{dks}, the climbing phenomenon only depends on the asymptotic behavior of the potential and forbids the scalar field to descend along steep enough exponential grades as it emerges from the initial singularity. We shall see a number of illustrations of this fact in the following.

The convenient gauge choice of eq.~\eqref{int8} has thus led to simple solutions in parametric time $t$, but the complications inherent in the problem have not disappeared. Rather, they have been moved to the actual link between $t$ and the cosmological time $t_c$. Indeed, in this class of models the relation between the ``parametric'' time $t$ and the cosmological time $t_c$ is determined
by eq.~\eqref{cosmic_parametric}, and considering for definiteness $\gamma<1$ and the climbing scalar solution determined by eq.~\eqref{int12}, it reads
\be
t_c \ = \ \int d t\ \left[\sinh\left(\frac{t}{2} \sqrt{1-\gamma^2} \right) \right]^\frac{\gamma}{1-\gamma} \ \left[\cosh\left(\frac{t}{2} \sqrt{1-\gamma^2} \right) \right]^{\ - \ \frac{\gamma}{1+\gamma}} \ , \label{int16}
\ee
or alternatively
\be
t_c \ = \ \frac{1}{\sqrt{1-\gamma^2}} \ \int d \tau \left( \tau \right)^\frac{\gamma-1/2}{1-\gamma} \ \left(1+\tau\right)^\frac{\,-\, \gamma - 1/2}{1+\gamma} \ , \label{int17}
\ee
after the substitution
\be
\tau \ = \ \sinh^2\left(\,\frac{t}{2} \,\sqrt{1-\gamma^2}\,\right) \ . \label{int18}
\ee

There are two elementary cases of eq.~\eqref{int17}, which correspond to $\gamma=\pm \frac{1}{2}$, or if you will to a climbing and a descending scalar in the same potential
with $\gamma=\frac{1}{2}$, by virtue of the ${\mathbb{Z}_2}$ symmetry that we have already elaborated upon. In general, however, the relation between $t$ and $t_c$ is considerably more complicated and reads
\be
t_c \ \sim \ \left[ \sinh^2\left(\,\frac{t}{2}\, \sqrt{1-\gamma^2}\,\right) \right]^{{\frac {1/2}{1-\gamma}}}
{\mbox{$_2$F$_1$}\left[{\frac {\gamma+1/2}{1+\gamma}},{\frac {1/2}{1-\gamma}};\,{\frac {3/2}{1-\gamma}};\,-\sinh^2\left(\,\frac{t}{2} \,\sqrt{1-\gamma^2}\,\right)\right]} \ , \label{int19}
\ee
where $_2$F$_1$ is a hypergeometric function. As we are about to see, some interesting FLRW generalizations of the elementary cases with $\gamma=\pm 1/2$ do exist also with potentials involving more than one exponential, although a wider class of exact solutions, which describe a number of interesting phenomena compatibly with the climbing regime that we have just reviewed, can only be found provided a non--trivial ${\cal B}$ is introduced.

\subsection{\sc Integrable scalar cosmologies involving more exponentials}\label{sec:morexp}

Let us now move on to search for more general potential functions ${\cal V}(\varphi)$ that lead to integrable scalar cosmologies. As one can anticipate from the preceding example, it is perhaps convenient to first tackle the systematic construction in a simplified context, the standard FLRW setting with $\mathcal{B}=0$, before considering more complicated systems. The corresponding exact solutions, as we are about to see, have the virtue of being simple, instructive and completely explicit.

\subsubsection{\sc Elementary systems with $\mathcal{B}=0$}\label{sec:B0}

The key idea underlying the construction is to bring the kinetic terms to their simplest form. In this case, starting from the Lagrangians \eqref{onefield}, this can be attained via the redefinitions
\be
{\cal A} \, =\, \log\left(x\,y \right) \ , \qquad \vf \, =\, \log\left(\frac{x}{y} \right)  \ , \label{int601}
\ee
where it should be understood that the independent variable $t$ will eventually take values in the portions of the real axis where the product $x y$ is positive. One is thus led to consider Lagrangians for the dynamical variables $x$ and $y$ of the type
\be
{\cal L} \ = \ - \ 2\ \dot{x} \, \dot{y} \ - \ x\, y \ {\cal V}\left( \frac{x}{y} \right) \ ,\label{int602}
\ee
and the cases of interest clearly correspond to potential functions such that $x\,y\,{\cal V}$ is at most a quadratic polynomial in $x$ and $y$, and thus to
\be
{\cal V} \ = \ C_{11} \ \frac{x}{y} \ + \  2\, C_{12} \ + C_{22} \ \frac{y}{x} \ , \label{int603}
\ee
with the $C_{ij}$ are arbitrary constants. The resulting equations of motion are
\bea
&& \ddot{x} \ - \ C_{12} \ x \ - \ C_{22} \ y  \ = \ 0 \ , \\
&& \ddot{y} \ - \ C_{11} \ x \ - \ C_{12} \ y \ = \ 0 \ , \label{int604}
\eea
and are to be supplemented with the Hamiltonian constraint
\be
\dot{x}\ \dot{y} \ = \ C_{12} \ x \, y  \ + \ \frac{1}{2} \ \Big( \, C_{11} \ x^2 \ + \
C_{22} \ y^2\, \Big) \ , \label{int605}
\ee
which could have been obtained varying $\mathcal{B}$ if this additional function had been retained in the space--time metric.

In terms of the original fields $a$ and $\vf$ these models correspond to the class of potentials
\be
{\cal V} \ = \ C_{11} \ e^{\, \vf} \ + \  2\, C_{12} \ + C_{22} \ e^{\, - \, \vf}  \ , \label{int7}
\ee
which combine in general a cosmological term with a pair of exponentials. Notice that these exponentials are always ``under--critical'', with $\gamma=\pm \frac{1}{2}$ in the notation of Section \ref{sec:climbing}, so that there is no climbing phenomenon in this class of models. Still, the inclusion of a cosmological constant and the possibility of dealing with potential wells entails a number of instructive lessons, as we shall see in Section \ref{sec:properties_exact}.

\subsubsection{\sc Triangular systems with ${\cal B} \neq 0$}\label{sec:Bnot0}

Let us now complicate the analysis that led to eq.~\eqref{int7}, allowing for a non--trivial $\mathcal{B}$. This yields a wider class of exact solutions for potentials involving combinations of exponentials with other values of $\gamma$. In a number of cases, our solutions will be identified letting
\be
 {\cal A} \ = \ \log\left(x^{\, \frac{1}{1-\alpha}}\ y^{\, \frac{1}{1-\beta}}\right) \ , \qquad \vf \ = \ \log\left(x^{\, \frac{1}{1-\alpha}} \ y^{\,- \,  \frac{1}{1-\beta}}\right)  \label{int20}
\ee
and allowing for a non--trivial ${\cal B}$ of the form
\be
{\cal B} \ = \ \log \left( x^{\, \frac{\alpha}{1-\alpha}}\ y^{\, \frac{\beta}{1-\beta}} \right) \ , \label{int21}
\ee
which makes it possible to replace the class of Lagrangians \eqref{onefield} with
\be
{\cal L} \ = \ - \ 4\, \dot{x} \ \dot{y}  \, -  \, 2\,(1-\alpha)(1-\beta) \ x^{\, \frac{1+\alpha}{1-\alpha}}\ y^{\, \frac{1+\beta}{1-\beta}} \ {\cal V}\left(x^{\, \frac{1}{1-\alpha}} \ y^{\,- \,  \frac{1}{1-\beta}}\right) \ . \label{int22}
\ee

Two interesting classes of potentials ${\cal V}(\varphi)$ can be readily identified in this way. They correspond to $\beta=-\alpha=\gamma$ in eq.~\eqref{int22}, with $\gamma \neq (-1,1)$, and combine eventually
in ${\cal L}$ a bilinear term in the two variables $x$ and $y$ with an algebraic power in one of them. The potentials belonging to the first class correspond to
\be
{\cal V} \ = \ C_1 \ e^{\,2\,\gamma\, \vf} \ +  C_2 \ e^{\,(\gamma\,+\,1)\, \vf} \ , \label{int23}
\ee
where the $C_i$ are arbitrary real constants, and in this case the relation between the parametric time $t$ and the cosmic time $t_c$ reads
\be
dt_c \ = \ x^{\,-\,\frac{\gamma}{1+\gamma}} \ y^{\,\frac{\gamma}{1-\gamma}} \ dt\ . \label{int24}
\ee
In terms of $x$ and $y$ the Lagrangians for this class of models take the form
\be
{\cal L} \ = \ - \ 4\, \dot{x} \ \dot{y}  \, -  \, 2\,(1\,-\,\gamma^2) \ \left[ C_1 \ x \, y \ + \ C_2 \ x^\frac{2}{1+\gamma} \right]  \ , \label{int25}
\ee
and the corresponding equations of motion thus define the following ``triangular'' systems
\bea
&& \ddot{x} \ - \ \frac{1-\gamma^2}{2} \ C_1 \ x \ = \ 0 \ , \\
&& \ddot{y} \ - \ \frac{1-\gamma^2}{2} \ C_1 \ y \ = \  C_2 \ (1-\gamma) \ x^{\, \frac{1-\gamma}{1+\gamma}} \ , \label{int255}
\eea
in which the first equation is elementary, so that the non--linear terms present in the second simply build up  a known source term. Finally, the equation of motion of ${\cal B}$ translates into the Hamiltonian constraint
\be
\dot{x} \ \dot{y}  \, =  \, \frac{1-\gamma^2}{2} \ \left[ C_1 \, x \, y \ + \ C_2 \, x^\frac{2}{1+\gamma} \right] \ . \label{int255a}
\ee
The second class of models would correspond to
\be
{\cal V} \ = \ C_1 \ e^{\,2\,\gamma\, \vf} \ + \ C_2 \ e^{\,(\gamma\,-\,1) \, \vf} \ , \label{int26}
\ee
where the $C_i$ are again arbitrary real constants. Notice, however, that these apparently new potentials can be mapped into those of eq.~\eqref{int23} by the combined redefinitions $(\vf,\gamma) \to (- \vf,-\gamma)$. Moreover, for $\gamma=0$ the potentials in eqs.~\eqref{int23} and \eqref{int26} become a special case of those in eq.~\eqref{int7}. As a result, the available phenomena can be explored referring only to the potentials of eq.~\eqref{int23}, letting $\gamma$ vary over the real axis, with the exclusion of the points $\gamma = \pm 1$ where the climbing phenomenon sets in. These ``critical'' potentials, however, can be reached via an asymmetric substitution, to which  now we turn  since it also allows the inclusion of a cosmological term. Performing  the redefinitions
\be
 {\cal A} \ = \ \frac{1}{1-\alpha} \ \log \,x  \ + \ v \ , \ \quad \vf = \ \frac{1}{1-\alpha} \ \log \,x \ - \ v \   \label{int30}
\ee
in eq.~\eqref{onefield} and making the gauge choice
\be
\mathcal{B} \ = \ \frac{\alpha}{1-\alpha} \ \log\,x \ + \ v \label{int301}
\ee
leads finally to
\be
{\cal L} \ = \ - \ \frac{4}{1-\alpha} \ \dot{x}\, \dot{v} \ - \ 2 \ e^{\,2\,v} \ x^{\,\frac{1+\alpha}{1-\alpha}} \ {\cal V}\left(x^\frac{1}{1-\alpha}\ e^{\,-\, v} \right) \ . \label{int31}
\ee
As a result, the most general potential ${\cal V}$ yielding a ``triangular'' system obtains in this case for $\alpha=-1$, and describes an arbitrary combination of an exponential potential that is  ``critical'' in the sense of Section \ref{sec:climbing} and a cosmological constant:
\be
{\cal V} \ = \ C_1 \, e^{\, 2\, \varphi} \ + \ C_2  \ . \label{int32}
\ee
In this case eq.~\eqref{int31} becomes
\be
{\cal L} \ = \ - \  2\, \dot{x}\, \dot{v} \ - \ 2\, C_1 \, x  \ - \ 2\, C_2 \, e^{\,2\,v} \ , \label{int33}
\ee
whose equations of motion,
\bea
&& \ddot{v} \ = \ C_1 \ , \\
&& \ddot{x} \ = \ 2\ C_2 \, e^{\,2\,v} \ , \label{int34}
\eea
form again a ``triangular'' system since the first is clearly elementary. As usual, the system \eqref{int34} is to be supplemented with the corresponding Hamiltonian constraint, which now reads
\be
\dot{x}\, \dot{v} \ = \ C_1 \, e^{\,2\,v} \ + \ C_2 \, x \ . \label{int35}
\ee
Finally, in this case the relation between the cosmological time $t_c$ and the parametric time $t$ follows from
\be
dt_c \ = \ e^{\,v} \ x^{\,-\,\frac{1}{2}} \ dt \ .
\ee

Actually, working again with $\alpha=-1$ there is another interesting gauge choice,
\be
\mathcal{B} \ = \ - \ {\cal A} \ - \ 2\, \varphi \
\ee
that together with the asymmetric redefinitions
\bea
&& {\cal A} \ = \ \frac{1}{4} \ \log x \ + \ v  \ , \nonumber \\
&& \varphi \ = \ \frac{1}{4} \ \log x \ - \ v
\eea
leads to the class of Lagrangians
\be
{\cal L} \ = \ - \ \frac{1}{2} \ \dot{x}\, \dot{v} \ - \ e^{\,-\,2\,v} \ x \ {\cal V}\left(x^\frac{1}{2}\ e^{\,-\, v} \right) \ . \label{int36}
\ee
Up to a shift of $\varphi$, one can thus associate a simple integrable dynamics to potential functions ${\cal V}$ of the form
\be
{\cal V} \ = \ C \ \varphi \ e^{\,2\,\varphi} \ . \label{int37}
\ee
For positive values of $C$ these are potential wells whose right end is an exponential wall that is ``critical'' in the sense of Section \ref{sec:climbing}. Therefore, the scalar can only emerge from the initial singularity proceeding toward it, and in the next section we shall see how in this type of system $\varphi$ can be (almost) stabilized as a result of the cosmological evolution.

With the potential function \eqref{int37}, the Lagrangian of eq.~\eqref{int36} becomes indeed
\be
{\cal L} \ = \ - \ \frac{1}{2} \ \dot{x}\, \dot{v} \ - \ C\, \left( \frac{1}{4} \ \log \,x \ - \ v \right)  \ , \label{int3655}
\ee
so that the resulting equations of motion,
\bea
&& \ddot{x} \ = \ - \ 2 \, C \ , \\
&& \ddot{v} \ = \ \frac{C}{2\,x}  \label{int38}
\eea
define once more a triangular system. Finally, the corresponding Hamiltonian constraint reads
\be
\dot{x} \ \dot{v} \ = \ C \, \left( \frac{1}{2} \ \log\,x \ - \ 2\, v \right) \ , \label{int39}
\ee
while in this case the relation between $t$ and the cosmological time $t_c$ is
\be
dt_c \ = \ x^{\, - \ \frac{3}{4}} \ e^{\,v} \ dt \ . \label{int395}
\ee
%
\subsubsection{\sc Systems integrable via quadratures}\label{sec:quadratures}

Two more classes of integrable potentials can actually be associated to eq.~\eqref{int22}. They are our first examples of a richer class of integrable scalar cosmologies, where the two--dimensional $({\cal A},\varphi)$ dynamics can be solved via quadratures since it becomes manifestly \emph{separable} in suitable coordinates.

The first class of integrable cosmologies of this type that we would like to describe belongs to the class of Lagrangians \eqref{int22} with $\alpha=\beta=-1$, which have the general form
\be
{\cal L} \ = \ - \ 4\, \dot{x} \ \dot{y}  \, -  \, 8\, {\cal V}\left[ \left(\frac{x}{y}\right)^{\, \frac{1}{2}} \right] \ , \label{int40}
\ee
and arises if
\be
{\cal V} \ = \  \ C \, \log\left( \frac{x+y}{x-y} \right) \ + \ D \ , \label{int401}
\ee
where $C$ and $D$ are two arbitrary constants. In terms of $\varphi$, it corresponds to the class of ``wall'' potentials
\be
{\cal V} \ = \ C \, \log\,\left( \coth \varphi \right) \ + \ D \ , \label{int402}
\ee
where the cosmological constant $D$ does not enter the equations for $\xi$ and $\eta$ introduced below but plays a role in the Hamiltonian constraint. Indeed, letting
\be
x \ = \ \frac{1}{2} \left(\xi \ + \ \eta\right) \ , \qquad y \ = \ \frac{1}{2} \left(\xi \ - \ \eta\right) \ , \label{int403}
\ee
eq.~\eqref{int40} takes  the separable form
\be
{\cal L} \ = \ - \ \dot{\xi}^{\,2} \ +  \ \dot{\eta}^{\,2}  \ -  \ 8\, C \, \log\left( \frac{\xi}{\eta} \right) \ - \ 8\, D \ , \label{int404}
\ee
whose equations of motion,
\bea
&& \ddot{\xi} \ = \ \frac{4\, C}{\xi} \ , \nonumber \\
&& \ddot{\eta} \ = \ \frac{4\, C}{\eta} \ , \label{int405}
\eea
can be turned into the conservation laws
\bea
&& \dot{\xi}^{\,2} \ -  \ 8\, C \, \log \xi \ = \ \epsilon_\xi \ , \nonumber \\
&& \dot{\eta}^{\,2} \ -  \ 8\, C \, \log \eta \ = \ \epsilon_\eta \ , \label{int406}
\eea
where $\epsilon_\xi$ and $\epsilon_\eta$ are integration constants.
The solutions by quadratures of these equations are to be subjected to the Hamiltonian constraint
\be
\dot{\xi}^{\,2} \ -  \ \dot{\eta}^{\,2}  \ =  \ 8\, C \, \log\left( \frac{\xi}{\eta} \right) \ + \ 8\, D \ , \label{int407}
\ee
which is tantamount to the condition
\be
\epsilon_\xi=\epsilon_\eta \,+\, 8\,D \ . \label{int4071}
\ee
Finally,
for this class of models the relation between the cosmological time $t_c$ and the parametric time $t$ takes the form
\be
dt_c \ = \ \frac{2\, dt}{\sqrt{\xi^{\,2} \ - \ \eta^{\,2}}} \ . \label{int408}
\ee

The second class of integrable cosmologies that we would like to describe possesses another instructive feature: its integrability rests on the recourse to complex combinations of the original variables. The
corresponding potential functions read
\be
{\cal V} \ = \ \mathrm{Im} \left[ \log\left( C \ \frac{e^{\,2\,\varphi}\ +\ i}{e^{\,2\,\varphi}\ -\ i}\right)\ + \ i \, \Lambda \right]\ , \label{Wcomplex}
\ee
where $\mathrm{Im}$ denotes the imaginary part and where the freedom of including the cosmological constant $\Lambda$ has the same origin as in the preceding example. Notice also that
\be
\mathrm{Im} \left[ \log\left( \frac{e^{\,2\,\varphi}\ +\ i}{e^{\,2\,\varphi}\ -\ i}\right)\right]\ = \ 2 \ \arctan\left(e^{\,-\, 2\,\varphi}\right) \ ,
\ee
so that we can confine our attention to \emph{real} values of $C$, since a phase would only bring about a shift of $\varphi$. These
potentials are essentially step functions that, as in the preceding example, result from a series of exponential terms. In terms of the new variables $x$ and $y$ above
the Lagrangians for this class of models read
\be
{\cal L} \ = \ - \ 4\, \dot{x} \ \dot{y}  \, -  \, 8\, \mathrm{Im} \left[ C\, \log\left( \frac{x\ + \ i\, y}{x\ - \ i\, y}\right) \ + \ i\, \Lambda \right] \ , \label{int4010}
\ee
so that, letting
\be
z \ = \ x \ + \ i\, y \ , \label{int4011}
\ee
eq.~\eqref{int4010} takes finally the form
\be
{\cal L} \ = \ 2\, \mathrm{Im} \Big[ \, - \ \dot{z}^2 \ - \ 8\, C \, \log z  \ - \ 4\, i \, \Lambda \Big] \ . \label{int4012}
\ee
As advertised, one is led to an integrable one--dimensional \emph{complex} dynamical system, whose equation of motion
\be
\ddot{z} \ = \ \frac{4\, C}{z} \label{int4013}
\ee
and Hamiltonian constraint
\be
\mathrm{Im} \Big[ \dot{z}^2 \ - \ 8\, C \, \log z  \ - \ 4\,i\, \Lambda \Big] \ = \ 0 \label{int40135}
\ee
are complex variants of eqs.~\eqref{int405} and \eqref{int407}.

Another interesting class of exact solutions can be identified looking for expressions of the type
\be
(\, x \ \pm \ y \,)^{\,p} \ = \ (xy)^{\,\frac{p}{2}} \ \left( \sqrt{\frac{x}{y}} \ \pm \ \sqrt{\frac{y}{x}}\right)^{\,p} \label{int3555}
\ee
and comparing them with the general Lagrangians of eq.~\eqref{onefield}. One can thus realize that the class of potentials
\be
 {\cal V}\ = \ C_1 \ \Big(\cosh\,\gamma\,\varphi \Big)^{\, \frac{2}{\gamma} \, - \, 2}\ + \ C_2 \ \Big( \sinh\,\gamma\,\varphi \Big)^{\,\frac{2}{\gamma} \, - \, 2}\ , \label{int36555}
\ee
which includes interesting examples of under--critical exponential wells, affords a relatively simple description in the class of gauges
\be
\qquad \mathcal{B} \ = \ {\cal A} \,(1\,-\,2\,\gamma) \ .
\ee
This can be  explicitly seen via the substitutions
\be
e^{\,2\, \gamma\,{\cal A}} \ = \ xy \ , \qquad e^{\,2\,\gamma\,\vf} \ = \ \frac{x}{y} \ , \label{int3755}
\ee
whose net result is the emergence of a pair of decoupled non--linear differential equations that exactly solvable in terms of Jacobi elliptic functions or generalizations thereof. Indeed, letting
\be
\xi \ = \ \frac{x \ + y}{2} \ , \qquad \eta \ = \ \frac{x \ - \ y}{2} \ , \label{int3656}
\ee
the Lagrangian \eqref{onefield} can be finally turned into the separable form
\be
{\cal L} \ = \ - \ \frac{1}{2\, \gamma^{\,2}} \left( \dot{\xi}^{\,2} \ - \ \dot{\eta}^{\,2} \right) \ - \ C_1 \ \xi^{\, \frac{2}{\gamma} \, - \, 2} \ - \ C_2 \ \eta^{\, \frac{2}{\gamma} \ - \ 2} \ ,
\ee
so that its equations of motion
\bea
&& \ddot{\xi} \ - \ 2\, C_1 \, \gamma(1\,-\,\gamma)\, \xi^{\, \frac{2}{\gamma} \,-\, 3}\ = \ 0 \ , \nonumber \\
&& \ddot{\eta} \ + \ 2\, C_2 \, \gamma(1\,-\,\gamma)\, \eta^{\, \frac{2}{\gamma} \,-\, 3}\ = \ 0 \label{int3657}
\eea
can be solved independently by quadratures. As usual, they are to be supplemented by the Hamiltonian constraint, which in this case takes the form
\be
\dot{\xi}^{\,2} \ - \ \dot{\eta}^{\,2} \ = \ 2\ \gamma^2 \left[ C_1 \ \xi^{\, \frac{2}{\gamma} \, - \, 2} \ + \ C_2 \ \eta^{\, \frac{2}{\gamma} \, - \, 2} \right] \ , \label{int3613}
\ee
while for this whole class of models the relation between the cosmological time $t_c$ and the parametric time $t$ reads
\be
dt_c \ = \ (\xi^2 \, - \, \eta^2)^{\,\frac{1}{2\,\gamma} \ - \ 1} \ dt \ .
\ee

There is also an interesting variant of this class of potentials along the lines of \eqref{Wcomplex}, which can be solved again in terms of complex combinations of the original variables. It rests again on the substitutions of eqs.~\eqref{int3755},
but now
\be
{\cal V} \ = \ \mathrm{Im} \,\Big[\,C \ \Big(\, i \ + \ \sinh\,2\,\gamma\,\varphi \, \Big)^{\, \frac{1}{\gamma} \ - \ 1}\,\Big] \ ,\label{Wcomplex2}
\ee
where $C$ is an arbitrary constant. Notice also that, if $C=|C|\,e^{\,i(\theta+\pi/2)}$,
\be
{\cal V}(\varphi) \ = \ |C|\,\Big(\cosh 2\,\gamma\,\varphi \Big)^{\frac{1}{\gamma} \ - \ 1}\, \cos \left[\Big(\frac{1}{\gamma} \ - \ 1\Big)\,\arccos\Big(\tanh\,2\,\gamma\,\varphi\Big)\ + \ \theta\,\right] \ .
\ee
Letting again
\be
z \ = \ x \ + \ i\, y \ ,
\ee
for this choice of ${\cal V}$ the Lagrangian of eq.~\eqref{onefield} takes  the final form
\be
{\cal L} \ = \ \mathrm{Im} \left[ \ - \ \frac{1}{2\, \gamma^{\,2}} \ \dot{z}^2 \ - \ \frac{C}{2} \ z^{\,\frac{2}{\gamma} \,-\, 2}\right] \ ,\label{int3615}
\ee
which can again be solved by quadratures, while the corresponding Hamiltonian constraint reads
\be
 \mathrm{Im} \left[ \frac{1}{2\, \gamma^{\,2}} \ \dot{z}^2 \ - \ \frac{C}{2} \ z^{\,\frac{2}{\gamma} \,-\, 2}\right] \, = \, 0 \, .
\ee
Interestingly, the potentials \eqref{int402}, \eqref{Wcomplex} and \eqref{Wcomplex2} involve infinite series of exponentials that can simulate, in String Theory, the resummation of an infinite number of loop corrections to tree--level exponential potentials. Indeed, as we shall see in Section \ref{sec:orientifolds}, in four dimensions loop corrections originating from the introduction of arbitrary numbers of handles in the world sheet can give rise precisely to corrections of a tree--level brane term by arbitrary powers of $e^{2\varphi}$, where $\varphi$ is a suitable combination of the ten--dimensional dilaton and of a scalar related to the internal volume of the compactification, up to an assumption concerning the stabilization of an orthogonal combination of the two fields that we shall discuss in detail there.

Another interesting class of potential functions is
\be {\cal V} \ = \ C_1 \ e^{2\,\gamma\,\varphi} \  + \ C_2 \
e^{\frac{2}{\gamma}\,\varphi} \ , \label{Liouville1}\ee
or more simply, if the $C_i$'s are both positive,
\be {\cal V} \ = \ \lambda \left( \ e^{2\,\gamma\,\varphi} \  + \
e^{\frac{2}{\gamma}\,\varphi} \right) \ , \label{Liouville11}\ee
with $\lambda>0$, since a relative factor between the two exponentials can clearly be absorbed into a shift of $\varphi$. One can also assume, without any loss of generality, that $0< \gamma< 1$, so that the first term is a mild exponential while the second is a steep one. This class of potentials can describe a scalar that can only emerge from an initial singularity while climbing them up to
then inject an inflationary phase during the subsequent descent in $d$ dimensions provided $\gamma<\frac{1}{\sqrt{d-1}}$.
\par
One can arrive at exact solutions for the class of potentials \eqref{Liouville1} working in the gauge
\be
{\cal B} \ = \ {\cal A}\ , \label{Liouville2}
\ee
which reduces eq.~\eqref{onefield} to
\be
{\cal L} \ = \ \frac{1}{2} \left( \dot{\varphi}^{\,2} \ - \ \dot{\cal A}^{\,2} \right) - e^{\,2\,{\cal A}} \ \left( C_1 \ e^{2\,\gamma\,\varphi} \  + \ C_2 \
e^{\frac{2}{\gamma}\,\varphi} \right) \ , \label{Liouville3}
\ee
and performing the ``Lorentz boost''
\bea
&& \widehat{\cal A}\ = \ \frac{1}{\sqrt{1-\gamma^{\,2}}} \ \Big( \cal A \ + \ \gamma \, \varphi \Big) \ ,\nonumber\\
&& \widehat{\varphi}\ = \ \frac{1}{\sqrt{1-\gamma^{\,2}}} \ \Big( \varphi \ + \ \gamma \, {\cal A} \Big)  \ , \label{Liouville4}
\eea
which turns the Lagrangian \eqref{Liouville3} into the separable form
\be
{\cal L} \ = \ \frac{1}{2} \left( {\dot{\widehat{\varphi}}}^{\,2} \ - \ {\dot{\widehat{\cal A}}}^{\,2} \right) \ - \ C_1\ e^{\,2\,\sqrt{1-\gamma^{\,2}}\, \widehat{\cal A}}\ - \ C_2\ e^{\,\frac{2}{\gamma}\,\sqrt{1-\gamma^{\,2}}\, \widehat{\varphi}}  \ . \label{Liouville5}
\ee
The resulting equations of motion are then
\bea
&& \ddot{\!\!\widehat{\cal A}} \ - \ 2\,C_1 \ \sqrt{1 - \gamma^{\,2}} \ e^{\, 2\,\sqrt{1-\gamma^{\,2}}\, \widehat{\cal A}} \ = \ 0 \ , \nonumber \\
&& \ddot{\!\widehat \varphi} \ + \ \frac{2\,C_2}{\gamma} \ \sqrt{1 - \gamma^{\,2}} \ e^{\,\frac{2}{\gamma}\,\sqrt{1-\gamma^{\,2}}\, \widehat{\varphi}} \ = \ 0 \ ,\label{Liouville6}
\eea
As usual these are to be supplemented with the Hamiltonian constraint, which in this case reads
\be
{\dot{\!\widehat{\varphi}}}^{\,2} \ - \ {\dot{\!\!\widehat{\cal A}}}^{\,2} \ = \ - \ 2\ C_1\ e^{\,2\,\sqrt{1-\gamma^{\,2}}\, \widehat{\cal A}}\ - \ 2\ C_2\ e^{\,\frac{2}{\gamma}\,\sqrt{1-\gamma^{\,2}}\, \widehat{\varphi}} \label{Liouville7} \ ,
\ee
while in this class of models the relation between the cosmological time $t_c$ and the parametric time $t$ takes the form
\be
dt_c \ = \ \exp\left[\, \frac{\widehat{\cal A} \, - \, \gamma \, \widehat{\varphi} }{\sqrt{1\,-\,\gamma^{\,2}}} \right]\ dt \ . \label{Liouville8}
\ee
%

\begin{table}[h]
\begin{center}
\caption{Families of integrable potential functions for the Lagrangians of eq.~\eqref{onefield}}
\begin{tabular}{l c c c}
\hline
\hline
& Potential function ${\mathcal V}$ & ${\cal A}, \varphi, {\cal B}$ & ${\cal L}$, Hamilt.~Constr., $dt_c$\\
\hline
\hline
\\
$\!\!\!\!\!(1)$ & {\tiny $\! C_{11} \, e^{\,\vf} \, + \, 2\, C_{12} \, + \, C_{22} \, e^{\, - \vf}$} & ${{\cal A} \, =\, {\log(x\, y)} \atop {\varphi \, = \, \log\left(\frac{x}{y}\right)}}\atop {{\mathcal B}\, =\, 0 \atop }$ & ${{{\mathcal L}\, =\, - \, 2 \dot{x} \, \dot{y} \, - \, C_{11} \, x^2 - 2\, C_{12} \, x \, y \, - \, C_{22} \, y^2 } \atop {2 \, \dot{x} \, \dot{y} \, = \, C_{11}\, x^2 \, +\, 2\, C_{12} \, x \, y \, + \, C_{22} \, y^2 }} \atop {{dt_c \, = \, dt}\atop }$ \\
\\

$\!\!\!\!\!(2)$ & {\tiny $\! C_1 \, e^{\,2\,\gamma \,\varphi}\, +\, C_2e^{\,(\gamma+1)\, \varphi} \ \ \ (\gamma^2 \neq 1)$} & ${{\cal A} \, = \, {\log\left(x^\frac{1}{1+\gamma}\, y^\frac{1}{1-\gamma}\right)} \atop {\varphi \, = \, \log\left(x^\frac{1}{1+\gamma}\, y^\frac{-1}{1-\gamma}\right)}}\atop {{\mathcal B} \, =\, \log\left(x^\frac{-\gamma}{1+\gamma}\, y^\frac{\gamma}{1-\gamma}\right) \atop }$ & ${{{\mathcal L}\, =\, - \, 4 \dot{x} \, \dot{y} \, - \, 2 (1 - \gamma^2) \left[ C_1 \, x \, y \, + \, C_2 \, x^\frac{2}{1+\gamma} \right]} \atop {2 \, \dot{x} \, \dot{y}  \, =  \, (1-\gamma^2) \left[ C_1 \, x \, y \, + \, C_2 \, x^\frac{2}{1+\gamma} \right]}} \atop {{dt_c \, = \, x^{- \frac{\gamma}{1+\gamma}} \, y^{\frac{\gamma}{1-\gamma}} \, dt} \atop }$\\

\\

$\!\!\!\!\!(3)$ & {\tiny $\! C_1 \, e^{\, 2\, \varphi} \ + \ C_2$} & ${{\cal A} \,= \,{\frac{1}{2} \log x \, + \, v } \atop {\varphi \,= \, \frac{1}{2} \log x - v}}\atop {{\mathcal B}\, =\, -\, \frac{1}{2} \log x + v \atop }$ & ${{\mathcal L}\, = \, - \,  2 \dot{x} \dot{v} \, - \, 2 \, C_1 x  \, - \, 2 \, C_2 \, e^{2 v} \atop {\dot{x}\, \dot{v} \, = \, C_1 \, e^{2v} \, + \, C_2 \, x}} \atop {{dt_c \, =\, e^{v} \, x^{-\frac{1}{2}} dt}\atop }$ \\
\\

$\!\!\!\!\!(4)$ & {\tiny $\! C \, \varphi \, e^{\, 2\, \varphi}$} & ${{\cal A}\, =\, \frac{1}{4} \log x \, + \, v \atop {\varphi \, = \, \frac{1}{4} \log x \, - \, v}}\atop {{{\cal B}\, =\, -\, \frac{3}{4} \log x \, + \, v} \atop }$ & ${{{\cal L} \, = \, - \, \frac{1}{2} \, \dot{x}\, \dot{v} \, - \, C\, \left( \frac{1}{4} \, \log \,x \, - \, v \right)} \atop {\dot{x} \, \dot{v} \, = \, C \, \left( \frac{1}{2} \, \log\,x \, - \, 2\, v \right)}}\atop {{dt_c \, = \, x^{\, - \, \frac{3}{4}} \, e^{\,v} \, dt} \atop }$\\
\\

$\!\!\!\!\!(5)$ & {\tiny $\! C \, \log (\coth \varphi) \, +\, D$} & ${{{\cal A}\, =\, \frac{1}{2} \, \log\left(\frac{\xi^2 \, - \, \eta^2}{2} \right) } \atop {\varphi \, = \, \frac{1}{2} \, \log \left( \frac{\xi\,+\,\eta}{\xi \, -\, \eta} \right) }} \atop {{{\cal B}\, =\, - \, \frac{1}{2} \, \log\left(\frac{\xi^2 \, - \, \eta^2}{2} \right)} \atop {}}$& ${{{\cal L} \, = \, - \, \dot{\xi}^{\,2} \, + \, \dot{\eta}^{\,2}  \, -  \, 8\, C \, \log\left( \frac{\xi}{\eta} \right) \, - \, 8\, D} \atop {\dot{\xi}^{\,2} \, -  \, \dot{\eta}^{\,2}  \, =  \, 8\, C \, \log\left( \frac{\xi}{\eta} \right) \, + \, 8\, D}} \atop {{dt_c \, = \, \frac{2\, dt}{\sqrt{\xi^{2} \, - \, \eta^{2}}}} \atop {}}$ \\
\\

$\!\!\!\!\!(6)$ & {\tiny $\! C \, \mathrm{Im} \left[ \log\left( \frac{e^{\,2\,\varphi}\, +\, i}{e^{\,2\,\varphi}\, -\, i}\right)\, + \, D\right]$} & ${{{\cal A}\, =\, \frac{1}{2} \, \log\left(\frac{\xi^2 \, - \, \eta^2}{2} \right)} \atop {\varphi \, = \, \frac{1}{2} \, \log \left( \frac{\xi\,+\,\eta}{\xi \, -\, \eta} \right) }} \atop {{{\cal B}\, =\, - \, \frac{1}{2} \, \log\left(\frac{\xi^2 \, - \, \eta^2}{2} \right)} \atop {z \, =\, \frac{1}{\sqrt{2}} \ \left( \xi \, e^{\,\frac{i\pi}{4}} \, + \, \eta \, e^{\,-\,\frac{i\pi}{4}} \right)}}$ & ${{{\cal L} \, = \, 2\, \mathrm{Im} \Big[ - \dot{z}^2 \, - \, 8\, C \, \log z  \, - \, 8\, D \Big]} \atop {\mathrm{Im} \Big[ \dot{z}^2 \, - \, 8\, C \, \log z  \, - \, 8\, D \Big] \, = \, 0}} \atop {{dt_c \, = \, \frac{2\, dt}{\sqrt{\xi^{2} \, - \, \eta^{2}}}} \atop {}}$ \\

\\
$\!\!\!\!\!(7)$ & {\tiny $\! C_1 \, \Big(\cosh\,\gamma\,\varphi \Big)^{\frac{2}{\gamma} \, - \, 2}\!\! + \, C_2 \Big( \sinh\,\gamma\,\varphi \Big)^{\frac{2}{\gamma} \, - \, 2}$} & ${{{\cal A}\, =\, \frac{1}{2\gamma} \, \log\left(\xi^2 \, - \, \eta^2\right)} \atop {\varphi \, = \, \frac{1}{2\gamma} \, \log \left( \frac{\xi\,+\,\eta}{\xi \, -\, \eta} \right) }} \atop {{{\cal B}\, =\, \left(\frac{1}{2\gamma}\,-\,1\right) \, \log\left(\xi^2 \, - \, \eta^2\right)} \atop {}}$ & ${{{\cal L} \, = \, - \, \frac{\dot{\xi}^{\,2} \, - \, \dot{\eta}^{\,2}}{2\, \gamma^{\,2}} \, - \, C_1 \ \xi^{\, \frac{2}{\gamma} \, - \, 2} \, - \, C_2 \, \eta^{\, \frac{2}{\gamma} \, - \, 2}} \atop {\dot{\xi}^{\,2} \, - \, \dot{\eta}^{\,2} \, = \, 2\, \gamma^2 \left[ C_1 \, \xi^{\, \frac{2}{\gamma} \, - \, 2} \, + \, C_2 \, \eta^{\, \frac{2}{\gamma} \, - \, 2} \right]}} \atop {{dt_c \, = \, (\xi^2 \, - \, \eta^2)^{\,\frac{1}{2\,\gamma} \ - \ 1} \, dt} \atop {}}$ \\
\\
$\!\!\!\!\!(8)$ & {\tiny $ \! \mathrm{Im} \,\Big[\,C \ \Big(\,i \, + \, \sinh\,2\,\gamma\,\varphi\, \Big)^{\, \frac{1}{\gamma} \ - \ 1}\,\Big]$} & ${{{\cal A}\, =\, \frac{1}{2\gamma} \, \log\left(\xi^2 \, - \, \eta^2\right)} \atop {\varphi \, = \, \frac{1}{2\gamma} \, \log \left( \frac{\xi\,+\,\eta}{\xi \, -\, \eta} \right) }} \atop {{{\cal B}\, =\, \left(\frac{1}{2\gamma}\,-\,1\right) \, \log\left(\xi^2 \, - \, \eta^2\right)} \atop {z \, =\, \frac{1}{\sqrt{2}} \ \left( \xi \, e^{\,\frac{i\pi}{4}} \, + \, \eta \, e^{\,-\,\frac{i\pi}{4}} \right)}}$ & ${{{\cal L} \, = \, \mathrm{Im} \left[ \, - \, \frac{1}{2\, \gamma^{\,2}} \, \dot{z}^2 \, - \, \frac{C}{2} \, z^{\,\frac{2}{\gamma} \,-\, 2}\right]} \atop {\mathrm{Im} \left[ \frac{1}{2\, \gamma^{\,2}} \ \dot{z}^2 \, - \, \frac{C}{2} \ z^{\,\frac{2}{\gamma} \,-\, 2}\right] \, = \, 0}} \atop {{dt_c \, = \, (\xi^2 \, - \, \eta^2)^{\,\frac{1}{2\,\gamma} \ - \ 1} \, dt} \atop {}}$ \\
\\
$\!\!\!\!\!(9)$ & {\tiny $\! C_1 \ e^{2\,\gamma\,\varphi} \  + \ C_2 \
e^{\frac{2}{\gamma}\,\varphi}\ \ \ (\gamma^2 \neq 1)$} & ${{{\cal A}\, = \, \frac{1}{\sqrt{1-\gamma^{\,2}}} \Big( \widehat{\cal A} \, - \, \gamma \, \widehat{\varphi} \Big)} \atop {\varphi \, = \, \frac{1}{\sqrt{1-\gamma^{\,2}}} \Big( \widehat{\varphi} \, - \, \gamma \, \widehat{\cal A} \Big)}} \atop {{{\cal B} \, = \, {\cal A}} \atop {}}$ & $\!\!\!{{{\cal L} \, = \, \frac{{\dot{\widehat{\varphi}}}^{\,2} \, - \, {\dot{\widehat{\cal A}}}^{\,2}}{2} \, - \, C_1\ e^{\,2\widehat{\cal A}\sqrt{1-\gamma^{\,2}}}\, - \, C_2\, e^{\,\frac{2\widehat{\varphi}\sqrt{1-\gamma^{\,2}}}{\gamma}} } \atop {\!\frac{{\dot{\!\widehat{\varphi}}}^{\,2} \, - \, {\dot{\widehat{\cal A}}}^{\,2}}{2} \, = \, - \, C_1\ e^{\,2\widehat{\cal A}\sqrt{1-\gamma^{\,2}}}\, - \, C_2\, e^{\,\frac{2\widehat{\varphi}\sqrt{1-\gamma^{\,2}}}{\gamma}}}} \atop {{dt_c \, = \, \exp\left[\, \frac{\widehat{\cal A} \, - \, \gamma \, \widehat{\varphi} }{\sqrt{1\,-\,\gamma^{\,2}}} \right]\, dt} \atop {}}$ \\
\\
\hline
\hline
\end{tabular}
\label{tab:families}
\end{center}
\end{table}
%


\section{\sc  Properties of the Exact Solutions}\label{sec:properties_exact}


In this section we describe in detail the solutions of the most significant integrable systems identified in Section \ref{sec:integrable}. Our emphasis will be on two main lessons that can be drawn from this study. The first is the emergence, in more general contexts, of the \emph{climbing phenomenon} that, as we have seen in Section \ref{sec:climbing}, first presented itself in a single sufficiently steep exponential potential ${\cal V}$ but only depends on the asymptotic behavior of the potential for large values of its argument. The second is the ubiquitous emergence of a \emph{Big Bang} followed by a \emph{Big Crunch} whenever the scalar field tries to settle at a negative extremum of ${\cal V}$. We begin in Section \ref{sec:ODEqualitative} with a detailed description of the variety of behaviors that are possible around critical points of the potentials and we apply the general theory of two--dimensional integrable systems to our examples of Section \ref{sec:integrable}, which are collected in Table \ref{tab:families}. We then describe in detail the solutions of the integrable families that we identified there, beginning in Section \ref{sec:B0_sol} from the relatively simple systems of Section \ref{sec:B0} that belong to the standard FLRW setting with ${\cal B}=0$. In Section \ref{sec:Bnot0_sol} we describe the solutions of the triangular systems of Section \ref{sec:Bnot0}, and we conclude in Section \ref{sec:quadratures_sol} with the systems solvable by quadratures of Section \ref{sec:quadratures}.

\subsection{\sc Qualitative analysis of fixed points: Subsystems I and II}
\label{sec:ODEqualitative}

It is instructive to spell out how one--scalar FLRW cosmologies are governed, in general, by nonlinear autonomous first--order dynamical systems on a two--dimensional plane. Our discussion moves from the redundancy of eqs.~\eqref{onefield_eqs}, a fact that we have already stressed in Section \ref{sec:effelag}, and for the sake of simplicity in this portion of the section we shall work in the gauge
\bea
&& {\cal B} \ = \ 0\ ,
\eea
although the exact solutions that will presented later on will rest, to a large extent, on more general gauge choices.
In this fashion ${\cal A}$ becomes a cyclic variable, since eqs.~(\ref{onefield_eqs}) become
\bea
&& \ddot{\varphi} \ + \  \dot{\cal A} \, \dot{\varphi} \ + \ \mathcal{V}^{\,\prime}(\varphi) \ = \ 0 \ , \nonumber \\
&& \ddot{\cal A} \ = \ - \ \dot{\varphi}^{\, 2} \  , \nonumber \\
&& \dot{\cal A}^{\,2} \ - \ \dot{\varphi}^{\,2} \ = \ 2\, \mathcal{V}(\varphi) \label{onefield_eqs_B0}\ ,
\eea
where ${\cal A}$ enters only via its time derivative, or equivalently via the (rescaled) Hubble function
\be
{\cal H} \ =\ \dot{\cal A} \ \equiv \ {H}\,(d-1) \ .
\ee
The solutions of this redundant system can indeed be generated starting from two irreducible sub--systems to which we now turn.
\begin{itemize}
\item {\bf Subsystem I} is
\bea
&& \dot{\varphi} \ = \ v \ ,\nonumber \\
&& \dot{v} \ = \ - \ \sigma \, v\, \sqrt{v^{\,2} \ + \ 2\, \mathcal{V}(\varphi)}\ - \
\mathcal{V}^{\,\prime}(\varphi)\ ,
\label{ODE3}
\eea
where the sign $\sigma =  \pm 1$ accounts for periods of \emph{expansion} or \emph{contraction} of the Universe, and where one should \emph{exclude} possible branches satisfying the conditions
\bea
&& v^{\,2} \ + \ 2\, \mathcal{V}(\varphi) \ = \ 0 \qquad {\rm if} \qquad
\mathcal{V}(\varphi) \ \neq \ 0 \ ,\nonumber \\
&& v^{\,2} \ + \ 2\, \mathcal{V}(\varphi) \ < \ 0 \ .
\label{ODE3a}
\eea
If eqs.~(\ref{ODE3}) are solved, one can also obtain
\bea
{\cal H} \ = \ \sigma \, \sqrt{v^{\,2} \ + \ 2\, \mathcal{V}(\varphi)}\ .
\label{ODE5}
\eea

\item{\bf Subsystem II} is
\bea
&& \dot{\varphi} \ = \ \sigma \, \Big({\cal H}^{\,2} \ - \ 2\, \mathcal{V}(\varphi)\,\Big)^{\frac{1}{2}}\ ,\nonumber \\
&& \dot{\cal H} \ = \ -\,\Big({\cal H}^{\,2} \ - \ 2\, \mathcal{V}(\varphi)\,\Big)\ ,
\label{ODE4}
\eea
where the sign $\sigma =  \pm 1$ accounts again for periods of expansion or contraction of the Universe, and where one should \emph{exclude} possible branches satisfying the conditions
\bea
&& {\cal H}^{\,2} \ - \ 2\, \mathcal{V}(\varphi) \ = \ 0 \qquad {\rm if} \qquad \
\mathcal{V}^{\,\prime}(\varphi) \ \neq \ 0 \ , \nonumber\\
&& {\cal H}^{\,2} \ - \ 2\, \mathcal{V}(\varphi) \ < \ 0 \ .
\label{ODE4a}
\eea

\end{itemize}
The branches that are compatible with eqs.~(\ref{ODE3a}) for Subsystem~I and with eqs.~(\ref{ODE4a}) for Subsystem~II will be referred to as ``admissible'' in what follows. One can verify that, within them, the complete original system \eqref{onefield_eqs_B0} can be recovered from any of the two nonlinear first--order systems, which are indeed autonomous on the two--dimensional Euclidean planes $(\varphi, \, v)$ or $(\varphi, \, {\cal H})$.

A well--developed theory of planar dynamical systems makes it possible to analyze qualitatively local and global properties of their phase portraits. Generic planar systems are indeed very regular and can have only a few different types of trajectories and limit sets. There are thus:
\begin{itemize}
\item{\bf fixed points} (critical or stationary);
\item{\bf periodic orbits} (cycles);
\item{\bf homoclinic orbits}: connecting a given fixed point with itself;
\item{\bf heteroclinic orbits}: connecting pairs of different fixed points.
\end{itemize}
An implication of these results is that generic planar systems cannot be chaotic \footnote{See, however, the peculiar behavior discussed in \cite{cd}: a scalar confined to a $\cosh$--well that is ``critical'' in the sense of Section \ref{sec:climbing} undergoes wild oscillations near the initial singularity. This occurs since, as the system proceeds toward a Big Crunch in the time reversal of the standard scenario, the scalar can never overcome the attractive pull of ``(over)critical'' exponential potentials.}, and are therefore very special if compared with dynamical systems in more than two dimensions, where chaotic regimes are frequently present.

Let us now turn to analyze in detail the nature of the fixed points present in the systems of Table \ref{tab:families}. This is quite instructive, although the more interesting mathematical structures present themselves in systems that are somewhat pathological from a physical viewpoint, since they involve potentials that are unbounded from below.

\subsubsection{\sc Qualitative Analysis of Subsystem I}
\label{subsystemI}
The fixed points of eqs.~(\ref{ODE3}) are determined by the conditions
\bea
v_c \ = \ 0\ , \quad \mathcal{V}^{\,\prime}(\varphi_c)\ = \ 0 \ ,
\label{ODE6}
\eea
and are admissible if
\bea
\mathcal{V}(\varphi_c)\ \geq \ 0\ .
\label{ODE6a}
\eea
If this condition is not fulfilled, Subsystem I does not possess fixed points and all points in its phase space are regular. A nonlinear system without admissible fixed points must possess only monotonic solutions, which can also blow up in a finite time. This behavior will show up repeatedly around the negative extrema present in some of our potentials: the scalar field $\varphi$ trying to settle down there will run off to infinity in a finite time while the Universe will experience a Big Crunch.
\par
If Subsystem I is linearized around a fixed point, the resulting equations read
\bea
&& \dot{\phi} \ = \ v \ ,\nonumber \\
&& \dot{v} \ = \ -\, \sigma \, \sqrt{ 2\, \mathcal{V}(\varphi_c)}\ v\ - \ \mathcal{V}^{\,\prime \prime}(\varphi_c) \ \phi \ ,
\label{ODE7}
\eea
where $\phi$ denotes the displacement of $\varphi$ from its critical value,
\be
\phi \ = \ \varphi \ - \ \varphi_c \ .
\ee
The corresponding eigenvalues
\be
 \lambda_{\pm} \ = \ - \ \sigma \, \sqrt{\frac{\mathcal{V}(\varphi_c)}{2}}\ \pm \  \sqrt{\frac{\mathcal{V}(\varphi_c)}{2}\ - \ \mathcal{V}^{\,\prime \prime }(\varphi_c)}
\label{ODE8}
\ee
characterize the critical points, and consequently define the phase portrait of the linearization. The nature of a fixed point is reflected in the corresponding eigenvalues. One can thus distinguish the following cases:
\begin{itemize}
\item{\bf Hyperbolic fixed point:}
\subitem{\bf - saddle:} if the two eigenvalues are real and have opposite signs;\,
\subitem{{\bf - node} {\it (attracting or repelling)}:} if the eigenvalues are real and have the same sign;
\subitem{{\bf - improper node} {\it (attracting or repelling)}:} if the two eigenvalues coincide;
\subitem{{\bf - focus} {\it (attracting or repelling)}:} if the two eigenvalues have the same real part;
\item{\bf Elliptic fixed point:} if the eigenvalues are purely imaginary.
\end{itemize}
An important result is that the phase portraits of a nonlinear system and of its linearization are qualitatively equivalent in a neighborhood of a \emph{hyperbolic fixed point}, where $Re\,(\lambda_{\pm})\,\neq\,0$. Let us add that Subsystem I does not possess periodic trajectories on account of Dulac's criterion, since the expression
\bea
\frac{\partial \dot{\varphi}}{\partial \varphi} \ + \
\frac{\partial \dot{v}}{\partial v} \ = \ -\,2\,\sigma \, \frac{v^{\,2} \ + \  \mathcal{V}(\varphi)}{\sqrt{v^{\,2} \ + \ 2\, \mathcal{V}(\varphi)}}
\label{ODE3per}
\eea
does not change sign on the whole two--dimensional plane. We are thus led to conclude that Subsystem I can only have \emph{fixed points}, \emph{heteroclinic orbits} or \emph{homoclinic orbits}.

In order to understand qualitatively the phase portrait in a neighborhood of an admissible fixed point one can analyze its structural stability. If the fixed point $\varphi_{c}$ is a local \emph{minimum} of the potential $\mathcal{V}(\varphi)$, one should define the weak Lyapunov function with the required properties,
\bea
&& f(\varphi,v)\ \equiv \  \sqrt{v^{\,2} \ + \ 2\, \mathcal{V}(\varphi)}\ - \ \sqrt{ 2\, \mathcal{V}(\varphi_{c})}\ > \ 0 \ , \quad f(\varphi_{c},v_{c})\ = \ 0\ ,\label{LyapGen1}\\
&& \dot{f}(\varphi,\,v)\ = \ -\,\sigma \, v^2 \ .\nonumber
\eea
By construction this function is positive definite in the domain of phase space delimited by the corresponding inequality (\ref{LyapGen1}) and vanish only at the fixed point, while its time derivative is negative or positive semi--definite depending on the sign of $\sigma \,=\,\pm 1$ and do not vanish identically on any trajectory other than the fixed point itself. From the constructed Lyapunov function one can conclude that this fixed point is unstable for $\sigma \,=\,-1$ and asymptotically stable for $\sigma \,=\,+1$. The inequality (\ref{LyapGen1}) defines explicitly the basin of attraction, i.e. the phase-space domain of asymptotic stability, and all trajectories crossing it approach asymptotically the fixed point as $t\, {\rightarrow} +\infty$.

The asymptotic behavior as $t\, {\rightarrow} +\infty$ of the Hubble function and of the scale factor that apply if the fixed point is asymptotically stable have the form
\bea
{\cal H} \ = \ \sqrt{2\,\mathcal{V}(\varphi_{c})}\ , \quad {\cal A} \ = \ \sqrt{2\,\mathcal{V}(\varphi_{c})}\,(t\,-t_0) \quad {\rm if} \quad \mathcal{V}(\varphi_{c}) \ > \ 0\ ,
\label{LyapGen3}
\eea
as pertains to an expanding de Sitter patch, while the exponential behavior leaves way to a power--like behavior if $\mathcal{V}(\varphi_{c}) \, = \, 0$. Let us also recall that in four dimensions $H=\frac{{\cal H}}{3}$ and $a = e^{\,{\cal A}/3}$.

\subsubsection{\sc Qualitative Analysis of Subsystem II and the ``separatrix''}
\label{subsystemII}
Let us now consider a class of potentials that can be represented in the form
\bea
\mathcal{V}(\varphi) \ = \ \frac{1}{2}\, \Big(\mathcal{P}(\varphi)^2 \ - \ \mathcal{P}^{\,\prime}(\varphi)^2\,\Big) \ , \label{sasha1}
\eea
where the function $\mathcal{P}(\varphi)$ is clearly defined modulo an overall sign.
This representation is actually not unique, and different functions $\mathcal{P}(\varphi)$
might correspond to the same potential $\mathcal{V}(\varphi)$. After all, eq.~(\ref{sasha1}) is a non--linear first--order differential equation for $\mathcal{P}(\varphi)$ that in principle can have both sporadic and continuous one--parameter families of solutions.

Resorting to this representation, we can now explain how to build the ``separatrix'' for a saddle, the solution that is not sensitive to the repulsive eigenvalue. To this end, let us note that for potentials that are compatible with eq.~\eqref{sasha1} the two simpler equations
\bea
\label{sasha2}
\dot{\varphi} & = & \sigma\  \mathcal{P}^{\,\prime}(\varphi) \ , \nonumber \\
{\cal H} & = & - \,\sigma \ \mathcal{P}(\varphi)\ ,
\label{sasha2a}
\eea
which can be integrated by quadratures, yield solutions of Subsystem II. However, if $\mathcal{P}(\varphi)$ denotes a particular solution of eq.~(\ref{sasha1}), the corresponding general solution of eqs.~(\ref{sasha2}) involves a single integration constant. On the contrary, the general solution of eqs.~(\ref{ODE4}) would involve two integration constants, so that only \emph{special solutions} of Subsystems I (\ref{ODE3}) and II (\ref{ODE4}) are captured in this fashion.
Regular fixed points $\varphi_c$ of $\mathcal{P}(\varphi)$ are regular fixed points of the one--dimensional gradient system (\ref{sasha2}) that are also admissible fixed points of the potential $\mathcal{V}(\varphi)$ (\ref{sasha1}). Moreover, the non--degenerate fixed points of the former are also non--degenerate fixed points of the latter, since
\bea
&&  \mathcal{P}^{\,\prime}(\varphi_c)\ = \ 0  \quad \Rightarrow \quad \mathcal{V}^{\,\prime}(\varphi_c) \ = \ \mathcal{P}^{\,\prime}(\varphi_c)\, \left(\,\mathcal{P}(\varphi_c) \ - \ \mathcal{P}^{\,\prime \prime}(\varphi_c)\,\right)\ = \ 0\ , \nonumber\\
&& \mathcal{V}^{\,\prime \prime}(\varphi_c) \ = \ \mathcal{P}^{\,\prime\prime}(\varphi_c)\, \left(\,\mathcal{P}(\varphi_c) \ - \ \mathcal{P}^{\,\prime \prime}(\varphi_c)\,\right)\ \neq \ 0 \quad \Rightarrow \quad \mathcal{P}^{\,\prime \prime}(\varphi_c)\ \neq \ 0 \ ,  \nonumber\\
&& \mathcal{V}(\varphi_c) \ = \ \frac{1}{2}\,(\mathcal{P}(\varphi_c))^2\ > \ 0\ .
\label{sasha1a}
\eea
\par
A suitable choice for $\sigma=\pm 1$ in (\ref{sasha2}) can always turn a fixed point of $\mathcal{V}$ into an attractor for the system (\ref{sasha2}), provided $\sigma$ is chosen in such a way that $\sigma\  \mathcal{P}^{\,\prime\prime}(\varphi_c)\ < \ 0$, so that if Subsystem I possesses a saddle point this becomes an attractor for eqs.~(\ref{sasha2}). The corresponding orbit approaches asymptotically this fixed point, and consequently it describes what is usually called the separatrix of the saddle point of Subsystem I (\ref{ODE3}). Eqs.~(\ref{sasha2}) are thus the \emph{separatrix equations}, a fact that we shall use repeatedly in what follows. Notice that a direct numerical approach would be bound to miss this curve, since the evolution near the fixed point would be dominated generically by the repelling eigenvalue.

The reader might have noticed some analogies between the representation (\ref{sasha1}) and the scalar potential of  $\mathcal{N}=1$  Supergravity coupled to a Wess--Zumino multiplet,
\begin{eqnarray}
    \mathcal{V} & = & 4 \, e^2 \,\exp \left[ \mathcal{K} \right]\left ( g^{z\bar{z}}\, \mathcal{D}_z W_h(z)\, \mathcal{D}_{\bar{z}} \overline{W_h}({\bar z}) \, - \, 3 \,  |W_h(z)|^2 \right )\ ,
    \label{frullini}
\end{eqnarray}
which is constructed  in terms of a K\"ahler potential ${\cal K}(z,\bar{z})$ and of a holomorphic superpotential $W_h(z)$ \cite{cfgv}. In eq.~\eqref{frullini} $z$ denotes a complex scalar field, whose kinetic term
\be
\sqrt{-g} \, g^{\,\mu\nu}\, \partial_z\partial_{\bar z} \mathcal{K} \ \partial_{\,\mu} z \ \partial_{\,\nu}\bar{z}
\ee
is determined by ${\cal K}$, while the two \emph{K\"ahler covariant derivatives} are
\begin{eqnarray}
  \mathcal{D}_z \, W&=& \partial_z W \, + \,  \partial_z \mathcal{K} \, W \nonumber \ , \\
  \mathcal{D}_{\bar z}\, { \overline{W}} &=& \partial_{\bar z} \overline{W} \, + \,
  \partial_{\bar z} \mathcal{K} \, \overline{W} \ . \label{felucide}
\end{eqnarray}
In \cite{prep} we shall advocate that the link between the potentials studied in this paper and $\mathcal{N}=1$ Supergravity rests on K\"ahler potentials of the form
\begin{equation}\label{calero}
    \mathcal{K}  \, = \, - \,\log \, \left[ \left(-{\rm i}\left(z-\bar{z}\right)\right)^q\right]
\end{equation}
and on the identifications
\begin{equation}\label{copperfildo}
    z \, = \, {\rm i} \, \exp\left( \frac{\varphi}{\sqrt{3q}} \right) \, + \, b \ .
\end{equation}
If the axion field $b$ can be consistently set to zero, the residual potential $\mathcal{V}(\varphi)$ acquires a form that becomes similar to eq.~(\ref{sasha1}) after the identification
\begin{equation}\label{mozzilla}
    \widetilde{\mathcal{P}}(\varphi) \, = \, \exp\left(-\frac{\varphi \sqrt{3 q}}{6} \right) \, \times \, W_h\left( {\rm i} \, \exp\left[ \frac{\varphi}{\sqrt{3q}} \right ] \, \, \right) \ .
\end{equation}
Assuming indeed that $W_h$ be a real function of $\varphi$ and parameterizing it as in eq.~(\ref{mozzilla}) leads to
\begin{equation}\label{cuprano}
    \mathcal{V} \, = \, \frac{3}{2^q} \, \left[ 4 \left(\widetilde{\mathcal{P}}^\prime(\varphi)\right)^2 \, - \, \left(\widetilde{\mathcal{P}}(\varphi)\right)^2\right]\ ,
\end{equation}
so that one might be tempted to foresee in $\mathcal{P}(\varphi) $ the imprint of a \textit{supersymmetric superpotential}. Yet, even leaving aside the factor $4$ in the first term, which could be absorbed rescaling $\varphi$, the crucial overall sign flip makes it impossible to identify ${\mathcal{P}}(\varphi)$ and $\widetilde{\mathcal{P}}(\varphi)$. Indeed, if for a given superpotential $W_h(z)$ the truncation of ${\cal V}$ to a vanishing axion allows the representation (\ref{sasha1}) in addition to the natural supersymmetric one (\ref{cuprano}), the link between $\mathcal{P}(\varphi)$ and $W_h\left( {\rm i} \, \exp\left[ \frac{1}{\sqrt{3q}} \, \varphi\right ] \, \, \right)$ is bound to be generally complicated and non local.  Hence the function $\mathcal{P}(\phi)$, if it exists, deserves the name \textit{fake superpotential} \cite{fakesup}, and this discussion should convey a flavor of the difficulties that are met when trying to fit integrable superpotentials in Supergravity.  Reverting the argument one can conclude that, given a potential that admits the representation (\ref{sasha1}) in terms of a real \textit{fake superpotential}, finding the corresponding holomorphic \textit{true superpotential} is generally a hard task.

The results collected in (\ref{sasha1a}) have a parallel in terms of the supersymmetric representation (\ref{cuprano}) and of the \textit{true superpotential} $\widetilde{\mathcal{P}}(\varphi)$. The main difference is that a critical point of the true superpotential is also an extremum of the scalar potential, but at a negative value of the potential:
\begin{equation}\label{crullostico}
    \mathcal{V}(\varphi_0) \ = \ - \, \frac{3}{2^q}\,(\widetilde{\mathcal{P}}(\varphi_0))^2\ < \ 0\ .
\end{equation}
Therefore, $\varphi_0$ physically corresponds to an anti de Sitter vacuum with unbroken supersymmetry, and critical points of the true superpotential are just what is required to preserve $\mathcal{N}=1$ supersymmetry. On the contrary, a critical point of the fake superpotential is a fixed point of the dynamical system, and from the physical viewpoint it corresponds to a de Sitter vacuum, which necessarily breaks all supersymmetries.
As it is well known, it is difficult to build de Sitter vacua in Supergravity. However, these types of vacua are guaranteed to exist in any given model whose supersymmetric scalar potential admits a representation like (\ref{sasha1}) in terms of a non--monotonic fake superpotential. A necessary, though not sufficient, condition for the existence of a fake superpotential is thus the existence of a de Sitter vacuum in the supergravity model under consideration.

Returning to eqs.~\eqref{sasha1}, let us notice that a new dependent variable $\mathfrak{P}(\varphi)$ defined by
\be
\mathcal{P}(\varphi)\ \equiv \ \sqrt{\frac{2\,{\cal V}(\varphi)}
{1\,-\,(\mathfrak{P}(\varphi))^{-2}}} \ ,
 \label{Abel_trans}
\ee
turns it into the product of two Abel equations of the first kind \cite{abel},
\bea
\mathfrak{P}^{\,\prime}(\varphi) \ = \  \left((\mathfrak{P}(\varphi))^2\ - \ 1 \right) \,\left(\frac{(\log|{\cal V}(\varphi)|)^{\,\prime}}{2}\ \mathfrak{P}(\varphi)\ \mp 1 \right)
\label{Abel_eq}
\eea
whose coefficient function depends on the potential ${\cal V}(\varphi)$.
Solving any of these two Abel equations provides a solution of the original problem, but neither of them is integrable for generic potentials. However, one can construct general solutions in some special cases.  These include the interesting potential well
\be
 {\cal V}(\varphi)\ = \ C\, \cosh\,\frac{2\,\varphi}{3} \ , \label{Ex3_1intintCOSHMainA}
\ee
which will show up again in Sections \ref{sec:Bnot0_sol} and \ref{sec:quadratures_sol}, and whose exact cosmological solution will be described in detail in Section \ref{sec:Bnot0_sol}. In
the new basis provided by $\mathfrak{P}(\varphi)$ and $\mathcal{U}(\varphi)$, where
\bea
\mathfrak{P}(\varphi) \ = \ \frac{1}{\sqrt{\mathcal{U}(\varphi)}}\ \pm\,\coth\frac{2\,\varphi}{3} \ ,
\label{AbelEq_basis1}
\eea
Abel's equation (\ref{Abel_eq}) reduces in this case to the linear equation
\bea
 \mathcal{U}^{\,\prime}(\varphi) \ = \
\left(\frac{2}{3}\,\tanh \frac{2\,\varphi}{3} \ + \ 2\,\coth \frac{2\,\varphi}{3} \right)\, \mathcal{U}(\varphi)\ - \ \frac{2}{3}\,\tanh \frac{2\,\varphi}{3}\ ,
\label{Abel_eqI}
\eea
and solving it one arrives at the general solution for $\mathfrak{P}(\varphi)$,
\bea
\mathfrak{P}(\varphi) \ = \
\frac{1}{|\sinh\frac{2\,\varphi}{3}|\,
\left(\,\alpha\, \sinh\frac{4\,\varphi}{3}\,+\,\cosh\frac{4\,\varphi}{3}\right)^{\frac{1}{2}}}\,\ \pm \ \,\coth\frac{2\,\varphi}{3} \ ,
\label{AbelEqSolut}
\eea
which is parameterized by an arbitrary integration constant $\alpha$. Let us stress that for $\alpha\,=\,\pm1$ this becomes a rational combination of exponential functions.
Using the same procedure one can also integrate Abel's equation (\ref{Abel_eq}) for the potential
\be
 {\cal V}(\varphi)\ = \ C\, \sinh\,\frac{2\,\varphi}{3} \ . \label{Ex3_1intintSINHainA}
\ee
Inserting the functions $\mathfrak{P}(\varphi)$ in eq.~(\ref{Abel_trans}) one then arrives at corresponding general solutions of eq.~(\ref{sasha1}) for $\mathcal{P}(\varphi)$. A number of $\mathcal{P}(\varphi)$ that solve eq.~(\ref{sasha1}) for other potentials will be constructed in the following subsections.	

\subsubsection{\sc Examples of fixed--point analysis} \label{sec:fixedpts_ana}
\noindent
Let us now turn to the detailed fixed--point analysis of an interesting class of potentials, not all integrable but whose choice is inspired the families of potentials in Table \ref{tab:families}. We shall occasionally distinguish various ranges of the relevant parameters, and for brevity we shall mostly leave out fixed points at infinity, unless they are the only ones present, as will be the case for the last examples. In the corresponding lists we shall reserve boldface characters to the physically more relevant cases of potentials bounded from below and we shall treat the two cases of systems evolving from a Big Bang (corresponding to $\sigma=1$ in eqs.~\eqref{ODE3} or \eqref{ODE4})  or evolving toward a Big Crunch (corresponding to $\sigma=-1$ in eqs.~\eqref{ODE3} or \eqref{ODE4}).
\begin{itemize}
\item[1. ] {\bf The potentials \ $\mathcal{V}(\varphi) \ = \ C \,\cosh(w\,\varphi) \ + \ \Lambda $}
\end{itemize}
\noindent
The class of potentials
\be
\mathcal{V}(\varphi) \ = \ C \,\cosh(w\,\varphi) \ + \ \Lambda\ , \qquad (C \ \neq \ 0\ , \quad w\ \neq \ 0) \ . \label{Ex1}
\ee
possesses an isolated fixed point,
\be
v_c \ = \ 0\ , \quad \varphi_c\ = \ 0 \ ,
\label{Ex2}
\ee
which is admissible provided
\be
\mathcal{V}(\varphi_c)\ = \ C \ + \ \Lambda \ \geq \ 0\ .
\label{Ex2a}
\ee
As we shall see in the following subsections, the condition \eqref{Ex2a} has an important physical consequence: the exact solutions for potential wells of this type will show indeed that when it is not fulfilled a scalar trying to settle at the extremum will readily run away. This behavior reflects, all in all, a familiar fact, the absence of spatially flat AdS slices.

The eigenvalues of eq.~\eqref{ODE8} for the potentials \eqref{Ex1} read
\be
\lambda_{\pm} \ = \ -\,\sigma \, \sqrt{\frac{C \ + \ \Lambda}{2}}\ \pm \  \sqrt{\frac{C \ + \ \Lambda}{2} \ - \ C \, w^2}\ \ ,
\label{Ex3}
\ee
so that in this case the admissible fixed point is simple (not degenerate) since $\lambda_{\pm}\ \neq \ 0$. Depending on the values of the parameters, these eigenvalues can correspond to a hyperbolic fixed point or alternatively to an elliptic one.

A hyperbolic fixed point obtains if
\bea
\label{hyperbolic1}
&h_1)&  \quad  \ \Lambda \ \geq \ -\,C \ ,\quad C \ < \ 0 \quad \\
&&(saddle: \ (\sigma \ = \ \pm \,1)) \ ,\\
&{\bf h_2)}&  \quad  \ \Lambda \ > \ -\,C \ ,\quad C \ > \ 0\ ,\quad w^2 \ < \ \frac{1}{2} \,\Big(1 \ + \ \frac{\Lambda}{C}\Big)  \nonumber\\
&& (node: \ repeller \ (\sigma \ = \ -1) {\rm \ or} \ attractor \ (\sigma \ = \ +1)) \ ,\\
&{\bf h_3)}&  \quad  \ \Lambda \ > \ -\,C \ ,\quad C \ > \ 0\ ,\quad w^2 \ = \ \frac{1}{2} \,\Big(1 \ + \ \frac{\Lambda}{C}\Big)  \nonumber\\
&& (improper \ node: \ repeller \ (\sigma \ = \ -1) {\rm \ or} \ attractor \ (\sigma \ = \ +1)) \ ,\\
&{\bf h_4)}& \quad  \ \Lambda \ > \ -\,C \ ,\quad C \ > \ 0\ ,\quad w^2 \ > \ \frac{1}{2} \,\Big(1 \ + \ \frac{\Lambda}{C}\Big)  \nonumber\\
&& (focus: \ repelling \ (\sigma \ = \ -1)\ {\rm or} \ attracting \ (\sigma \ = \ +1) \ spirals) \nonumber  \ ,
\label{hyperbolic4}
\eea
while an \emph{elliptic} fixed point obtains if
\bea
{\bf e)} \quad C \ + \ \Lambda \ = \ 0\ , \quad C \ > \ 0 \quad (center) \ .
\label{Ex3b}
\eea

The phase portraits of the nonlinear system \eqref{ODE3} and of its linearization \eqref{ODE7} are qualitatively equivalent in a neighborhood of a \emph{hyperbolic} fixed point.
A stable node, an improper stable node and a stable focus are asymptotically stable as well, so that every trajectory approaches the fixed point as $t\, {\rightarrow} +\infty$.
On the other hand, the nature of an \emph{elliptic} fixed point can change in the nonlinear system, but one can show nonetheless that it is unstable for $\sigma \,=\,-1$ and asymptotically stable at $\sigma \,=\,+1$, due to the existence of a weak Lyapunov function $f(\varphi\,,v)$ with the required properties,
\bea
&& f(\varphi,v)\ = \  \sqrt{v^{\,2} \ + \ 2\, \mathcal{V}(\varphi)}\ > \ 0\ , \nonumber\\&& f(0\,,0)\ = \ 0\ , \qquad \dot{f}(\varphi,\,v)\ = \ -\,\sigma \, v^2\ .
\label{Lyap1}
\eea
This is a positive definite function in the whole admissible domain of phase space that vanishes only at the fixed point, while its time derivative is negative (positive) semi--definite for the case $\sigma \,=\,+1 \, (-1$) and does not vanish identically on any trajectory other than the fixed point itself. We shall see explicitly these types of behavior in the relatively simple exact solutions for the case $w=1$ that will be discussed in Section \ref{sec:B0_sol} for the case $\sigma=1$, which corresponds to a Big Bang singularity in the past.

Let us conclude this discussion of the potentials (\ref{Ex1}) by noticing that the asymptotic behavior as $t\, {\rightarrow} +\infty$ of the Hubble function and of the scale factor that apply if the fixed point is asymptotically stable are simply
\bea
{\cal H}\ = \  \sqrt{2(C \ + \ \Lambda)}\ , \quad {\cal A} \ = \ \sqrt{2(C \ + \ \Lambda)}\,(t\,-t_0) \qquad {\rm if} \quad C \ + \Lambda\ \geq \ 0\ , \quad C \ > \ 0\ ,
\label{HubAsym}
\eea
as pertains to an expanding de Sitter patch.

The subclass of potentials (\ref{Ex1}) with
\bea
\label{sasha3}
&& C\ = \  +\, \Lambda\ \frac{4\, - \, w^2}{4\, + \, w^2} \ , \quad  \Lambda\ > \ 0 \ , \quad w \ \neq 2\ , \\
&& C\ = \   - \, \Lambda\ \frac{4\, - \, w^2}{4\, + \, w^2} \ ,  \quad  \Lambda\ < \ 0 \ , \quad w \ \neq 2
\label{sasha3abcd}
\eea
can equivalently be represented in the form (\ref{sasha1}) with
\bea
\label{sasha4}
&& \mathcal{P}(\varphi)\ = \ 4 \,\sqrt{\frac{\Lambda}{4\, + \, w^2}} \ \cosh\Big(\,\frac{w}{2}\,\varphi\,\Big) \ , \quad  \Lambda\ > \ 0 \ , \\
&& \mathcal{P}(\varphi)\ = \ 4 \,\sqrt{\frac{-\, \Lambda}{4\, + \, w^2}} \ \sinh\Big(\,\frac{w}{2}\,\varphi\,\Big) \ , \quad  \Lambda\ < \ 0 \ .
\label{sasha4abcdef}
\eea
Only the first subclass (\ref{sasha3}) possesses the admissible fixed point (\ref{Ex2}), which in this case is characterized by
\bea
\lambda_{\pm} \ = \ 2\,\sqrt{\frac{\Lambda}{4\, + \, w^2}}\,\Big(-\,\sigma \ \pm \ \frac{1}{2}\,|w^2\ - \ 2 |\ \Big)\ ,
\label{sasha5}
\eea
so that it is of the following types:
\bea
\label{sasha6}
& h_1)&  saddle \ (\sigma \ = \ \pm \,1) \ {\rm if} \
\ |w| \ > \ 2 \ ,\\
&{\bf h_2)}& node: \ repeller(attractor) \ {\rm for} \ \sigma \ = \ -1 (+1) : w \ < \ 2 \ {\rm and} \ w \ \neq \  \pm\, \sqrt{2}\ ,\\
&{\bf h_3)}& improper \ node: \ repeller(attractor) \ {\rm for} \ \sigma \ = \ -1 (+1) : w \ = \ \pm\, \sqrt{2}\ .
\eea
Since for the choice of $\mathcal{P}(\varphi)$ in eq.~(\ref{sasha4}) $\mathcal{P}^{\,\prime \prime}(\varphi_c) > 0$, the discussion after eq.~\eqref{sasha1a} implies that the \emph{separatrix equations} (\ref{sasha2}) for the saddle (\ref{sasha6}) are found letting $\sigma=-1$, and thus read
\bea
&& \dot{\varphi} \ = \ -\, 2\, w \,\sqrt{\frac{\Lambda}{4\, + \, w^2}}\ \sinh\Big(\,\frac{w}{2}\,\varphi\,\Big) \ , \nonumber \\
&& {\cal{H}} \ = \ 4 \,\sqrt{\frac{\Lambda}{4\, + \, w^2}} \ \cosh\Big(\,\frac{w}{2}\,\varphi\,\Big)\ ,
\label{ssaddle1}
\eea
so that the corresponding solutions are
\bea
&& \varphi \ = \ \pm \ \frac{2}{w} \,\log \left\{ \coth\left[\,\frac{w^2}{4}\,\sqrt{\frac{\Lambda}{4\, + \, w^2}}\,(t\ - \ t_0)\right]\right\} \ , \nonumber \\
&& {\cal A} \ = \ {\cal A}_0 \,\Big|\sinh\Big(\,\frac{w^2}{2}\,\sqrt{\frac{\Lambda}{4\, + \, w^2}}\,(t\ - \ t_0)\,\Big)\Big|^{\frac{8}{w^2}}\ .
\label{sasha2Solut_1}
\eea
Here $t_0$ and $a_0$ are integration constants, while the reader should appreciate that $\varphi$ in eq.~\eqref{sasha2Solut_1} possesses two distinct branches.
\vskip 24pt
\begin{itemize}
\item[2. ]{\bf The integrable potentials (1) of Table \ref{tab:families}}
\end{itemize}
The potentials
\be
\mathcal{V}(\varphi)\ = \ \epsilon_1\,e^{\varphi} \ + \ \epsilon_2\,e^{-\,\varphi} \ + \ \Lambda\ , \quad \epsilon_1 \ = \ \pm \, 1\ , \quad \epsilon_2 \ = \ \pm \, 1\ ,
\label{Ex2_1New}
\ee
where more general values of the parameters $\epsilon_1$ and $\epsilon_2$ can always be reached by rescalings of the time variable and constant shifts the scalar field, possess the fixed point
\be
v_c \ = \ 0\ , \quad \varphi_c\ = \ 0\ ,\qquad {\rm if} \quad \epsilon_1\ = \ \epsilon_2 \ ,
\ee
\be
\mathcal{V}(\varphi_c)  \ = \   2\,\epsilon_1 \ + \ \Lambda \ , \quad \mathcal{V}^{\,\prime \prime}(\varphi_c)\ = \ 2\, \epsilon_1 \ ,
\label{Ex2_3New}
\ee
which is admissible if
\be
2\, \epsilon_1 \  + \ \Lambda \ \geq 0\ .
\label{Ex2_4New}
\ee
The corresponding eigenvalues \eqref{ODE8} are
\bea
\lambda_{\pm} \ = \ \left(-\,\sigma \, \sqrt{\frac{\Lambda}{2} \ + \ \epsilon_1}\ \pm \  \sqrt{\frac{\Lambda}{2} \ - \ \epsilon_1}\ \right)\ ,
\label{Ex3New}
\eea
and can correspond to a hyperbolic fixed point, or alternatively to an elliptic one. A hyperbolic fixed point obtains if
\bea
\label{hyperbolic1New}
&h_1)&   \Lambda \ \geq \ 2 \ ,\quad \epsilon_1 \ = \ -1 \quad (saddle\ (\sigma \ = \ \pm\, 1)) \ ,\\
&{\bf h_2})&    \Lambda \ \geq \ 2 \ ,\quad \epsilon_1 \ = \ 1  \quad (node: \ repeller/attractor \ (\sigma \ = \ -1/+1)) \ ,\\
&{\bf h_3)}&     \Lambda \ = \ 2 \ , \quad \epsilon_1 \ = \ 1   \quad (improper \ node: \, repeller/attractor \, (\sigma \, = \, -1/+1))\, ,\\
&{\bf h_4)}&  |\Lambda| \ < \ 2 \ ,\quad \epsilon_1 \ = \ 1 \quad (focus: \ repelling/attracting\ (\sigma \ = \ -1/+1))\ ,
\label{hyperbolic4New}
\eea
while an elliptic one obtains if
\bea
{\bf e)} \quad \Lambda \ = \ -\,2\, , \quad \epsilon_1 \ = \ 1 \quad (center: unstable/asymptotically \, stable\, (\sigma \,=\,-1/+1))\ .
\label{Ex3bNew}
\eea

The potential (\ref{Ex2_1New}) can equivalently be represented in the form (\ref{sasha1}) with
\bea
\label{sasha4New}
\mathcal{P}_{\pm}(\varphi)\ = \ \frac{1}{\Gamma_{\pm}}\,\Big(\epsilon_1\,e^{\varphi} \ + \ \epsilon_2\,e^{-\,\varphi}\, \Big) \ + \ \Gamma_{\pm} \ , \quad \Gamma_{\pm} \ \equiv \ \sqrt{\Lambda \ \pm \ \sqrt{\Lambda^2 \ - \ 4\,\epsilon_1\,\epsilon_2}}\ .
\eea

For the saddle (\ref{hyperbolic1New}) $\mathcal{P}_{\pm}(\varphi)$ becomes
\bea
\label{sasha4Newsaddle}
&& \mathcal{P}_{s\pm}(\varphi)\ = \ -\,\frac{2}{\Gamma_{s\pm}}\,\cosh{\varphi} \ + \ \Gamma_{s\pm} \ , \quad \Gamma_{s\pm} \ \equiv \ \sqrt{\Lambda \ \pm \ \sqrt{\Lambda^2 \ - \ 4}}\ , \nonumber\\
&& \mathcal{P}^{\,\prime \prime}_{s\pm}(\varphi_c)\ = \ -\,\frac{2}{\Gamma_{s\pm}} \ < \ 0\ ,
\eea
and the separatrix equations (\ref{sasha2})--(\ref{sasha2a}) read
\bea
\dot{\varphi} \ = \ \mathcal{P}^{\,\prime}_{s\pm}(\varphi) \ , \quad
{\cal H} \ = \ - \, \mathcal{P}_{s\pm}(\varphi)  \quad (\Lambda \ \geq \ 2) \ .
\label{sasha2a2gammaNew}
\eea
These are essentially eqs.~\eqref{ssaddle1}, so that we can refrain from displaying their solutions, which can be deduced from eqs.~\eqref{sasha2Solut_1} with simple substitutions.
\vskip 24pt
\begin{itemize}
\item[3. ]{\bf The potentials \ $\mathcal{V}(\varphi)\ = \ \epsilon_1\,e^{\gamma_1\,\varphi} \ + \ \epsilon_2\,e^{\gamma_2\,\varphi}$}
\end{itemize}
\noindent
The class of potentials
\bea
 \mathcal{V}(\varphi)\ = \ \epsilon_1\,e^{\gamma_1\,\varphi} \ + \  \epsilon_2\,e^{\gamma_2\,\varphi}\ , \quad \epsilon_1 \ = \ \pm \, 1\ , \quad \epsilon_2 \ = \ \pm \, 1\ , \quad \gamma_1 \ \neq \ \gamma_2\ , \quad \gamma_i \ \neq \ 0 \ ,
\label{Ex2_1}
\eea
where more general values of the parameters $\epsilon_1$ and $\epsilon_2$ can always be reached by rescalings of the time variable and constant shifts the scalar field, possesses the fixed point

\bea
v_c \ = \ 0\ , \quad \varphi_c\ = \ \frac{1}{\gamma_1\, - \, \gamma_2}\ \log \Big( -\, \frac{\epsilon_2\, \gamma_2}{\epsilon_1\, \gamma_1} \Big)\ , \quad  \frac{\epsilon_2\, \gamma_2}{\epsilon_1\, \gamma_1}\ < \ 0\ ,
\label{Ex2_2}
\eea
such that
\bea
&& \mathcal{V}(\varphi_c)  \ = \ \Big( -\, \frac{\epsilon_2\, \gamma_2}{\epsilon_1\, \gamma_1} \Big)^{\frac{\gamma_2}{\gamma_1 \, - \, \gamma_2}}\, \epsilon_2 \,\Big( 1 \, -\, \frac{\gamma_2}{\gamma_1}\Big)\ ,
\label{Ex2_3}\\
&& \mathcal{V}^{\,\prime \prime}(\varphi_c)\ = \ -\, \gamma_1\,\gamma_2\,\mathcal{V}(\varphi_c)\ .
\label{Ex2_3sasha}
\eea
The fixed point (\ref{Ex2_2}) is admissible in the ranges
\bea
\label{Ex2_4}
&1)&  \quad  \ -\,\epsilon_1\ = \ \epsilon_2 \ = \ 1\ , \quad 0 \ < \ \frac{\gamma_2}{\gamma_1} \ < \ 1 \ ,\\
\label{Ex2_5}
&2)&   \quad  ~ \epsilon_1\ = \ \epsilon_2 \ = \ 1\ , \quad \frac{\gamma_2}{\gamma_1} \ < \ 0 \ ,\\
&3)&    \quad ~ ~  \epsilon_1\ = \ -\, \epsilon_2 \ = \ 1\ , \quad \frac{\gamma_2}{\gamma_1} \ > \ 1 \ ,
\label{Ex2_6}
\eea
and the corresponding eigenvalues \eqref{ODE8} read
\bea
\lambda_{\pm} \ = \ \sqrt{\frac{\mathcal{V}(\varphi_c)}{2}}\,\Big(-\,\sigma \ \pm \  \sqrt{1 \ + \ 2\, \gamma_1 \, \gamma_2}\ \Big)\ .
\label{Ex2_7}
\eea
In the ranges of eqs.~(\ref{Ex2_4})--(\ref{Ex2_6}) these eigenvalues correspond to a hyperbolic fixed point of the following types:
\bea
\label{Ex2_8}
&1) \ {\rm and} \ 3)&  \quad    saddle \quad (unstable)\ , \quad \gamma_1 \, \gamma_2\ > \ 0 \ ,\\
\label{Ex2_10}
&\textbf{2a)}&   \quad -\,1\ < \ 2\,\gamma_1 \, \gamma_2\ < \ 0\nonumber\\
&& node \ (repeller \ (\sigma \ = \ -1), \ attractor  \ (\sigma \ = \ +1)) \ ,\\
&\textbf{2b)}&   \quad -\,1\ = \ 2\,\gamma_1 \, \gamma_2\nonumber\\
&& improper \ node \ (repeller \ (\sigma \ = \ -1), \ attractor \ (\sigma \ = \ +1)) \ ,\\
&\textbf{2c)}&   \quad -\,1\ > \ 2\,\gamma_1 \, \gamma_2\ ,\nonumber\\
&& focus \ (repelling \ (\sigma \ = \ -1)\ {\rm or} \ attracting \ (\sigma \ = \ +1) \ spirals)\ . \\
\label{Ex2_11}
\eea
\begin{itemize}
\item[4. ]{\bf The integrable potentials (2) of Table \ref{tab:families}}
\end{itemize}
\noindent
There is a first integrable subset of the potentials just discussed,
\bea
\mathcal{V}(\varphi)\ = \ \epsilon_1\,e^{2\,\gamma\,\varphi} \ + \ \epsilon_2\,e^{({\gamma}\ + \ 1)\,\varphi}\ ,
\label{Ex2_7Integrgg}
\eea
This class is particularly interesting since the case $\gamma=\frac{2}{3}$ corresponds to a supersymmetric integrable model. In \cite{prep} we shall discuss in detail this model, which is obtained coupling to $\mathcal{N}=1$ Supergravity a single Wess-Zumino multiplet with the kinetic term of the $S^3$ model (case $q=3$ in eq.(\ref{calero}) ) and a superpotential
\begin{equation}\label{fornicatore}
    W_h(z) \, = \, \frac{1}{\sqrt{5}} \ \big(24 \, z^5 \, + \,{\rm i} \, 2 \, \omega \, z^3 \big) \ ,
\end{equation}
where $\omega$ is a constant. After a consistent truncation to vanishing axion ($b=0$), and upon the identification (\ref{copperfildo}), the scalar potential generated by (\ref{fornicatore}) becomes (up to an overall positive constant)
\begin{equation}\label{federalissimo}
    \mathcal{V}_{SUSY}(\varphi)\, = \, \frac{6\omega}{5} \ \exp\left(\frac{4\varphi}{3}\right) \, +\, \frac{24}{5}\ \exp\left(\frac{5 \varphi}{3} \right) \ ,
\end{equation}
which can be turned into the form (\ref{Ex2_7Integrgg}) with $\gamma=\frac{1}{3}$ by a shift of $\varphi$.

The family of potentials \eqref{Ex2_7Integrgg} displays a variety of behaviors that will be discussed in detail in Section \ref{sec:Bnot0_sol} and possesses the fixed point
\bea
v_c \ = \ 0\ , \quad \varphi_c\ = \ \frac{1}{\gamma\, - \, 1}\ \log \left[ -\, \frac{\epsilon_2}{2\,\epsilon_1} \Big(1\,+\,\frac{1}{\gamma}\Big)\right]\ , \quad \frac{\epsilon_2}{\epsilon_1} \Big(1\,+\,\frac{1}{\gamma}\Big)\ < \ 0\ ,
\label{Ex2_2mnmnmn}
\eea
with eigenvalues
\bea
\lambda_{\pm} \ = \ \sqrt{\frac{\mathcal{V}(\varphi_c)}{2}}\,\Big(-\,\sigma \ \pm \  ( 2\, \gamma \ + \ 1)\ \Big)\ .
\label{Ex2_7acbd}
\eea
As a result, one can distinguish the following cases:
\bea
\label{Ex2_12}
&1)&  \quad  \ -\,\epsilon_1\ = \ \epsilon_2 \ = \ 1\ ,\quad    |\gamma| \ > \ 1\  \quad (saddle) \ ,\\
\label{Ex2_10bis}
&\textbf{2a)}&  \quad  ~ \epsilon_1\ = \ \epsilon_2 \ = \ 1\ , \quad -\,1\ < \ \gamma\ < \ 0\ , \quad  \gamma \ \neq \ -\,\frac{1}{2} \quad  (node)  \ ,
\label{Ex2_10nodeCOS}\\
&\textbf{2b)}&   \quad  ~ \epsilon_1\ = \ \epsilon_2 \ = \ 1\ ,\quad  \gamma \ = \ -\,\frac{1}{2} \quad
(improper \ node) \ ,\\
&3)&    \quad  \epsilon_1\ = \ -\, \epsilon_2 \ = \ 1\ ,\quad 0\ < \ \gamma \ < \ 1 \quad (saddle) \ .
\label{Ex2_13ddd}
\eea

Let us also note that the subclass of  potentials (\ref{Ex2_7Integrgg}) with
\bea
\epsilon_1\, = \, \sign(1\, - \,\gamma^2)\ ,
\label{Ex2_corr}
\eea
which is consistently correlated with the constraints (\ref{Ex2_12})--(\ref{Ex2_13ddd}) on the type of fixed point, can equivalently be represented in the form (\ref{sasha1}) with
\bea
\label{sasha4Newbis}
\mathcal{P}(\varphi)\ = \ \sqrt{\frac{2}{|1\, - \,\gamma^2|}}\,\,e^{\gamma\,\varphi} \ + \ \epsilon_2\, \sqrt{\frac{1}{2}\,\Big|\frac{1\, + \,\gamma}{1\, - \,\gamma}\Big|}\,\sign(1\, - \,\gamma)\,e^{\varphi}\ .
\eea
$\mathcal{P}(\varphi)$ possesses the non--degenerate fixed point (\ref{Ex2_2mnmnmn}) with the properties
\bea
&& \mathcal{P}^{\,\prime}(\varphi_c)\ = \  0\ , \quad \mathcal{P}^{\,\prime \prime}(\varphi_c)\ = \ \left[ -\, \frac{\epsilon_2}{2\,\epsilon_1} \Big(1\,+\,\frac{1}{\gamma}\Big)\right]^{\frac{\gamma}{\gamma\,-\,1}} \, \frac{\sqrt{2}\,\gamma\,(\gamma\,-\,1)}{\sqrt{|1\,-\,\gamma^2|}}\ \neq \ 0\ , \nonumber\\
&& \mathcal{P}^{\,\prime \prime}(\varphi_c)\ > \ 0 \quad {\rm for} \quad |\gamma| \ > \ 1 \quad {\rm and} \quad \mathcal{P}^{\,\prime \prime}(\varphi_c)\ < \ 0 \quad for \quad 0\ < \ \gamma \ < \ 1\ ,
\label{sasha4W}
\eea
so that the \emph{separatrix} equations (\ref{sasha2}) adapted to the saddles (\ref{Ex2_12}) and (\ref{Ex2_13ddd}) are
\bea
\label{sasha22gamma}
&& \dot{\varphi} \ = \ -\,  \mathcal{P}^{\,\prime}(\varphi) \ , \quad
{\cal H} \ = \  + \, \mathcal{P}(\varphi) \quad  {\rm for} \quad |\gamma| \ > \ 1 \ , \\
&& \dot{\varphi} \ = \ +\,  \mathcal{P}^{\,\prime}(\varphi) \ , \quad
{\cal H} \ = \ - \, \mathcal{P}(\varphi) \quad   {\rm for} \quad 0\ < \ \gamma \ < \ 1\ .
\eea
Let us further remark that the potential well
\be
 {\cal V}(\varphi)\ = \ 2\,\epsilon_1\, \cosh\,\frac{2\,\varphi}{3} \label{Ex3_1intintCOSH} \ ,
\ee
belongs to the series (\ref{Ex2_7Integrgg}), where it emerges for $\gamma\,=\,-\,\frac{1}{3}$ and $\epsilon_2\,=\,\epsilon_1$, and that its fixed point is the node (\ref{Ex2_10nodeCOS}).

\vskip 24pt
\begin{itemize}
\item[5. ]{\bf The integrable potentials (9) of Table \ref{tab:families}}
\end{itemize}
\noindent
The integrable potentials
\bea
 \mathcal{V}(\varphi)\ = \ \epsilon_1\,e^{\,2\,\gamma\,\varphi} \ + \  \epsilon_2\,e^{\,\frac{2}{\gamma}\,\varphi}\ , \quad \gamma \ \neq \ 0 \ ,
\eea
possess the fixed point
\bea
v_c \ = \ 0\ , \quad \varphi_c\ = \ \frac{\gamma}{1\, - \, \gamma^2}\ \log |\gamma|\ ,
\label{Ex2_2sasha}
\eea
\bea
\lambda_{\pm} \ = \ \sqrt{\frac{\mathcal{V}(\varphi_c)}{2}}\,\Big(-\,\sigma \ \pm \ 3\ \Big)\ ,
\label{Ex2_7acbdf}
\eea
and there are two cases:
\bea
\label{Ex2_13}
&1)&  \quad  \ -\,\epsilon_1\ = \ \epsilon_2 \ = \ 1\ \quad {\rm for} \quad    |\gamma| \ > \ 1\  \quad (saddle) \ ,\\
&2)&   \quad  ~ \epsilon_1\ = \ -\, \epsilon_2 \ = \ 1\ \quad {\rm for} \quad |\gamma| \ < \ 1 \ \quad (saddle) \ .
\label{Ex2_14}
\eea
The potentials admitting the fixed point (\ref{Ex2_13})--(\ref{Ex2_14}) can be recast in the form
\bea
 \mathcal{V}(\varphi)\ = \ \sign(1\ - \ \gamma^2) \ e^{2\,\gamma\,\varphi} \ + \  \sign(\gamma^2 \ - \ 1)\ e^{\frac{2}{\gamma}\,\varphi}\ ,
\label{Ex2_1_equiv}
\eea
which admits the representation (\ref{sasha1}) with
\bea
\mathcal{P}_{\pm}(\varphi)\ = \  \sqrt{\frac{2}{|1\ - \ \gamma^2|}} \ \Big(e^{\gamma\,\varphi} \ \pm \  |\gamma|\ e^{\frac{1}{\gamma}\,\varphi}\Big)\ .
\label{sasha4Wbis}
\eea
Since $\mathcal{P}_{+}(\varphi)$ has no fixed point while $\mathcal{P}_{-}(\varphi)$ possesses the non--degenerate fixed point (\ref{Ex2_2sasha}) with the properties
\bea
&& \mathcal{P}^{\,\prime}_{-}(\varphi_c)\ = \  0\ , \quad \mathcal{P}^{\,\prime \prime}_{-}(\varphi_c)\ = \  \sqrt{|1\, -\, \gamma^2|}\, |\gamma|^{\frac{\gamma^2}{1\,-\,\gamma^2}} \, \sign(\gamma^2 \, -\, 1)\ \neq \ 0\ , \nonumber\\
&& \mathcal{P}^{\,\prime \prime}_{-}(\varphi_c)\ > \ 0 \quad {\rm for} \quad |\gamma| \ > \ 1 \quad {\rm and} \quad \mathcal{P}^{\,\prime \prime}_{-}(\varphi_c)\ < \ 0 \quad {\rm for} \quad |\gamma| \ < \ 1\ ,
\label{sasha4Wter}
\eea
the \emph{separatrix} equations (\ref{sasha2}) are in this case
\bea
\label{sasha22gammabis}
&& \dot{\varphi} \ = \ -\,  \mathcal{P}^{\,\prime}_{-}(\varphi) \ , \quad
{\cal H} \ = \  + \, \mathcal{P}_{-}(\varphi) \quad  {\rm for} \quad |\gamma| \ > \ 1 \ , \\
&& \dot{\varphi} \ = \ +\,  \mathcal{P}^{\,\prime}_{-}(\varphi) \ , \quad
{\cal H} \ = \ - \, \mathcal{P}_{-}(\varphi) \quad   {\rm for} \quad |\gamma| \ < \ 1 \ .
\eea
\begin{itemize}
\item[6. ]{\bf The integrable potential (4) of Table \ref{tab:families}}
\end{itemize}
\noindent
The potential
\bea
\mathcal{V}(\varphi)\ = \  C\,\varphi\,e^{2\,\varphi}\ ,
\label{New_pot_1}
\eea
possesses the fixed point
\bea
v_c \ = \ 0\ , \quad \varphi_c\ = \ -\,\frac{1}{2}\ ,
\label{New_pot_1a}
\eea
which is admissible if
\bea
C \ \leq \ 0\ ,
\label{New_pot_3}
\eea
but this choice makes it unbounded from below as $\varphi \to \infty$. The eigenvalues \eqref{ODE8} read
\bea
\lambda_{\pm} \ = \ \sqrt{\frac{\mathcal{V}(\varphi_c)}{2}}\ (-\,\sigma \ \pm \  3) \ ,
\label{New_pot_4}
\eea
and correspond in this case to a \emph{saddle}.

The potential (\ref{New_pot_1}) with the admissible saddle point (\ref{New_pot_1a}) -- (\ref{New_pot_3}) can equivalently be represented in the form (\ref{sasha1}), with
\bea
&& \mathcal{P}(\varphi)\ = \  \sqrt{-\,C}\ \Big(\,\varphi\,-\frac{1}{2}\,\Big)\,e^{\varphi}\ ,
\label{New_pot_5}\\
&& \mathcal{P}^{\,\prime}(\varphi_c)\ = \ 0\ , \quad \mathcal{P}^{\,\prime \prime}(\varphi_c)\ = \
\sqrt{-\,C}\, e^{-1} \ > \ 0 \ ,
\eea
so that its separatrix eqs.~(\ref{sasha2})--(\ref{sasha2a}) read
\bea
\dot{\varphi} \ = \ -\,\mathcal{P}^{\,\prime}(\varphi) \ , \quad
{\cal H} \ = \ \mathcal{P}(\varphi) \ .
\label{New_pot_6}
\eea
\vskip 24pt
\begin{itemize}
\item[7. ]{\bf The potentials \ ${\cal V}(\varphi)\ = \ C_1 \ \Big(\, \cosh\,\gamma\,\varphi \, \Big)^{\beta}\ + \ C_2 \ \Big(  \, \sinh\,\gamma\,\varphi \,\Big)^{\beta}$}
\end{itemize}
\noindent
The potentials
\be
 {\cal V}(\varphi)\ = \ C_1 \ \Big(\, \cosh\,\gamma\,\varphi \, \Big)^{\beta}\ + \ C_2 \ \Big(  \, \sinh\,\gamma\,\varphi \,\Big)^{\beta}\ , \quad \beta \ \neq \ 0 \ ,\quad \gamma \ \neq \ 0 \label{Ex3_1}
\ee
possess the fixed points
\bea
&& v_{1c} \ = \ 0\ , \quad \varphi^{+}_{1c}\ = \ \frac{1}{2\,\gamma}\ \log \left( ~\frac{1 \,+\, \Big(-\,\frac{C_1}{C_2}\,\Big)^{\frac{1}{\beta\, - \, 2}}}{1 \,-\, \Big(-\,\frac{C_1}{C_2}\,\Big)^{\frac{1}{\beta\, - \, 2}}} ~\right)\ , \quad \beta \ \neq \ 2 \ ,
\label{Ex3_2}\ \\
&& {\rm for \ odd \ \beta:} \ \Big|\frac{C_1}{C_2}\Big| \ < \ 1 \quad {\rm if} \quad \beta \ >  \ 2\ \quad {\rm or} \quad \Big|\frac{C_1}{C_2}\Big| \ > \ 1 \quad {\rm if} \quad \beta \ <  \ 2 \ ,\nonumber\\
&& {\rm for \ even \ \beta:} \ additionally \ \varphi^{-}_{1c}\ = \ - \, \varphi^{+}_{1c}\ , \quad \frac{C_1}{C_2}\ < \ 0\label{Ex3_2aaa}
\eea
and
\be
v_{2c} \ = \ 0\ , \quad \varphi_{2c}\ = \ 0 \ , \quad \beta \ > \ 1
\label{Ex3_3}\ .
\ee
The fixed--point values of the potential are
\be
{\cal V}(\varphi^{\pm}_{1c}) \ = \ C_1\,
\Big(~1 \,-\, \Big(-\,\frac{C_1}{C_2}\,\Big)^{\frac{2}{\beta\, - \, 2}}~ \Big)^\frac{2\,-\,\beta}{2}
\label{Ex3_4}\
\ee
and
\be
{\cal V}(\varphi_{2c}) \ = \ C_1 \ ,
\label{Ex3_5}\
\ee
respectively, so that they are admissible if
\bea
 C_1 \ \geq \ 0\ .
\eea

The eigenvalues \eqref{ODE8} are
\bea
\lambda_{1\pm} \ = \ \sqrt{\frac{\mathcal{V}(\varphi^{\pm}_{1c})}{2}}\,\Big(-\,\sigma \ \pm \  \sqrt{1 \ + \ 2\, \gamma^2 \, \beta\,(\beta \, - \,2)}\ \Big)\ ,
\label{Ex3_6}
\eea
so that the corresponding types of hyperbolic fixed points are
\bea
\label{Ex3_8}
&1)&    saddle \quad (unstable), \ {\rm if} \ \beta \ < \ 0 \quad {\rm or}
\quad \beta \ > \ 2\ ,\\
\label{Ex3_10a}
&\textbf{2)}&    node \ (repeller \ (\sigma \ = \ -1)\ {\rm or} \ attractor  \ (\sigma \ = \ +1)): \nonumber\\
&&  |\gamma| \ < \ \frac{1}{\sqrt{2}}\ , \quad 0 \ < \ \beta \ < \ 2\ ,\\
&&  |\gamma| \ \geq \ \frac{1}{\sqrt{2}}\ , \quad
0 \ < \ \beta \ < \ 1 \ +\ \sqrt{1 \ - \ \frac{1}{2\, \gamma^2}}
\quad or \quad
1 \ +\ \sqrt{1 \ - \ \frac{1}{2\, \gamma^2}} \ < \ \beta \ < \ 2
\ , ~~~~~~~~\\
&\textbf{3)}& improper \ node \ (repeller \ (\sigma \ = \ -1)\ {\rm or} \ attractor \ (\sigma \ = \ +1)): \nonumber\\ &&  |\gamma| \ \geq \ \frac{1}{\sqrt{2}}\ , \quad \beta \ = \ 1 \ \pm \ \sqrt{1 \ - \ \frac{1}{2\, \gamma^2}}\ ,\\
&\textbf{4)}&   focus \ (repelling \ (\sigma \ = \ -1)\ {\rm or} \ attracting \ (\sigma \ = \ +1) \ spirals): \nonumber\\
&& |\gamma| \ > \ \frac{1}{\sqrt{2}}\ , \quad 1 \ -\ \sqrt{1 \ - \ \frac{1}{2\, \gamma^2}} \ < \ \beta \ < \  1 \ +\ \sqrt{1 \ - \ \frac{1}{2\, \gamma^2}}
\label{Ex3_11a}
\eea
for the fixed point of eqs.~(\ref{Ex3_2})--(\ref{Ex3_2aaa}). In addition, for the fixed point of eq,~\eqref{Ex3_3}
\bea
&& \lambda_{2\pm} \ = \ \sqrt{\frac{\mathcal{V}(\varphi_{2c})}{2}}\,\Big(-\,\sigma \ \pm \  \sqrt{1 \ - \ 2\, \gamma^2 \, \beta} \,\Big) \ {\rm for} \quad \beta \ > \ 2\ , \\
&&  \lambda_{2\pm} \ = \ \sqrt{\frac{\mathcal{V}(\varphi_{2c})}{2}}\,\Big(-\,\sigma \ \pm \  \sqrt{1 \ - \ 4\, \gamma^2 \, \Big(1 \ + \ \frac{C_2}{C_1}\Big)} \,~\Big) \ {\rm for} \quad \beta \ = \ 2\ ,
\label{Ex3_7}
\eea
\bea
&1)&    saddle \quad (unstable): \quad  \beta \ = \ 2\ ,  \quad \frac{C_2}{C_1} \ < \ -1\ ,\\
\label{Ex3_8b}
&\textbf{2)}&    node \ (repeller \ (\sigma \ = \ -1), \ attractor  \ (\sigma \ = \ +1)): \nonumber\\
&&  |\gamma| \ < \ \frac{1}{\sqrt{2\,\beta}}\quad {\rm for}  \quad  \beta \ > \ 2\ ,\\
&&  -1 \ < \ \frac{C_2}{C_1}\ < \ \frac{1}{4\,\gamma^2}\ - \ 1\ \quad {\rm for} \quad  \beta \ = \ 2\ ,\\
&\textbf{3)}& improper \ node \ (repeller \ (\sigma \ = \ -1), \ attractor \ (\sigma \ = \ +1)): \nonumber\\ &&  \gamma \ = \ \pm \, \frac{1}{\sqrt{2\,\beta}}\quad {\rm for} \quad  \beta \ > \ 2\ ,\\
&&  \frac{C_2}{C_1}\ = \ \frac{1}{4\,\gamma^2}\ - \ 1\quad {\rm for} \quad  \beta \ = \ 2\ ,\\
&\textbf{4)}&   focus \ (repelling \ (\sigma \ = \ -1)\ {\rm or} \ attracting \ (\sigma \ = \ +1) \ spirals): \nonumber\\
&&  |\gamma| \ > \ \frac{1}{\sqrt{2\,\beta}}\quad {\rm for} \quad  \beta \ > \ 2\ , \\
&&  \frac{C_2}{C_1}\ > \ \frac{1}{4\,\gamma^2}\ - \ 1\quad {\rm for} \quad  \beta \ = \ 2 \ .
\label{Ex3_11ab}
\eea
\vskip 24pt
\begin{itemize}
\item[8. ]{\bf The integrable potentials (7) of Table \ref{tab:families}}
\end{itemize}
\noindent
If $\beta \ = \ 2\, \left(\frac{1}{\gamma} \ - \ 1 \right)$ the potentials (\ref{Ex3_1}) become
\be
 {\cal V}(\varphi)\ = \ C_1 \ \Big(\, \cosh\,\gamma\,\varphi \, \Big)^{2\, \left(\frac{1}{\gamma} \ - \ 1 \right)}\ + \ C_2 \ \Big(  \, \sinh\,\gamma\,\varphi \,\Big)^{2\, \left(\frac{1}{\gamma} \ - \ 1 \right)}\label{Ex3_1intint}
\ee
which are integrable and are characterized by the following types of hyperbolic fixed points:
\bea
\lambda_{1\pm} \ = \ \sqrt{\frac{\mathcal{V}(\varphi^{\pm}_{1c})}{2}}\,\Big(-\,\sigma \ \pm \  (4\,\gamma \ -\ 3) \Big)
\label{Ex3_6c}
\eea
\bea
\label{Ex3_8c}
&1)&    saddle \quad (unstable): \ \gamma \ < \ \frac{1}{2} \quad {\rm or} \quad \gamma \ > \ 1\ ,\\
\label{Ex3_10ac}
&\textbf{2)}&    node \ (repeller \ (\sigma \ = \ -1), \ attractor  \ (\sigma \ = \ +1)): \nonumber\\
&& \frac{1}{2} \ < \ \gamma \ < \ \frac{3}{4} \quad {\rm or} \quad
\frac{3}{4} \ < \ \gamma \ < \ 1\ ,\\
&\textbf{3)}& improper \ node \ (repeller \ (\sigma \ = \ -1), \ attractor \ (\sigma \ = \ +1)): \quad  \gamma \ = \ \frac{3}{4}
\label{Ex3_11ac}
\eea
and
\bea
&& \lambda_{2\pm} \ = \ \sqrt{\frac{\mathcal{V}(\varphi_{2c})}{2}}\,\Big(-\,\sigma \ \pm \  ( 2\, \gamma \ - \ 1)\Big) \quad {\rm for} \quad 0 \ < \ \gamma \ < \ \frac{1}{2}	\ , \\
&&  \lambda_{2\pm} \ = \ \sqrt{\frac{\mathcal{V}(\varphi_{2c})}{2}}\,\left(-\,\sigma \ \pm \  \sqrt{-\, \frac{C_2}{C_1}} \,~\right) \quad {\rm for}  \quad \gamma \ = \ \frac{1}{2}\ .
\label{Ex3_7a}
\eea
\bea
\label{Ex3_8cd}
&1)&    saddle \quad (unstable): \quad \gamma \ = \ \frac{1}{2}\quad {\rm for} \quad \frac{C_2}{C_1} \ < \ -1\ ,\\
\label{Ex3_10acd}
&\textbf{2)}&    node \ (repeller \ (\sigma \ = \ -1), \ attractor  \ (\sigma \ = \ +1)): \nonumber\\
&& {\rm for} \ 0 \ < \ \gamma \ < \ \frac{1}{2}	 \quad {\rm or} \quad
 \gamma \ = \ \frac{1}{2}\ , \quad -1 \ < \ \frac{C_2}{C_1} \ < \ 0\ ,\label{Ex3_10acdnode}\\
&\textbf{3)}& improper \ node \ (repeller \ (\sigma \ = \ -1), \ attractor \ (\sigma \ = \ +1)): \nonumber\\ && {\rm for}\ \gamma \ = \ \frac{1}{2}\ , \quad C_2 \ = \ 0\ ,
\nonumber\\
&\textbf{4)}&   focus \ (repelling \ (\sigma \ = \ -1)\ {\rm or} \ attracting \ (\sigma \ = \ +1) \ spirals): \nonumber\\
&& {\rm for}\quad \gamma \ = \ \frac{1}{2}\ , \quad \frac{C_2}{C_1} \ > \ 0\ .
\label{Ex3_11acd}
\eea

The potentials (\ref{Ex3_1intint}) with
\bea
C_1\, > \,0\ , \quad C_2\,<\,0\ ,
\label{Ex2_corrAAA}
\eea
can equivalently be represented in the form (\ref{sasha1}) with
\be
 \mathcal{P}_{\pm}(\varphi)\ = \ \sqrt{2\,C_1} \ \Big(\, \cosh\,\gamma\,\varphi \, \Big)^{\frac{1}{\gamma}}\ \pm \ \sqrt{-2\,C_2} \ \Big(  \, \sinh\,\gamma\,\varphi \,\Big)^{\frac{1}{\gamma}}\ ,
\label{Ex3_1intintW}
\ee
where $\mathcal{P}_{+}(\varphi)$ possesses both fixed points (\ref{Ex3_2})--(\ref{Ex3_2aaa}) and (\ref{Ex3_3}), while $\mathcal{P}_{-}(\varphi)$ possesses only the fixed point (\ref{Ex3_3}).

Finally the potentials
\bea
&& {\cal V}(\varphi)\ = \ C_1\, \cosh\left(\frac{2\,\varphi}{3}\right) \ , \label{Ex3_1intintCOSHbis}\\
&&{\cal V}(\varphi)\ = \ C_1\, \cosh\,\varphi \ ,
\label{Ex3_1intintCOSH1}
\eea
which we have already discussed,
belong to the class (\ref{Ex3_1intint}) and correspond to $\gamma\,=\,\frac{1}{3}\,, \ C_2\,=\,-\,C_1$ and $\gamma\,=\,\frac{1}{2}\,, \ C_2\,=\,C_1$, so that their fixed points are the node (\ref{Ex3_10acdnode}) and the focus (\ref{Ex3_11acd}), respectively.

\vskip 24pt
\begin{itemize}
\item[9. ]{\bf The integrable potentials (8) of Table \ref{tab:families}}
\end{itemize}
\noindent
Let us now turn to consider the class of potentials of eq.~\eqref{Wcomplex2} (letting $C=|C|\,e^{\,i(\theta+\pi/2)}$)
\bea
{\cal V}(\varphi) & = & |C|\,\mathrm{Im} \,\left[\,e^{\,i(\theta+\pi/2)} \ \Big(\,i \ + \ \sinh\,2\,\gamma\,\varphi \, \Big)^{\, \frac{1}{\gamma} \ - \ 1}\,\right] \nonumber\\ & \equiv & |C|\,\Big(\cosh 2\,\gamma\,\varphi \Big)^{\frac{1}{\gamma} \ - \ 1}\, \cos \left[\Big(\frac{1}{\gamma} \ - \ 1\Big)\,\arccos\Big(\tanh\,2\,\gamma\,\varphi\Big)\ + \ \theta\,\right]
\label{Exemple4_1}
\eea
where $0\ < \ \theta \ \leq \ 2\,\pi$ is a real parameter. This expression has no definite symmetry in general with respect to the inversion $\varphi \, \rightarrow \ -\,\varphi$, and it is bounded both from below and above, $|{\cal V}(\varphi)| \ \leq \ |C|$, for $\gamma \ < \ 0$ and for $\gamma\,>\,1$. On the other hand, if $0\,<\,\gamma \, < \, 1$, the two conditions
\bea
\cos \left[\pi \,\Big(1\ - \ \frac{1}{\gamma} \Big)\ + \ \theta\,\right]\ > \ 0\ , \quad \cos \theta \ > \ 0\ ,
\label{Exemple4_1cond1}
\eea
guarantee that the potential is bounded from below, ${\cal V}(\varphi) \ \geq \ -\,|C|$ asymptotically as $\varphi \ \rightarrow \ \pm\,\infty$. For the case $\gamma \,=\, \frac{1}{n}\ (n\,\in\,{\mathbb{N}})$, the potentials of this class become polynomials in the hyperbolic functions $\sinh\,(2\,\gamma\,\varphi)$ and $\cosh\,(2\,\gamma\,\varphi)$. Moreover, for $\gamma \,=\, \frac{1}{2n}\ (n\,\in\,\mathbb{N})$ and $\theta=\frac{\pi}{2}, \frac{3\pi}{2}\ (\theta=0,\pi)$ they are even (odd) functions of $\varphi$, and for $\gamma \,=\, \frac{1}{2n+1}\ (n\,\in\,\mathbb{N})$ and $\theta= 0,\pi \left(\theta=\frac{\pi}{2},\frac{3\pi}{2}\right)$ they are even (odd) functions of $\varphi$.

The potential (\ref{Exemple4_1}) possesses the fixed points
\bea
v_{kc} \, = \, 0\ , \quad \varphi_{kc}\, = \, \frac{1}{4\,\gamma}\,\log\left(\,\frac{1\, + \, \cos\frac{\frac{\pi}{2}\,+\, \pi\,k\, - \, \theta}{\frac{1}{\gamma} \, - \, 2}}{1\ - \ \cos\frac{\frac{\pi}{2}\,+\, \pi\,k\, - \, \theta}{\frac{1}{\gamma} \, - \, 2}}\,\right)\ ,
\label{Example_fixed_points}
\eea
where the possible values of $k\, \in \, Z$ are constrained by the range of the principal value of the $\arccos$ function, ($0\, \leq \,\arccos \,\leq \,\pi$):
\bea
&&-\,\frac{1}{2} \,+\,\frac{\theta}{\pi} \, \leq \,k\,\leq\,
\frac{1}{\gamma}\,-\, \frac{5}{2} \, + \, \frac{\theta}{\pi} \quad {\rm if} \quad \frac{1}{\gamma}\,>\,2\ ,\\
&& \frac{1}{\gamma}\,-\, \frac{5}{2} \, + \, \frac{\theta}{\pi}\, \leq \,k\,\leq\,-\,\frac{1}{2} \,+\,\frac{\theta}{\pi} \quad {\rm if} \quad \frac{1}{\gamma}\,<\,2\ ,
\label{Example_fixed_points_constraint}
\eea
The fixed--point values of the potential and of its second derivative
\bea
\label{Exemple4_2}
&&{\cal V}(\varphi_{kc}) \ = \  C\,(-1)^{k+1}\,\left(\sin\frac{\frac{\pi}{2}\,+\, \pi\,k\, - \, \theta}{\frac{1}{\gamma} \ - \ 2}\,\right)^{\, 2 \ - \ \frac{1}{\gamma}}\ ,\\
&& {\cal V}^{\,\prime \prime}(\varphi_{kc}) \ = \ -\,4\,(1 \, - \, \gamma)\,(1 \, - \, 2\,\gamma)\,{\cal V}(\varphi_{kc})\ ,
\label{Exemple4_2a}
\eea
define the admissible fixed points
\bea
C\,(-1)^{k+1}\ > \ 0
\label{Exemple4_3}
\eea
and determine their types
\bea
\lambda_{k\pm} \ = \ \sqrt{\frac{\mathcal{V}(\varphi_{kc})}{2}}\,\Big(-\,\sigma \ \pm \  (4\,\gamma \ -\ 3) \Big)\ ,
\label{Exemple4_4}
\eea
\bea
\label{Exemple4_5}
&1)&    saddle \quad (unstable): \ \gamma \ < \ \frac{1}{2} \quad {\rm or} \quad \gamma \ > \ 1\ ,\\
\label{Exemple4_6}
&\textbf{2)}&    node \ (repeller \ (\sigma \ = \ -1), \ attractor  \ (\sigma \ = \ +1)): \nonumber\\
&& \frac{1}{2} \ < \ \gamma \ < \ \frac{3}{4} \quad {\rm or} \quad
\frac{3}{4} \ < \ \gamma \ < \ 1\ ,
\label{Exemple4_7nodeex}\\
&\textbf{3)}& improper \ node \ (repeller \ (\sigma \ = \ -1), \ attractor \ (\sigma \ = \ +1)): \quad  \gamma \ = \ \frac{3}{4}\ .
\label{Exemple4_7}
\eea

It is instructive to take a closer look at some representatives of the different types of admissible fixed points.

1. The symmetric potential with parameters $C\, > \, 0\,,\, \theta \,=\,0\,,\, \gamma\,=\,\frac{1}{5}$ that is bounded from below possesses one admissible fixed point (\ref{Example_fixed_points}) with $k\,=\,1$, the saddle of eq.~(\ref{Exemple4_5}):
\bea
v_{1c} \, = \, 0\ , \quad \varphi_{1c}\, = \,0 \ .
\label{Example_fixed_points_res}
\eea

2. The potential with parameters $C\, > \, 0\,,\,\theta \,= \,0\,,\, \gamma\,=\,\frac{5}{6}$ that is bounded from below possesses one admissible fixed point (\ref{Example_fixed_points}) with $k\,=\,-1$, the node of eq.~(\ref{Exemple4_7nodeex}):
\bea
v_{-1c} \, = \, 0\ , \quad \varphi_{-1c}\, = \, \frac{3}{15}\,\log \left( \,\frac{1\, + \, \cos \frac{5\,\pi}{8}}{1\ - \ \cos\frac{5\,\pi}{8}}\,\right)\ .
\eea

3. The potential with parameters $C\, > \, 0\,,\,\theta \,= \,0\,,\, \gamma\,=\,\frac{3}{4}$ that is bounded from below possesses one admissible fixed point (\ref{Example_fixed_points}) with $k\,=\,-1$, the improper node of eq.~(\ref{Exemple4_7}):
\bea
v_{-1c} \, = \, 0\ , \quad \varphi_{-1c}\, = \, \frac{1}{3}\,\log\left(\,\frac{\sqrt{2}\, - \, 1}{\sqrt{2}\, + \, 1}\,\right)\ .
\eea

The potential (\ref{Exemple4_1}) with
\bea
C\, > \,0
\label{Ex2_corrBBB}
\eea
can equivalently be represented in the form (\ref{sasha1}), with
\bea
\mathcal{P}(\varphi) \ = \ \sqrt{C}\,\Big(\cosh 2\,\gamma\,\varphi \Big)^{\frac{1}{2\,\gamma}}\, \cos \left[\frac{1}{2\,\gamma}\,\arccos\Big(\tanh\,2\,\gamma\,\varphi\Big)\, + \, \frac{\theta}{2} \, - \, \frac{\pi}{4}\,\right] \ ,
\label{Exemple4_1WWW}
\eea
where $\mathcal{P}(\varphi)$ possesses the fixed points (\ref{Example_fixed_points}) for odd values of $k$.

\vskip 24pt
\begin{itemize}
\item[10. ]{\bf The integrable potentials (6) of Table \ref{tab:families}}
\end{itemize}
\noindent
Let us now consider the potential
\bea
\mathcal{V}(\varphi)\ = \ 2 \,C\, \arctan\left(e^{\,-\, 2\,\varphi}\right)\ + \ D \ ,
\label{Exemple10_1}
\eea
which possesses the admissible fixed points at \emph{infinity}
\bea
&& v^{+}_{c} \, = \, 0\ , \quad \varphi^{+}_{c}\, = \, +\,\infty \quad {\rm if} \quad D\ \geq \ 0\ ,\label{Example_fixed_points10a}\\
&& v^{-}_{c} \, = \, 0\ , \quad \varphi^{-}_{c}\, = \, -\,\infty \quad {\rm if} \quad C\,\pi\ + \  D\ \geq \ 0 \ ,
\label{Example_fixed_points10}
\eea
with
\bea
&& {\cal V}(\varphi^{+}_{c}) \ = \ D\ , \quad \quad \quad \quad \ \,{\cal V}^{~\prime \prime}(\varphi^{+}_{c}) \ = \ 0 \ ,\\
&& {\cal V}(\varphi^{-}_{c}) \ = \ C\,\pi\ + \ D\ , \quad {\cal V}^{~\prime \prime}(\varphi^{-}_{c}) \ = \ 0 \ .
\label{Ex10_4}\
\eea
The corresponding eigenvalues \eqref{ODE8},
\bea
\lambda^{\pm}_{\pm} \ = \ \sqrt{\frac{\mathcal{V}(\varphi^{\pm}_{c})}{2}}\,(-\,\sigma \ \pm \  1)\ ,
\label{Ex10_6}
\eea
include zeroes, so that the fixed points (\ref{Example_fixed_points10a}) and (\ref{Example_fixed_points10}) are \emph{not} hyperbolic. As a result, the linearization of this dynamical system around these fixed points must be combined with an analysis of their structural stability properties in order to determine the phase portrait. To this end, one can consider different domains for the admissible parameter space $\{C\,,D\}$ and define the following weak Lyapunov functions $f^{\pm}(\varphi,v)$ with the required properties in these domains:
\bea
&& \quad \quad \quad \quad \quad \quad \quad \quad {\rm 1) \ for\ }\ C\ \geq \ 0\ , \quad  D\ \geq \ 0 \ ,
\label{LyapEx10_a} \\
&& f^{+}(\varphi,v)\ \equiv \  \sqrt{v^{\,2} \ + \ 2\, \mathcal{V}(\varphi)}\ - \ \sqrt{ 2\, \mathcal{V}(\varphi^{+}_{c})}\ > \ 0 \ {\rm for} \ -\infty \ < \{v,\varphi\} \ < \ +\infty \, ,\label{LyapEx10bbb} \\
&&f^{+}(\varphi^{+}_{c},v^{+}_{c})\ = \ 0\ , \quad \dot{f}^{+}(\varphi,\,v)\ = \ -\,\sigma \, v^2 \ , \nonumber
\eea
and
\bea
&& \quad \quad \quad \quad \quad \quad \quad \quad \quad {\rm 2) \ for\ }\  -\,\frac{D}{\pi}\ \leq C\ < \  0 \ , \quad  D\ \geq \ 0 \ ,
\label{LyapEx10_c} \\
&& f^{-}(\varphi,v)\ \equiv \ \sqrt{v^{\,2} \ + \ 2\, \mathcal{V}(\varphi)}
\ - \ \sqrt{ 2\, \mathcal{V}(\varphi^{-}_{c})} \ > \ 0 \quad {\rm for} \quad -\infty \ < \{v,\varphi\} \ < \ +\infty \ ,~~~~~\label{LyapEx10d} \\
&&
f^{-}(\varphi^{-}_{c},v^{-}_{c})\ = \ 0\ , \quad \dot{f}^{-}(\varphi,\,v)\ = \ -\,\sigma \, v^2\ .\nonumber
\eea
These functions are positive definite in the whole phase space and vanish only at the fixed points, while their time derivative is negative or positive semi--definite depending on the sign of $\sigma \,=\,\pm 1$ and does not vanish identically on any trajectory other than the fixed points themselves. From these Lyapunov functions one can conclude that the fixed points (\ref{Example_fixed_points10a}) and (\ref{Example_fixed_points10}) are \emph{unstable} for $\sigma \,=\,-1$ and asymptotically \emph{stable} for $\sigma \,=\,+1$ in the parameter domain (\ref{LyapEx10_a}) and (\ref{LyapEx10_c}), respectively. Eqs.~(\ref{LyapEx10bbb}) and (\ref{LyapEx10d}) define explicitly the basin of attraction (stability domain), which is the whole phase space in this example, so that all trajectories approach asymptotically as $t\, {\rightarrow} +\infty$ the corresponding fixed points.

The asymptotic behavior as $t\, {\rightarrow} +\infty$ of the Hubble function and of the scale factor that apply if the fixed points (\ref{Example_fixed_points10a}-- \ref{Example_fixed_points10}) are asymptotically stable have as usual the form
\bea
{\cal H} \ = \ \sqrt{2\,\mathcal{V}(\varphi^{\pm}_{c})}\ , \quad {\cal A} \ = \  \sqrt{2\,\mathcal{V}(\varphi^{\pm}_{c})}\,(t\,-t_0) \quad {\rm if} \quad \mathcal{V}(\varphi^{\pm}_{c}) \ > \ 0\ ,
\label{HubAsymEx10}
\eea
as pertains to an expanding de Sitter patch, while the exponential behavior leaves way to a power--like behavior if $\mathcal{V}(\varphi^{\pm}_{c}) \, = \, 0$. Let us also recall that in four dimensions $H=\frac{{\cal H}}{3}$ and $a = e^{\,{\cal A}/3}$. Similar considerations apply to the last two examples, to which we now turn.
\vskip 24pt
\begin{itemize}
\item[11. ]{\bf The integrable potentials (5) of Table \ref{tab:families}}
\end{itemize}
\noindent
Let us now consider the potential
\bea
\mathcal{V}(\varphi)\ = \ C\, \log |\coth \varphi| \ + \ D \ ,
\label{Exemple11_1}
\eea
which possesses the admissible fixed points at \emph{infinity}
\bea
v^{\pm}_{c} \, = \, 0\ , \quad \varphi^{\pm}_{c}\, = \, \pm\,\infty \quad at \quad D\ \geq \ 0\ ,
\label{Example_fixed_points11}
\eea
with
\bea
{\cal V}(\varphi^{\pm}_{c}) \ = \ D\ , \quad {\cal V}^{~\prime \prime}(\varphi^{\pm}_{c}) \ = \ 0 \ .
\eea
The corresponding eigenvalues \eqref{ODE8}
\bea
\lambda^{\pm}_{\pm} \ = \ \sqrt{\frac{\mathcal{V}(\varphi^{\pm}_{c})}{2}}\ (-\,\sigma \ \pm \  1)
\label{Ex11_6}
\eea
include zero, so that these fixed points (\ref{Example_fixed_points11}) are \emph{not} hyperbolic, and in order to understand qualitatively the phase portrait in their vicinity one needs to analyze their structural stability properties. To this end, one can define the following weak Lyapunov functions $f^{\pm}(\varphi,v)$ with the required properties:
\bea
&& \quad \quad \quad \quad \quad \quad \quad \quad \quad \quad \quad \quad C\ \geq \ 0\ , \quad  D\ \geq \ 0 \ ,
\label{LyapEx11_a} \\
&& f^{\pm}(\varphi,v)\ \equiv \  \sqrt{v^{\,2} \ + \ 2\, \mathcal{V}(\varphi)}\ - \ \sqrt{ 2\, \mathcal{V}(\varphi^{\pm}_{c})}\ > \ 0  \ {\rm for} \ -\infty \ < \{v,\varphi\} \ < \ +\infty \ ,\label{LyapEx11b}\\
&& f^{\pm}(\varphi^{\pm}_{c},v^{\pm}_{c})\ = \ 0\ , \quad \dot{f}^{\pm}(\varphi,\,v)\ = \ -\,\sigma \, v^2 \nonumber
\eea
These functions are positive definite in the whole phase space and vanish only at the fixed points, while their time derivative is negative or positive semi--definite depending on the sign of $\sigma \,=\,\pm 1$ and does not vanish identically on any trajectory other than the fixed points themselves. From these Lyapunov functions one can conclude that the fixed points (\ref{Example_fixed_points11}) are \emph{unstable} for $\sigma \,=\,-1$ and asymptotically \emph{stable} for $\sigma \,=\,+1$ in the parameter domain (\ref{LyapEx11_a}). Eqs.~(\ref{LyapEx11b}) define explicitly the basin of attraction (stability domain), which is the whole phase space in this example, so that all trajectories approach asymptotically as $t\, {\rightarrow} +\infty$ the corresponding fixed points.
\begin{itemize}
\item[12. ] {\bf The potential \ $\mathcal{V}(\varphi) \ = \ {\cal V}_0  \,e^{2\,\gamma\,\varphi} \ + \ \Lambda $}
\end{itemize}
\noindent
Our last analysis concerns the potential
\be
\mathcal{V}(\varphi) \ = \ {\cal V}_0  \,e^{2\,\gamma\,\varphi} \ + \ \Lambda \ , \quad \gamma \ > \ 0 \ ,
\label{Ex12_1New}
\ee
which possesses an admissible fixed point \emph{at infinity},
\be
v_c \ = \ 0\ , \quad \varphi_c\ = \ -\,\infty \ ,\label{Ex12_3aaaaNew}
\ee
\be
\mathcal{V}(\varphi_c)  \ = \    \ \Lambda \ , \quad \mathcal{V}^{\,\prime \prime}(\varphi_c)\ = \ 0\ ,
\label{Ex12_3New}
\ee
provided
\be
\Lambda \ \geq 0\ .
\label{Ex12_4New}
\ee
The corresponding eigenvalues \eqref{ODE8}
\bea
\lambda_{\pm} \ = \ \sqrt{\frac{\mathcal{V}(\varphi_{c})}{2}}\,(-\,\sigma \ \pm \  1)
\label{Ex12_6}
\eea
include zero, so that one is confronted with a degenerate non--hyperbolic fixed point.

As in the preceding subsection, one can define a Lyapunov function,
\bea
&& {\rm if} \ {\cal V}_0\ \geq \ 0\ , \quad  \Lambda \ \geq \ 0 \ :
\label{LyapEx12_a} \\
&& f(\varphi,v)\ \equiv \  \sqrt{v^{\,2} \ + \ 2\, \mathcal{V}(\varphi)}\ - \ \sqrt{ 2\, \mathcal{V}(\varphi_{c})}\ > \ 0 \nonumber \\
&& {\rm for} \ -\infty \ < \{v,\varphi\} \ < \ +\infty \ ,\label{LyapEx12b}\\
&& f(\varphi_{c},v_{c})\ = \ 0\ , \quad \dot{f}(\varphi,\,v)\ = \ -\,\sigma \, v^2 \nonumber
\eea
and as a result one can conclude that the fixed point (\ref{Ex12_3aaaaNew}) is \emph{unstable} for $\sigma \,=\,-1$ and asymptotically \emph{stable} for $\sigma \,=\,+1$ in the parameter domain (\ref{LyapEx12_a}). Eqs.~(\ref{LyapEx12b}) characterize the basin of attraction, which is the whole phase space in this example, so that all trajectories approach asymptotically as $t\, {\rightarrow} +\infty$ the fixed point (\ref{Ex12_3aaaaNew}).

In order to clarify how the system approaches the fully degenerate non--hyperbolic fixed point (\ref{Ex12_3aaaaNew}) with $\lambda_{\pm}\,=\,0$ in eq. (\ref{Ex12_6}) with $\Lambda \, =\, 0$ and ${\cal V}_0\,>\,0$ and to specify the corresponding directions in the phase portrait, one can represent Subsystem I of eq.~(\ref{ODE3}) in polar coordinates $(r\,,\theta)$, introducing the new evolution parameter $\tau$ defined as
\bea
&& \frac{d r}{d \tau} \ = \ -\ \sigma \, r \, \sin^2 \theta \ ,\nonumber \\
&& \frac{d \theta}{d \tau} \ = \ - \ \cos \theta \,( \sigma\, \sin \theta \ + \ \gamma)\ ,
\label{ODE3Ex12}
\eea
where
\be
r\,d t \ \equiv \  {d \tau}\ , \quad
\sqrt{2\, {\cal V}_0} \, e^{\,\gamma\,\varphi} \ \equiv  \ r\, \cos \theta\ , \quad
v \  \equiv  \  r\, \sin \theta\ .
\label{ODE3Ex12Coordinates}
\ee
For $0 \, < \, \gamma \, < \, 1$ this transformation (\ref{ODE3Ex12Coordinates}) blows up the degenerate fixed point (\ref{Ex12_3aaaaNew}) into the circle $(0\,, \ 0 \leq \ \theta \ < 2\,\pi)$, where the resulting dynamical system (\ref{ODE3Ex12}) possesses the four non--degenerate hyperbolic fixed points
\bea
&& \left(0\,,\frac{\pi}{2}\right)\ , \quad \left(0\,,\frac{3\,\pi}{2}\right)\ ,\label{gammafixedpoints}\\
&& (0\,,\mp\arcsin\gamma)\ , \quad (0\,,\pi \pm \arcsin\gamma)
\label{ODE3Ex12FixedPolar}
\eea
for $\sigma=\pm 1$, with the eigenvalues
\bea
&& \lambda_1 \ = \ -\,\sigma\,\sin^2\theta\ , \quad \lambda_2 \ = \ -\,\sigma\,\cos (2\theta)\ + \ \gamma\,\sin\theta \ .
\label{ODE3Ex12FixedPolarEigen}
\eea
Only the two fixed points (\ref{gammafixedpoints}) survive for $\gamma \, \geq \, 1$, so that the system undergoes a bifurcation at $\gamma \,=\,1$ where the climbing phenomenon sets in. The angles in eqs.~(\ref{gammafixedpoints}) and (\ref{ODE3Ex12FixedPolar}) define the discrete set of phase portrait directions along which the trajectories can approach asymptotically the original degenerate fixed point (\ref{Ex12_3aaaaNew}).
%
\subsection{\sc Exact solutions of the models of Table \ref{tab:families}}\label{sec:mode_tab1}

We now turn to a detailed discussion of the solutions of the most significant models listed in Table \ref{tab:families}.

\subsubsection{\sc Solutions of the elementary systems with ${\cal B}=0$}\label{sec:B0_sol}

Let us open our discussion of the exact solutions with a relatively simple special case in which only $C_{12}$ is not zero in the class of potentials \eqref{int7}. Only a cosmological constant is then present, and if we further assume initially that $\Lambda >0$ and let
\be
C_{12}\ = \ \frac{\Lambda}{2} \ , \qquad \lambda \ = \ \sqrt{\frac{\Lambda}{2}}  \label{B0_sol_01}
\ee
the solution takes the form
\bea
&& \varphi \ = \ \log\left( \frac{c_1 \, \cosh\left(\lambda \, t_c \right) \ + \ d_1 \, \sinh\left(\lambda \, t_c \right)}{c_2 \, \cosh\left(\lambda \, t_c \right) \ + \ d_2 \, \sinh\left(\lambda \, t_c \right)} \right) \ , \\
&& e^{\,\cal A} \ = \ \left( c_1 \, \cosh\left(\lambda \, t_c \right) \ + \ d_1 \, \sinh\left(\lambda \, t_c \right)\right)\, \left(c_2 \, \cosh\left(\lambda \, t_c \right) \ + \ d_2 \, \sinh\left(\lambda \, t_c \right)\right) \ , \label{B0_sol_1}
\eea
while the Hamiltonian constraint reduces to the condition
\be
c_1 \, c_2 \ = \ d_1 \, d_2 \
\ee
on the four integration constants.

\begin{figure}[h]
\begin{center}$
\begin{array}{cc}
\epsfig{file=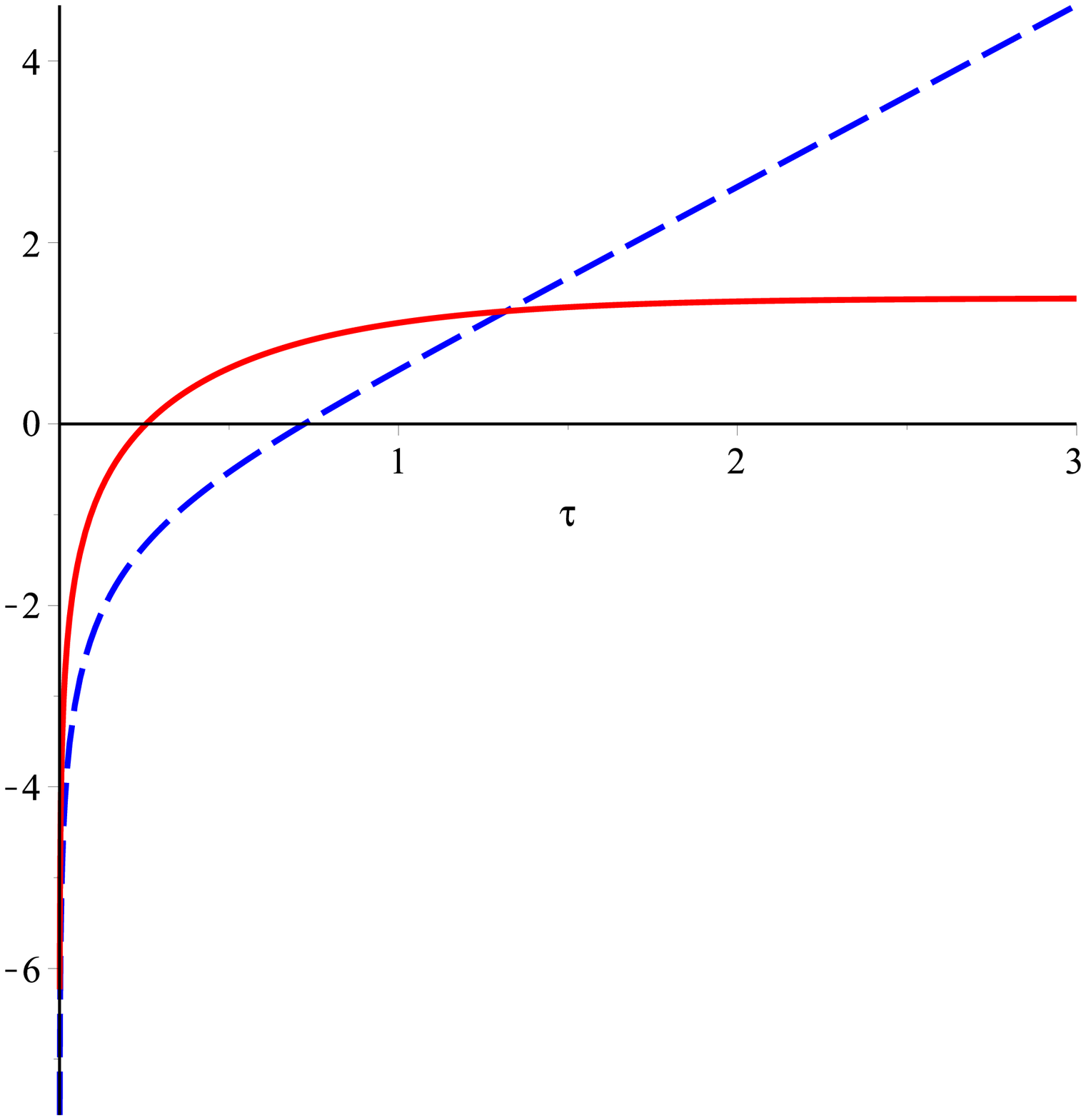, height=1.5in, width=1.5in} & \qquad \qquad
\epsfig{file=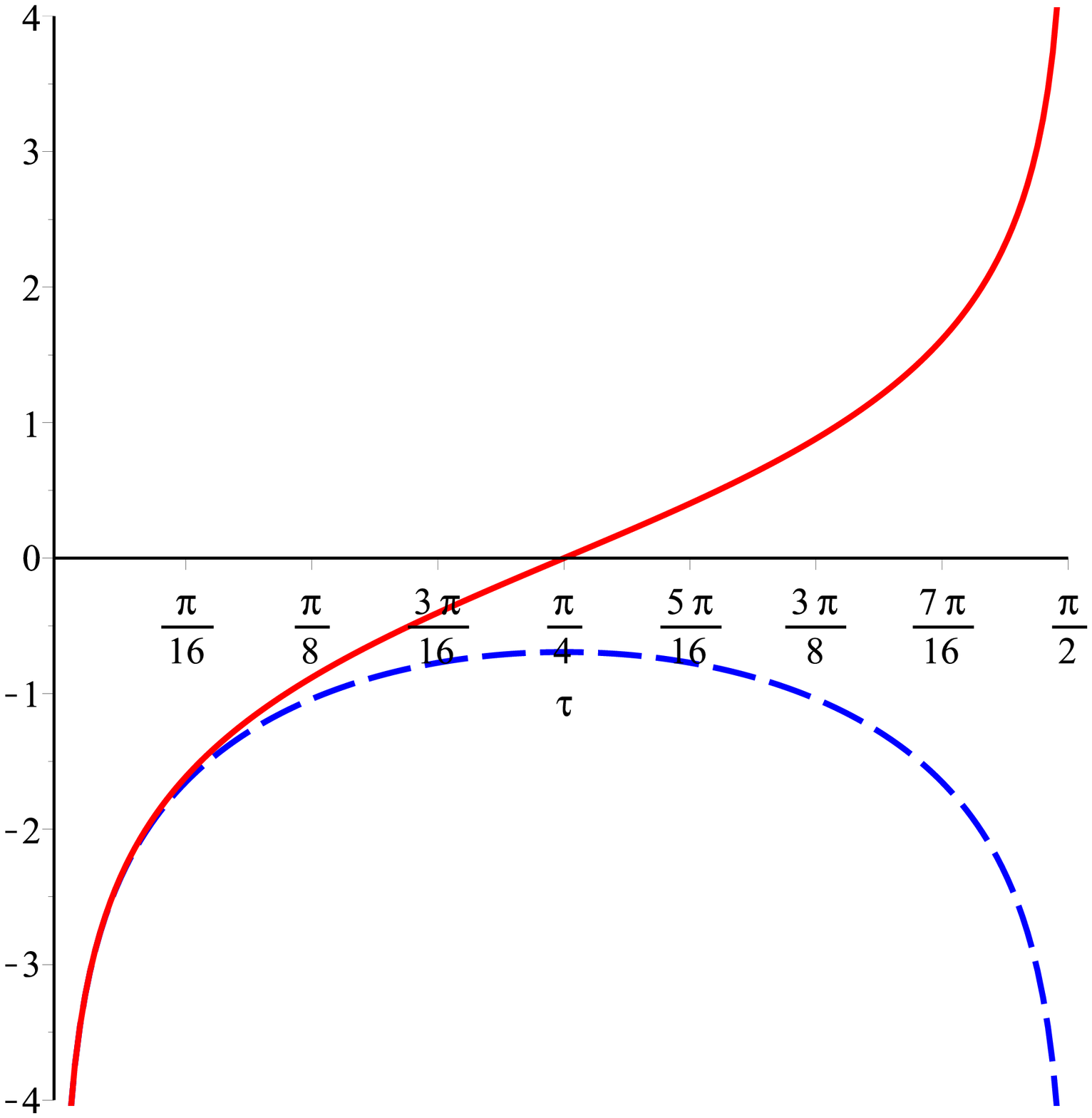, height=1.5in, width=1.5in}
\end{array}$
\end{center}
\caption{$\varphi(\tau)$ and ${\cal A}(\tau)$ (dashed), with $\tau=\sqrt{|\Lambda|/2} \ t_c$, according to eqs.~\eqref{B0_sol_101} and eqs.~\eqref{B0_sol_11}. With $\Lambda>0$ the scalar is brought to rest by cosmological friction as the Universe begins an exponential expansion, while with $\Lambda<0$ the scalar runs off to infinity as the Universe undergoes an eventual Big Crunch.}
\label{onlylambda}
\end{figure}

Together with the familiar de Sitter patch, which is recovered if the four constants are all equal, this setting includes two other interesting solutions that entail a non--trivial cosmological evolution of the scalar field. They differ only in the direction of its motion, and hence it suffices to illustrate one of them, which obtains if $c_1$ and $d_2$ vanish and reads
\bea
&& \varphi \ = \ \log\left[ \frac{d_1}{c_2} \, \tanh\left(\lambda \, t_c \right)\right] \ , \\
&& e^{\,\cal A} \ = \ \Big[ c_2 \, d_1 \, \sinh\left(\lambda \, t_c \right) \, \cosh\left(\lambda \, t_c \right) \Big] \ . \label{B0_sol_101}
\eea
If $c_2$ and $d_1$ have identical signs one can work for $t_c \geq 0 $, and then the scalar emerges from the initial singularity at $t=0$ from $-\infty$ (fig.~\ref{onlylambda}), moves toward larger values and when it is brought eventually to rest by cosmological friction the Universe enters an epoch of exponential expansion. On the other hand, if $c_2$ and $d_1$ have opposite signs one can capture the time reversal of this evolution. This option presents itself in all of our examples, but we shall leave it aside for brevity in the following.

The solutions of eq.~\eqref{B0_sol_1} admit an interesting continuation to imaginary values of $t$. What we have stressed in Section \ref{sec:effelag} is manifest in this case: the end result can be regarded as a solution for $\Lambda>0$ with Euclidean signature, or alternatively as a solution for $\Lambda < 0$ with Minkowski signature. Abiding to the latter interpretation, one can present it in the form
\bea
&& \varphi \ = \ \log\left[ \frac{d_1}{c_2} \, \tan\left(|\lambda| \, t_c \right)\right] \ , \\
&& e^{\,\cal A} \ = \ c_2 \, d_1 \, \sin\left(|\lambda| \, t_c \right) \, \cos\left(|\lambda| \, t_c \right) \ , \label{B0_sol_11}
\eea
where $c_2$ and $d_1$ were also analytically continued in order that the arguments remain real and their product stays positive. This solution illustrates a key feature of this class of models, since it describes a Universe that emerges from a Big Bang at $t_c=0$ and ends up in a Big crunch at
\be
t_c^{\,\star} \ = \ \frac{\pi}{2} \ \sqrt{\frac{2}{|\Lambda|}} \ ,
\ee
when the scalar is rapidly driven to infinity. Notice in fact that during the collapsing phase cosmological damping leaves way to cosmological amplification, so that in this case the scalar field sweeps out the whole real axis during its evolution. This type of behavior reflects the attractive nature of a negative cosmological constant and the absence of a flat $AdS$ slicing, and will surface time and over again in more complicated examples.

Our next class of integrable potentials illustrates the crucial role of the allowed range for the independent variable $t$. It concerns a case that was already discussed in Section \ref{sec:climbing}, an exponential potential with $\gamma=\frac{1}{2}$, that we write
\be
{\cal V} \ = \ C \ e^{\, \varphi} \ , \label{B0_sol_1111}
\ee
letting for brevity $C_{11}=C$. However, now the solution will be given directly in terms of cosmic time, since after all, as we already noticed there, this value of $\gamma$ brings about an elementary relation between parametric time $t$ and cosmic time $t_c$. Assuming to begin with that $C$ be \emph{positive}, as pertains to a stable potential, the solution reads
\bea
&& \varphi \ = \ \log\left[ \frac{d_1 \, t_c \ + \ c_1}{\frac{C \, d_1}{6} \ t_c^{\,3} \ + \ \frac{C \, c_1}{2} \ t_c^{\,2} \ + \ d_2 \, t_c \ + \ c_2} \right] \ , \\
&& e^{\,\cal A} \ = \ \left[ \left(d_1 \, t_c \ + \ c_1\right)\, \left(\frac{C \, d_1}{6} \ t_c^{\,3} \ + \ \frac{C \, c_1}{2} \ t_c^{\,2} \ + \ d_2 \, t_c \ + \ c_2 \right) \right] \ , \label{B0_sol_12}
\eea
where the constants are not independent since the Hamiltonian constraint \eqref{int605} yields the quadratic relation
\be
C\, c_1^{\,2} \ = \ 2\, d_1\, d_2 \ . \label{B0_sol_2}
\ee

Notice that $d_1$ cannot vanish, otherwise the Hamiltonian constraint would force $c_1$ to vanish as well. As a result, shifting $t$ one can always remove $c_1$ and then $d_2$ is bound to vanish on account of eq.~\eqref{B0_sol_2}, so that all in all the solution can be reduced in general to the form
\bea
&& \varphi \ = \ \log\left[ \frac{6}{C} \ \frac{t_c}{\left(t_c^3 + \Delta \right)} \right] \ , \nonumber \\
&& e^{\,\cal A} \ = \ \left[ \frac{C\, d_1^{\,2}}{6} \ t_c \, \left(t_c^3 \ + \ \Delta \right) \right] \ , \label{B0_sol_4}
\eea
where $\Delta$ is a combination of the remaining constants. Actually, a further rescaling of the time variable $t_c$ to
\be
\tau \ = \ \frac{t_c}{|\Delta|}
\ee
gives rise if $\Delta>0$ to
\bea
&& \varphi \ = \ \log\left[ \frac{6}{C} \ \frac{\tau}{\left(\tau^3 \ + \ 1 \right)} \right] \ , \nonumber \\
&& e^{\,\cal A} \ = \ \left[ \frac{C\, d_1^{\,2}}{6} \ \tau \, \left(\tau^3 \ + \ 1 \right) \right] \ , \label{B0_sol_41}
\eea
which describes for $\tau>0$ the evolution in the presence of a \emph{climbing} scalar, and if $\Delta<0$ to
\bea
&& \varphi \ = \ \log\left[ \frac{6}{C} \ \frac{\tau}{\left(\tau^3 \ - \ 1 \right)} \right] \ , \nonumber \\
&& e^{\,\cal A} \ = \ \left[ \frac{C\, d_1^{\,2}}{6} \ \tau \, \left(\tau^3 \ - \ 1 \right) \right] \ , \label{B0_sol_42}
\eea
which describes for $\tau>1$ the evolution in the presence of a \emph{descending} scalar. Notice also that these two expressions, with their excluded regions, are mapped into each other by a time--reversal operation.

How about the region that we excluded in the standard case $C>0$ ? It also plays a role, albeit in the case $C<0$ that we excluded since it involves an inverted potential. In this pathological case the solutions behave again along the lines of what happened for $\Lambda<0$ in the preceding example: the Universe emerges from a Big Bang to end up in a Big Crunch while the scalar sweeps out the whole real axis.

We can now move one step forward, combining the preceding potentials into
\be
{\cal V}\ = \ C\, e^{\,\varphi} \ + \ \Lambda  \ , \label{B0_sol_6}
\ee
where both $C$ and the cosmological constant $\Lambda$ are initially assumed to be positive. In this case the solution can be cast in the form
\bea
&& x \ = \ c_1 \, \cosh \lambda t_c \ + \ d_1 \, \sinh \lambda t_c \ , \nonumber \\
&&  y \ = \ \frac{C\,t_c}{2\,\lambda} \ \big(c_1 \ \sinh \lambda t_c \ + \ d_1 \ \cosh \lambda t_c\big) \ + \ c_2 \, \cosh \lambda t_c \ + \ d_2 \, \sinh \lambda t_c \ , \label{B0_sol_7}
\eea
where $\lambda$ is given again in eq.~\eqref{B0_sol_01}, while the Hamiltonian constraint yields the quadratic relation
\be
\left( d_1^{\,2} \ - \ c_1^{\,2}\right) \ = \ \frac{\Lambda}{C} \ \left( c_1 \, c_2 \ - \ d_1 \, d_2 \right) \ . \label{B0_sol_8}
\ee
\begin{figure}[h]
\begin{center}$
\begin{array}{ccc}
\epsfig{file=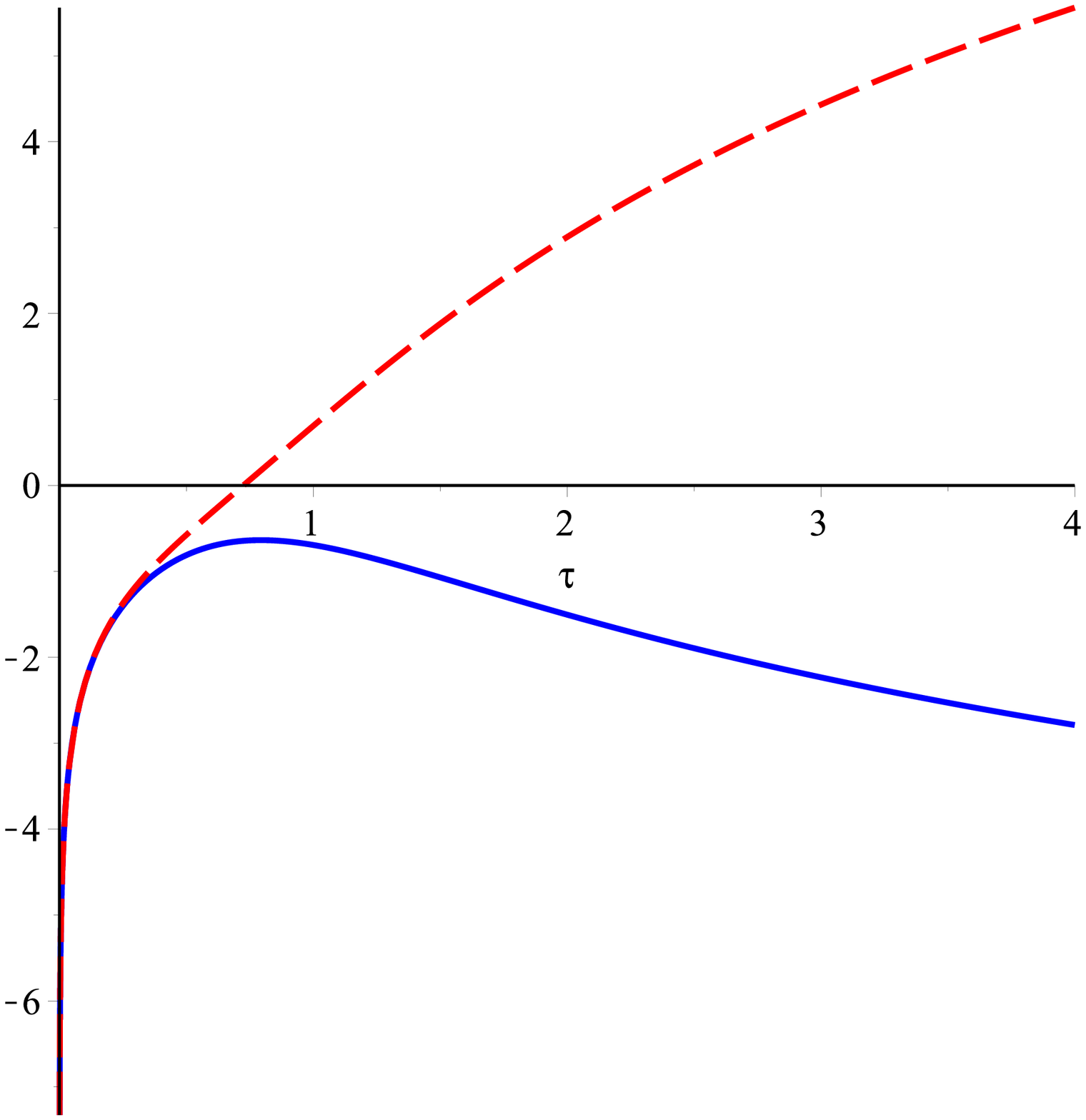, height=1.5in, width=1.5in} \qquad &
\epsfig{file=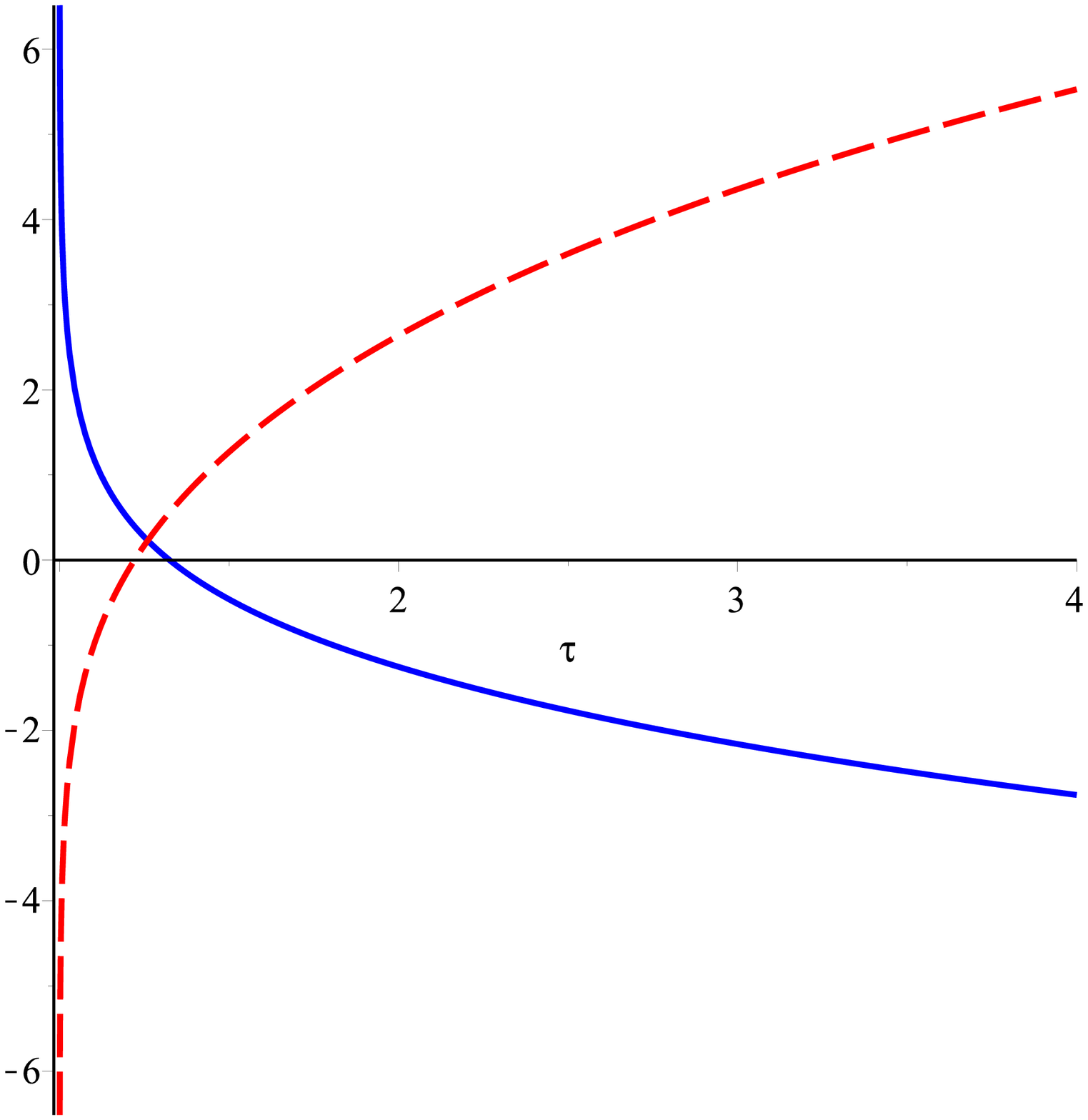, height=1.5in, width=1.5in} \qquad &
\epsfig{file=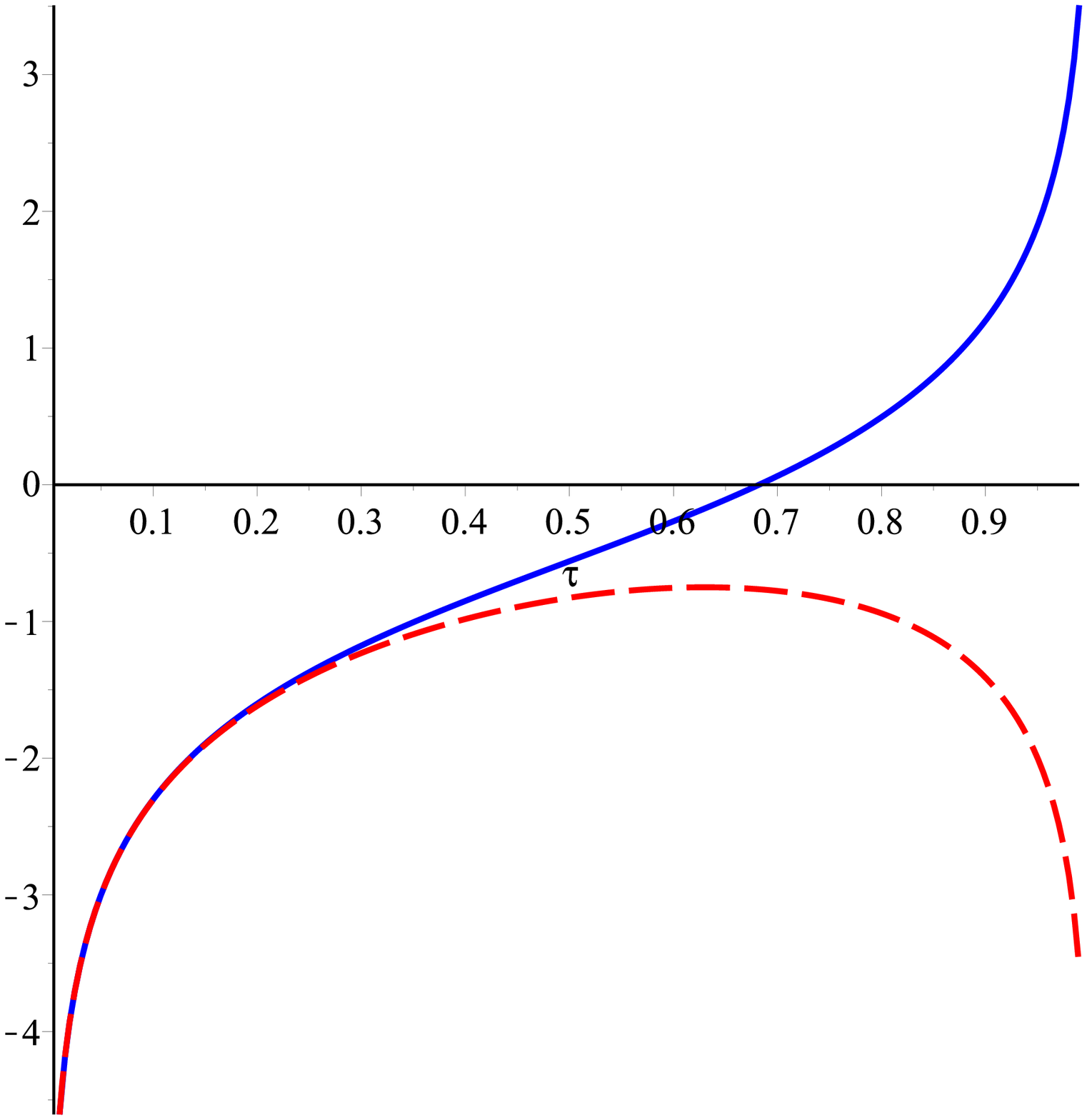, height=1.5in, width=1.5in}
\end{array}$
\end{center}
\caption{The solutions for $\varphi(\tau)$ and ${\cal A}(\tau)$ (dashed) in the potential \eqref{B0_sol_1111}, with $\tau=\sqrt{|\Lambda|/2} \ t_c$: \emph{descending} or \emph{climbing} if $C>0$, with an eventual Big Crunch if $C<0$.}
\label{desclimbcrunch}
\end{figure}
There are clearly two classes of independent solutions, the first of which, a deformation of the \emph{descending} solution, obtains for $d_1=0$ and, taking into account the Hamiltonian constraint \eqref{B0_sol_8}, can be cast in the form\footnote{Notice that some of the coefficients present in these solutions and in eqs.~\eqref{B0_sol_12_1} are bound to have a singular limit as $\Lambda \to 0$. As a result, as $t_c\to \infty$ $\varphi$ approaches a limiting speed that is \emph{half} of the value determined by eqs.~\eqref{B0_sol_41} and \eqref{B0_sol_42} in the absence of a cosmological term. In other words, the $\Lambda \to 0$ limit of the late time behavior exhibits a \emph{discontinuity} at $\Lambda=0$.}
\bea
\varphi &=& - \ \log \, \frac{C}{\Lambda} \Big[ \left(\lambda\,t_c \,+\, \Delta_1 \right) \, \tanh(\lambda t_c) \ - \ 1 \Big] \ , \nonumber \\
e^{\,{\cal A}} &=&  c_1^{\,2}\, \frac{C}{\Lambda} \ \cosh^2(\lambda t_c) \, \Big[ \left(\lambda\,t_c \,+\, \Delta_1 \right) \, \tanh(\lambda t_c) \ - \ 1 \Big] \ , \label{B0_sol_9}
\eea
where $\Delta_1$ is a combination of the other constants, and the corresponding range of $t_c$ is defined by the condition
\be
C \ \Big[ \left(\lambda\,t_c \,+\, \Delta_1 \right) \, \tanh(\lambda t_c) \ - \ 1 \Big] \ > \ 0 \ .
\ee
In a similar fashion, a deformation of the \emph{climbing} solution obtains for $c_1=0$ and takes the form
\bea
\varphi &=& - \ \log \frac{C}{\Lambda} \Big[ \left(\lambda\,t_c \,+\, \Delta_2 \right) \, \coth(\lambda t_c) \ - \ 1 \Big] \ , \nonumber \\
e^{\,{\cal A}} &=&  d_1^{\,2}\, \frac{C}{\Lambda} \ \sinh^2(\lambda t_c) \, \Big[ \left(\lambda\,t_c \,+\, \Delta_2 \right) \, \coth(\lambda t_c) \ - \ 1 \Big] \label{B0_sol_12_1}
\eea
if $\Delta_2$ is a \emph{positive} combination of the other constants. In general, however, the range of $t_c$ is defined by the condition
\be
\Big[ \left(\lambda\,t_c \,+\, \Delta_2 \right) \, \coth(\lambda t_c) \ - \ 1 \Big] \ > \ 0 \ , \label{B0_sol_13}
\ee
so that this solution describes a climbing scalar only if the inequality \eqref{B0_sol_13} is satisfied for all $t \geq 0$, but otherwise it describes a descending scalar.
In all cases, the presence of a positive cosmological constant in eq.~\eqref{B0_sol_6} drives the system to approach more rapidly the attractor behavior
\be
\varphi \ \sim \ - \log \lambda\, t_c \ ,
\ee
a result that affords an intuitive explanation since its net effect is to flatten the potential, reducing the effective value of $\gamma$, in the language of Section \ref{sec:climbing}, that is felt by the scalar field during the descent.

As in the preceding example, the ranges thus identified depend on the sign of $C$, so that for $C<0$ one is confronted again with the type of behavior encountered in eqs.~\eqref{B0_sol_11}: the scalar sweeps the whole real axis during an epoch following a Big Bang and preceding a Big Crunch.
\begin{figure}[h]
\begin{center}$
\begin{array}{ccc}
\epsfig{file=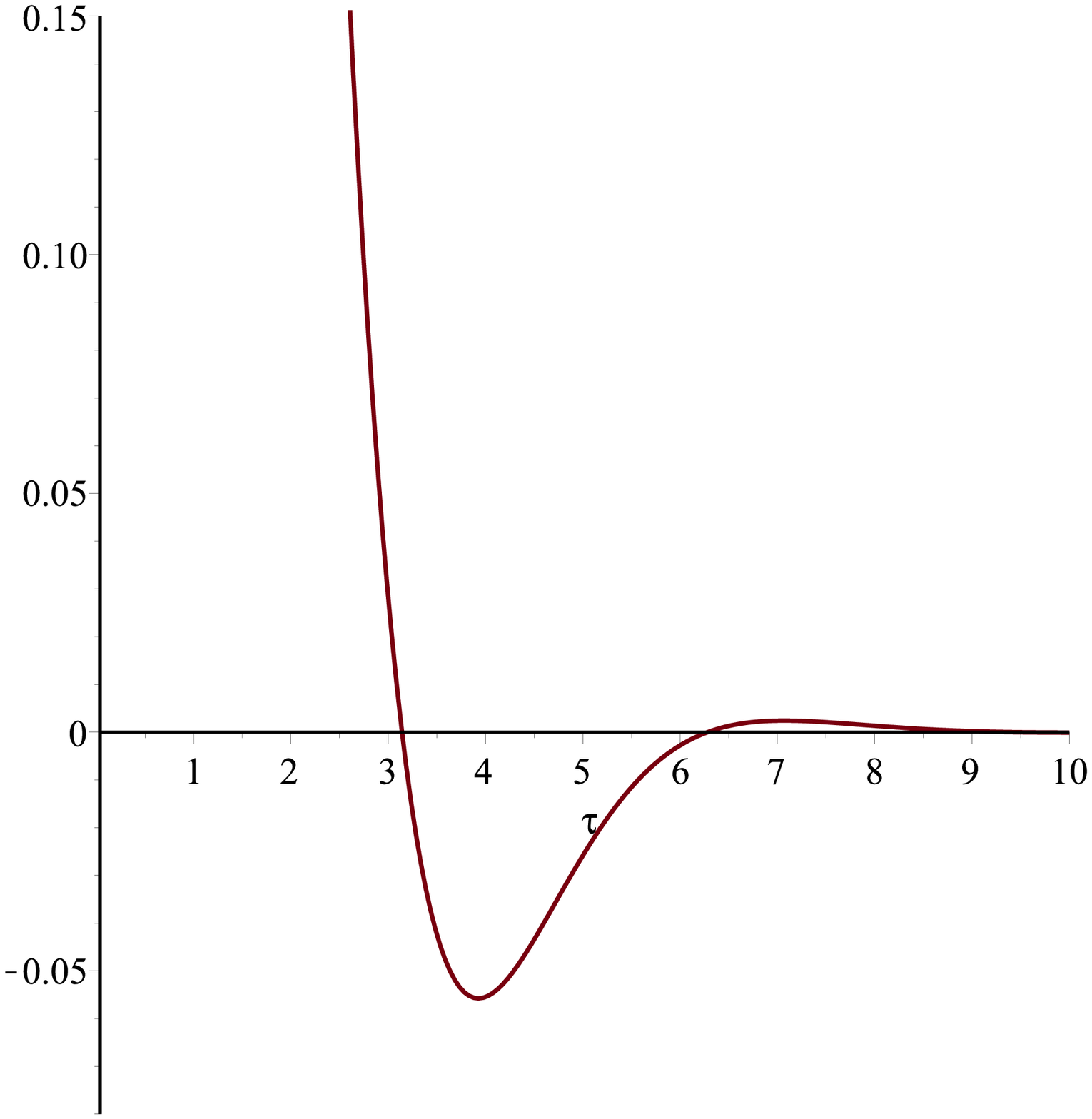, height=1.5in, width=1.5in} \qquad &
\epsfig{file=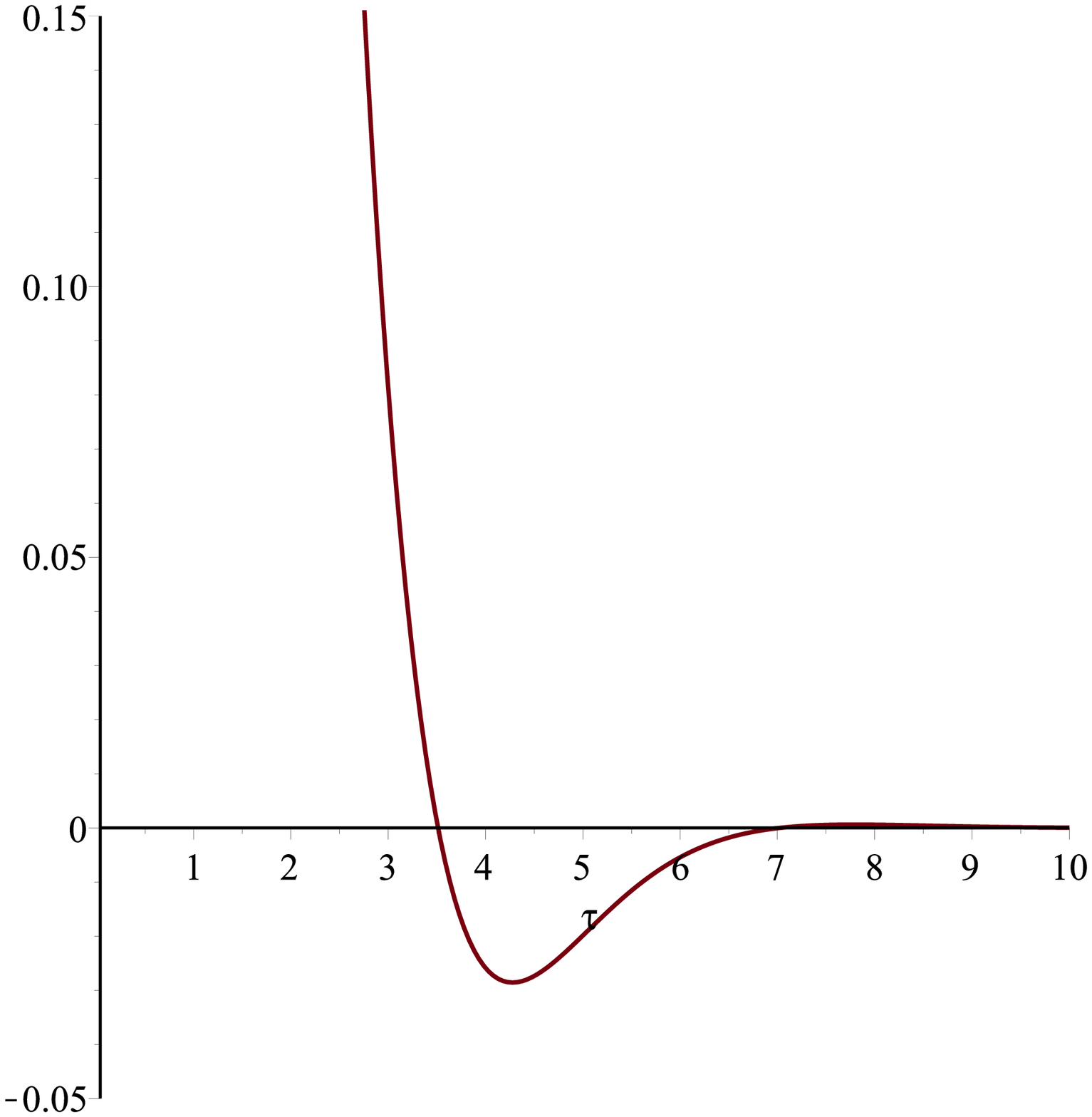, height=1.5in, width=1.5in} \qquad &
\epsfig{file=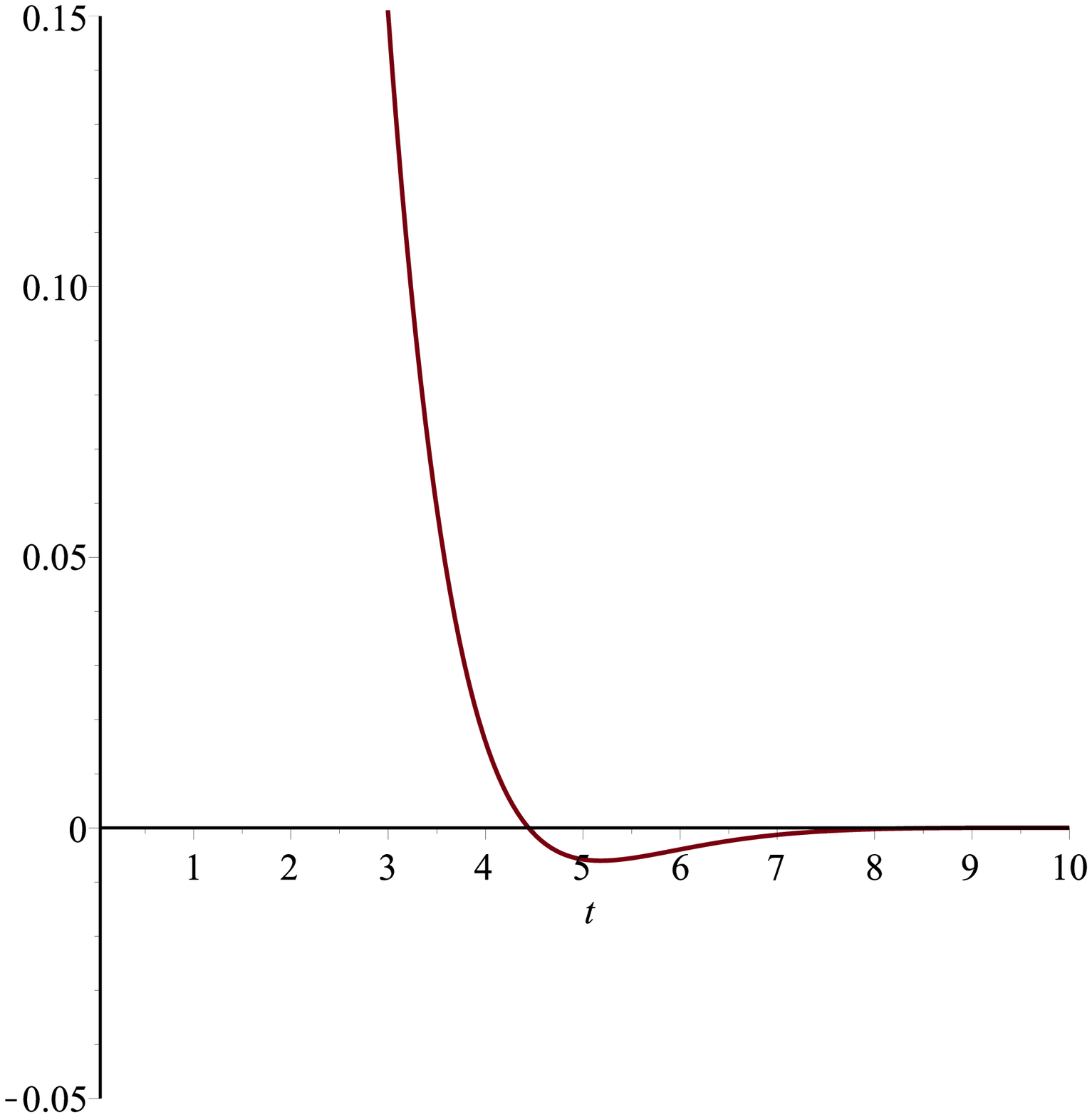, height=1.5in, width=1.5in}
\end{array}$
\end{center}
\caption{The evolution of $\varphi$ in the potentials \eqref{B0_sol_14}, with $\epsilon$ equal to 0 (left), to 0.1 (center) and to 0.5 (right), as a function of $\tau$. The extent of the oscillations is quickly reduced as $\epsilon$ increases.}
\label{cosh-Lambda}
\end{figure}

Similar types of behavior are encountered in the class of potential wells
\be
{\cal V} \ = \ C \ \cosh \varphi \ + \ \Lambda \ \ . \label{B0_sol_14}
\ee
The preceding discussion makes it possible to confine our analysis, for brevity, to the class of models where $\Lambda$ and
$C$ are both positive, which exhibit nonetheless an interesting crossover between two qualitatively different types of behavior. For all these models eq.~\eqref{B0_sol_14} describes potential wells whose ends are under--critical in the sense of section \ref{sec:climbing}, so that the scalar $\varphi$ can descend from either side after the initial singularity, but there are two types of subsequent behavior. They reflect the eigenvalues of the system \eqref{int604}, that for this class of potentials are determined by the relation
\be
\lambda^2 \ = \ \frac{\Lambda\, \pm\, C}{2} \ ,
\ee
so that they are all real if $\Lambda > C$ while two are real and two are imaginary if $\Lambda < C$. Letting
\be
\tau \ = \ t \ \sqrt{\frac{C}{2}} \ , \qquad \epsilon \ = \ \frac{\Lambda}{C}
\ee
the interesting behavior for $\Lambda < C$ is captured for instance by the exact solution
\be
\varphi \ = \ \log \left[ \frac{\sqrt{1 - \epsilon}\ \sinh\left(\tau\, \sqrt{1 + \epsilon} \right) \  + \ \sqrt{1 + \epsilon}\ \sin\left(\tau\, \sqrt{1 - \epsilon} \right)}{\sqrt{1 - \epsilon}\ \sinh\left(\tau\, \sqrt{1 + \epsilon} \right) \  - \ \sqrt{1 + \epsilon}\ \sin\left(\tau\, \sqrt{1 - \epsilon} \right)} \right] \ ,
\ee
which describes a scalar field that emerges from the initial singularity climbing down the right end of the well and readily approaches its minimum performing damped oscillations about it (fig.~\ref{cosh-Lambda}).

Briefly stated, the two sub--classes of systems thus identified behave, respectively, as an over--damped and an under--damped oscillator, so that if $\Lambda > C$ cosmological friction dominates and the scalar descends from one of the walls and comes to rest at the bottom, while in the opposite case it also undergoes damped oscillations. One can understand intuitively why this should be the case: as we have seen, the net effect of a positive $\Lambda$ is to enhance the effect of cosmological damping relative to the potential drive, so that only if it is small enough can the potential have a say close to the origin. In the language of Section \ref{sec:fixedpts_ana}, the two types of behavior reflect the transition between a fixed point of type ``node'' and one of type ``focus''.

\subsubsection{\sc Solutions of the triangular systems with ${\cal B} \neq 0$}\label{sec:Bnot0_sol}
The class of potentials of eqs.~\eqref{int26} is rather rich and has the virtue of displaying neatly a number of effects, including the onset of the climbing phenomenon when the potential becomes sufficiently steep. Let us notice, to begin with, that in the most interesting case in which both $c_i$ are not zero in the potential \eqref{int23}, shifting $\varphi$ one can make them identical up to a sign. Therefore, we shall work generically with
\be
{\cal V} \ = \ C \left( \epsilon_1 \ e^{\,2\,\gamma\,\varphi} \ + \ \epsilon_2 \ e^{\,(\gamma\,+\,1)\,\varphi} \right)\ ,
\ee
with $\gamma \neq \pm 1$,
and we shall actually set $C=1$ in the following, absorbing its positive value in the time scale. In all cases, as we have seen in Section \ref{sec:integrable},
\be
 {\cal A} \ = \ \log\left(x^{\, \frac{1}{1+\gamma}}\ y^{\, \frac{1}{1-\gamma}}\right) \ , \qquad \vf \ = \ \log\left(x^{\, \frac{1}{1+\gamma}} \ y^{\,- \,  \frac{1}{1-\gamma}}\right) \ , \label{Bnot0_3_001}
\ee
and
\be
dt_c \ = \ x^{\,-\,\frac{\gamma}{1+\gamma}} \ y^{\,\frac{\gamma}{1-\gamma}} \ dt\ . \label{Bnot0_3_002}
\ee
Some care should be exercised when interpreting the dynamics in terms of $t$, since in certain ranges for the parameter $\gamma$ large values of $t$ translate into small values of $t_c$ and vice versa.

For brevity, here we shall restrict our attention to the cases where the potential does not become arbitrarily large and negative, since we know already from preceding examples what this type of pathology would imply (see \emph{e.g.} the last of figs.~\ref{desclimbcrunch}), but nonetheless different ranges of $\gamma$ bring about different types of interesting phenomena:
\begin{itemize}
\item[1. ] for $\gamma>1$, both exponents in the potential are \emph{positive} and \emph{over--critical} in the sense of Section \ref{sec:climbing}. In this region the potential is dominated asymptotically by the first term in eq.~\eqref{int23}, so that we shall confine our attention to the case $\epsilon_1=1$, allowing however for $\epsilon_2=\pm 1$, since negative values give rise to shallow potential wells terminating on an over--critical potential, an interesting situation that will also be captured by a subsequent class of examples;
\item[2. ] for $0<\gamma<1$ both exponents are \emph{positive} and \emph{under--critical}, and here one can play similarly with $\epsilon_1=\pm 1$ while fixing $\epsilon_2=1$, since the second term now dominates asymptotically;
\item[3. ] for $-1<\gamma<0$ both exponents are \emph{under--critical} but have \emph{opposite signs}, so that eq.~\eqref{int23} describes \emph{stable} potential wells only if $\epsilon_1=\epsilon_2=1$;
\item[4. ] for $-3<\gamma<-1$ both exponents are \emph{negative}, one is \emph{over--critical} and one is \emph{under--critical}, so that for $\epsilon_2=1$ eq.~\eqref{int23} captures the onset on an inflationary phase spurred by the climbing phenomenon, as will be also the case for one of the models belonging to the last class of Section \ref{sec:quadratures}. In this region one can also allow for $\epsilon_2=- 1$ compatibly with the existence of a \emph{negative} lower bound for ${\cal V}$;
\item[5. ] for $\gamma<-3$ both exponents are \emph{negative} and \emph{over--critical}, so that this last range has the same qualitative features as the first one.
\end{itemize}

To begin with, the solutions apply in regions where $x(t)$ and $y(t)$ are both positive and possess widely different features for $|\gamma|<1$, where the climbing phenomenon is not present, and in the complementary interval $|\gamma|>1$, where it occurs. We have chosen to present them in an explicit form, although some contributions can be related to hypergeometric functions.
\begin{figure}[h]
\begin{center}$
\begin{array}{cc}
\epsfig{file=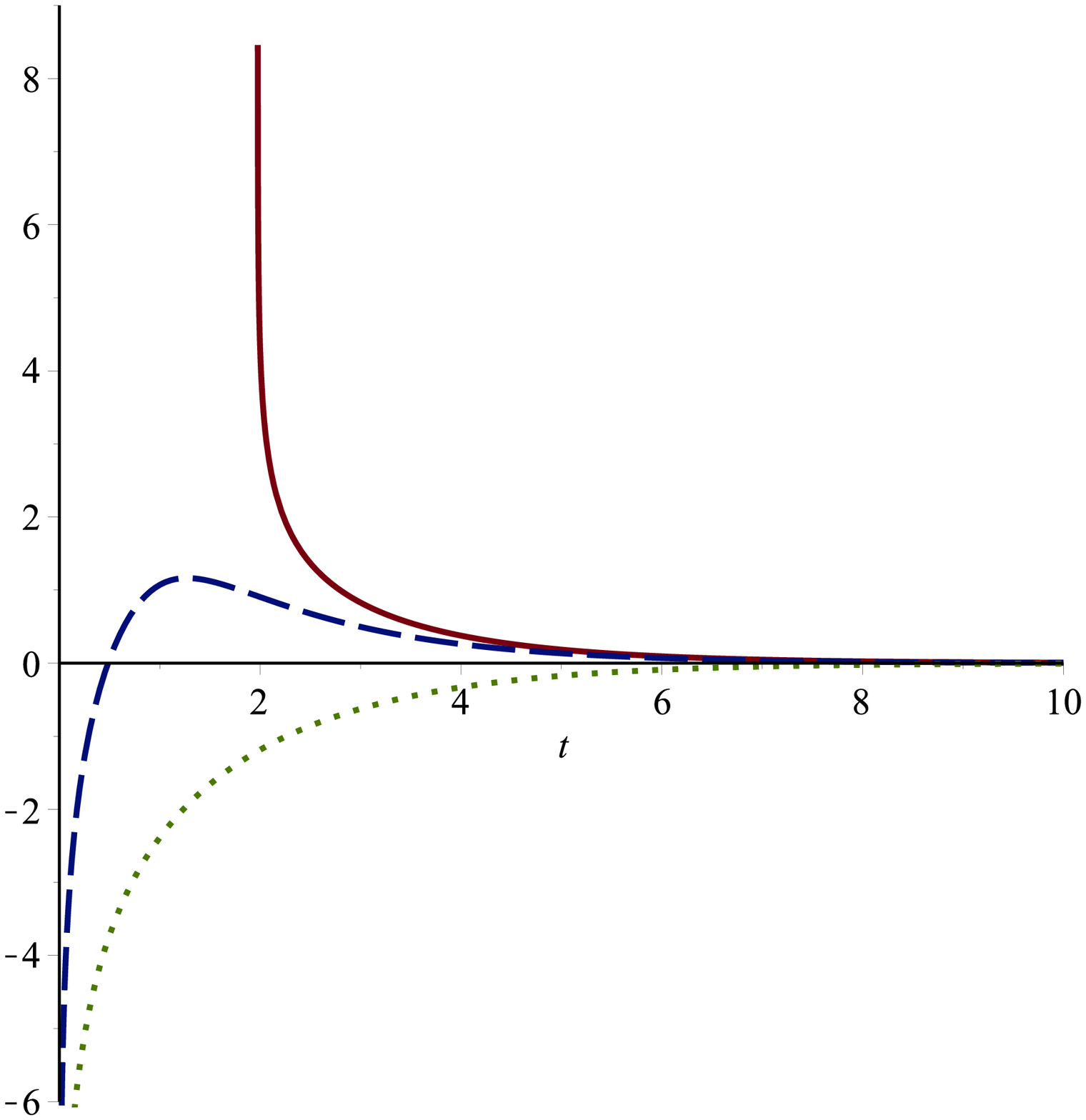, height=1.5in, width=1.5in} \qquad \quad &
\epsfig{file=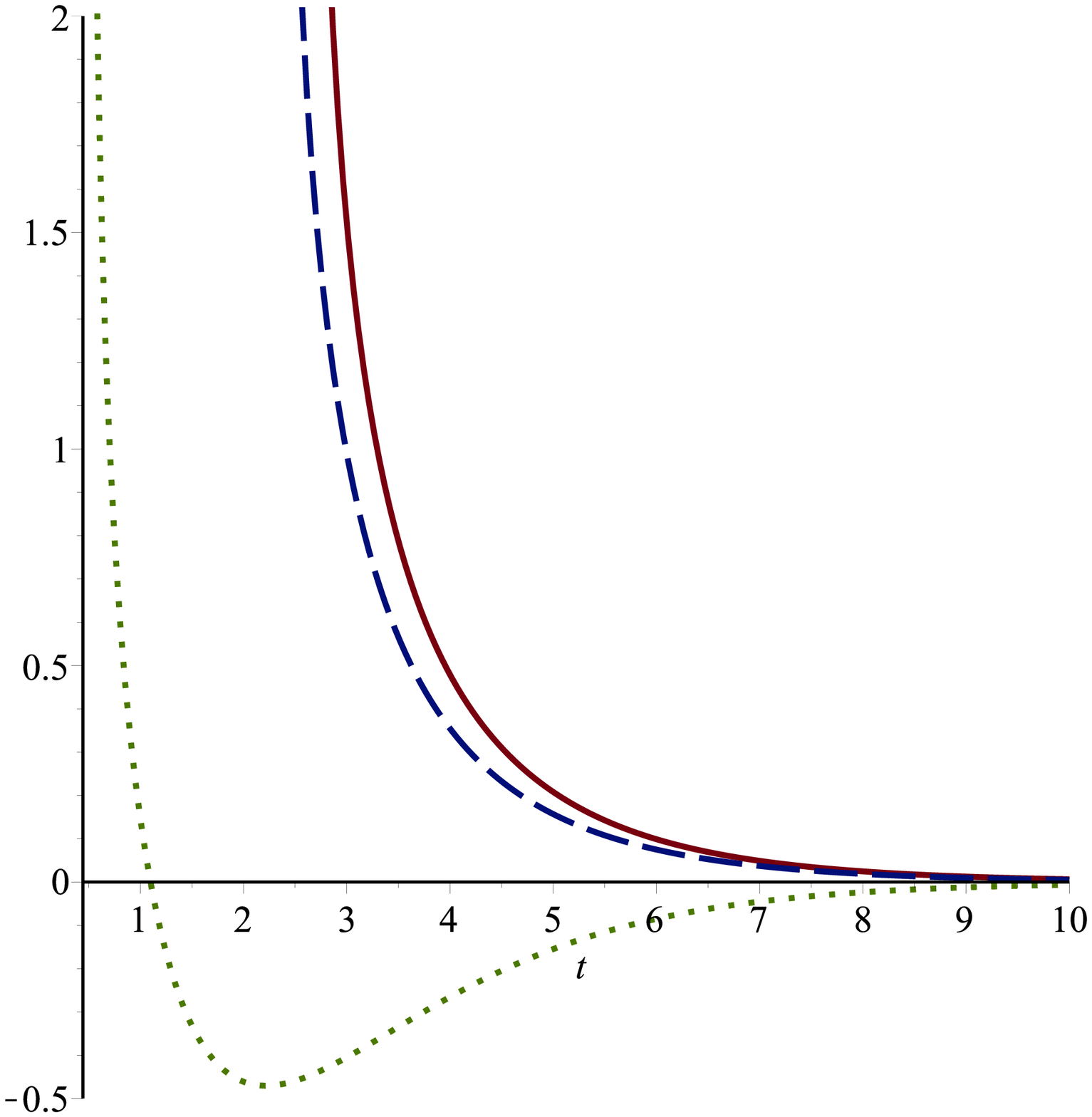, height=1.5in, width=1.5in} \qquad
\end{array}$
\end{center}
\caption{Cosmological evolution of the scalar field in the potential well \eqref{Bnot0_3_0051}: the solutions \eqref{Bnot0_3_0052} for $a-2=-2.5,-1.9,2.5$ (left: continuous, dashed, dotted), and the solution \eqref{Bnot0_3_0053} for $a=-2.5,-1.9,2.5$ (right: continuous, dashed, dotted). In the former case an interesting transition occurs at the point $a=-2$, below which the solutions actually emerge from the right.}
\label{crit_well}
\end{figure}

For $|\gamma| < 1$ and $\epsilon_1=1$ the homogeneous solutions are hyperbolic, and letting
\be
\Omega \ = \ \sqrt{\frac{1\, - \, \gamma^{\,2}}{2}} \label{Bnot0_3_003}
\ee
there are actually two distinct classes of behavior. They are captured by
\bea
x(t) &=& \sinh \left(\Omega t \right) \ , \label{Bnot0_3_004} \\
y(t) &=& \left[ a \ - \ \frac{1}{1+\gamma} \int_{0}^{\sinh^2\left(\Omega t \right)} \!\!\!\!\!du \ u^{\, \frac{1-\gamma}{2(1+\gamma)}} \ (1+u)^{\, -\, \frac{1}{2}} \right] \cosh\left(\Omega t \right) \ + \ \Big[\sinh \left(\Omega t \right) \Big]^{\, \frac{3+\gamma}{1+\gamma}} \ , \nonumber
\eea
which typically describes for $0 < \gamma < 1$ a \emph{climbing} scalar, and by
\bea
x(t) &=& \cosh \left(\Omega t \right) \ , \label{Bnot0_3_005} \\
y(t) &=& \left[ a \ + \ \frac{1}{1+\gamma} \int_{1}^{\cosh^2\left(\Omega t \right)} \!\!\!\!\!du \ u^{\, \frac{1-\gamma}{2(1+\gamma)}} \ (u-1)^{\, -\, \frac{1}{2}} \right] \sinh\left(\Omega t \right) \ - \  \Big[\cosh \left(\Omega t \right) \Big]^{\, \frac{3+\gamma}{1+\gamma}} \ , \nonumber
\eea
which typically describes for $0< \gamma < 1$ a \emph{descending} scalar. For instance, in the region $-1<\gamma<0$ the names of climbing and descending solutions should perhaps be interchanged, since the potential becomes steeper on its left end. More importantly, the behavior undergoes a sharp change when $a$ endows $y(t)$ with a zero for positive $t$, thus setting a positive lower bound on the admissible region for the cosmological evolution, since eq.~\eqref{Bnot0_3_004} then describes a descending scalar. Aside from taking into account the Hamiltonian constraint, in these solutions we have also absorbed an overall constant originating from $x(t)$ into the definition of the spatial scale.

The case $\gamma \, = \, - \, \frac{1}{3}\,$ is particularly instructive, since for the corresponding stable potential,
\be
V \ = \ 2\, \cosh \left( \frac{2 \, \varphi}{3} \right) \ , \label{Bnot0_3_0051}
\ee
the integrals of eqs.~\eqref{Bnot0_3_004} and \eqref{Bnot0_3_005} become particularly simple and the solutions read
\bea
&& {\cal A} \ = \ \frac{3}{4} \ \log\left\{\sinh^2\left(\frac{2 \, t}{3} \right) \, \left[\cosh^2\left(\frac{2 \, t}{3}\right) \,+\, (a-2)\, \cosh\left(\frac{2 \, t}{3}\right) \,+\, 1\right] \right\} \ , \\
&&\varphi \ = \ \frac{3}{4} \ \log\left\{\frac{\sinh^2\left(\frac{2 \, t}{3} \right)}{\cosh^2\left(\frac{2 \, t}{3}\right) \,+\, (a-2)\, \cosh\left(\frac{2 \, t}{3} \right)\,+\,1} \right\}
\label{Bnot0_3_0052}
\eea
and
\bea
&& {\cal A} \ = \ \frac{3}{4} \ \log\left\{\cosh^2\left(\frac{2 \, t}{3} \right) \, \left[\sinh^2\left(\frac{2 \, t}{3}\right) \,+ \, a\, \sinh\left(\frac{2 \, t}{3} \right)\,-\,1 \right] \right\} \ , \\
&&\varphi \ = \ \frac{3}{4} \ \log\left\{\frac{\cosh^2\left(\frac{2 \, t}{3} \right)}{\sinh^2\left(\frac{2 \, t}{3}\right) \,+\, a\, \sinh\left(\frac{2 \, t}{3} \right)\,-\, 1} \right\} \ .
\label{Bnot0_3_0053}
\eea
As we have stressed, these expressions apply in regions where $x(t)$ and $y(t)$ are both positive, where they describe Universes where the scalar approaches the critical point $\varphi=0$ of the potential after emerging from the initial singularity from negative or positive values. The transition that we have mentioned after eq.~\eqref{Bnot0_3_005} manifests itself in eq.~\eqref{Bnot0_3_0052} as the parameter $a$ entering the solution becomes less than -2.

On the other hand for $|\gamma| > 1$, letting
\be
\omega \ = \ \sqrt{\frac{\gamma^{\,2}\,-\,1}{2}} \label{Bnot0_3_006}
\ee
and up to a shift of the origin of $t$ there is the single class of solutions
\bea
&& x(t) \ = \ \sin \left(\omega t \right) \ , \label{Bnot0_3_007} \\
&& y(t) \ = \ \left[ a \ - \ \frac{\epsilon}{\gamma+1} \int_{\sin^2\left(\omega t \right)}^{1} \!\!\!\!\!du \ u^{\,\frac{1-\gamma}{2(\gamma+1)}} \ (1-u)^{\, -\, \frac{1}{2}} \right] \cos\left(\omega t \right) \ - \ \epsilon \, \Big[\sin \left(\omega t \right) \Big]^{\, \frac{\gamma+3}{\gamma+1}} \ . \nonumber
\eea
For $\epsilon=1$ these describe, for $0<\omega \, t < \omega^\star\, t < \frac{\pi}{2}$, where $y(t)$ vanishes at the upper end, a climbing scalar in an expanding Universe, and the constant $a$ determines the extent of the climbing phase. On the other hand, for $\epsilon=-1$ this solution applies in the wider range $0<\omega \, t < \omega^\star\, t < {\pi}$ and its nature changes drastically, as one could have anticipated from the preceding examples, since the extremum corresponds to a negative cosmological constant. As a result, the scalar lingers about it for a while but this drives the Universe to collapse again, so that the scalar is accelerated and eventually climbs up the potential all the way until an eventual Big Crunch takes place. There is another interesting region for $-3<\gamma<-1$, where the climbing is followed by a ``hard kick'' that leads to an inflationary phase during the subsequent descent. The behavior of this class of potentials is along the lines of what we shall see for the models of eq.~\eqref{Liouville3}, so that we can postpone a more detailed discussion of this case for the sake of brevity, but let us note that for the sequence of rational values $\gamma=- 1-\frac{1}{2n}$ the integrals become elementary and are expressible in terms of trigonometric functions. Eq.~\eqref{Bnot0_3_007}, and the previous considerations for $\epsilon=-1$, also apply to the potential wells obtained choosing $\epsilon_1=-1$ in the region $0<\gamma<1$, provided
\be
\omega \ = \ \sqrt{\frac{1\,-\,\gamma^{\,2}}{2}} \ .
\ee

We can forego a detailed description of the solutions of eqs.~\eqref{int34}, which can be expressed in terms of error functions, since the key effect of a cosmological constant was already illustrated in the preceding section and this model adds to that discussion only the inevitable presence of the climbing phenomenon brought about by its ``critical'' exponential potential.
\begin{figure}[h]
\begin{center}$
\begin{array}{cc}
\epsfig{file=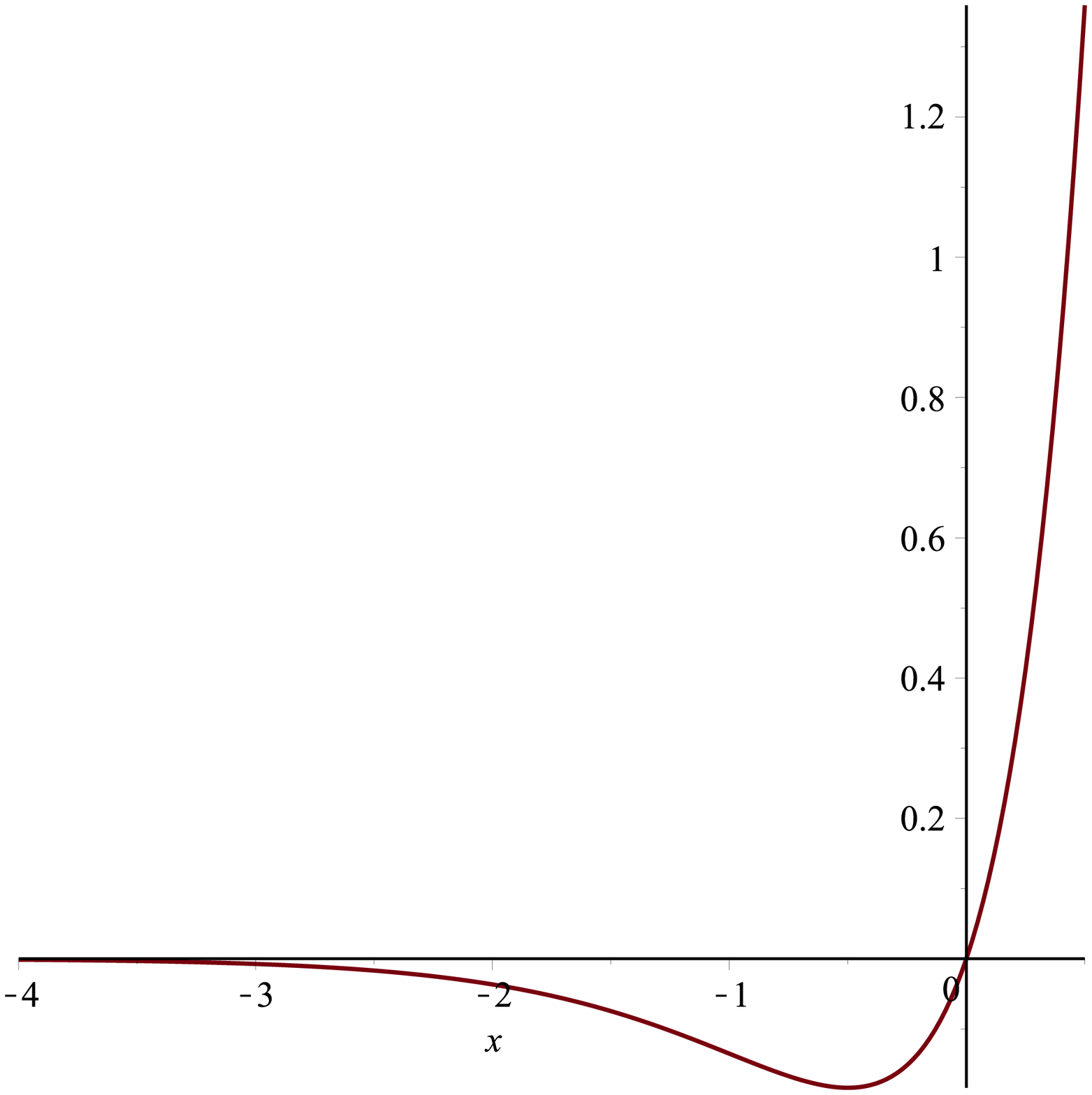, height=1.5in, width=1.5in} \qquad \quad &
\epsfig{file=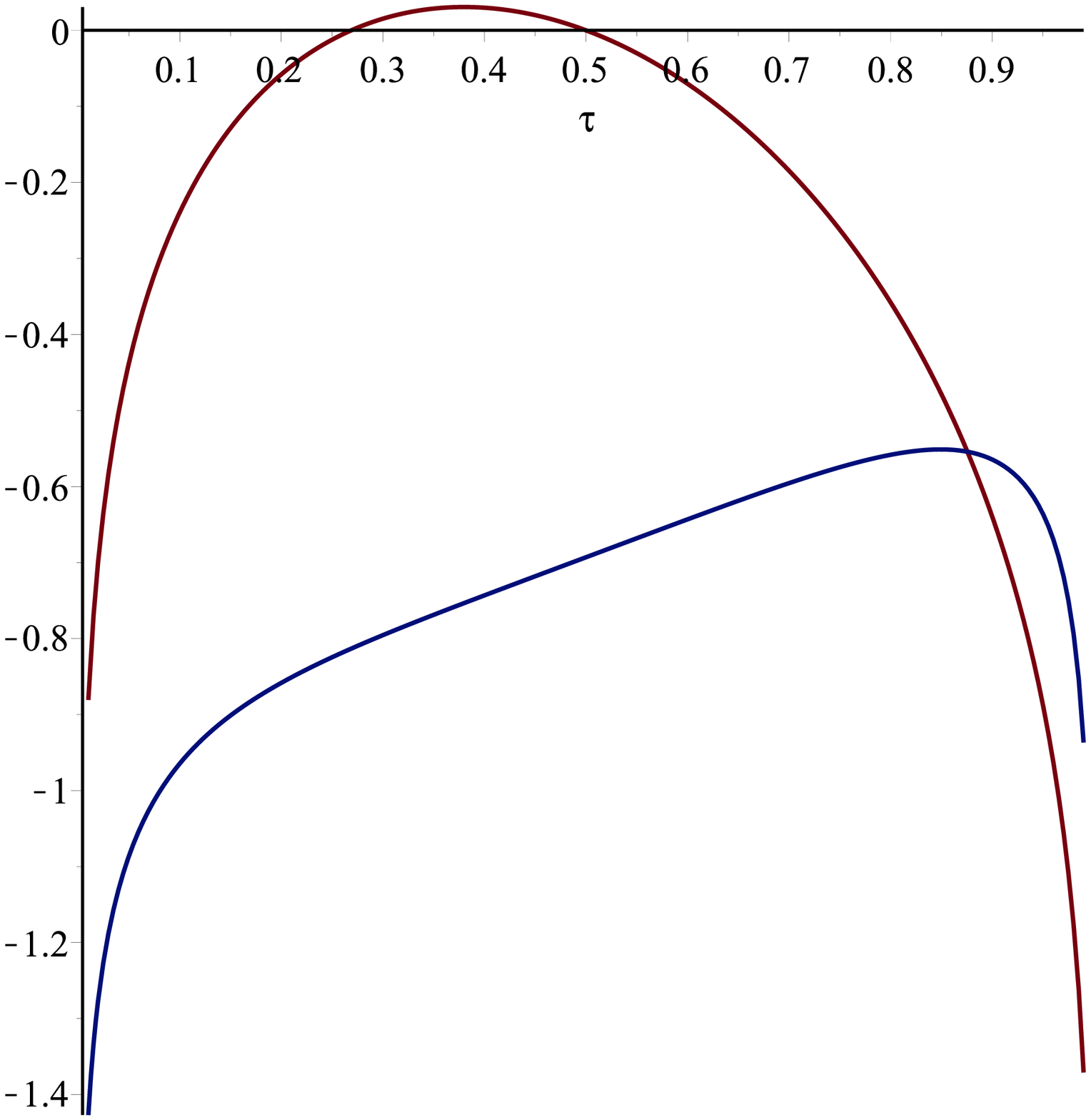, height=1.5in, width=1.5in} \qquad
\end{array}$
\end{center}
\caption{The potential well \eqref{int37} and the corresponding $\varphi(\tau)$ and ${\cal A}(\tau)$ (dashed).}
\label{crit_wellbis}
\end{figure}
Therefore, we can now turn to describe the solutions of the dynamical system of eqs.~\eqref{int3655}, which corresponds to the potential function of eq.~\eqref{int37}. As we have stressed, for $C>0$ this describes a shallow potential well whose right end leaves way to a ``critical'' exponential in the sense of Section \ref{sec:climbing}. Hence, one might expect the scalar to emerge from an initial singularity while climbing up the potential and to try to approach the bottom of the well for suitable initial conditions.

Up to a choice for the origin of the parametric time $t$, the solution of the first of eqs.~\eqref{int38} reads
\be
x \ = \ C \, t \, \left( t_2 \ - \ t \right) \ , \label{Bnot0_3_1}
\ee
and in the spirit of the preceding section we work in the interval $0<t<t_2$ where $x$ is positive. It is thus convenient to perform the rescaling
\be
\tau \ = \ \frac{t}{t_2} \label{Bnot0_3_2}
\ee
and taking into account the Hamiltonian constraint \eqref{int39} the complete solution takes finally the form
\bea
x &=& C \ t_2^2 \ \tau \, \left( 1 \ - \ \tau \right) \\
v &=& \frac{1}{2} \, \bigg[ \tau\, \log \tau \ + \ \left( 1 \ - \ \tau \right) \, \log\left( 1 \ - \ \tau \right)\bigg] \ + \ \frac{1}{4} \ \log (C\,t_2^{\,2}) \nonumber \\ &-& \frac{C}{2}\left( 1 \ - \ 2\, \tau \right) \ ,
\eea
so that
\bea
\varphi &=& \frac{1}{4} \ \left( 1 \ - \ 2\, \tau \right) \  \left[\log\left(\frac{\tau}{1 \ - \ \tau}\right) \ + \ 2\, C \right] \ , \nonumber \\
{\cal A} &=& \frac{1}{4} \  \bigg[\left( 1 \ + \ 2\, \tau \right)\, \log \tau \ + \ \left( 3 \ - \ 2\, \tau \right)\, \log (1\,-\,\tau) \bigg] \nonumber \\
 &+& \ \frac{C}{2} \ \left( 2\, \tau \ - \ 1 \right) \ + \ \frac{1}{2} \ \log (C\,t_2^{\,2}) \ .
\eea
Finally
\be
dt_c \ = \ C^{\, -\, \frac{1}{2}} \ dt \ e^{\,\frac{C}{2}\left( 2\, \tau \ - \ 1 \right)} \
\left[ \tau^{\, \tau \,-\, \frac{3}{2}}\, (1\,-\,\tau)^{\, 1\,-\, \tau \,-\, \frac{3}{2}} \right]^{\,\frac{1}{2}} \ .
\ee

An example of this class of solutions is displayed in fig.~\ref{crit_wellbis}. They follow the general trend of the spacetimes of eq.~\eqref{FLRWgen} with a negative cosmological constant, so that a Big Bang is always followed by a Big Crunch, and in this example the two are separated by an interval of cosmological time that grows exponentially with $C$. As expected, $\varphi$ does emerge from the Big Bang from the milder end of the potential, the only option allowed by its ``critical'' logarithmic slope, and larger positive values of $C$ bring about larger penetrations into the right repulsive region. However, for $\varphi<0$ the potential is mostly very small and negative, so that the scalar lingers around the extremum trying to settle there, but this is impossible since it would result in an expanding solution driven by a mere negative cosmological constant, which as we have stressed does not exist in the class of metrics of \eqref{FLRWgen}. As a result, the Universe starts to contract, cosmological damping turns into cosmological amplification and the scalar eventually runs off to infinity while the Universe experiences a Big Crunch.

\subsubsection{\sc Solutions of the systems integrable via quadratures}\label{sec:quadratures_sol}

Another wide class of the integrable models that we have identified is solvable by quadratures. The solutions bring about in some cases a new complication, the need to invert simultaneously a pair of functions, but nonetheless one can extract relatively simply some information about the behavior close to the initial singularity and at late times.

Error functions have already surfaced in Section \ref{sec:Bnot0_sol}, and play a prominent role also in the solutions for the potentials of eqs.~\eqref{int402} and \eqref{Wcomplex}, which deserve a closer look since they possess the novel feature of involving infinite series of exponentials. Let us begin from the first case, which rests on a pair of independent dynamical systems, and let us concentrate on the non--pathological case $C>0$, where the potential describes an infinite logarithmic barrier at $\varphi=0$ that is accompanied, for $\varphi >0$, by a region that is flat up to exponentially small corrections. Rescaling the time variable according to
\be
\tau \ = \ \sqrt{8\,C} \, t \label{quad_1}
\ee
and letting
\be
\epsilon_{\,\xi} \ = \ - \ \log \xi_{\,0}  \ , \qquad \epsilon_{\,\eta} \ = \ - \ \log \eta_{\,0} \, \label{quad_2}
\ee
the system reduces to the two identical--looking equations
\bea
&& \dot{\xi}^2 \ = \ \log\left(\frac{\xi}{\xi_{\,0}}\right) \ , \\
&& \dot{\eta}^2 \ = \ \log\left(\frac{\eta}{\eta_{\,0}}\right) \ , \label{quad_3}
\eea
while the Hamiltonian constraint \eqref{int4071} translates into the condition
\be
\eta_{\,0} \ = \ \xi_{\,0}\, e^{\,\rho} \ , \label{quad_4}
\ee
where
\be
\rho \ = \ \frac{D}{C} \ . \label{quad_5}
\ee

Eqs.~\eqref{quad_3} can now be solved by quadratures, so that one readily arrives at the
two relations
\bea
&& t \ = \ 2\, \xi_0 \, \int_{u_0}^{\sqrt{{\cal A}\, + \, \log\left(\frac{2\, \cosh \varphi}{\xi_{\,0}}\right)}} e^{\,u^{\,2}}\, du \ , \nonumber \\
&& t \ = \ 2\, \xi_0 \, e^{\,\rho} \, \int_{v_0}^{\sqrt{{\cal A}\, + \, \rho \, +\, \log\left(\frac{2\, \sinh \varphi}{\xi_{\,0}}\right)}} e^{\,v^{\,2}}\, dv \ , \label{quad_6}
\eea
where $u_0$ and $v_0$ are a pair of positive lower bounds, but this implicit form is clearly
not very handy. Still, one can extract from it the main features of the dynamics.

Let us first analyze qualitatively the behavior of these expressions in the case $\rho=0$. The key observation is that for large $t$ the upper limits must also grow, while the integrands are dominated by the regions near them. As a result, the two expressions are compatible if $\varphi$ also grows in such a way that the two upper bounds become essentially, in both cases, $\sqrt{{\cal A} + \varphi}$. This means effectively that $x$ grows faster than $y$, so that
\be
{\cal A} \ \sim \ \varphi \  \sim \ \frac{1}{2} \ \log t \ . \label{quad_7}
\ee
As we have seen in Section \ref{sec:climbing}, this type of expression describes the late--time descent along a ``critical'' exponential tail, which is clearly the dominant behavior for large positive values of $\varphi$ also for the potential \eqref{int402}. On the other hand, for small values of $t$ one finds in general the limiting relations
\bea
&& {\cal A} \ \sim \  \frac{1}{2} \ (-\rho + u_0^{\,2}+v_0^{\,2}) \ + \ \frac{1}{2} \ \log\left( \frac{\xi_{\,0}^{\,2}\, \sinh(\rho + u_0^{\,2}-v_0^{\,2})}{2}\right) \ , \nonumber \\
&& \log\left(\coth{\varphi}\right) \ \sim \ \rho \, +\, u_0^{\,2} \, - \, v_0^{\,2} \ , \label{quad_8}
\eea
which we shall specialize again momentarily to the case $\rho=0$.
It should then be clear that the only consistent way to set the Big Bang time at $t=0$ is to identify the two lower bounds, choosing $v_0=u_0$, but the scalar is then bound to emerge after the initial singularity from large positive values, and thus while \emph{climbing up} the exponential wall. This is indeed the expected result, since the logarithmic slope of $W$ diverges as $\varphi \to 0^+$, so that we are well beyond the ``critical'' value there. And indeed one can come readily to this conclusion from the differential equation for $\varphi$ near $t=0$ in the gauge of Section \ref{sec:climbing}, which is a viable choice for this model since $W$ never vanishes.

As in some preceding examples, there is a again a sharp difference between the cases of positive and negative $\rho$, so let us begin by illustrating the former, which is simpler, referring again to eqs.~\eqref{quad_8}. The behavior near the original singularity does not change, although now $u_0$ is reduced with respect to $v_0$ according to
\be
u_0 \, = \, \sqrt{v_0^{\,2} \ - \ \rho} \ . \label{quad_9}
\ee
This compensates, for low values of $t$, the combined effects of the positive $\rho$ in the exponential pre--factor and in the upper bound of the second of eqs.~\eqref{quad_6}. However, as $t$ increases the integrals become dominated by their upper ends, so that the presence of $\rho$ is to be compensated by the difference between the two hyperbolic functions. In other words, as in preceding examples, a positive $\rho$ enhances cosmological friction and $\varphi$ attains a \emph{finite} limiting value for large $t$. For $\rho<0$ we expect, as for the preceding examples, an eventual collapse of the Universe. The behavior for small $t$ remains the same, so that the system emerges again from the initial singularity with the scalar climbing up, as can be seen from eqs.~\eqref{quad_8}, but when the integrals become dominated by their upper bounds at some point the two eqs.~\eqref{quad_6} become incompatible, due to the combined effects of a negative $\rho$ in the exponent and in the upper bound, and this signals in this formulation the expected Big Crunch.
\begin{figure}[h]
\begin{center}$
\begin{array}{ccc}
\epsfig{file=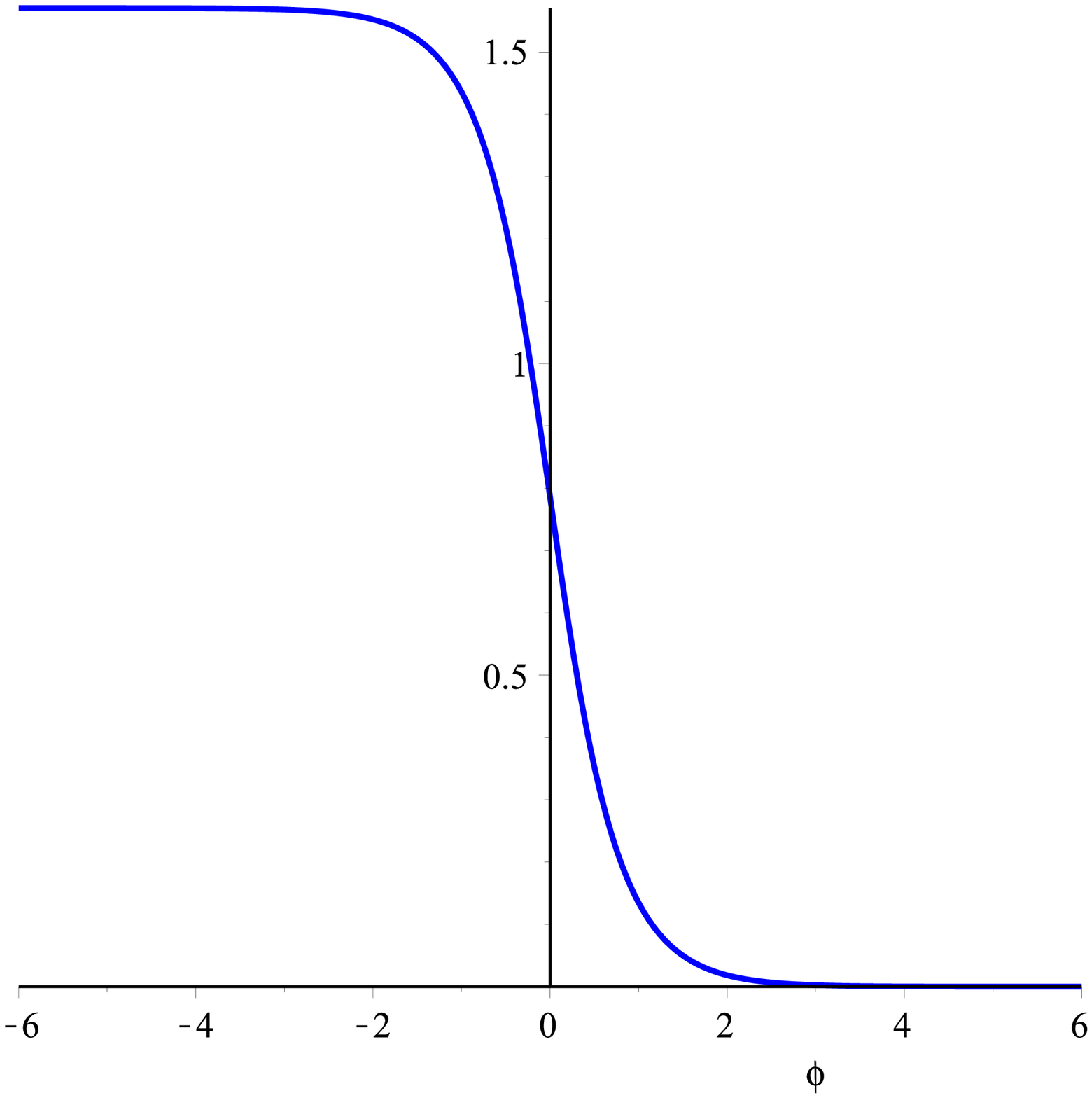, height=1.5in, width=1.5in}\qquad \quad &
\epsfig{file=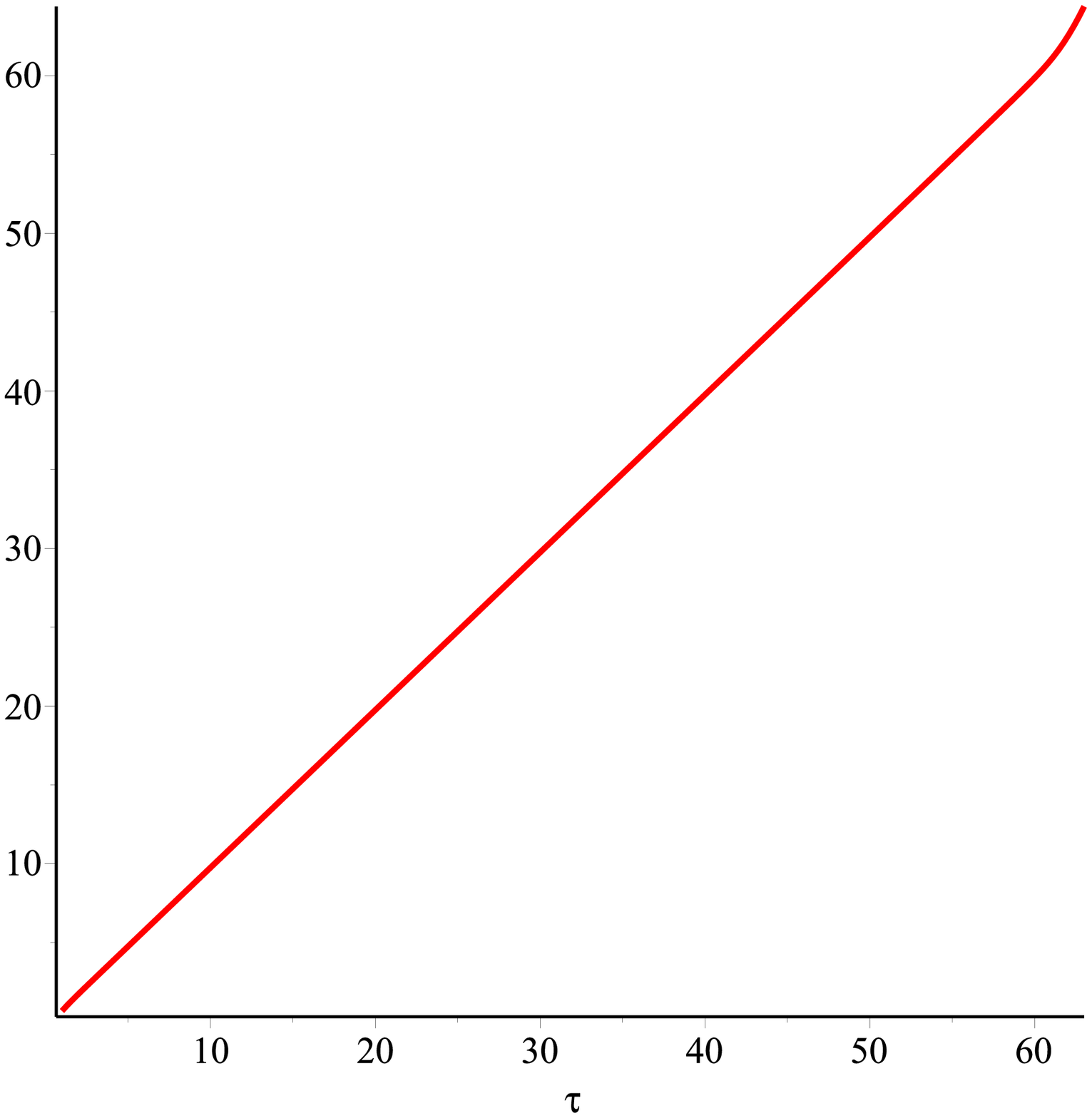, height=1.5in, width=1.5in}\qquad \quad &
\epsfig{file=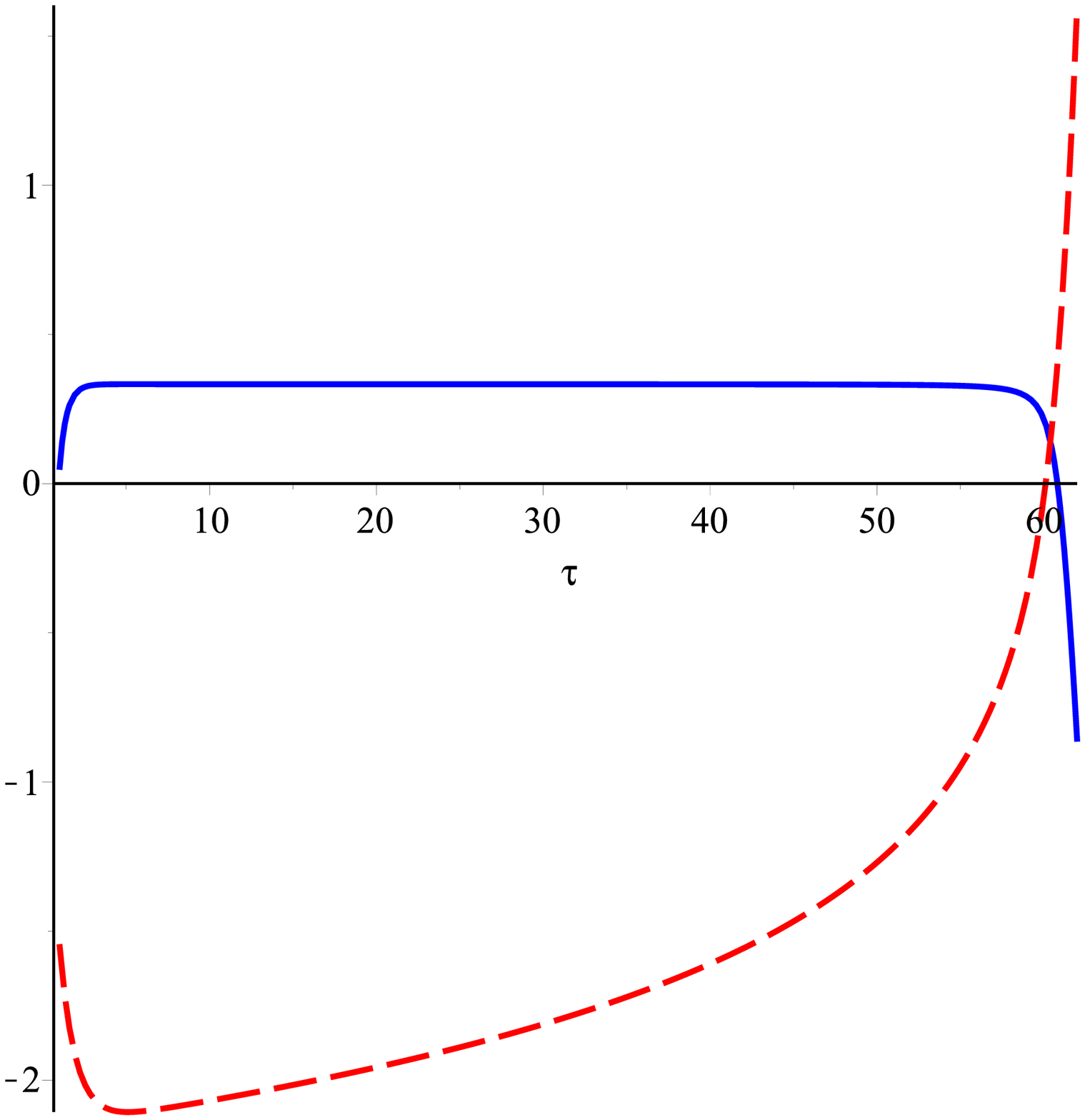, height=1.5in, width=1.5in}
\end{array}$
\end{center}
\caption{A typical numerical \emph{climbing} solution for the potential \eqref{quad_911} (left), in the gauge of Section \ref{sec:climbing}. The figure displays the scale factor ${\cal A}$ (center), the scalar field $\varphi$ (right, dashed) and the cosmic acceleration (right, continuous). In this four--dimensional example 60 $e$--folds of accelerated expansion terminate in a graceful exit when $\varphi$ climbs down and moves away toward $+\infty$.}
\label{step_potential}
\end{figure}

Let us now turn to the potential of eq.~\eqref{Wcomplex}, which we shall specialize to the form
\be
 {\cal V} \ = \ \arctan\left( e^{\,-\,2\,\varphi} \right) \ .\label{quad_911}
\ee
This potential never vanishes and describes essentially a finite potential step, which the scalar should be able to climb or descend during the cosmological evolution, and there are indeed two distinct classes of solutions that realize these options. Let us add that the effective $\gamma$,
\be
\gamma_{\,eff} \ = \ \frac{1}{2\, {\cal V}}\ \frac{\partial {\cal V}}{\partial \varphi} \label{quad_10}
\ee
tends to zero from negative values as $\varphi \to - \infty$ and to $-1$ from larger negative values as $\varphi \to \infty$. In other words, in the convenient gauge of Section \ref{sec:climbing} the scalar would be subject to an effective potential drive ${\cal V}^{\,\prime}/2\,{\cal V}$ that is essentially constant on both sides of the step, with a $\gamma_{\,eff}$ that is slightly negative on the left and negative and essentially critical on the right. Hence, depending on the initial conditions the scalar $\varphi$ can indeed emerge from $-\infty$ moving toward larger values, or alternatively it can emerge from $+\infty$ trying to climb up and then revert its motion somewhere, but it has no other option than proceeding indefinitely to the right at late epochs.

These types of behavior can thus be anticipated proceeding along the lines of Section \ref{sec:climbing}, but they can also be foreseen from the implicit solution by quadratures of the system \eqref{int4010} that, up to a rescaling of the time variable, takes the form of the contour integral
\be
t \ = \ \int_{a}^{z} \frac{dz}{\sqrt{\log\left(\frac{z}{z_0}\right)}} \  \label{quad_11} \ ,
\ee
to be defined in the complex plane cut along the line connecting the origin to the point $z_0$. The freedom in the choice of the lower limit $a$, where one of the two variables $x$ and $y$ is to vanish if the initial singularity occurs at $t=0$, reflects the behavior of $\varphi$ as it emerges from it. Recalling from Section \ref{sec:quadratures} that for this model
\be
\varphi \ = \ \frac{1}{2} \ \log \left( \frac{x}{y} \right) \ , \qquad {\cal A} \ = \ \frac{1}{2} \ \log \left( {x}{y} \right) \label{quad_1155} \ ,
\ee
one can readily conclude that a real (imaginary) $a$ implies that the scalar emerges from $+\infty$ ($-\infty$). In a similar fashion, the asymptotic slope of the contour for large $|z|$ determines the eventual fate of $\varphi$ at late epochs.

For brevity, in the following let us take $z_0=1$, up to a rescaling of the spatial metric in eq.~\eqref{FLRWgen}. One could thus envisage, a priori, scalar trajectories emerging from $\pm \infty$ and terminating at any point of the real axis, but the Hamiltonian constraint should be imposed, which in this case translates into the condition that the contour be chosen in such a way that $t$ remain real throughout the cosmological evolution. One can then show that, at any point of the $z$--plane, the slope of the integration contour is to be
\be
\frac{dy}{dx} \ = \ \frac{\arctan\left(\frac{y}{x} \right)}{\log\sqrt{x^2+y^2} \ + \ \sqrt{\left(\log\sqrt{x^2+y^2}\right)^2 \ + \ \big(\arctan\left(\frac{y}{x} \right)\big)^2}} \ , \label{quad_11555}
\ee
which clearly tends to zero for large values of $x$ and $y$, so that indeed $\varphi \to +\infty$. Notice that a cosmological term $\Lambda$ would displace $\frac{y}{x}$ in the argument of the inner $\arctan$ to $\frac{y}{x} + \alpha\, \Lambda$, with $\alpha$ positive. If $\Lambda>0$ the contour still flattens so that $x/y$ grows and $\varphi \to \infty$ at late times. However, if $\Lambda<0$ the slope at some point changes sign and $y$ is driven to zero, which signals the expected Big Crunch.

Alternatively, starting from the contour integral one can generate the asymptotic series
\be
t \ = \ z \, \int_{0}^{1} \frac{du}{\sqrt{\log(u)+\log\left(z\right)}}\ = \ \frac{z}{\sqrt{\log\left(z\right)}} \ \sum_{n=0}^\infty \ \frac{\Gamma\left(n+\frac{1}{2}\right)}{\Gamma\left(\frac{1}{2}\right)} \ \log\left(z\right)^{\,-\,n} \ , \label{quad_12}
\ee
which is clearly dominated by the first term for large values of $|z|$. Even in this fashion, one is led to conclude that real values of $t$ obtain at late times \emph{only} for contours that tend to become parallel to the real axis of the $z$-plane, which is tantamount to saying that $\varphi$ must inevitably tend to $+\infty$ at late epochs. This is precisely as expected from the shape of the potential.

One can actually turn the (complex) integrable system into the set of coupled first--order differential equations
\bea
\frac{dx}{dt} &=& \frac{\sqrt[4]{\left(\log\sqrt{x^2+y^2}\right)^2 \ + \ \big(\arctan\left(\frac{y}{x} \right)\big)^2}}{\sqrt{1 \ + \ \left( \frac{\arctan\left(\frac{y}{x} \right)}{\log\sqrt{x^2+y^2} \ + \ \sqrt{\left(\log\sqrt{x^2+y^2}\right)^2 \ + \ \big(\arctan\left(\frac{y}{x} \right)\big)^2}} \right)^2}} \ , \label{quad_13} \\
\frac{dy}{dt} &=& \frac{\arctan\left(\frac{y}{x} \right) \ \sqrt[4]{\left(\log\sqrt{x^2+y^2}\right)^2 \ + \ \big(\arctan\left(\frac{y}{x} \right)\big)^2}}{\sqrt{\left(\log\sqrt{x^2+y^2} \ + \ \sqrt{\left(\log\sqrt{x^2+y^2}\right)^2 \ + \ \big(\arctan\left(\frac{y}{x} \right)\big)^2} \right)^2 \ + \ \big(\arctan\left(\frac{y}{x} \right)\big)^2}} \, , \nonumber
\eea
the second of which is obtained combining the first with eq.~\eqref{quad_11555}, or alternatively one can recast this system in terms of the cosmic time $t_c$.

This model is remarkable, since it combines naturally an inflationary phase with an eventual graceful exit. Indeed, if the scalar emerges from the initial singularity moving in from $+ \infty$ and manages to climb up the step, it slows down on the plateau as a result of cosmological friction in the almost flat potential that it finds there, reverts its motion and proceeds eventually in fast roll toward $+\infty$. A similar behavior also occurs if the scalar moves in from $-\infty$ and undergoes slow roll before descending the step, and in both cases, as $\varphi$ lingers in the left region, the Universe can undergo several $e$--folds of accelerated expansion. A typical numerical solution displaying this type of behavior for a scalar climbing up from the right is shown in fig.~\ref{step_potential}. Interestingly, as we shall briefly explain in Section \ref{sec:orientifolds}, in four dimensions a power series in $e^{\,2\,\varphi}$ as in eq.~\eqref{quad_911} can emerge precisely from closed--string loop corrections, so that in principle these types of potentials can have a role in String Theory. Moreover, since
\be
\arctan\left( e^{\,-\,2\,\varphi} \right) \ = \ \frac{\pi}{2} \ - \arctan\left( e^{\,2\,\varphi} \right) \label{duality}
\ee
this combination of loop corrections enjoys a remarkable behavior under the inversion of its argument, which in String Theory would be naturally related to the string coupling, as we shall see in Section \ref{sec:orientifolds}.

Out next class of examples refers to the potentials of eq.~\eqref{int36555}, which we present here again for the reader's convenience,
\be
 {\cal V}\ = \ C_1 \ \Big(\, \cosh\,\gamma\,\varphi \, \Big)^{\, 2\, \left(\frac{1}{\gamma} \ - \ 1 \right)}\ + \ C_2 \ \Big(  \, \sinh\,\gamma\,\varphi \,\Big)^{\, 2\, \left(\frac{1}{\gamma} \ - \ 1 \right)}\ . \label{quad_36555}
\ee
Depending on the choice made for the real exponent $\gamma$, these potentials can describe barriers or wells of different shapes, and the presence of the second term restricts in general the domain to the region $\varphi>0$. For the sake of brevity and simplicity, we shall concentrate on a special but very significant case of potential wells, with $\gamma=\frac{1}{3}$, which affords relatively handy solutions in terms of elliptic functions. The potentials that we would like to discuss here in detail are thus
\be
W \ = \ C_1 \, \left( \cosh \frac{\varphi}{3} \right)^4 \ + \ C_2 \, \left( \sinh \frac{\varphi}{3} \right)^4 \ = \ C_1 \left[ \left( \cosh \frac{\varphi}{3} \right)^4 \ + \ \epsilon\, \left( \sinh \frac{\varphi}{3} \right)^4 \right] \ , \label{quad_15}
\ee
with $|\epsilon|<1$ in order to exclude a pathological behavior for large $|\varphi|$. Notice that positive values of $\epsilon$ tend to squeeze the well, while negative ones tend to widen it. In terms of $\xi$ and $\eta$
\be
(x\,y)^{\, 2}\ W \ = \ C_1 \left( \xi^{\,4} \ + \ \epsilon \ \eta^{\,4}\right) \ , \label{quad_16}
\ee
so that the starting point is provided in this case by the two conservation laws
\bea
&& \dot{\xi}^2 \ - \ \frac{2}{9} \ C_1 \ \xi^4 \ = \ A \ , \nonumber \\
&& \dot{\eta}^2 \ + \ \frac{2}{9} \ C_1 \ \epsilon \ \eta^4 \ = \ A \ , \label{quad_17}
\eea
where the two ``total energies'' are bound to coincide because of the Hamiltonian constraint \eqref{int3613}. The relations
\be
e^{\,2\,\gamma\, {\cal A}} \ = \ \xi^{\,2} \ - \ \eta^{\,2} \ , \qquad e^{\,2\,\gamma\, {\varphi}} \ = \ \frac{\xi \ + \ \eta}{\xi \ - \ \eta}
\ee
were already presented in Section \ref{sec:quadratures}, and connect $\xi$ and $\eta$ to the physical variables ${\cal A}$ and $\varphi$.

The solutions for $\xi$ and $\eta$ can be expressed in terms of Jacobi elliptic functions with imaginary modulus $k$, and for $\epsilon>0$ a possible choice corresponding to the scalar initially climbing down the right end of the well is
\bea
&& \xi \ = \ e^{\,-\,i\frac{\pi}{4}}\ \left( \frac{9\,A}{2\,C_1}\right)^\frac{1}{4} \ {\rm sn} \left[ t\, e^{\,i\frac{\pi}{4}}\, \left(\frac{2\,A\,C_1}{9}\right)^\frac{1}{4} \, , \, i \right] \ , \nonumber \\
&& \eta \ = \ \left( \frac{9\,A}{2\,\epsilon\,C_1}\right)^\frac{1}{4} \ {\rm sn} \left[ t\, \left(\frac{2\,A\,\epsilon\,C_1}{9}\right)^\frac{1}{4} \, , \, i \right]  \ , \label{quad_18}
\eea
where the second function generalizes a $\sin$ function, while the first is also real and generalizes a $\tan$ function. On the other hand for $\epsilon<0$ both $\xi$ and $\eta$ behave as $\xi$ above, so that
\bea
&& \xi \ = \ e^{\,-\,i\frac{\pi}{4}}\ \left( \frac{9\,A}{2\,C_1}\right)^\frac{1}{4} \ {\rm sn} \left[ t\, e^{\,i\frac{\pi}{4}}\, \left(\frac{2\,A\,C_1}{9}\right)^\frac{1}{4} \, , \, i \right] \ , \nonumber \\
&& \eta \ = \ e^{\,-\,i\frac{\pi}{4}}\ \left( \frac{9\,A}{2\,|\epsilon|\,C_1}\right)^\frac{1}{4} \ {\rm sn} \left[ t\, e^{\,i\frac{\pi}{4}}\, \left(\frac{2\,A\,|\epsilon|\,C_1}{9}\right)^\frac{1}{4} \, , \, i \right]  \ , \label{quad_18_neg}
\eea
Let us also recall that, in terms of Jacobi $\vartheta$--functions,
\be
{\rm sn}(u,k) \ = \ \frac{\vartheta_3(0|\tau)}{\vartheta_2(0|\tau)} \ \frac{\vartheta_1\left( \frac{u}{\vartheta_3^2(0|\tau)}\Big| \tau \right)}{\vartheta_4\left( \frac{u}{\vartheta_3^2(0|\tau)}\Big| \tau \right)} \ , \label{int3659}
\ee
where the modulus $\tau$ of the associated torus is defined implicitly in terms of $k$ by the relation
\be
k \ = \ \frac{\vartheta_2^2(0|\tau)}{\vartheta_3^2(0 | \tau)} \ . \label{int3612}
\ee

The solutions that we have displayed describe a scalar $\varphi$ that emerges from the initial singularity descending the right end of the wells and comes to the extremum, which lies at $\varphi=0$ for the whole class of potentials at stake, within a finite interval of the parametric time $t$, or asymptotically
as $t_c\to\infty$. This is the familiar behavior for extrema corresponding to positive values of the potentials, and of course another type of solutions, where the scalar descends along the left end, also exists, which can be simply obtained from these reversing the sign of $\eta$.
\begin{figure}[h]
\begin{center}$
\begin{array}{ccc}
\epsfig{file=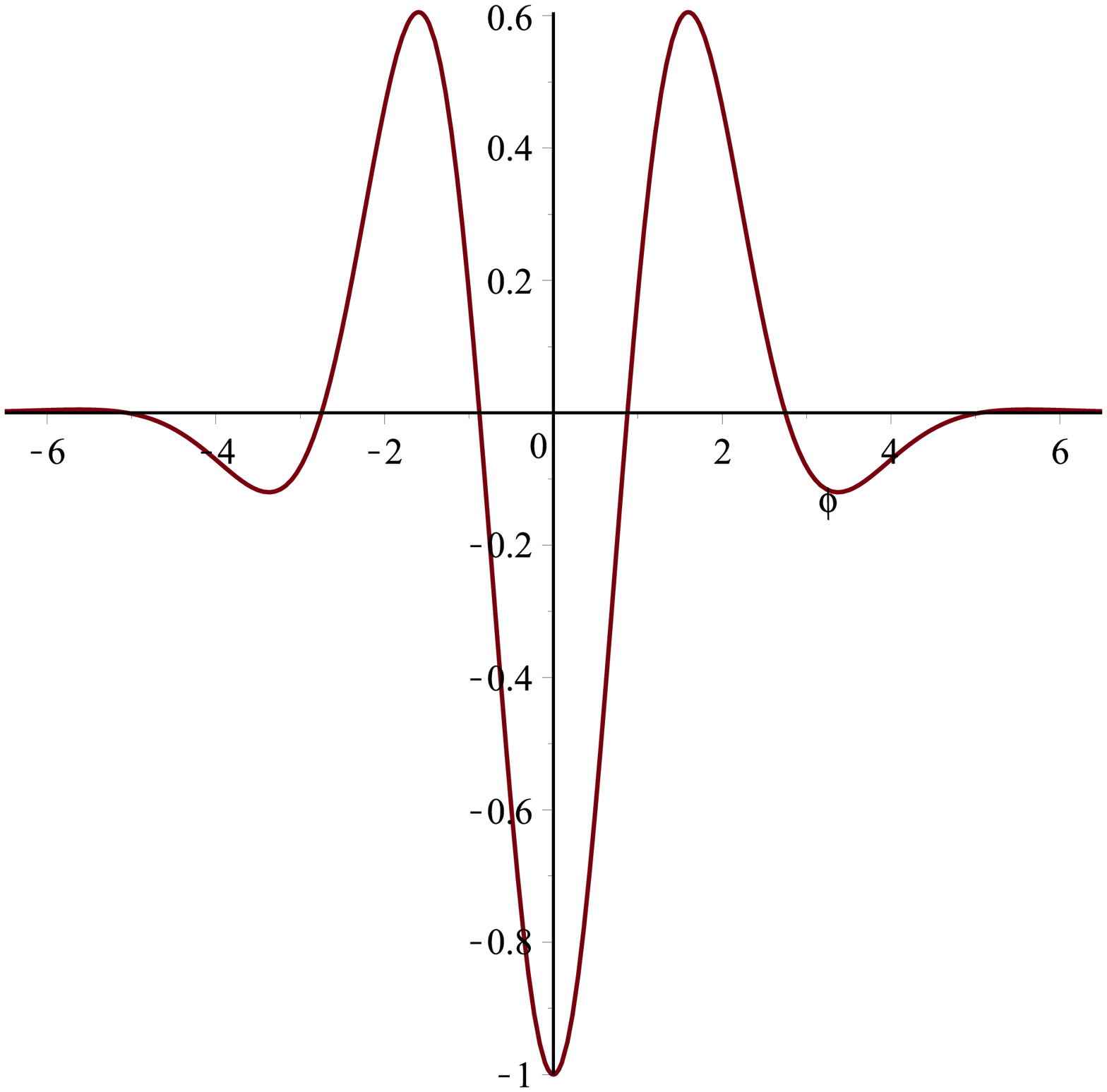, height=1.5in, width=1.5in}\qquad \quad &
\epsfig{file=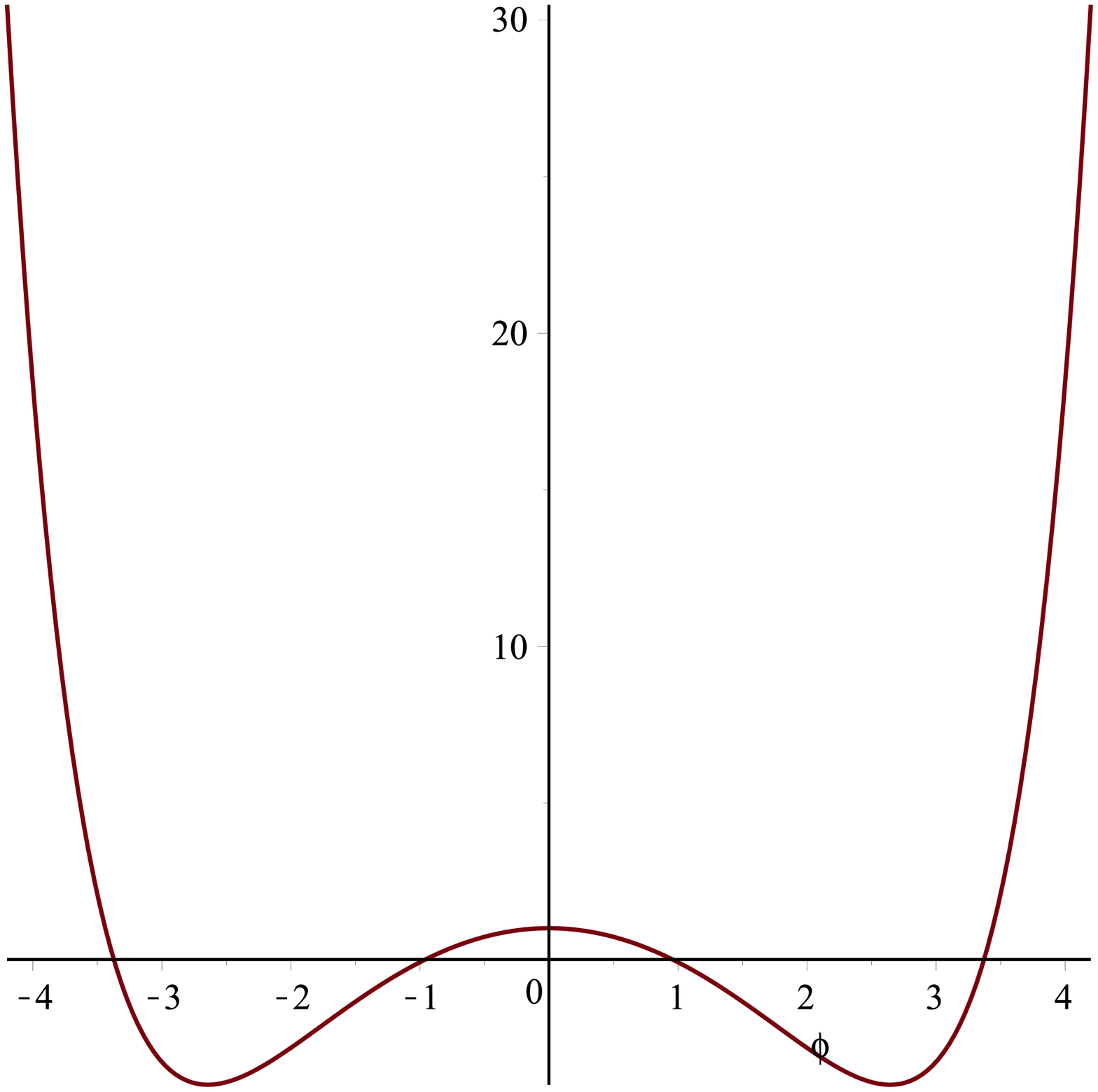, height=1.5in, width=1.5in}\qquad \quad &
\epsfig{file=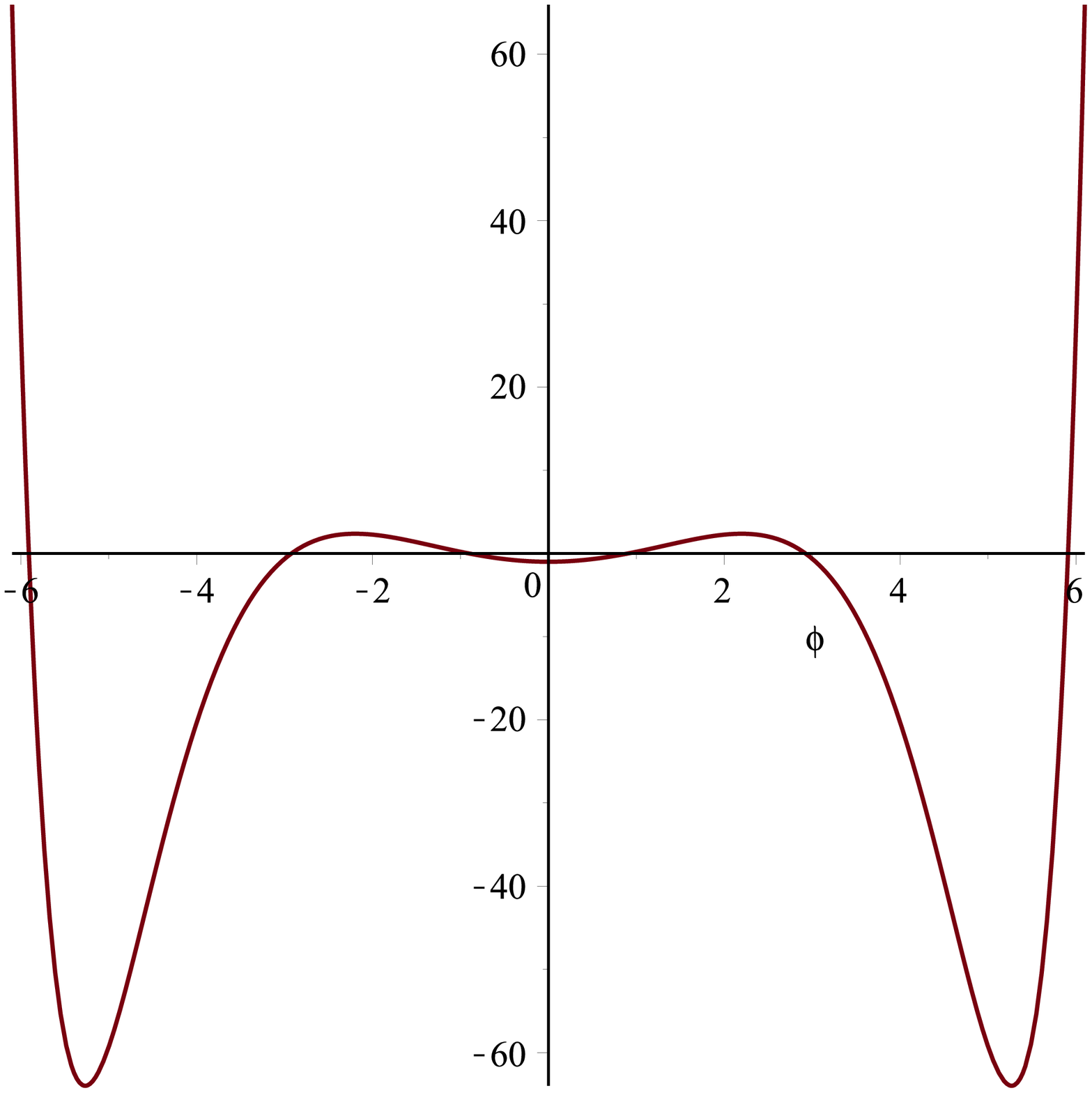, height=1.5in, width=1.5in}
\end{array}$
\end{center}
\caption{Some instances of the stable even potentials that can emerge from eq.~\eqref{Wcomplex2} for $\gamma=\frac{1}{2\,n}$, displayed here for the three cases $n=-5,3,4$. There are always negative extrema where the systems will try to settle before ending up in a Big Crunch.}
\label{multicritical}
\end{figure}

In discussing the class of potentials described by eq.~\eqref{Wcomplex2}, we shall leave aside once more pathological cases in which they are unbounded from below as the field grows to arbitrarily large (positive or negative) values. The interesting subclass of potentials that fulfill this condition are then even functions that arise for special ranges of $\gamma$, where generically they have a number of critical points with \emph{negative} cosmological constant. We refrain from analyzing these examples in detail, since we already know that whenever such extrema are present, as the scalar tries to settle there the expansion stops and leaves way to an eventual Big Crunch, but some of these interesting potentials are displayed in fig.~\ref{multicritical}.

Our last class of examples is very instructive, since their potentials
\be
W = \lambda \left[ e^{\,2\,\gamma\,\vf} \ + \ e^{\,\frac{2}{\gamma}\,\vf}\right] \ , \label{quad_25}
\ee
with $\lambda>0$, where one can clearly restrict the attention to the case $0<\gamma<1$, combine an (over)critical term with an (under)critical one and the solutions are relatively simple.  These are the most general potentials in eq.~\eqref{Liouville3} whenever the two coefficients $C_1$ and $C_2$ are both positive, up to shifts of $\varphi$, and we confine our attention to this sub--class for brevity, since other choices bring about the types of pathologies that we have already come across. For low enough $\gamma$, the first term in eq.~\eqref{quad_25} can drive an inflationary phase, so that the key qualitative features of the models are along the lines of the semi--analytic treatment in \cite{dkps}, or of the class of models discussed at the beginning of Section \ref{sec:Bnot0_sol}.

The solution by quadratures of this class of models rests on the gauge choice \eqref{Liouville2} and on a ``boost'' that turns the Lagrangian into a separable form. The resulting equations of motion lead to the ``energy conservation'' relations
\bea
&& {\dot{\widehat{\cal A}}}^{\,2} \ = \ 2\, \lambda \, \left( e^{\,2\,\widehat{\cal A}\, \sqrt{1\,-\,\gamma^{\,2}}} \ + \ e^{\,2\,\widehat{\cal A}_0\, \sqrt{1\,-\,\gamma^{\,2}}} \right) \ , \nonumber \\
&& {\dot{\widehat{\varphi}}}^{\,2} \ = \ 2\, \lambda\, \left( e^{\,2\,\widehat{\varphi}_0 \, \sqrt{\frac{1}{\gamma^{\,2}}\,-\,1}} \ - \ e^{\,2\,\widehat{\varphi} \, \sqrt{\frac{1}{\gamma^{\,2}}\,-\,1}} \right) \ , \label{quad_26}
\eea
so that $\widehat{\varphi}$ is bound to be less than $\widehat{\varphi}_0$. This is the manifestation of the climbing phenomenon: $\widehat{\varphi}$ cannot proceed along the steep potential beyond a certain point. As in previous examples, however, the two constants ${\widehat{\cal A}}_0$ and $\widehat{\varphi}_0$ are not independent, but
\be
\widehat{\cal A}_0 \ = \ \frac{1}{\gamma} \ \widehat{\varphi}_0 \label{quad_27}
\ee
on account of the Hamiltonian constraint \eqref{Liouville7}. Letting
\bea
&& X \, = \, (\widehat{\cal A} \, - \, \widehat{\cal A}_0) \, \sqrt{1 \,- \, \gamma^{\,2}} \ , \nonumber \\
&& Y \, = \, (\widehat{\varphi} \, - \, \widehat{\varphi}_0) \, \sqrt{\frac{1}{\gamma^{\,2}}\,-\,1} \ , \nonumber \\
&& \omega^2 \, =\, 2\, \lambda \, e^{\,2\, \widehat{\cal A}_0 \, \sqrt{1 \,- \, \gamma^{\,2}}}\ \left( \frac{1}{\gamma^{\,2}}\,-\,1\right) \ , \label{quad_28}
\eea
eqs.~\eqref{quad_26} take the neat form
\bea
&& \dot{X}^2 \, = \, \omega^2\, \gamma^{\,2} \left( 1 \ + \ e^{\,2\,X} \right) \ , \nonumber \\
&& \dot{Y}^2 \, = \, \omega^2\left( 1 \ - \ e^{\,2\,Y} \right) \ . \label{quad_28_1}
\eea
while the Hamiltonian constraint becomes
\be
\dot{X}^2 \ - \ \gamma^{\,2}\, \dot{Y}^2 \ = \ \omega^{\,2}\, \gamma^{\,2} \Big( e^{\,2\,X} \ + \ e^{\,2\,X} \Big) \ . \label{quad_29}
\ee
The solutions of eqs.~\eqref{quad_28_1} can be obtained by quadratures, which can be conveniently inverted so that finally
\be
e^{\,{\cal A}} \ = \ e^{\, {\cal A}_0}\ \frac{\cosh\big[\omega\,\left(t\, - \, t_{\widehat{\varphi}}\right)\big]^{\,\frac{\gamma^{\,2}}{1 \,- \, \gamma^{\,2}}}}{\sinh\big[\omega\,\gamma\,\left(t\, - \,t_{\widehat{\cal A}}\right)\big]^{\,\frac{1}{{1 \,- \, \gamma^{\,2}}}}} \label{quad_33}
\ee
and
\be
e^{\,\vf} \ = \ e^{\, \vf_0}\ \frac{\sinh\big[\omega\,\gamma\,\left(t\, - \, t_{\widehat{\cal A}}\right)\big]^{\,\frac{\gamma}{1 \,- \, \gamma^{\,2}}}}{\cosh\big[\omega\,\left(t\, - \, t_{\widehat{\varphi}}\right)\big]^{\,\frac{\gamma}{{1 \,- \, \gamma^{\,2}}}}} \ , \label{quad_34}
\ee
where one can choose $t_{\widehat{\cal A}}=0$ with no loss of generality.
\begin{figure}[h]
\begin{center}$
\begin{array}{cc}
\epsfig{file=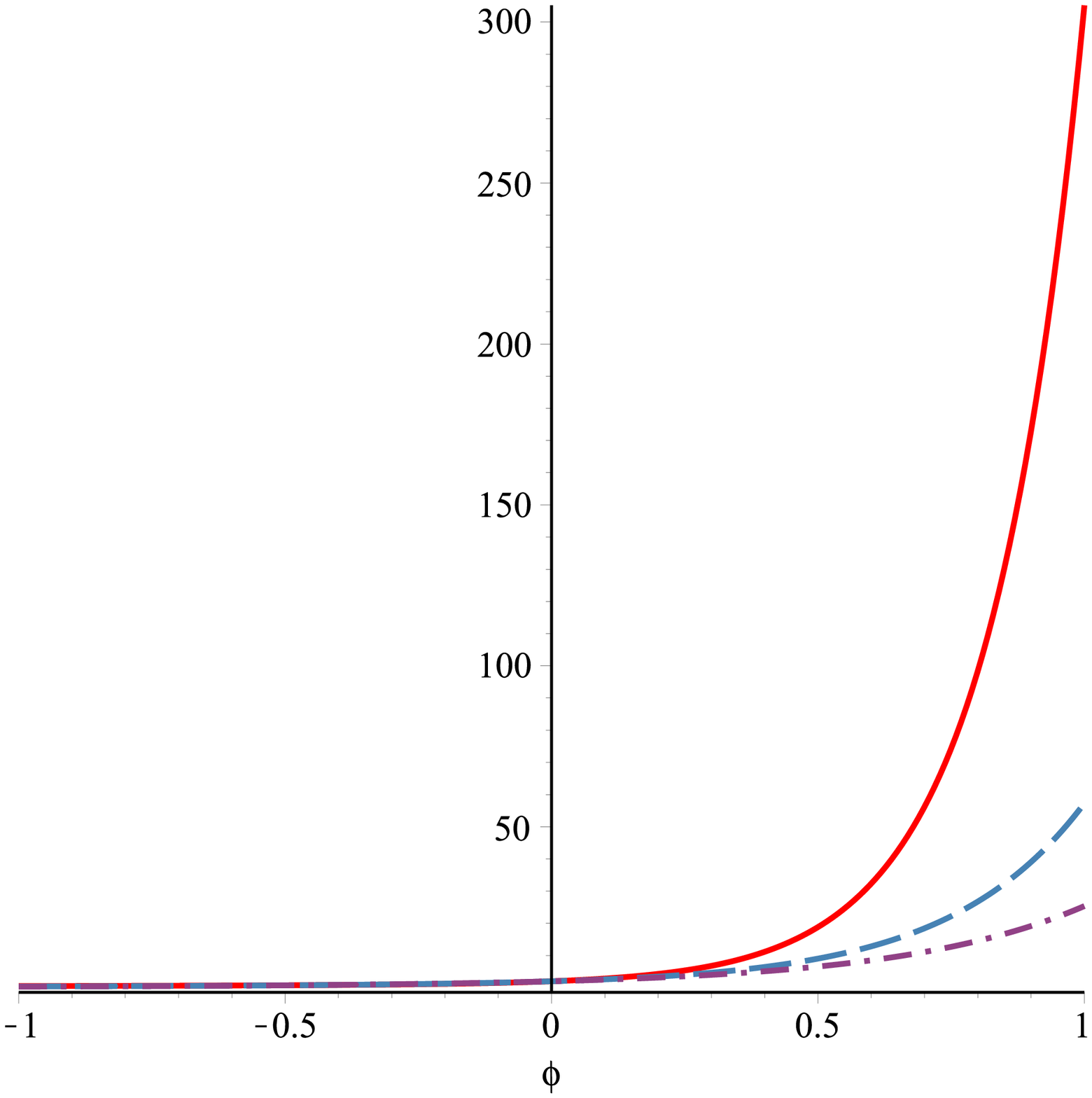, height=1.5in, width=1.5in}\qquad \qquad &
\epsfig{file=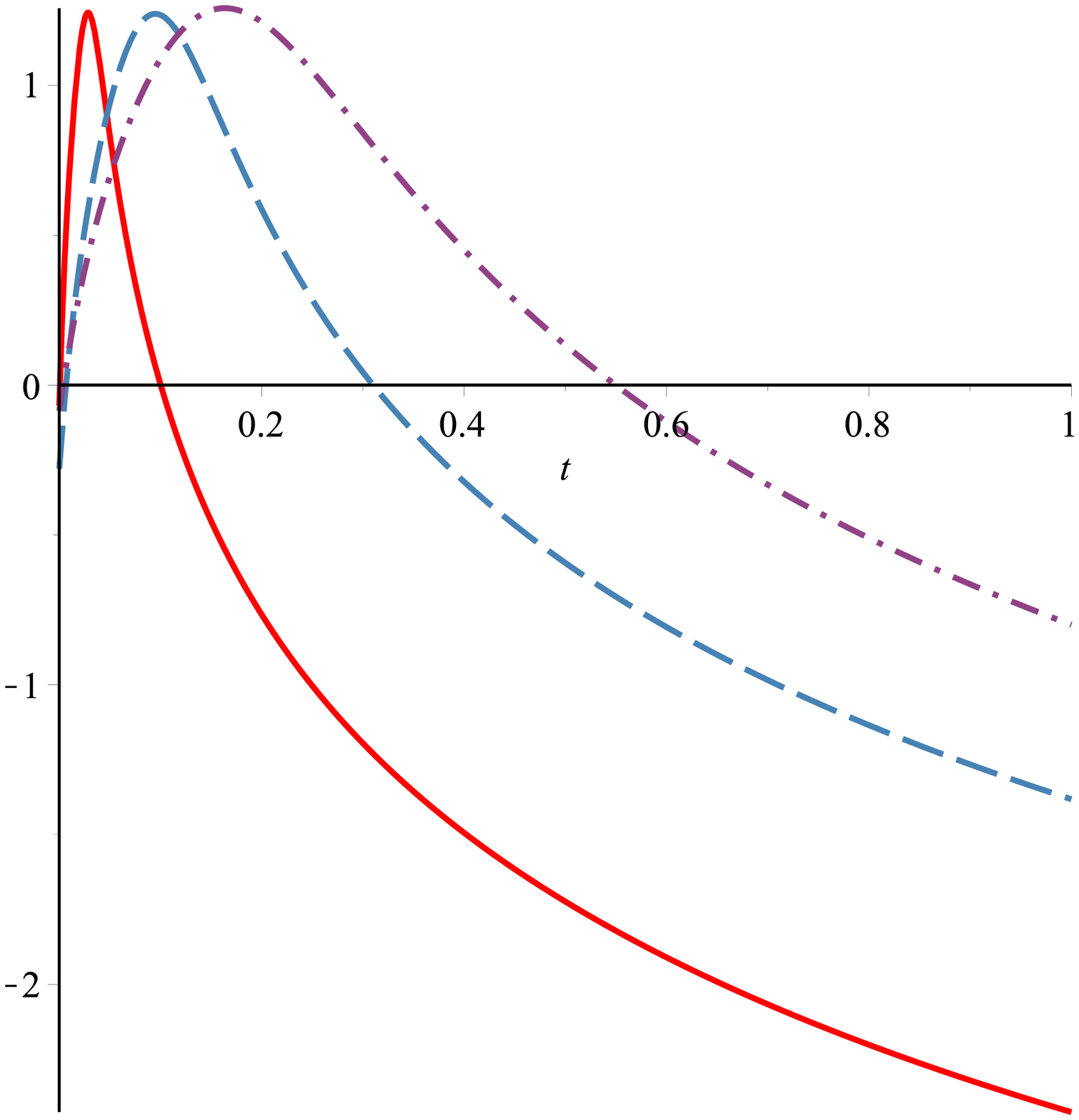, height=1.5in, width=1.5in}
\end{array}$
\end{center}
\caption{Three potentials of eq.~\eqref{quad_25} (left diagram), with $\gamma=0.35$ (red, continuous), $\gamma=0.5$ (blue, dashed) and $\gamma=0.65$ (purple, dashed-dotted) and the corresponding behavior of $\varphi$ in cosmic time for the same penetration in the exponential barrier (right diagram). The scalar is bound to emerge from the initial singularity while climbing up the potential, but in the presence of a steeper barrier it equilibrates more slowly to its attractor curve $\varphi \sim \frac{1}{\gamma} \log t_c$ during the descent, which in the first two examples supports an inflationary phase in four dimensions.}
\label{liouville_potentials}
\end{figure}

In order to properly interpret these expressions, let us take a closer look at the relation between $t$ and the actual cosmological time $t_c$, determined by
\be
dt_c \ = \ \pm e^{\,{\cal A}}\, dt \ = \ \pm \ e^{\, {\cal A}_0}\ \frac{\cosh\big[\omega\,\left(t\, - \, t_{\widehat{\varphi}}\right)\big]^{\,\frac{\gamma^{\,2}}{1 \,- \, \gamma^{\,2}}}}{\sinh\big[\omega\,\gamma\,t \big]^{\,\frac{1}{{1 \,- \, \gamma^{\,2}}}}}\ dt \ . \label{quad_37}
\ee
There is a subtlety here: one must pick the ``minus'' sign, so that large positive values of $t$ translate into early cosmological epochs, close to the initial singularity, while small values of $t$ translate into large values of $t_c$, and thus into late epochs. Indeed with the sign choice in \eqref{quad_37}, for small values of $t$ where
\be
e^{\,{\cal A}} \ \sim \ t^{\,-\, \frac{1}{1 \,- \, \gamma^{\,2}}} \ , \quad e^{\,\vf} \ \sim \ t^{\,\frac{\gamma}{1 \,- \, \gamma^{\,2}}} \ ,
\ee
one finds that
\be
t_c \ \sim \ - \ \int e^{\,{\cal A}} \, dt \ \sim \ t^{\,-\, \frac{\gamma^{\,2}}{1 \,- \, \gamma^{\,2}}} \ .
\ee
Expressing $e^{\,{\cal A}}$ and $e^{\,\vf}$ in terms of $t_c$ now gives
\be
e^{\,{\cal A}} \ \sim \ t_c^{\,\frac{1}{\gamma^{\,2}}} \ , \quad e^{\,\vf} \ \sim \ t_c^{\, -\, \frac{1}{\gamma}} \ ,
\ee
which are precisely the equations describing the behavior at large times in a ``mild'' exponential potential $e^{\,2\,\gamma\,\vf}$ reviewed in Section \ref{sec:climbing}, once one recalls that ${\cal A}=(d-1)A$ and specializes the discussion to the $d=4$ case. In a similar fashion the opposite limit, $t \to \infty$, corresponds to the behavior of the scalar as it climbs up the ``mild'' potential $e^{\,2\gamma\,\vf}$ right after the initial singularity. Indeed, the limiting behavior is now
\be
e^{\,{\cal A}} \ \sim \ e^{\,\vf} \ \sim e^{\,-\,\frac{\gamma\,\omega\,t}{\gamma\,+\,1}} \ ,
\ee
so that
\be
t_c \ \sim \ - \ \int e^{\,-\,\frac{\gamma\,\omega\,t}{\gamma\,+\,1}} \ \sim \ e^{\,-\,\frac{\gamma\,\omega\,t}{\gamma\,+\,1}} \ ,
\ee
where the integration constant was chosen so that as $t \to \infty$ $t_c \to 0$. Hence
\be
e^{\,{\cal A}} \ \sim \ e^{\,\vf} \ \sim t_c \ ,
\ee
a result that we saw already in Section \ref{sec:climbing}.

All in all, this class of models describes, in a relatively handy analytic form but at the price of a counter--intuitive relation between $t$ and $t_c$, spatially flat cosmologies with a climbing scalar that impinges on a ``hard'' exponential, reverts its motion and, if $\gamma < \frac{1}{\sqrt{d-1}}$, drives an inflationary phase during its final descent.
\section{\sc  Additional Integrable Potentials}

In addition to the cases introduced in Section \ref{sec:integrable}, the most significant of which were solved in detail in Section \ref{sec:properties_exact}, we have identified 26 sporadic one--field potentials that lead to integrable cosmologies. Their emergence rests on the transformations
\be
\xi \ = \ \frac{1}{2} \left( e^{\,\gamma\,(\mathcal{A}+\vf)} \ + \ e^{\,\gamma\,(\mathcal{A}-\vf)} \right) \ , \quad \eta \ =\ \frac{1}{2} \left( e^{\,\gamma\,(\mathcal{A}+\vf)} \ - \ e^{\,\gamma\,(\mathcal{A}-\vf)} \right) \ , \quad \mathcal{B} \ = \ \mathcal{A}\,(1\,-\,2\,\gamma)  \label{spor_1}
\ee
introduced in Section \ref{sec:quadratures} and on special choices for the parameter $\gamma$. The corresponding Lagrangians (\ref{onefield}) can always be recast in the form
\begin{equation}\label{carolinus}
    \mathcal{L} \ = \ - \ \frac{1}{2} \ \left( \dot{\xi}^2 \, - \, \dot{\eta}^2\right) \ - \ \mathcal{V}_c(\xi, \eta) \ ,
\end{equation}
where the potential functions
\be
\mathcal{V}_c(\xi,\eta) \ = \ \gamma^{\,2} \, \left( \xi^2 \, - \, \eta^2 \right)^{\left(\frac{1}{\gamma} \, - \, 1 \right)} \ \mathcal{V}(\varphi) \ ,
\ee
correspond to a list of sporadic two--field dynamical systems that are Liouville integrable thanks to the existence of a conserved charge $Q$ in addition to the Hamiltonian $H$. The compilation of this list was a major mathematical achievement resulting from the work of several authors (see the review papers \cite{hietarinta, perelomov, conte} and references therein), and here we are translating these results into the cosmological setting. It is indeed remarkable that these integrable dynamical systems can be mapped into cosmological models with a potential ${\mathcal V}(\varphi)$ that depends only on exponentials of the scalar field $\varphi$.

In Section \ref{sec:sporadic_pot} we list the 26 $\mathcal{V}(\varphi)$, the corresponding $\mathcal{V}_c(\xi,\eta)$ and the additional conserved charge that exists in each of these models and guarantees its Liouville integrability. We also display plots of the most significant potentials and add some qualitative remarks on the corresponding solutions that can be anticipated on the basis of the results of Section \ref{sec:properties_exact}. We conclude in Sections \ref{sec:trigonometric_pot} and \ref{sec:toda_pot}, where we display some links that exist, up to analytic extensions, between the cosmological systems and trigonometric potentials or Toda systems.

\subsection{\sc Sporadic integrable potentials}\label{sec:sporadic_pot}

Let us now turn to a description of the 26 sporadic potentials and of the corresponding integrable cosmological models. The intuition provided by the explicit analysis performed in Section \ref{sec:properties_exact} will suffice to capture the key qualitative features of all the interesting models whose potentials are bounded from below, although obtaining explicit solutions would be much harder in general for these sporadic models. For the sake of order and clarity, we shall split the 26 cases into groups that can be turned into the canonical form of eq.~\eqref{carolinus} with the same value of the parameter $\gamma$ by the transformation \eqref{spor_1}.
\begin{figure}[h]
\begin{center}$
\begin{array}{ccc}
\epsfig{file=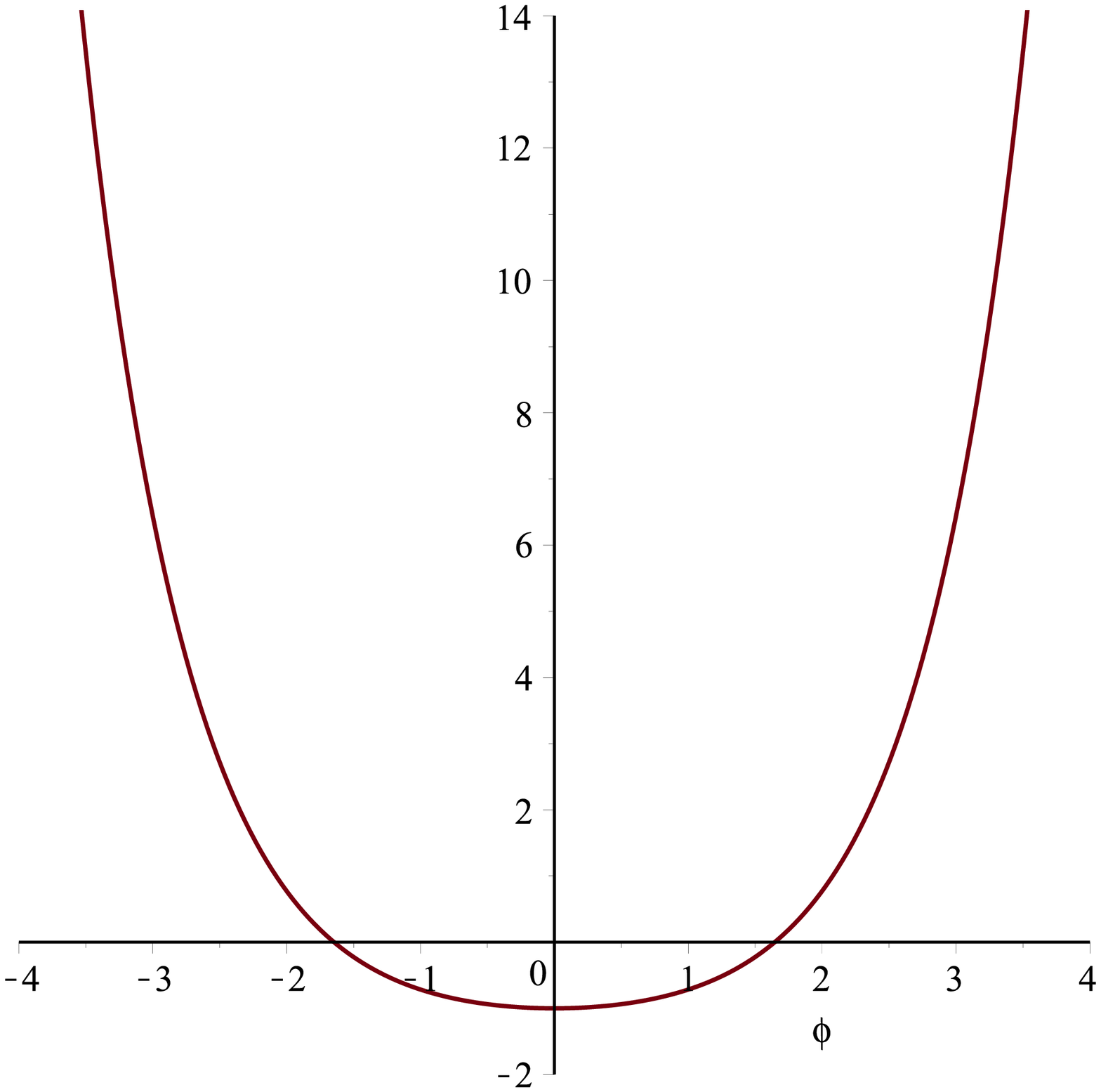, height=1.5in, width=1.5in}\qquad \quad &
\epsfig{file=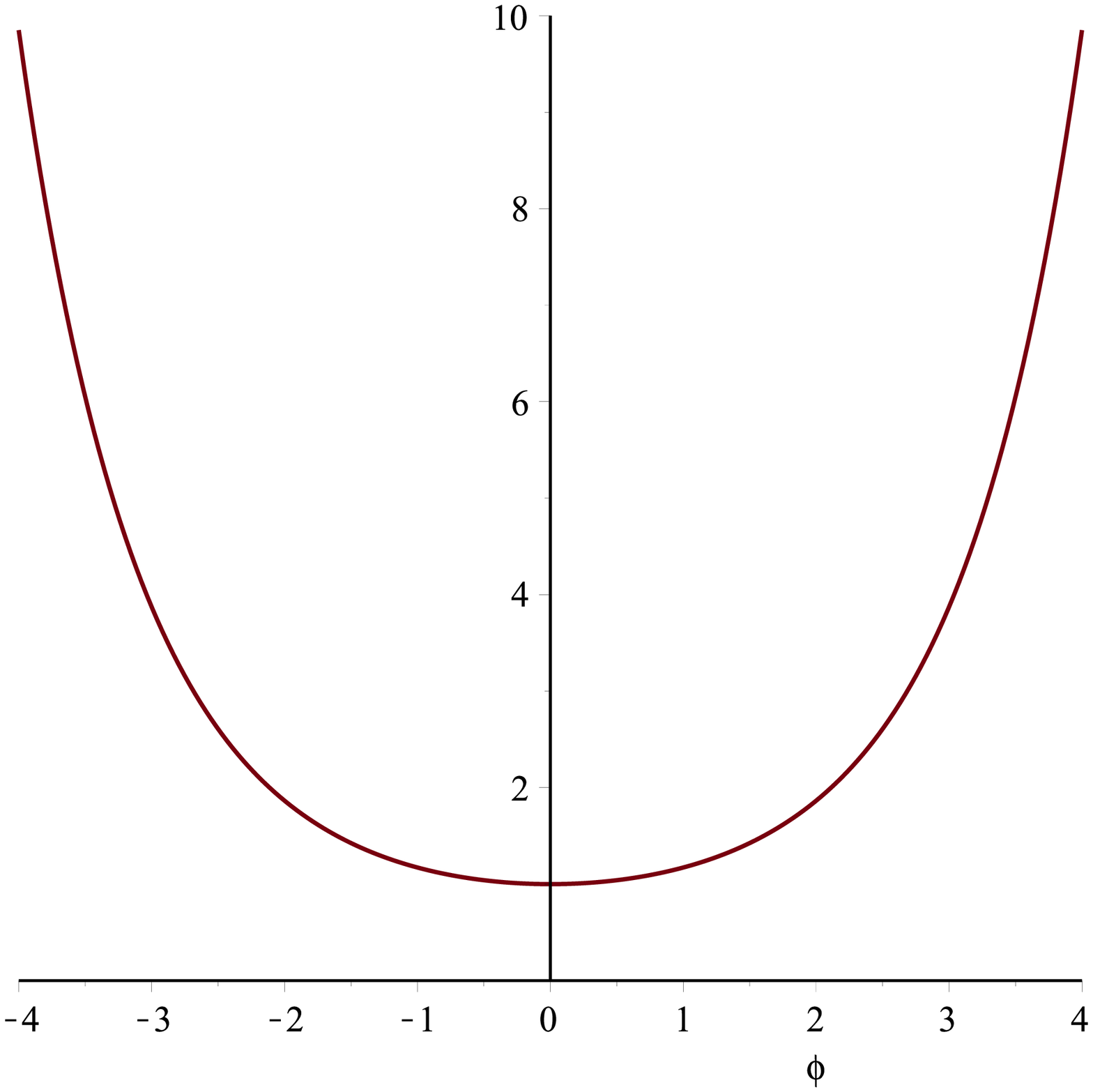, height=1.5in, width=1.5in}\qquad \quad &
\epsfig{file=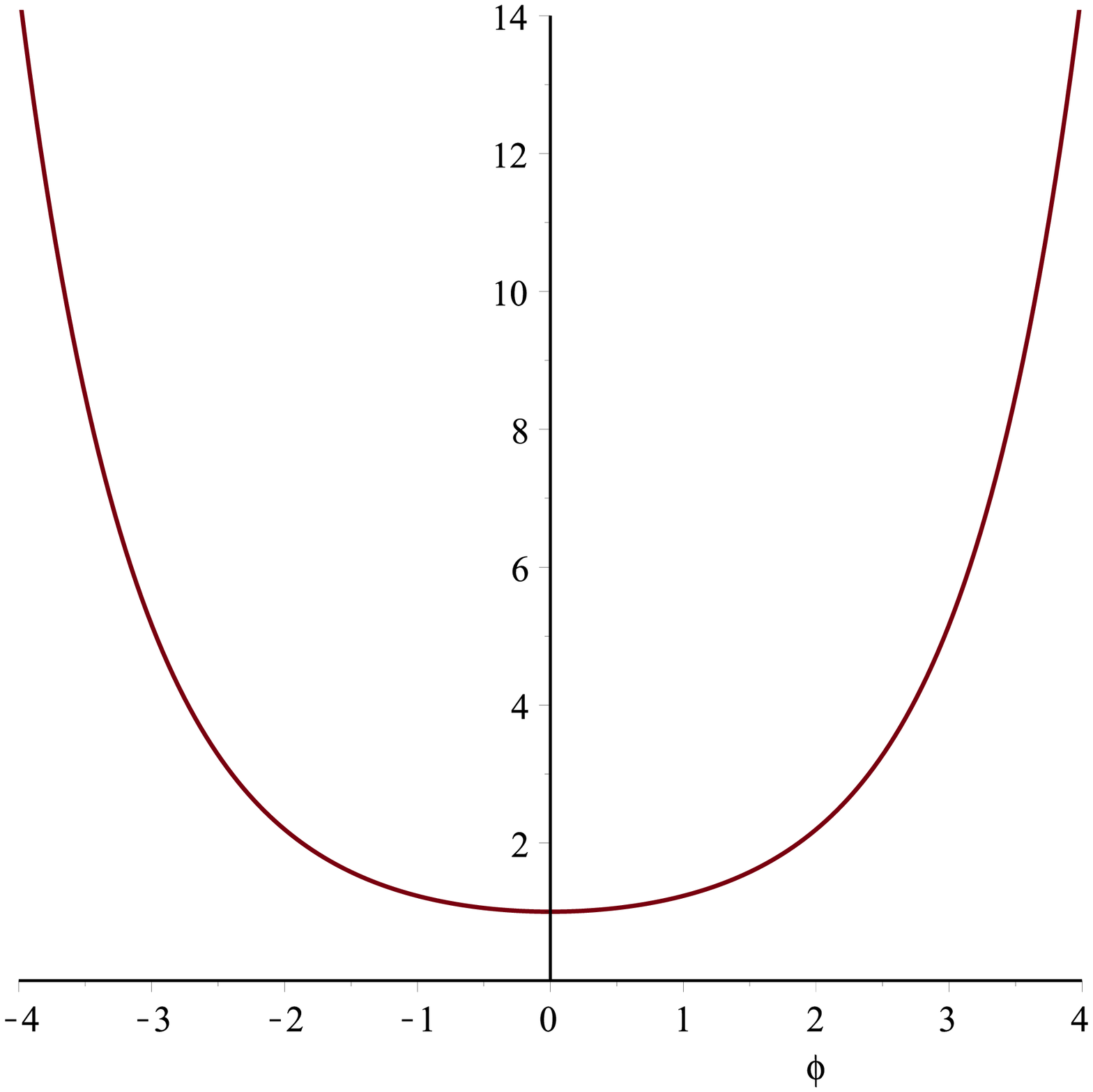, height=1.5in, width=1.5in}
\end{array}$
\end{center}
\caption{The three potentials of eq.~\eqref{GIa} with $\lambda < 0$ in the first case and $\lambda>0$ in the others. The analysis of Section \ref{sec:properties_exact} indicates that in the first model the Universe will experience a Big Crunch while in the others it will readily enter a de Sitter phase with the scalar stabilized at the bottom.}
\label{v1a_potentials}
\end{figure}
 \paragraph{Group Ia $\left(\gamma \, = \, \frac{2}{5}\right)$} In this case the scalar potentials are
 \begin{eqnarray}\label{GIa}
  \mathcal{V}_{Ia}(\varphi) & = &  \lambda \left[ a \ \cosh ^3\left(\frac{2 \varphi
   }{5}\right)\ +\ b\ \sinh
   ^2\left(\frac{2 \varphi
   }{5}\right) \cosh \left(\frac{2
   \varphi }{5}\right)\right] \nonumber\\
   & = & \frac{\lambda}{4} \left[(a+b)
   \cosh\left(\frac{6\varphi}{5}\right)+(3 a-b)
   \cosh\left(\frac{2\varphi}{5}\right)\right]
    \ ,
 \end{eqnarray}
where $\lambda$ is an overall scale, so that
 \begin{equation}\label{GIac}
  {\mathcal V}_{c,\, Ia}(\xi,\eta) \, = \,  \frac{4}{25}\ \lambda \, \Big[ a\ \xi^3 \ + \ b\ \xi\ \eta^2 \Big] \ .
 \end{equation}
There are three independent choices for the parameters $a$ and $b$,
 \begin{equation}\label{GIapar}
    \left\{a,b\right\} \, = \, \left\{
\begin{array}{cc}
 1  & -3  \\
 1  & -\frac{1}{2} \\
 1  & -\frac{3}{16}
\end{array}
\right\} \ ,
 \end{equation}
and selecting appropriately the sign of $\lambda$ the even potentials $\mathcal{V}_{Ia}$, displayed in fig.~\ref{v1a_potentials}, are always bounded from below.

The Liouville integrability of the systems corresponding to eq.~(\ref{GIapar}) is guaranteed by the existence of an additional conserved charge (see \emph{e.g.} \cite{hieta2} and references therein), which in the three cases takes the form
 \bea\label{AddInt1}
&& {\mathcal Q}_{\,Ia}^{(1)}(\xi,\eta) \, = \, 25\,\dot{\xi}\, \dot{\eta} \ + \ 4\, \lambda \, \eta\,\Big( \eta^2 \ - \ 3\, \xi^2 \Big)\ , \nonumber\\
&& {\mathcal Q}_{\,Ia}^{(2)}(\xi,\eta) \, = \, 50\, \dot{\eta}\,
\Big(\eta \,\dot{\xi} \ - \ \dot{\eta}\,\xi \Big) \ + \ \lambda \, \eta^2 \, \Big( \eta^2 \ - \ 4\, \xi^2\ \Big)\ , \\
&& {\mathcal Q}_{\,Ia}^{(3)}(\xi,\eta)) \, = \,  \dot{\eta}^4 \ + \  \frac{\lambda}{25} \, \dot{\eta}\, \eta^2\,\Big(
\eta \,\dot{\xi} \ - \ 3\, \dot{\eta}\,\xi  \Big)\ + \ \frac{\lambda^2}{5000} \, \eta^4\,\Big( \eta^2 \ - \ 6\, \xi^2 \Big)\ . \nonumber
\eea

\paragraph{Group Ib $\left(\gamma \, = \, \frac{2}{5}\right)$} In this case the scalar potentials are
 \begin{eqnarray}\label{GIb}
  \mathcal{V}_{Ib}(\varphi) & = &  \lambda \left[ a \sinh ^3\left(\frac{2 \varphi
   }{5}\right)+b \cosh
   ^2\left(\frac{2 \varphi
   }{5}\right) \sinh \left(\frac{2
   \varphi }{5}\right)\right] \nonumber\\
   & = & \frac{\lambda}{4} \left[(a+b)
   \sinh\left(\frac{6\varphi}{5}\right)-(3 a-b)
   \sinh\left(\frac{2\varphi}{5}\right)\right] \ ,
 \end{eqnarray}
so that
 \begin{equation}\label{GIbc}
  {\mathcal V}_{c,\,Ib}(\xi,\eta) \, = \, - \, \frac{4}{25}\  \lambda \, \Big[ a\ \eta^3 \ + \ b\ \xi^2 \ \eta \Big] \ .
 \end{equation}
One has again the three options in eq.~(\ref{GIapar}) for the parameters $a$ and $b$, but these potentials are less interesting, since they are clearly all unbounded from below. Notice that the potentials (\ref{GIbc}) and the corresponding additional conserved charges, which guarantee the Liouville integrability of these systems, can be obtained from eqs.~(\ref{GIac}) and (\ref{AddInt1}) via the transformations
 \begin{equation}\label{GIIIbcTransForm}
\xi \ \rightarrow \ i\,\eta \ , \qquad \eta \ \rightarrow \ i\,\xi \ , \qquad
\lambda \ \rightarrow - \ i\,\lambda \ .
 \end{equation}
\begin{figure}[h]
\begin{center}
$\begin{array}{ccc}
\epsfig{file=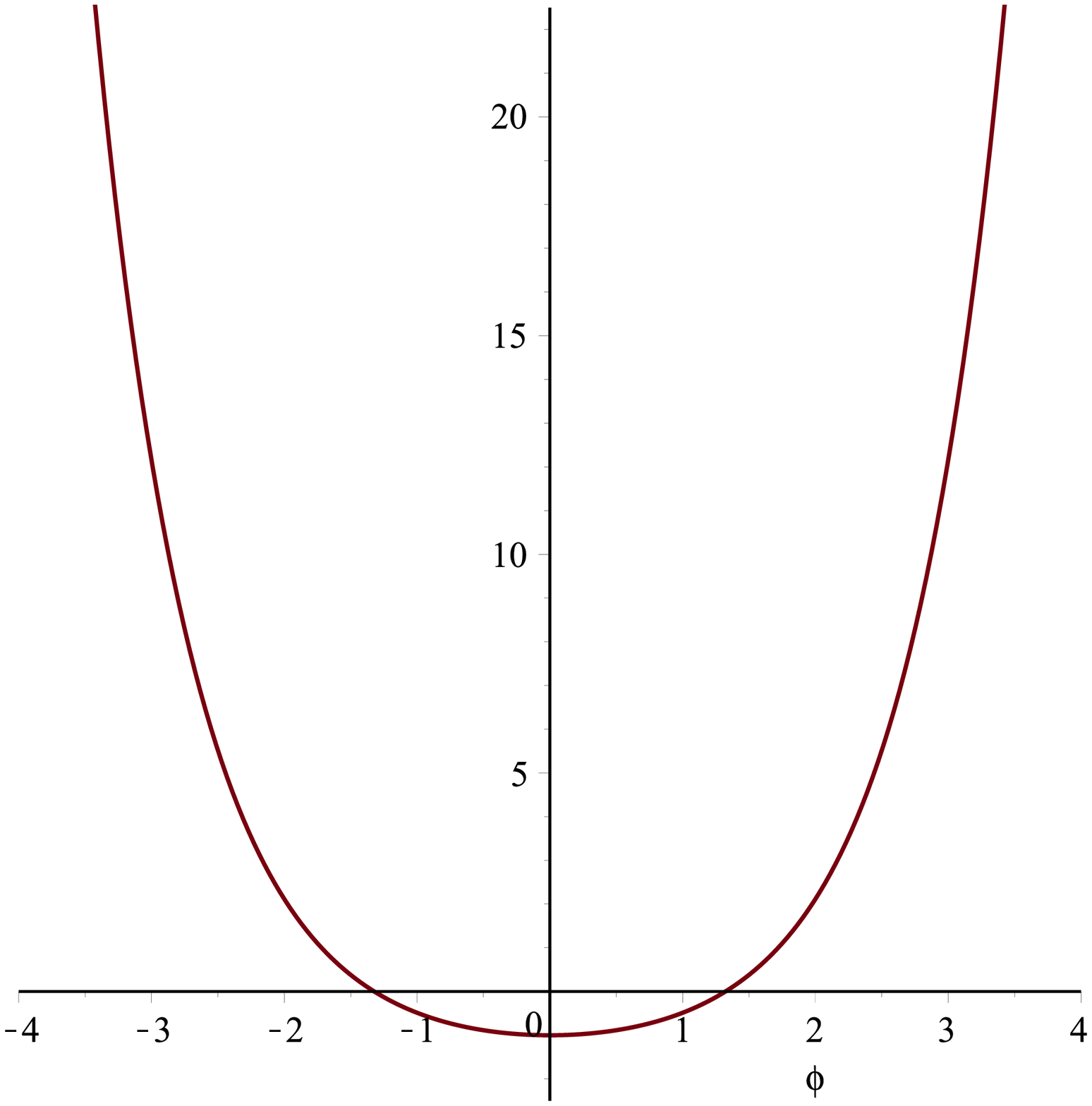, height=1.5in, width=1.5in}\qquad \quad &
\epsfig{file=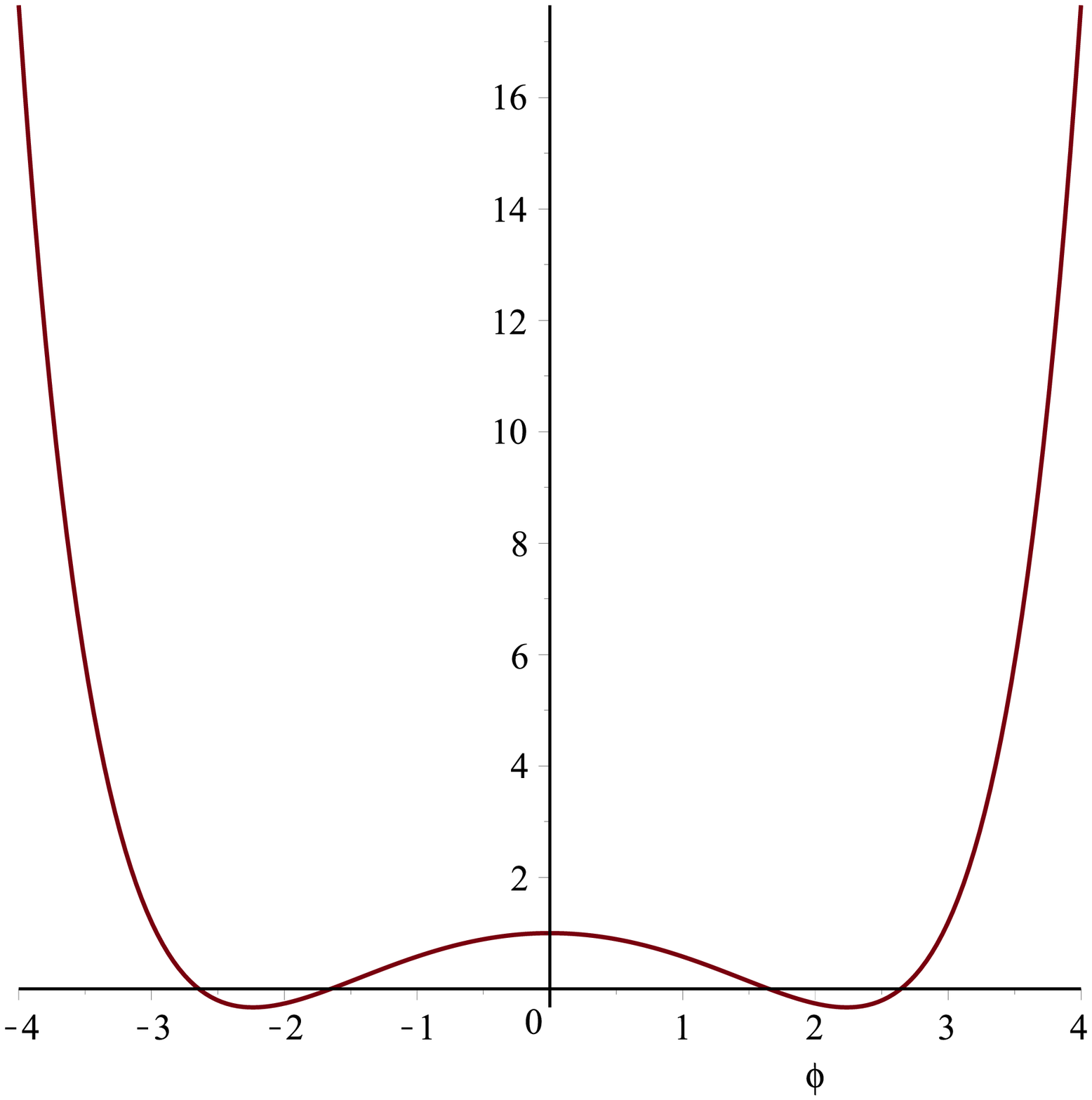, height=1.5in, width=1.5in}\qquad \quad &
\epsfig{file=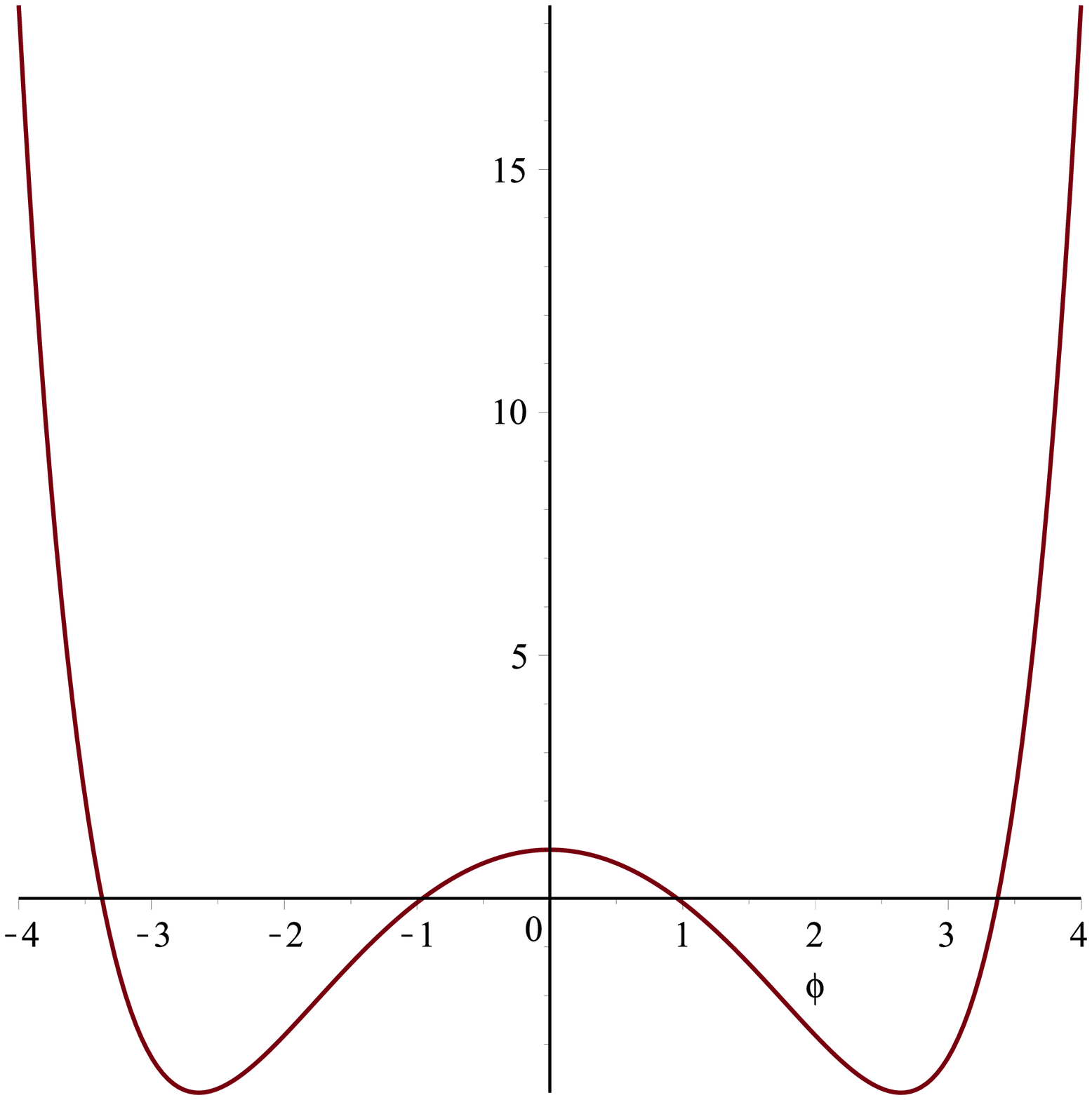, height=1.5in, width=1.5in}
\end{array}$
\vskip 10pt
$\begin{array}{ccc}
\epsfig{file=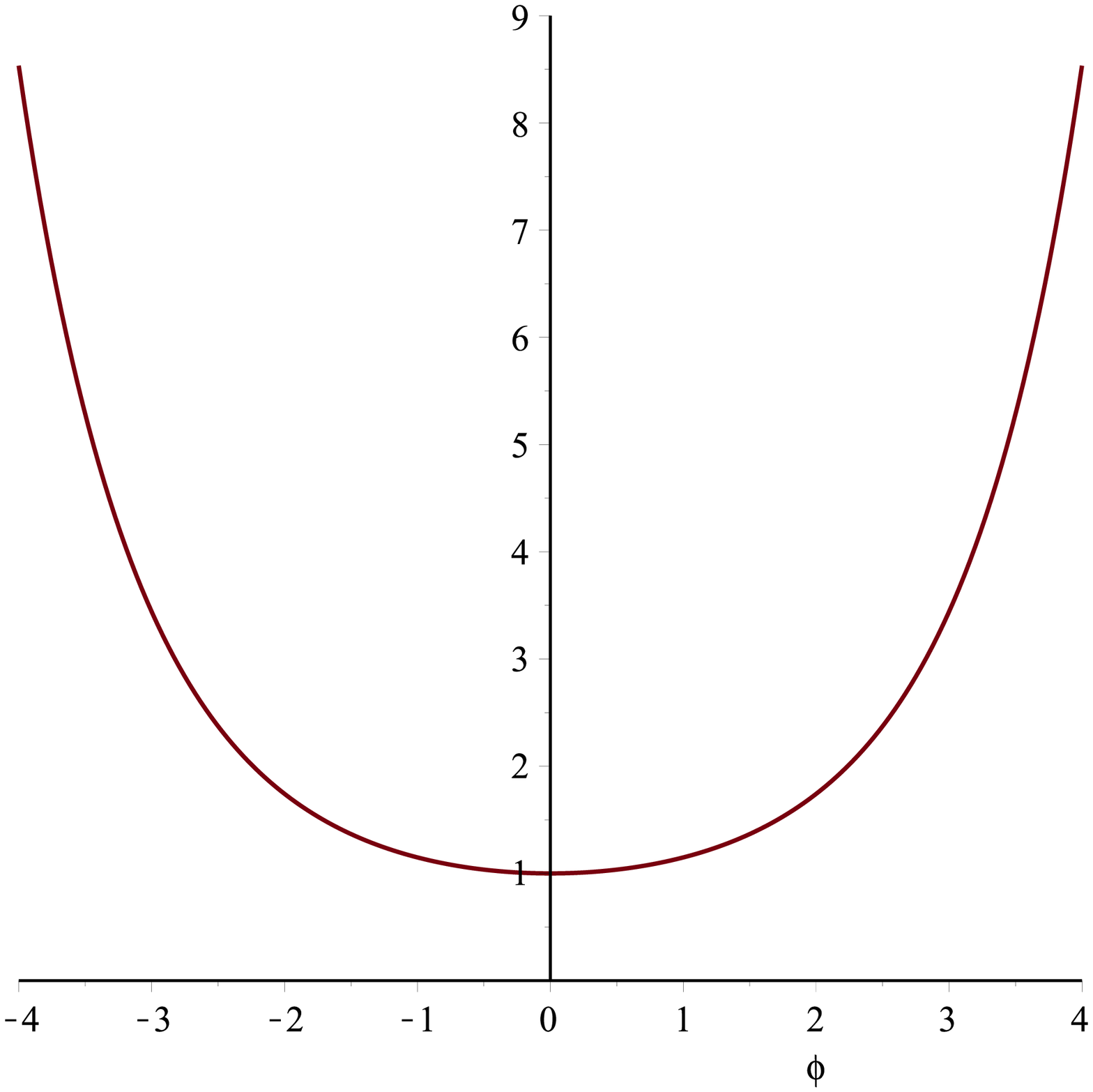, height=1.5in, width=1.5in}\qquad \quad &
\epsfig{file=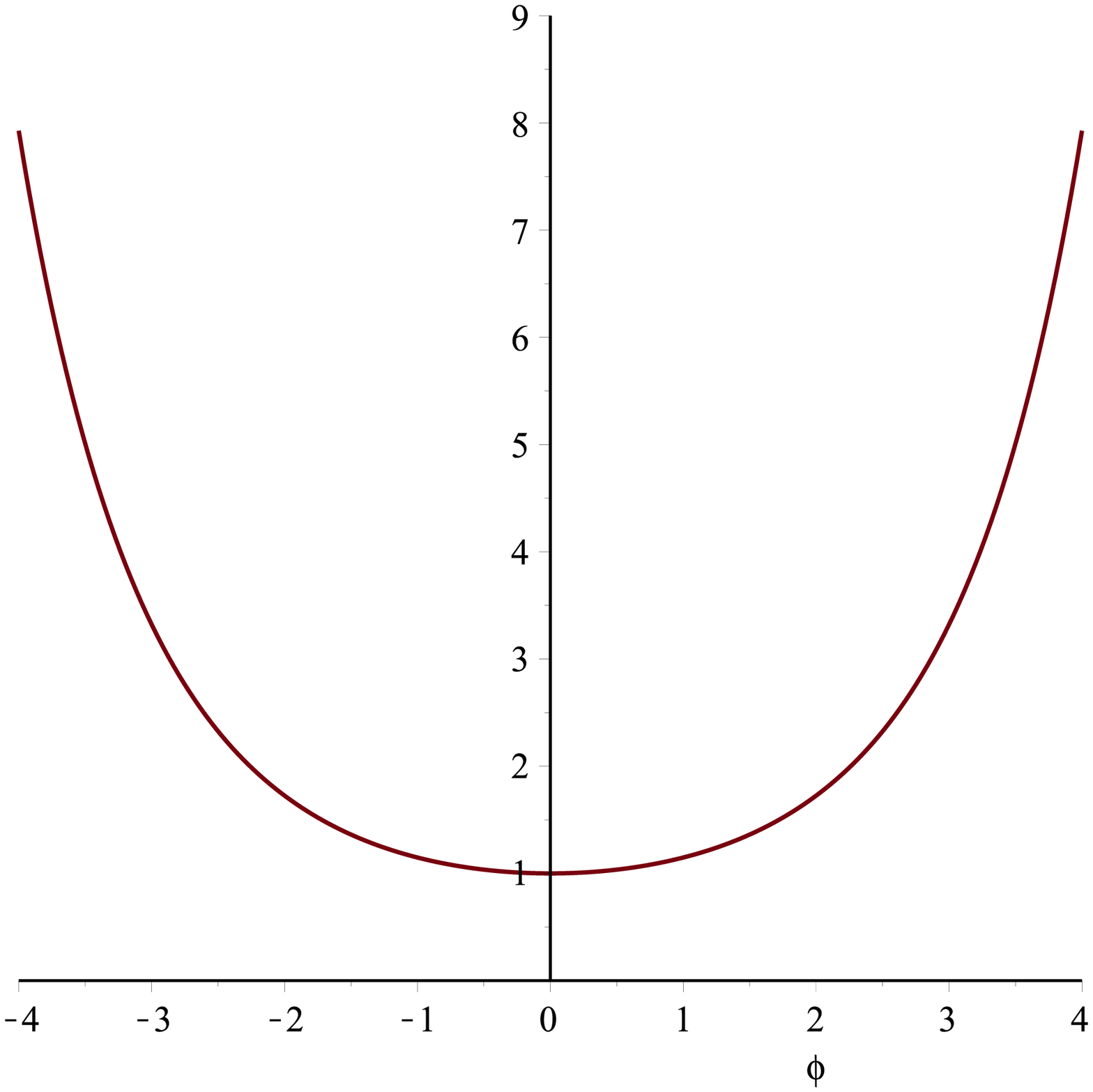, height=1.5in, width=1.5in}\qquad \quad &
\end{array}$
\end{center}
\caption{The five non--trivial potentials of eq.~\eqref{GII}, which correspond to cases 2--6 since the first is constant, with $\lambda<0$ in the second case. The analysis of Section \ref{sec:properties_exact} indicates that in cases 1, 2 and 3 above the Universe will experience a Big Crunch while in the others it will readily enter a de Sitter phase with the scalar stabilized at the bottom.}
\label{v23_potentials}
\end{figure}
\paragraph{Group II $\left(\gamma \, = \, \frac{1}{3}\right)$} In this case the scalar potentials are
 \begin{eqnarray}\label{GII}
  \mathcal{V}_{II}(\varphi) & = & \lambda \left[ a \, \cosh ^4\left(\frac{\varphi
   }{3}\right)\ +\ b \, \sinh
   ^4\left(\frac{\varphi }{3}\right)\ +\ c \, \sinh
   ^2\left(\frac{\varphi }{3}\right)
   \cosh ^2\left(\frac{\varphi
   }{3}\right)\right] \nonumber\\
   & = & \frac{\lambda}{8} \left[3 a+3 b- c+4( a- b)
   \cosh\left(\frac{2\varphi}{3} \right)+(a+b+c) \cosh\left(\frac{4\varphi}{3} \right)\right]\nonumber\ ,
 \end{eqnarray}
so that
 \begin{equation}\label{GIIc}
  {\mathcal V}_{c,\,II}(\xi,\eta) \, = \,  \frac{1}{9} \ \lambda \, \Big[ a \ \xi^4 \ + \ b\ \eta^4 \ + c\ \xi^2 \ \eta^2 \Big] \ .
 \end{equation}
with six independent options for the parameters $a$, $b$ and $c$ in eq.~(\ref{GII}),
 \begin{equation}\label{GIIpar}
    \left\{a,b,c\right\} \, = \, \left\{
\begin{array}{ccc}
 1  & 1  & -2
   \\
 1  & 1  & -6
   \\
 1  & 8  & -6
   \\
 1  & 16  & -12
    \\
 1  & \frac{1}{8} &
   -\frac{3}{4} \\
 1  & \frac{1}{16} &
   -\frac{3}{4}
\end{array}
\right\} \ ,
 \end{equation}
and choosing appropriately the sign of $\lambda$ these even potentials are always bounded from below.

The Liouville integrability of the systems corresponding to eq.~(\ref{GIIpar}) is guaranteed by the existence of an additional conserved charge (see \emph{e.g.} \cite{hieta2} and references therein), which in the six cases takes the form
 \bea\label{AddInt2}
&& {\mathcal Q}_{II}^{(1)}(\xi,\eta) \, = \,
\eta \,\dot{\xi} \ - \ \dot{\eta}\,\xi \ , \nonumber\\
&& {\mathcal Q}_{\,II}^{(2)}(\xi,\eta) \, = \, \frac{9}{4} \,\dot{\xi}\, \dot{\eta} \ + \  \lambda \, \eta\,\xi\,\Big( \eta^2 \ - \ \xi^2 \Big)\ , \nonumber\\
&& {\mathcal Q}_{\,II}^{(3)}(\xi,\eta) \, = \,  \Big[\frac{9}{2}\ \dot{\xi}^2\ + \
\lambda\,\xi^2\,(2\,\eta^2\ - \ \xi^2) \Big]^2 \ + \ 9\, \lambda\, \xi^2\,\, \Big(\xi \,\dot{\eta} \ - \ 2\,\dot{\xi}\,\eta \Big)^2 \ , \nonumber\\
&& {\mathcal Q}_{\,II}^{(4)}(\xi,\eta) \, = \, (\xi \,\dot{\eta} \ - \ \dot{\xi}\,\eta)\,\dot{\xi} \ - \ \frac{4}{9}\ \lambda\,\eta\, \xi^2\,(\xi^2 \ - \  2\,\eta^2)\ , \\
&& {\mathcal Q}_{\,II}^{(5)}(\xi,\eta) \, = \,  \left[ \frac{9}{2} \dot{\eta}^2\, +\,\frac{1}{8} \lambda  \left(\eta
   ^2-2 \xi ^2\right) \eta^2\right]^2-\frac{9}{8} \lambda
   \eta ^2 \left(2 \xi  \dot{\eta}-\eta  \dot{\xi}\right)^2 , \nonumber\\
&& {\mathcal Q}_{\,II}^{(6)}(\xi,\eta) \, = \, (\eta \,\dot{\xi} \ - \ \dot{\eta}\,\xi)\,\dot{\eta} \ + \ \frac{1}{36}\, \lambda\,\xi\, \eta^2\,(\eta^2 \ - \  2\,\xi^2)\ . \nonumber
\eea

If $\eta$ is continued to purely imaginary values, these potentials correspond to the cubic and quartic generalized H\'enon--Heiles integrable systems with meromorphic genus--two hyperelliptic general solutions (see \cite{hietarinta, perelomov, conte} and references therein).
\begin{figure}[h]
\begin{center}
\epsfig{file=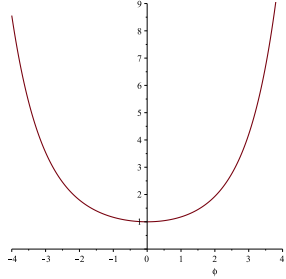, height=1.5in, width=1.5in}
\end{center}
\caption{The bounded Ramani potential corresponding to eq.~\eqref{GIIIa}. The analysis of Section \ref{sec:properties_exact} indicates that the Universe will readily enter a de Sitter phase with the scalar stabilized at the bottom.}
\label{ramani_potentials}
\end{figure}
\paragraph{Group III $\left(\gamma \, = \, \frac {2}{5}\right)$ -- (Ramani potentials)}
There are two scalar potentials in this class. The first potential is bounded from below and reads
 \begin{eqnarray}\label{GIIIa}
  \mathcal{V}_{IIIa}(\varphi) & = &  \lambda \left[ a \, \cosh ^3\left(\frac{2 \varphi
   }{5}\right)\ + \ b \, \sinh
   ^2\left(\frac{2 \varphi
   }{5}\right) \cosh \left(\frac{2
   \varphi }{5}\right)\ +\ c \, \sinh
   ^3\left(\frac{2 \varphi
   }{5}\right)\right] \ ,
  \end{eqnarray}
so that
 \begin{equation}\label{GIIIac}
  {\mathcal V}_{c,\,IIIa}(\xi,\eta) \, = \,  \frac{4}{25}\  \lambda \, \Big[ a\ \xi^3 \ + \ b\ \xi \ \eta^2 \ + \ c \ \eta^3 \Big] \ ,
 \end{equation}
and in this case
 \begin{equation}\label{GIIIapar}
    \left\{a,b,c\right\} \, = \, \left\{
 1\,, \ -\frac{1}{2}\,, \
   \frac{1}{6 \sqrt{3}} \right\} \ \ .
 \end{equation}
so that the potential (\ref{GIIIa}) can be recast in the form
\begin{eqnarray}\label{expGIIIa}
    \mathcal{V}_{IIIa}(\varphi) & = & \frac{\lambda}{16} \left[\left(1-\frac{1}{3
   \sqrt{3}}\right) e^{-6 \varphi
   /5}+\left(7+\frac{1}{\sqrt{3}}\right)
   e^{-2 \varphi
   /5} \right. \\
   && \left. +\left(7-\frac{1}{\sqrt{3}}\right)
   e^{2 \varphi /5}+\left(1+\frac{1}{3
   \sqrt{3}}\right) e^{6 \varphi
   /5}\right]\ .
\end{eqnarray}

The Liouville integrability of the system corresponding to eq.~(\ref{GIIIa}) is guaranteed by the existence of the additional conserved charge (see \emph{e.g.} \cite{hieta2} and references therein)
 \bea\label{AddInt3}
{\mathcal Q}_{\,IIIa}(\xi,\eta)& = & \dot{\eta}^4 \ + \ \frac{2}{\sqrt{3}}\,\dot{\xi}\,\dot{\eta}^3\ - \ \frac{4\, \lambda}{25 \sqrt{3}}\,\dot{\xi}^2\,\eta^3\ + \ \frac{4\, \lambda}{25}\,(\eta^3 \ + \ \sqrt{3}\,\xi\,\eta^2)\,\dot{\xi}\,\dot{\eta}\nonumber\\
& + & \frac{4\, \lambda}{25}\,
\Big(-2\sqrt{3}\, \xi^2\,\eta \ + \ \frac{1}{\sqrt{3}}\,\eta^3 \ - \ 2\, \xi\, \eta^2\Big)\,\dot{\eta}^2\nonumber\\
& + & \frac{4\, \lambda^2}{25^2}\Big(\frac{4}{\sqrt{3}}\,\xi^3\,\eta^3\ - \
\frac{2}{\sqrt{3}}\, \xi\,\eta^5\ - \ \xi^2\,\eta^4\ + \ \frac{5}{9} \, \eta^6\Big)\ .
\eea
On the other hand, the second potential is not bounded from below and reads
 \begin{equation}\label{GIIIb}
  \mathcal{V}_{IIIb}(\varphi) \, = \,  a \, \cosh ^3\left(\frac{2 \varphi
   }{5}\right)\ +\ b \, \sinh \left(\frac{2
   \varphi }{5}\right) \cosh
   ^2\left(\frac{2 \varphi
   }{5}\right)\ +\ c \, \sinh
   ^3\left(\frac{2 \varphi
   }{5}\right)\ ,
  \end{equation}
so that
 \begin{equation}\label{GIIIbc}
 {\mathcal V}_{c,\,IIIb}(\xi,\eta)\, = \, - \,  \frac{4}{25}\  \lambda \, \Big[ a\ \xi^3 \ + \ b\ \xi^2  \ \eta  \ + \ c \ \eta^3 \Big] \ ,
 \end{equation}
and in this case
 \begin{equation}\label{GIIIapar2}
    \left\{a,b,c\right\} \, = \, \left\{
 1\,, \ -3 \sqrt{3}\,,\   6 \sqrt{3}\right\}  .
 \end{equation}
 With these values of the coefficients, eq.~(\ref{GIIIb}) can be recast in the form
 \begin{eqnarray}
  \mathcal{V}_{IIIb}(\varphi) &=& \frac{\lambda}{16} \left[\left(2-18
   \sqrt{3}\right) e^{-6 \varphi
   /5}+\left(6+30 \sqrt{3}\right) e^{-2
   \varphi /5}\right.\\
   &&\left. +\left(6-30
   \sqrt{3}\right) e^{2 \varphi
   /5}+\left(2+18 \sqrt{3}\right) e^{6
   \varphi /5}\right] \ .
   \label{ExpGIIIb}
 \end{eqnarray}
The potential (\ref{GIIIbc}), and the additional charge that guarantees the Liouville integrability of the system, can be derived from eqs.~(\ref{GIIIac}) and (\ref{AddInt3}) via the transformations
 \begin{equation}\label{GIIIbcTrans}
\xi \ \rightarrow \ i\,\eta \ , \qquad \eta \ \rightarrow \ i\,\xi \ , \qquad
\lambda \ \rightarrow \ - \, i\,6 \sqrt{3}\,\lambda \ .
 \end{equation}
\begin{figure}[h]
\begin{center}
$\begin{array}{ccc}
\epsfig{file=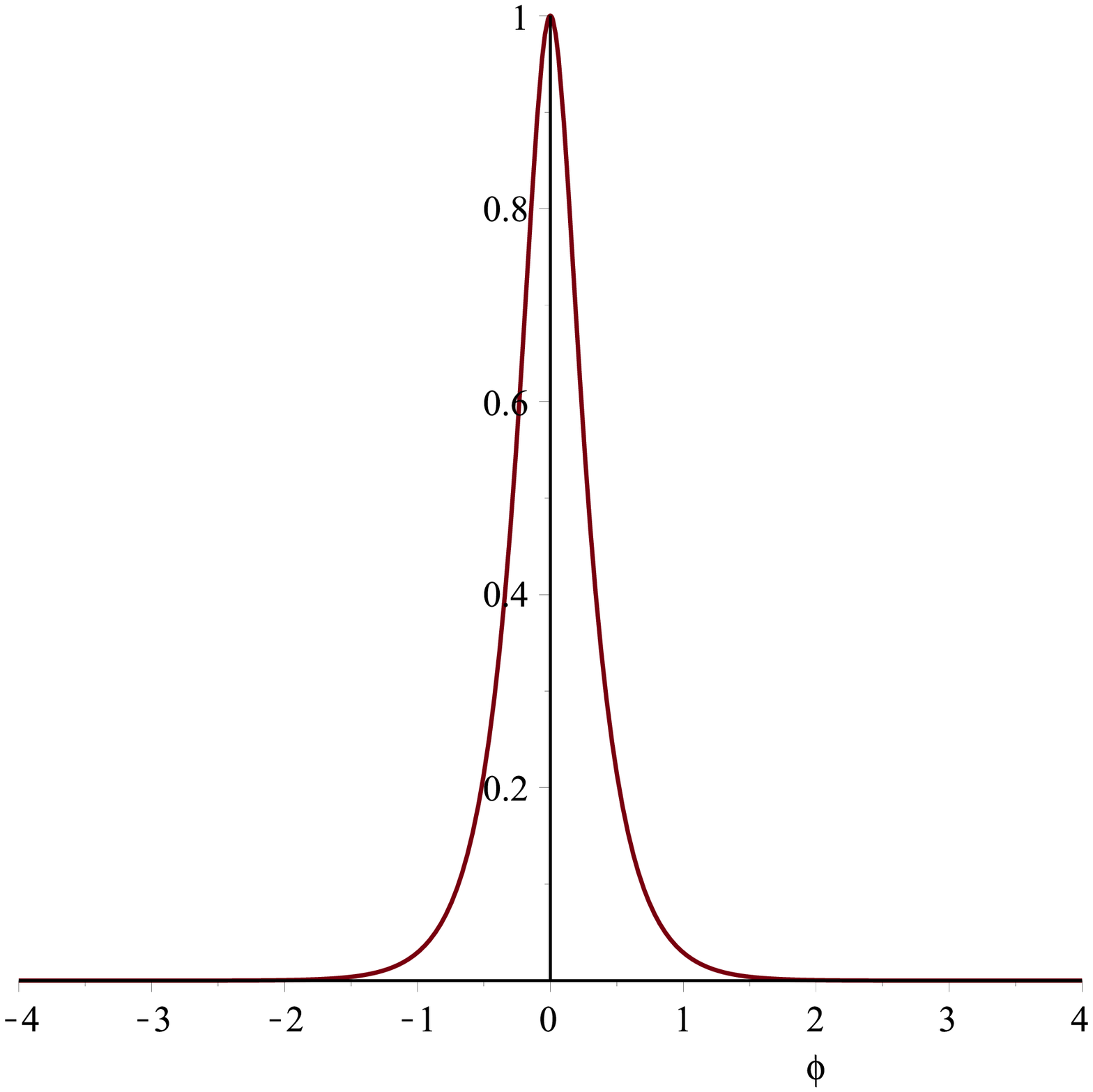, height=1.5in, width=1.5in}\qquad \quad &
\epsfig{file=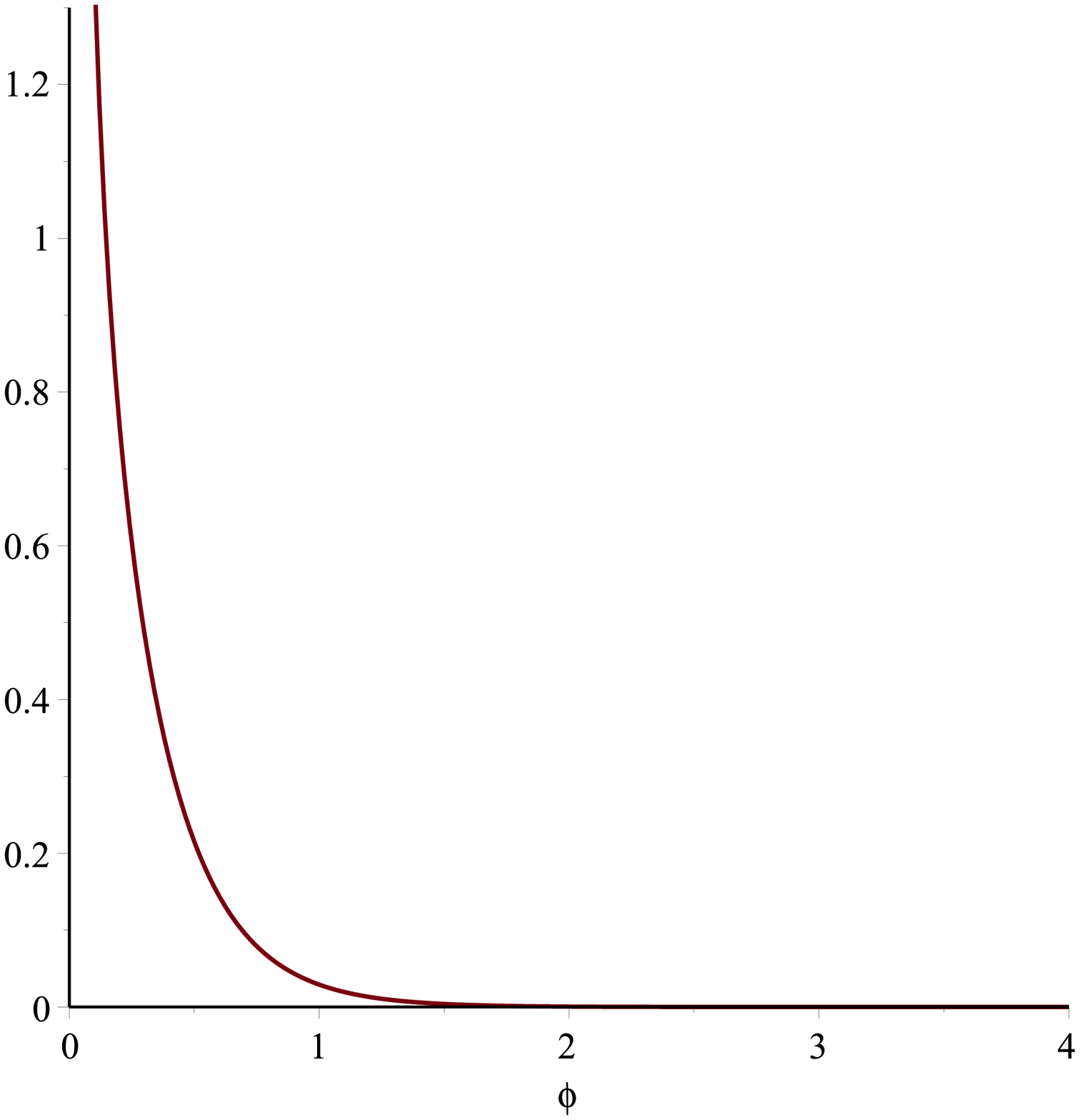, height=1.5in, width=1.5in}
\end{array}$
\end{center}
\caption{The potentials of Group IV in their real domain. In both cases the Universe will end up in a decelerated phase with the scalar running off to infinity, but in the first case it can either bounce back or overcome the wall, depending on the initial conditions.}
\label{v4_potentials}
\end{figure}
\paragraph{Group IV $\big(\gamma=3\big)$ --  (Fokas, Lagestrom, Inozemtsev potentials)} This group comprises the two potentials
\begin{equation}\label{ciccius4}
    \begin{array}{ccc}
       \mathcal{V}_{IVa} & = & \frac{\lambda }{\cosh
   ^{\frac{2}{3}}(6 \varphi )}  \ , \\
   \null &\null & \null \\
       \mathcal{V}_{IVb} & = &  \frac{\lambda }{\sinh
   ^{\frac{2}{3}}(6 \varphi )}  \ ,
     \end{array}
\end{equation}
which are displayed in fig.~\ref{v4_potentials}, so that
\begin{equation}\label{ciccius4c}
    \begin{array}{ccc}
       \mathcal{V}_{IVac} & = & 9 \ \frac{\lambda }{\left(\xi^2 \ + \ \eta^2\right)^{\frac{2}{3}}}  \ , \\
   \null &\null & \null \\
       \mathcal{V}_{IVbc} & = &  9\  \frac{\lambda }{\left(2\, \xi \, \eta \right)^{\frac{2}{3}}} \ .
     \end{array}
\end{equation}
\begin{figure}[h]
\begin{center}
$\begin{array}{ccc}
\epsfig{file=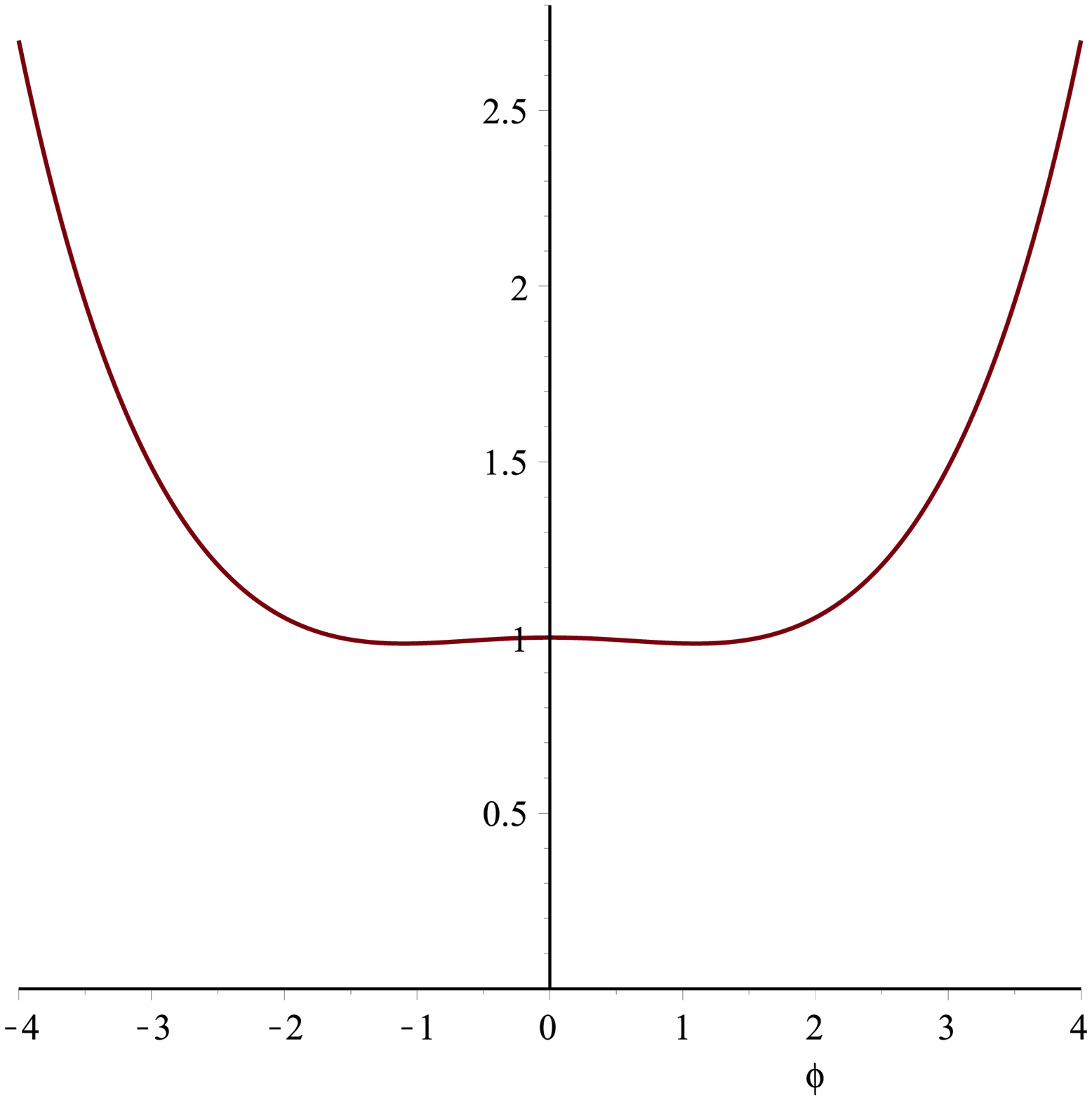, height=1.5in, width=1.5in}\qquad \quad &
\epsfig{file=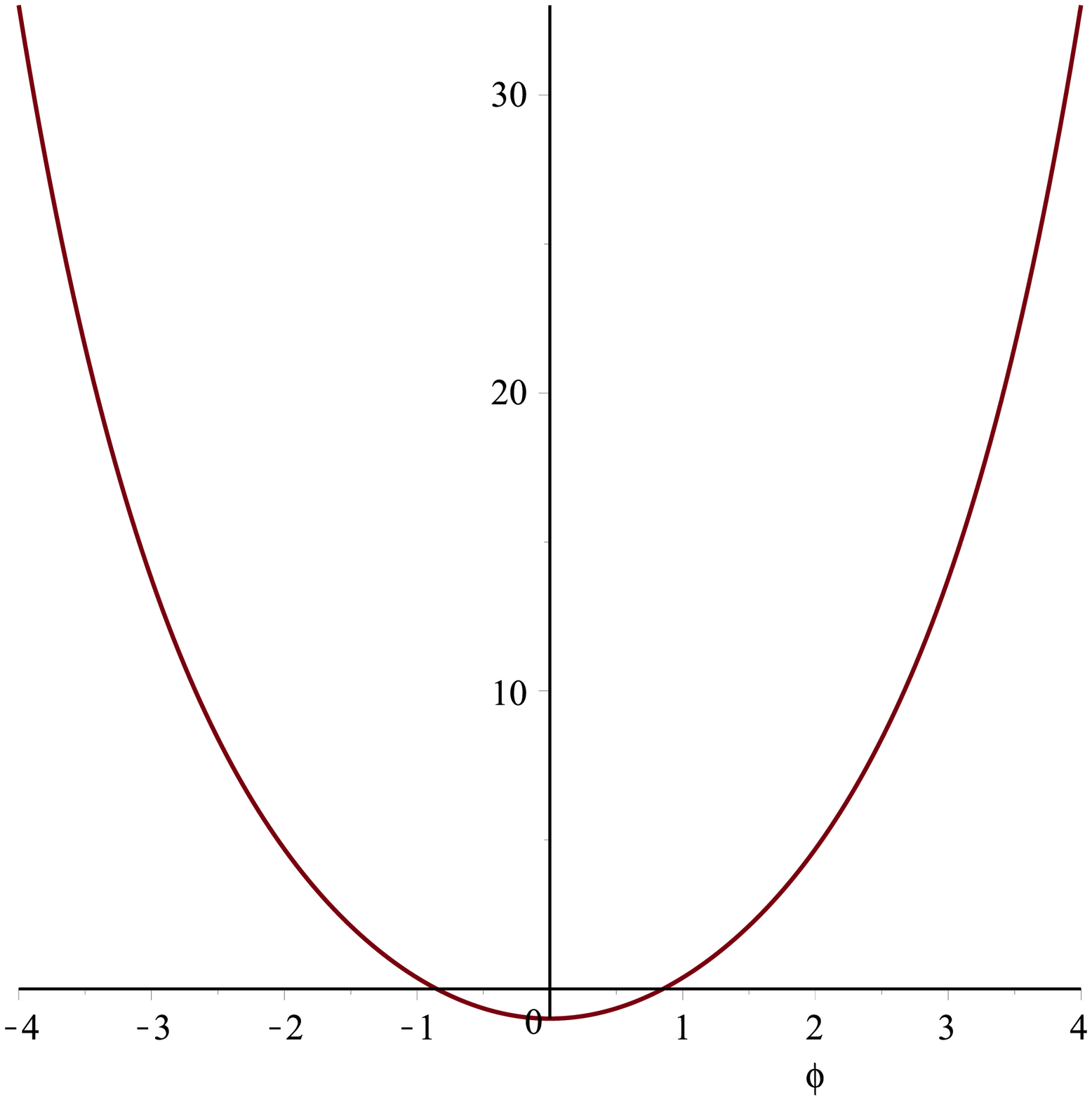, height=1.5in, width=1.5in}\qquad \quad &
\epsfig{file=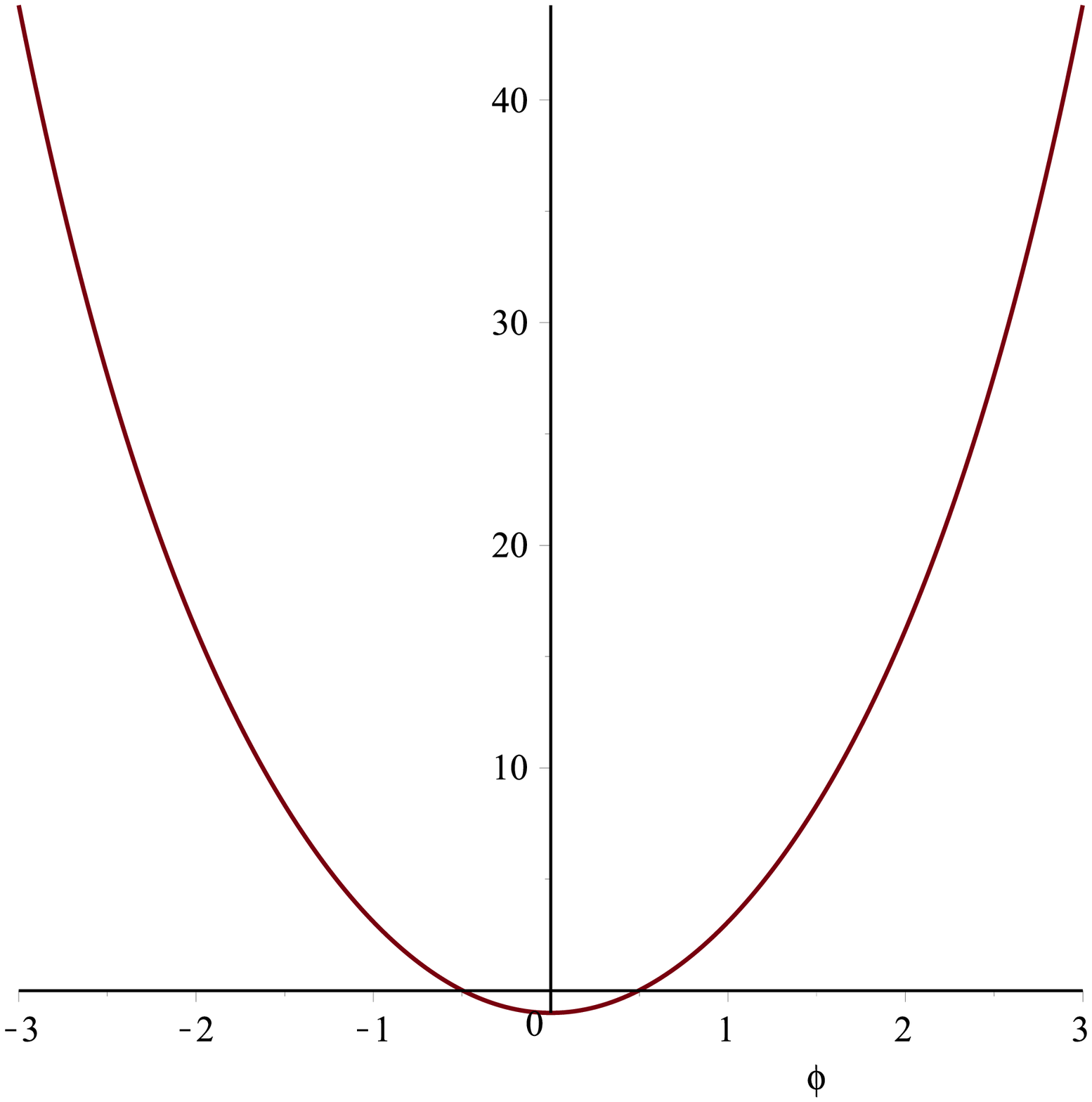, height=1.5in, width=1.5in}
\end{array}$
\vskip 10pt
$\begin{array}{ccc}
\epsfig{file=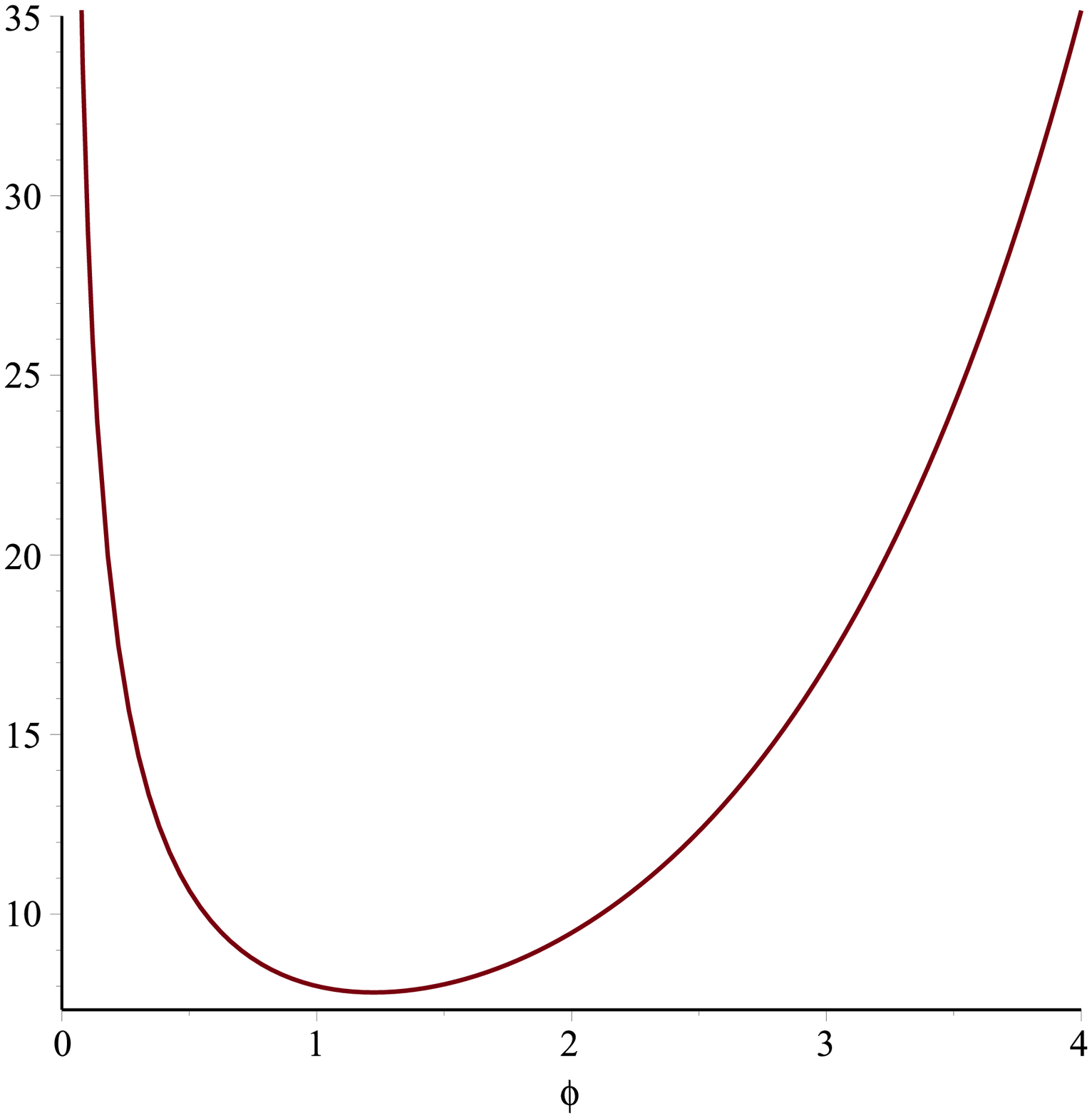, height=1.5in, width=1.5in}\qquad \quad &
\epsfig{file=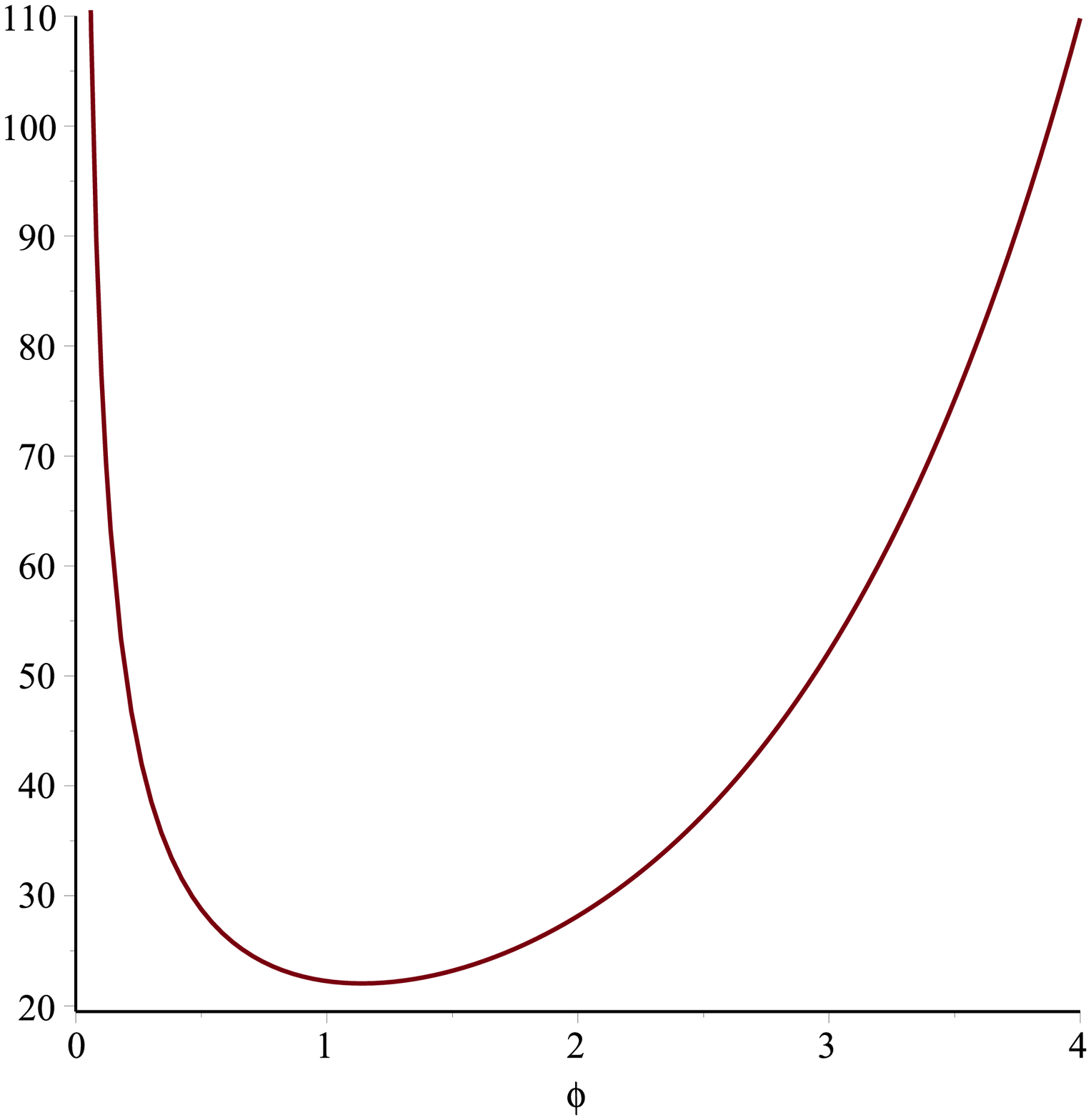, height=1.5in, width=1.5in}\qquad \quad &
\end{array}$
\end{center}
\caption{The bounded Holt--Drach potentials of eqs.~\eqref{GVa} and \eqref{GVb} in their real domains. They are all the potentials of eq.~\eqref{GVa}, with negative $\lambda$ in the last two cases, and the last two potentials of eq.~\eqref{GVb}, again with negative $\lambda$. The analysis of Section \ref{sec:properties_exact} indicates that in cases 2 and 3 above the Universe will experience a Big Crunch while case 1 will enter readily a de Sitter phase with the scalar stabilized at one of the critical points. In cases 4 and 5 the scalar will descend from the right end after the initial singularity and will readily end at the bottom as the Universe enters a de Sitter phase.}
\label{v5_potentials}
\end{figure}

The Liouville integrability of the systems corresponding to eqs.~\eqref{ciccius4c} is guaranteed by the existence of an additional conserved charge \cite{inozem}, which for the two potentials at stake reads
 \bea\label{AddInt4}
{\mathcal Q}_{\,IVa}(\xi,\eta) & = & (\dot{\eta}^2 \ + \ \dot{\xi}^2)\,
(\eta\,\dot{\xi}\ - \ \dot{\eta}\,\xi) \ - \ 4\,\lambda\, (\eta\,\dot{\xi}\ + \ \dot{\eta} \, \xi)\,(\eta^2 \ + \ \xi^2)^{-\frac{2}{3}}\ , \nonumber\\
{\mathcal Q}_{\,IVb}(\xi,\eta) & = & (\eta\,\dot{\xi}\ - \ \dot{\eta} \, \xi)\, \dot{\xi}\,\dot{\eta} \ - \ 2\,\lambda\, (\eta\,\dot{\eta}\ + \  \xi\,\dot{\xi})\,(2\,\eta\,\xi)^{-\frac{2}{3}}\ .
\eea
\paragraph{Group Va $\left(\gamma\ = \ \frac{3}{5}\right)$ -- (Holt-Drach potentials)} In this case the scalar potentials are
 \begin{eqnarray}\label{GVa}
  \mathcal{V}_{Va}(\varphi) & = &  \lambda \left[a \, \cosh
   ^{\frac{4}{3}}\left(\frac{3
   \varphi }{5}\right)\ + \ b\, \frac{\sinh ^2\left(\frac{3
   \varphi }{5}\right)}{\cosh
   ^{\frac{2}{3}}\left(\frac{3
   \varphi }{5}\right)}\right] \nonumber\\
   & = & \frac{a-b+(a+b) \cosh
   \left(\frac{6 \varphi
   }{5}\right)}{2 \cosh
   ^{\frac{2}{3}}\left(\frac{3
   \varphi }{5}\right)}\ ,
 \end{eqnarray}
so that
 \begin{equation}\label{GVac}
  \mathcal{V}_{c,Va}(\xi,\eta) \, = \,  \frac{9}{25}\ \lambda \Big[a \ \xi^\frac{4}{3} \ + \ b\ \xi^{\,-\, \frac{2}{3}} \ \eta^2 \Big] \ ,
 \end{equation}
with the three options
 \begin{equation}\label{GVapar}
    \left\{b,a\right\} \, = \, \left\{ \begin{array}{cc}
1  & -\frac{3}{4} \\
1  & -\frac{9}{2} \\
1  & -12
\end{array}
\right\} \ \ .
 \end{equation}

The Liouville integrability of the systems of eq.~(\ref{GIapar}) is guaranteed by the existence of an additional conserved charge (see \emph{e.g.} \cite{hieta2} and references therein) that in the three cases takes the form
 \bea\label{AddInt5}
{\mathcal Q}_{\,Va}^{(1)}(\xi,\eta) & = &  \Big[\dot{\eta}^2 \ - \ \frac{3}{2}\,\dot{\xi}^2\ + \ \frac{9}{25}\,\lambda \Big( \frac{9}{2}\, \xi^\frac{4}{3}\ + \ 3\,  \xi^{-\frac{2}{3}}\,\eta^2\Big)\Big]\,\dot{\eta}\ - \ \frac{81}{25}\,\lambda\, \eta\,\xi^{\frac{1}{3}}\,\dot{\xi}\ ,\nonumber\\
{\mathcal Q}_{\,Va}^{(2)}(\xi,\eta) & = &  \Big(\dot{\eta}^2 \ - \ 2\,\dot{\xi}^2\ + \ \frac{36}{25}\,\lambda\, \xi^{-\frac{2}{3}}\,\eta^2\Big)\,\dot{\eta}^2\ - \ \frac{216}{25}\,\lambda\, \eta\,\xi^{\frac{1}{3}}\,\dot{\eta}\,\dot{\xi}\ -\ \frac{5832}{625}\,\lambda\, \eta^2\,\xi^{\frac{2}{3}}\ ,~~~\\
{\mathcal Q}_{\,Va}^{(3)}(\xi,\eta) & = &  \Big[\dot{\eta}^2 \ - \ 3\,\dot{\xi}^2\ - \ \frac{54}{25}\,\lambda \Big(3\,\xi^\frac{4}{3}\ - \  \xi^{-\frac{2}{3}}\,\eta^2\Big)\Big]\,\dot{\eta}^4 \ - \ \frac{648}{25}\,\lambda\, \eta\,\xi^{\frac{1}{3}}\,\dot{\eta}^3\,\dot{\xi}
\nonumber\\ & - & \Big(\frac{9}{25}\,\lambda\Big)^2\, 648\, \eta^2\,\xi^{\frac{2}{3}}\,\dot{\eta}^2\ - \ \Big(\frac{9}{25}\,\lambda\Big)^3\,648\,\eta^4\ .\nonumber
\eea
\paragraph{Group Vb $\left(\gamma\ = \ \frac{3}{5}\right)$ -- (Holt--Drach potentials)} In this case the scalar potential is
 \begin{eqnarray}\label{GVb}
  \mathcal{V}_{Vb}(\varphi) & = &  \lambda \left[a \, \sinh
   ^{\frac{4}{3}}\left(\frac{3
   \varphi }{5}\right)\ + \ b\, \frac{\cosh ^2\left(\frac{3
   \varphi }{5}\right)}{\sinh
   ^{\frac{2}{3}}\left(\frac{3
   \varphi }{5}\right)}\right] \nonumber\\
   &= &\frac{-a+b+(a+b) \cosh
   \left(\frac{6 \varphi
   }{5}\right)}{2 \sinh
   ^{\frac{2}{3}}\left(\frac{3
   \varphi }{5}\right)} \ ,
 \end{eqnarray}
so that
 \begin{equation}\label{GVbc}
  \mathcal{V}_{Va}(\varphi) \, = \, - \,  \frac{9}{25}\ \lambda \Big[a \ \eta^\frac{4}{3} \ + \ b\ \xi^2 \ \eta^{\,-\, \frac{2}{3}} \Big] \ ,
 \end{equation}
again with the three options of eq.~(\ref{GVapar}) for the parameters $a$ and $b$, the first of which leads to a result that is unbounded from below.

The potentials (\ref{GVbc}) and the corresponding second integrals, which guarantee the Liouville integrability of the systems, can be derived from eqs.~(\ref{GVac}) and (\ref{AddInt5}) via the transformations
 \begin{equation}\label{GIIIbcTransFormDrach}
\xi \ \rightarrow \ i\,\eta \ , \quad \eta \ \rightarrow \ i\,\xi \  \quad \lambda \, \rightarrow \, - \, \lambda\ .
 \end{equation}
\paragraph{Group VI $\left(\gamma\ = \ \frac {6}{7}\right)$ --  (Drach potentials)} This group comprises the two potentials
\begin{equation}\label{ciccius4bis}
    \begin{array}{ccc}
       \mathcal{V}_{VIa} & = & \frac{\lambda  \sinh \left(\frac{6
   \varphi }{7}\right)}{\cosh
   ^{\frac{2}{3}}\left(\frac{6
   \varphi }{7}\right)} \ ,  \\
       \null & \null & \null \\
       \mathcal{V}_{VIb} & = &   \frac{\lambda  \cosh \left(\frac{6
   \varphi }{7}\right)}{\sinh
   ^{\frac{2}{3}}\left(\frac{6
   \varphi }{7}\right)} \ ,
     \end{array}
\end{equation}
so that
\begin{equation}\label{ciccius4cbis}
    \begin{array}{ccc}
       \mathcal{V}_{VIac} & = & \frac{36}{49} \ \lambda  \ \eta \ \xi^{\,-\,\frac{2}{3}}\ ,  \\
       \null & \null & \null \\
       \mathcal{V}_{VIbc} & = &   \frac{36}{49} \ \lambda  \ \xi \ \eta^{\,-\,\frac{2}{3}}\ .
     \end{array}
\end{equation}
The first potential, $\mathcal{V}_{VIa}(\varphi)$, is unbounded from below, while $\mathcal{V}_{VIb}(\varphi)$ describes a well with a positive minimum that is qualitatively similar to the last example in fig.~\ref{v5_potentials}.

The Liouville integrability of these models is guaranteed by the existence of an additional conserved charge ${\cal Q}$ \cite{tsiganov} that in the two cases takes the form
\bea\label{AddInt5DrachSys}
{\mathcal Q}_{\,VIa}(\xi,\eta) & = &  \dot{\xi}\,(3\,\dot{\eta}^2 \ - \ 2\,\dot{\xi}^2)\ + \ \frac{18}{49}\,\lambda \,\eta\,\xi^{-\frac{5}{3}}\Big( \frac{2}{3}\, \eta \,\dot{\eta}\ - \ \xi\,\dot{\xi}\Big)\ ,\\
{\mathcal Q}_{\,VIb}(\xi,\eta) & = &  \dot{\eta}\,(3\,\dot{\xi}^2 \ - \ 2\,\dot{\eta}^2)\ -\ \frac{18}{49}\,\lambda \,\xi\,\eta^{-\frac{5}{3}}\Big( \frac{2}{3}\, \xi \,\dot{\xi}\ - \ \eta\,\dot{\eta}\Big)\ .
\eea
\begin{figure}[h]
\begin{center}
$\begin{array}{cc}
\epsfig{file=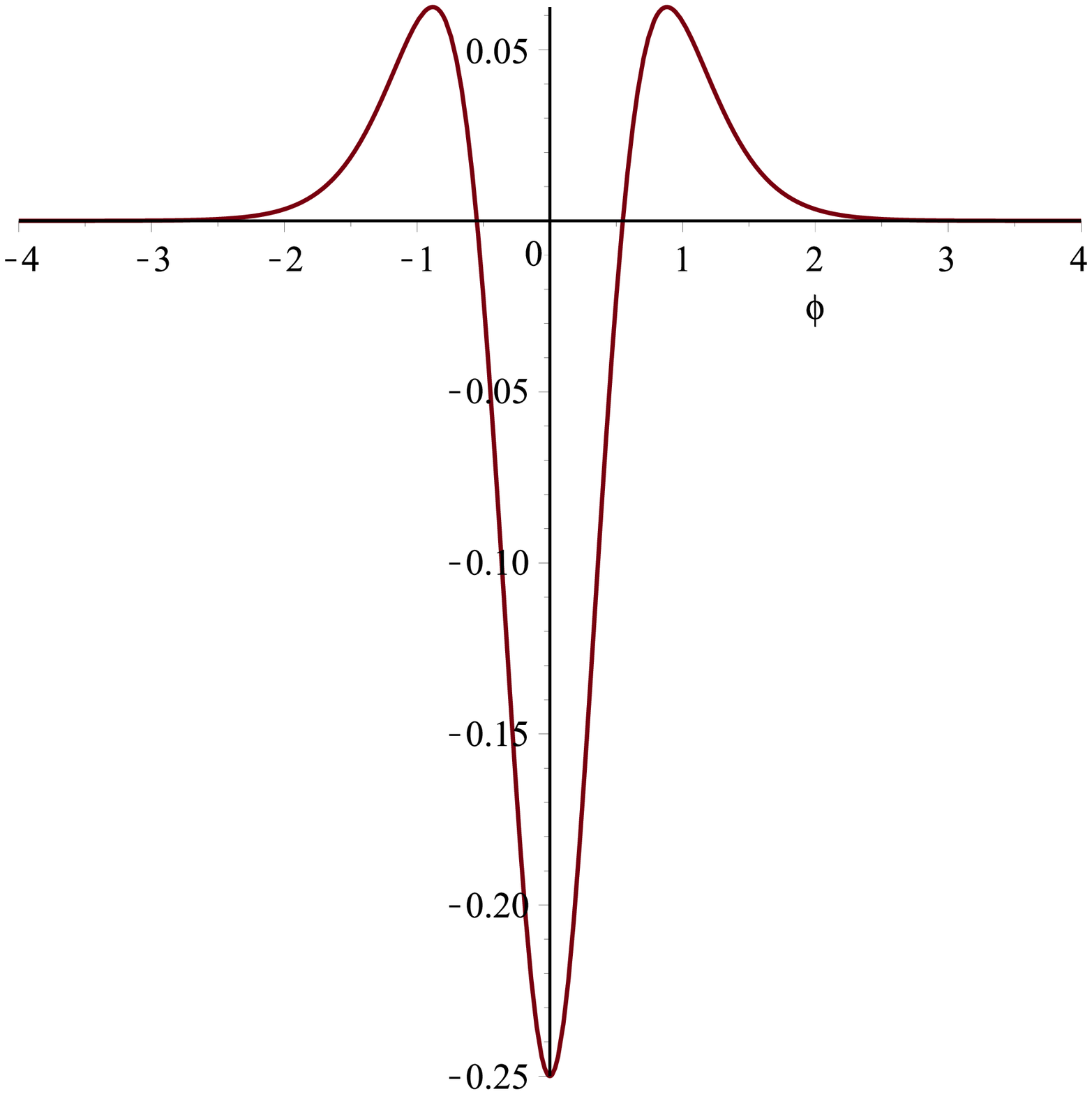, height=1.5in, width=1.5in}\qquad \quad &
\epsfig{file=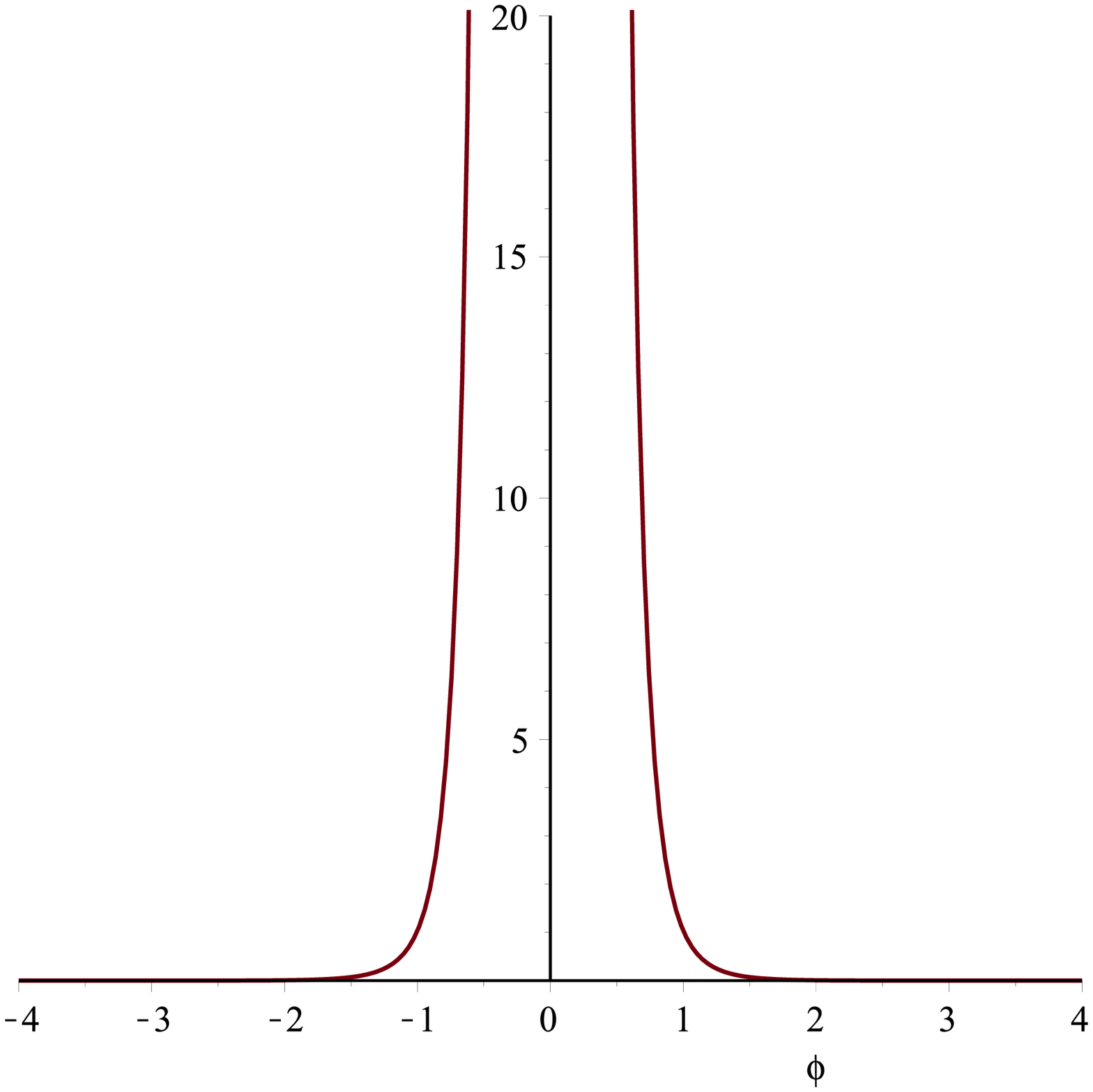, height=1.5in, width=1.5in}
\end{array}$
\end{center}
\caption{The two Hietarinta potentials of eqs.~\eqref{Sporadic1}. The analysis of Section \ref{sec:properties_exact} indicates that in case 1 special initial conditions will lead to a Big Crunch. In case 2 the behavior will be qualitatively similar to that illustrated in fig.~\ref{v4_potentials}.}
\label{v7_potentials}
\end{figure}
\paragraph{Group VII $\left(\gamma\ = \ - \ 1 \right)$ --  (Hietarinta potentials)} This group comprises the two potentials
\bea
&& {\mathcal{V}_{VIIa}(\varphi)}\ = \ \lambda \left[ - \,\frac{1}{4}\, {(\cosh \varphi)}^{-4} \ + \ (\sinh \varphi)^{2}\, (\cosh \varphi)^{-6}\right] \ , \\
&& {\mathcal{V}_{VIIb}(\varphi)}\ = \ \lambda \left[ - \,\frac{1}{4}\, (\sinh \varphi)^{-4} \ +\ (\cosh \varphi)^{2} \,(\sinh \varphi)^{-6}\right] \ .
\label{Sporadic1}
\eea
Applying the transformations of eqs.~(\ref{int3755}) and (\ref{int3656}) with $\gamma = -1$, these expressions become
\bea
&& {\mathcal{V}}_{c,VIIa}(\xi,\eta) \ = \ \lambda \left[ - \,\frac{1}{4}\, {\xi}^{-4} \ + \ {\eta}^{2}\, {\xi}^{-6}\right] \ , \\
&& {\mathcal{V}}_{c,VIIb}(\xi,\eta) \ = \ \lambda \left[ - \,\frac{1}{4} \, {\eta}^{-4}\ + \ {\xi}^{2}\, {\eta}^{-6}\right] \ .
\label{Sporadic2}
\eea
The Liouville integrability of each of these systems is guaranteed by the existence of an additional conserved charge (see \emph{e.g.} \cite{hieta2} and references therein) that in the two cases take the form
 \bea\label{AddInt4a}
{\mathcal Q}_{\,VIIa}(\xi,\eta) & = &  (\dot{\eta} \, \xi\ - \
\eta\,\dot{\xi})\, \dot{\xi} \ - \ \lambda\, \eta\,\xi^{-4} \ + \ 2\, \lambda\, \eta^3\,\xi^{-6}\ , \\
{\mathcal Q}_{\,VIIb}(\xi,\eta) & = & (\dot{\xi}\,\eta \ - \
\xi \,\dot{\eta} )\, \dot{\eta} \ + \ \lambda\, \xi\,\eta^{-4} \ - \ 2\, \lambda\, \xi^3\,\eta^{-6}
\ .
\eea
%

\subsection{\sc An integrable trigonometric potential}\label{sec:trigonometric_pot}

It is perhaps worth mentioning another simple and yet potentially instructive case that does not belong to the class of potentials that constitute the main subject of this paper. This is the simple trigonometric potential
\be
\mathcal{V}_{trig} (\varphi)\ = \ \lambda \ \cos{2\,\varphi} \ ,  \label{trig_1}
\ee
where $\lambda$ is a real constant. With this potential, taking ${\cal B}={\cal A}$, one can reduce the general lagrangian (\ref{onefield}) to the separable form
\bea {\cal L}_t \ = \  - \ \dot{\mathcal{U}}_{\,1}^{\,2} \ - \  \dot{\mathcal{U}}_{\,2}^{\,2} \ - \
 \frac{\lambda}{2} \ \exp{\Big(2\, \sqrt{2} \ \mathcal{U}_{\,1}}\Big)\ - \
 \frac{\lambda}{2} \ \exp{\Big(2\, \sqrt{2} \ \mathcal{U}_{\,2}}\Big) \label{trig_2} \ , \eea
at the price of moving to complex variables via the transformation
\bea
&& \mathcal{U}_{\,1} \ = \ \frac{1}{\sqrt2}\ \Big({\mathcal A} \ + \ i\ \varphi \Big) \ ,\nonumber\\
&& \mathcal{U}_{\,2} \ = \ \frac{{\rm i}}{\sqrt2}\ \Big({\mathcal A} \ - \ i\ \varphi \Big) \ . \label{trig_3} \eea
This potential is closely related to the Liouville potentials of eq.~\eqref{Liouville1}, since it obtains formally setting $\gamma=i$ in them, and of course one could also consider, along similar lines, a combination of trigonometric $\sin$ and $\cos$ functions, allowing for a complex $\lambda$.

\subsection{\sc Relations to Toda systems}\label{sec:toda_pot}
We can conclude this section with some cursory remarks on the explicit link between some special Toda systems and potentials that yield integrable cosmological models.

The starting point is provided by the Toda chain Lagrangians associated with the $A_2$ simple roots,
\begin{eqnarray}
{\cal L}_{A_2} & = & \frac{1}{2} \ \Big(\dot{x_1}^{2} \ + \
\dot{x_2}^{2} \ + \ \dot{x_3}^{2} \Big) \  + \ \lambda \ \Big( e^{x_2\
- \ x_1}\ +\ \epsilon \ e^{x_3\ - \ x_2}\Big) \ , \nonumber\\
x_3& \equiv & - \ x_1 \ - \ x_2 \ ,
\label{TodaA2}
\end{eqnarray}
where the two choices $\epsilon=1$ and $\epsilon=-1$ correspond to the $SL(3)/SO(3)$ and $SL(3)/SO(1,2)$ chains.
It is well known that this Lagrangian describes a completely integrable dynamical system in the complex hyperplane $\mathbb{C}^2$ where the $x_i \,\,\,(i=1,2,3)$ take values. The transformations
\be
x_1\ = \ \frac{2\,{\rm i}}{\sqrt{3}} \ \varphi \ - \ 2\, \mathcal{A}\,,
\qquad x_2\ = \ -\,\frac{4\,{\rm i}}{\sqrt{3}} \ \varphi  \label{TodaA3}
\ee
suffice to map eq.~(\ref{TodaA2}) into a Lagrangian proportional to (\ref{onefield}) with the gauge choice
\be
\qquad \mathcal{B} \ = \ \mathcal{A} \label{TodaA4}
\ee
and
\be
\mathcal{V}_{A_2}(\varphi)\ = \ \lambda \, \cos\left ( 2\, \sqrt{3} \, \varphi \right ) \  \label{trigo2}
\ee
As a result, even this trigonometric potential defines an integrable cosmological model.

Finally, let us remark that the trigonometric potential \eqref{trig_1}, which was already related to a Liouville system with $\gamma=i$, can be also retrieved starting the Toda--chain Lagrangian based on the simple roots of the $D_2$ system,

\be {\cal L}_{D_2} \ = \ \frac{1}{2} \ \Big(\dot{x_1}^{2} \ + \
\dot{x_2}^{2} \Big) \ + \ \lambda \ \Big( e^{x_2\ - \ x_1}\ + \
e^{-\ x_2\ - \ x_1}\Big) \ , \label{TodaD2}
\ee
working in the gauge
\be
\mathcal{B} \ = \ \mathcal{A}\  , \label{TodaF2}
\ee
via the redefinitions
\be
x_1\ = \ - \ 2\, {\cal A} \ , \qquad x_2\ = \ 2 \, i \ \varphi\ . \label{TodaE2}
\ee
%
\section{\sc  Orientifold Vacua and Exponential Potentials}\label{sec:orientifolds}

In orientifold vacua \cite{orientifolds}, non--conventional projections and charge neutrality conditions can conspire to force classically stable non--BPS combinations
of branes and orientifolds into a bulk that was originally supersymmetric.  This phenomenon is usually termed ``brane supersymmetry breaking'' (BSB) \cite{bsb}, and its simplest
manifestation occurs in the ten--dimensional Sugimoto model in \cite{bsb}, a ``false friend'' of the type-I superstring. Whereas the latter has an $SO(32)$ gauge
group resulting from $D9$ branes and a conventional $O9-$ orientifold with \emph{negative} tension and charge, the Sugimoto model has an $O9+$ orientifold with \emph{positive}
tension and charge, so that charge neutrality forces $\overline{D}9$ branes into its vacuum. As a result, \emph{supersymmetry is broken at the string scale} leaving behind no order
parameter to restore it in ten dimensions, and the low--energy effective field theory is consequently the Volkov--Akulov supergravity \cite{va} described in \cite{dmpr}, to be contrasted with
the more conventional $N=(1,0)$ coupled system of Supergravity and supersymmetric Yang--Mills theory that accompanies the type--I superstring. All in all, these models are characterized by runaway potentials that leave no other option in flat space than complicated resummations \cite{resummations}. Lower--dimensional vacua are largely uncharted, although the nine--dimensional solution of \cite{dm9d} is clearly a first significant step.

Our focus in this paper has been on cosmological models with background metrics of the type
\be
ds^{\,2} \ = \ - \ e^{\,2\,{\cal B}(t)}  \, dt^2 \ + \ e^{\,2\,A(t)}  \, d{\bf x} \cdot d{\bf x} \ . \label{st11}
\ee
In order to keep an eye on the possible role of BSB in the Early Universe, we can forego a detailed discussion of low--energy couplings, but for one key fact: in models of this type, the tensions of
orientifolds and branes conspire to generate tree--level exponential potentials whose string--frame form is uniquely determined by their origin from the (projective) disk. Similar, if more intricate
constructions, involving non--BPS combinations of orientifolds and (anti--)$D9$ and $D5$ branes, first show up in six dimensions \cite{bsb}. No tachyons are again present, to lowest order,
in these more complicated settings, although supersymmetry is again broken in their open sectors at the string scale from the outset. In lower dimensions, or after $T$--duality transformations, geometric
moduli or lower--dimensional branes can also intervene, and bring about singular limits where supersymmetry can be recovered and, correspondingly, more conventional low--energy supergravities that clearly
deserve a closer look, but the very existence of the Sugimoto model of \cite{bsb} is somehow a conceptual puzzle for String Theory, on the par with the celebrated option of recovering eleven non--compact dimensions \cite{11dsugra} via string dualities \cite{11d}.

Despite the aesthetic appeal of settings where supersymmetry breaking is an inevitable feat, the applications of BSB in conventional Minkowski backgrounds have been seriously hampered by the corresponding
exponential potentials, which for one matter have no critical points, so that conventional perturbative techniques are inapplicable without resummations, which lie well beyond current
technology \cite{resummations}. On the other hand, the string and Planck scales are typically close enough to make one wonder whether this class of string vacua is trying to provide
interesting clues on the Early Universe. This expectation drove the analysis of \cite{dks}, where the Sugimoto model of \cite{bsb} was shown to result in intriguing spatially flat cosmologies where a scalar (the dilaton in the ten--dimensional case) can only emerge from the initial singularity while climbing a steep exponential potential. This behavior was shown to persist even in the presence of the non--minimally coupled axion partner present in supergravity--inspired \cite{noscale} scenarios as in \cite{kklt}, and actually we should stress that BSB provides a natural origin for their uplift. These considerations were stimulated by exact solutions that have actually a long history \cite{lm,exp_sol}, and may be regarded as the simplest members of the cosmologies driven by combinations of exponential potentials that we have explored in the preceding sections.

One can actually identify a few interesting types of exponential potentials that may emerge at tree level in models where BSB is at work. This entails following closely two scalar modes that are generally present in these orientifolds, up to dualities, the dilaton and the scalar associated to the volume associated to compact extra dimensions. The ensuing derivation is inspired by the four--dimensional analysis in \cite{wit4} and will also provide an opportunity to correct eq.~(2.17) of \cite{dks}, which conflicts with the correct four--dimensional analysis presented there.

In the string frame and in a mostly negative signature, the ten--dimensional low--energy effective field theory of the Sugimoto model of \cite{bsb} includes the terms
\be
{\cal S}_{10} \,=\, \frac{1}{{2\kappa_{10}^2}} \ \int d^{\,10} x \sqrt{-\widetilde{g}}
\, e^{\,-\,2\,\phi}\,  \Big[ \, R(\widetilde{g}) \, - \, 4 \, (\partial \phi)^2 \, - \, T_{9}  \, e^{\,-\, \phi} \, +\, \ldots\, \Big]  \, , \label{st1}
\ee
which suffice for our purposes since, as in the preceding section, we shall confine our attention to the metric, whose string--frame form we call here $\widetilde{g}$, and to the dilaton $\phi$. The Weyl rescaling
\be
\widetilde{g}_{MN} \ = \ g_{MN} \ e^{\,\frac{\phi}{2}} \label{st112}
\ee
of the ten--dimensional metric turns the action \eqref{st1} into its Einstein--frame form
\be
{\cal S}_{10} \,=\, \frac{1}{{2\kappa_{10}^2}} \ \int d^{\,10} x \sqrt{-g}
\, \Big[ \, R \, + \, \frac{1}{2} \, (\partial \phi)^2 \, - \, T_{9}  \, e^{\,\frac{3}{2}\, \phi} \, +\, \ldots\, \Big]  \, , \label{st2}
\ee
while a subsequent Kaluza--Klein reduction to $d$ dimensions on the metric
\be
ds^2 \ \equiv \ {g}_{MN} \, dx^M\,dx^N \ = \ e^{\,-\, \frac{(10-d)}{(d-2)} \,\sigma} \ g_{\mu\nu}\, dx^\mu\, dx^\nu \ - \ e^{\,\sigma} \ \delta_{ij}\, dx^i \, dx^j \ , \label{st3}
\ee
where $\sigma$ reflects the internal volume, results in the Einstein--frame $d$--dimensional effective action
\bea
{\cal S}_{d} &=& \frac{1}{2\kappa_{d}^2} \ \int d^{\,d} x \sqrt{-g}
\  \Big[  \, R \ + \ \frac{1}{2} \ (\partial \phi)^2  \ + \ 2\ \frac{(10-d)}{(d-2)} \ (\partial \sigma)^2 \nonumber \\
 &-& \ T_{9}  \  e^{\,\frac{3}{2}\, \phi} \ e^{\,-\, \frac{(10-d)}{(d-2)}\, \sigma} \Big] \, . \label{st4}
\eea
It is thus convenient to define the two fields
\bea
&& \Phi_t \ = \ \sqrt{\frac{d-2}{2(d-1)}}\ \left( \frac{3}{2} \ \phi\ - \  {\frac{10-d}{d-2}} \ \sigma \, \right)\ , \\
&& \Phi_s \ = \ \sqrt{\frac{10-d}{2(d-1)}}\ \left( \frac{1}{2}\ \phi \ + \ 3 \, \sigma \,\right) \ ,  \label{st5}
\eea
only one of which enters the tadpole potential, so that eq.~\eqref{st4} takes finally the form
\be
S_{d} \ = \ \frac{1}{2\kappa_{d}^2} \ \int d^{\,d} x \sqrt{-g}
\  \Big[  \, R \ + \ \frac{1}{2} \ (\partial \Phi_s)^2  \ + \ \frac{1}{2} \, (\partial \Phi_t)^2 \ -  \ T_{9}  \  e^{\, \sqrt{\frac{2(d-1)}{d-2}}\ \Phi_t} \Big] \, . \label{st6}
\ee
Following standard practice, in the ensuing discussion we shall also begin by assuming that $\Phi_s$ is somehow stabilized, so that the dynamics is dominated by $\Phi_t$ alone. As a result, $S_d$ reduces for our purposes to
\be
{\cal S}_{d} \ = \ \frac{1}{2\kappa_{d}^2} \ \int d^{\,d} x \sqrt{-g}
\  \Big[  \, R \ + \ \frac{1}{2} \, (\partial \Phi_t)^2 \ -  \ T_{9}  \  e^{\,\sqrt{\frac{2(d-1)}{d-2}}\ \Phi_t} \Big] \ , \label{st7}
\ee
so that the scalar potential takes the form
\be
V_9 \ = T_9 \ e^{\, \gamma_9 \, \sqrt{\frac{2(d-1)}{d-2}}\ \Phi_t} \ , \label{gamma9}
\ee
and taking into account the choice of normalization for ${\cal S}_d$ made in String Theory the comparison between eqs.~\eqref{st7} and \eqref{redefin2} results in the redefinition
\be
\varphi \ = \ \sqrt{\frac{d-1}{2(d-2)}} \ \Phi_t \ , \label{varphiphit}
\ee
and thus in the striking fact that $\gamma_9 = 1$ \emph{for all dimensions} $4 \leq d \leq 10$.

In Sections \ref{sec:integrable} and \ref{sec:properties_exact} we have elaborated at length on the special meaning of this value, generalizing the results of \cite{dks,dkps}. Suffice it to repeat here that, in models involving exponential potentials as in \eqref{gamma9}
and with background metrics as in eq.~\eqref{st11}, if $\gamma_9 \geq 1$ $\Phi_t$ deserves to be called a ``climbing scalar'', since it is forced to emerge from the initial singularity while
climbing up the potential, while if $\gamma_9 < 1$ a ``descending'' solution also exists \footnote{One can show that this type of asymptotic behavior persists even in extensions of \eqref{st11} with curved spatial slices, but in this paper we shall confine our attention to the spatially flat case.}.

Notice also that if $\Phi_s$ is somehow stabilized, letting for brevity $\Phi_s = 0$, eqs.~\eqref{st5} and \eqref{varphiphit} imply that
\be
\varphi \ = \ \frac{2(d-1)}{3(d-2)} \ \phi \ .
\ee
The original dilaton and the rescaled field $\varphi$ that enters the cosmological equations of Section \ref{sec:effelag} thus coincide in this case in our dimensions, so that closed--string loop factors contribute indeed integer powers of $\exp(2\,\varphi)$ to the potential, as anticipated before eq.~\eqref{duality}.

Lower--dimensional vacua of this type contain in general other ingredients, to which we can now turn. These include the $D5$ (or $\overline{D}5$) branes, which are widely present in
the models of \cite{bsb}, and also the non--BPS $D3$ identified in \cite{dms} along the lines of Sen's approach \cite{sen}, which is stable also in the presence of $D9$ branes. Proceeding as above, under the assumption that $\Phi_s$
remains somehow stabilized despite the fact that it enters these other contributions, one can show that in dimensions $d \leq 6$, where they are space filling, $D5$ branes give rise to an additional exponential potential with
\be
\gamma_5 \ = \ \frac{d}{2(d-1)} \ ,
\ee
which lies below the critical value $\gamma_5=1$ but above the upper bound for slow--roll inflation,
\be
\gamma_{s.r.} \ = \ \frac{1}{\sqrt{d-1}} \ . \label{uppergamma}
\ee
A similar reasoning \cite{dks} shows that in four dimensions the non--BPS $D3$ brane of \cite{dms} would bring about a space--filling contribution with
$\gamma_3 = \frac{1}{2}$, which can drive an inflationary phase since this value lies below the upper bound \eqref{uppergamma}.
On the other hand, taking to account both $\Phi_s$ and $\Phi_t$, in the general four--dimensional case when $D9$ (anti--)branes and the stable non--BPS brane of \cite{dms} are present, one
would be confronted with models of the type
\bea
{\cal S}_{4} &=& \frac{1}{{2\kappa_{4}^2}} \ \int d^{\,4} x \sqrt{-g}
\, e^{\,-\,2\,\phi}\,  \Big[ \, R \ + \ \frac{1}{2} \ (\partial \Phi_s)^2  \ + \ \frac{1}{2} \, (\partial \Phi_t)^2 \nonumber \\ &-& T_{9}  \  e^{\, \sqrt{3}\ \Phi_t} \ - \ T_5 \ e^{\, \sqrt{3}\left( \frac{2}{3} \ \Phi_t \ - \ \frac{1}{\sqrt{3}} \ \Phi_s
\right)} \ - T_3 \ e^{\, \sqrt{3}\left( \frac{1}{2} \ \Phi_t \ - \ \frac{\sqrt{3}}{2} \ \Phi_s \right)}  \ +
\ \ldots\, \Big]  \, . \label{st4c}
\eea
This expression has a very interesting structure, and confining the attention to case where $D5$--branes are not present, so that the $D3$--brane is stable, it is instructive to take a closer look at the
two--scalar dynamics lifting the assumption that various sorts of corrections that we have neglected stabilize $\Phi_s$.

To begin with, one can notice that independent shifts of
$\Phi_s$ and $\Phi_t$ can turn the general scalar potential resulting from $D9$ and $D3$ branes into the universal form
\be
V \ = \ V_0 \left( e^{\, \sqrt{3}\ \Phi_t} \  +  \ e^{\, \sqrt{3}\left( \frac{1}{2} \ \Phi_t \ - \ \frac{\sqrt{3}}{2} \ \Phi_s \right)} \right) \ ,
\ee
which comprises ``critical'' values of $\gamma$ along a pair of directions that are 60 degrees apart. There is rich set of solutions, and indeed ``climbing'' still takes place near the initial
singularity for a range of initial directions, although it is no more inevitable, while the late--time dynamics is dominated by a (fast--roll) attractor along the valley between the two peaks.

The integrable spatially--flat cosmologies that we have described include the two classes of examples of eqs.~\eqref{int26} and \eqref{Liouville11}, whose potentials convey interesting lessons for this type of compactifications in the presence of BSB. Indeed, as we have seen they result in a relatively simple
integrable dynamics, but are really close in spirit to what has just emerged from BSB models with $D9$ and $D3$ branes once one assumes that $\Phi_s$ is somehow stabilized by higher--order corrections.
Both cases involve indeed a climbing scalar that bounces essentially against a hard wall and subsequently injects an inflationary phase, in qualitatively similar fashions.

One can add to this discussion a further degree of freedom, allowing for an off-critical bulk of dimension $d$. Confining our attention to the case $d>10$, let us add some cursory remarks on the
resulting potential after a compactification to four dimensions. For simplicity, let us confine our attention to the contributions arising from $D9$ branes and from the conformal anomaly originally
described by Polyakov in \cite{polyakov}. Up to shifts of the two fields $\Phi_s$ and $\Phi_t$, the resulting potential contains again two terms with identical normalizations,
and assuming again that $\Phi_s$ is somehow stabilized, one is finally confronted with
\be
V \ = \ V_0 \left( e^{\, \sqrt{3}\ \gamma_9 \ \Phi_t} \  +  \ e^{\, \sqrt{3} \ \gamma_\Lambda \ \Phi_t} \right) \ ,
\ee
where
\be
\gamma_9 \ = \  \sqrt{\frac{d^2 \,- \, 14\,d \,+\, 184}{24\,(d\,-\, 4)}} \ , \quad \gamma_\Lambda \ = \ - \ \frac{10}{3} \ \frac{(d\,-\,4)(d\,-\,10)}{\sqrt{2(d^2 \,- \, 14\,d \,
+\, 184)}}\
\ee
Interestingly, for $d$ slightly larger than ten $\gamma_\Lambda$ is \emph{small and negative} while $\gamma_9$ is very close to one, so that one has a potential well which combines a steep wall with a rather flat one. As a result, the scalar is essentially bound to emerge from the initial singularity with the scalar descending along the mild wall and to stabilize readily at the bottom as the Universe enters a de Sitter phase. We met potentials of this type in Sections \ref{sec:Bnot0_sol} and \ref{sec:sporadic_pot}.

We would like to conclude this section with a brief look into the possible link between $\gamma$ and other fundamental branes. This step is motivated in part by the wide scan carried out in \cite{branescan}, and the computation involved is a slight generalization of what we presented so far, the main new ingredient being a more general coupling to the dilaton of a $p$--brane in its natural presentation, in the string frame and in $p+1$ dimensions, namely
\be
S_{p} \ \sim \ - \ \int d^{\,p+1} x \, \sqrt{-g}\, e^{\,-\,\alpha \, \phi} \ .
\ee
The following steps parallel what we have seen so far, and thus involve the Weyl rescaling that puts the bulk portion of the effective action in the Einstein frame, a wrapping of the $p$--brane from the original $(p+1)$--dimensional space to four dimensions, the field redefinitions \eqref{st5} of $\phi$ and the breathing mode $\sigma$ into $\Phi_t$ and $\Phi_s$ and, again, the \emph{assumption} that $\Phi_s$ is somehow stabilized. All in all, one is thus led to
\be
\gamma \ = \ \frac{1}{12} \ \big(p \ + \ 9 \ - \ 6\, \alpha \big) \ , \label{gammabranes}
\ee
so that these $\gamma$'s are amusingly all multiples of $\frac{1}{12}$. The allowed values include those that we already ran across for the $D9$ and $D3$ branes, while the value $\frac{1}{12}$ appears related to an NS fivebrane, which is interestingly \emph{unstable} in these orientifold models, with one of its dimensions wrapped on a compact internal cycle. This value is tantalizingly close to those investigated in \cite{dkps} in connection with the class of potentials
\be
V \ = \ V_0 \left( e^{\,2\varphi} \ +\  e^{\,2\,\gamma\,\varphi}\right)\ ,
\ee
minimal modifications of the $D9$--brane term that embody the climbing mechanism and can account qualitatively for CMB power suppression at large scales within a perturbative regime for the string coupling. However, more than a word of caution is needed. As the Referee correctly stressed to us, these top--down considerations cannot afford a quantitative comparison with the CMB in the absence of well--motivated completions of potential, which are necessary for instance to grant a proper tensor--to--scalar ratio and whose presence will affect in general the precise link between the available $\gamma$'s and the spectral index $n_s$.

\section{\sc  Conclusions}\label{sec:conclusion}

This paper was largely devoted to the search for exact cosmological solutions involving a single scalar field $\phi$ with a standard kinetic term in spatially flat metrics. Most of our analytic solutions, however, were obtained within a mild generalization of the FLRW setting that rests on a gauge function ${\cal B}(t)$, as in eq.~\eqref{FLRWgen}, so that they are expressed in terms of a ``parametric time'' $t$. Although in most cases the actual relation between $t$ and the cosmic time measured by comoving observers is somewhat implicit, we showed how to extract the key physical indications of these models. Hopefully, these exact solutions will stimulate further progress in the development of the inflationary scenario, which can nowadays begin to afford detailed comparisons with the new data provided by the PLANCK experiment \cite{XCMB}.

The objective of our search was the scalar potential $V(\phi)$ of the models, whose nature determines whether the corresponding equations are integrable, and in particular whether they can be integrated explicitly. Our search for integrable potentials was driven by the reduced action principle of eq.~\eqref{onefield}, and the corresponding equations of motion \eqref{onefield_eqs} always include the Freedman constraint, the condition that the Hamiltonian vanish for the allowed solutions. Canonical transformations were identified that could map the reduced Lagrangians into known types of integrable systems with two degrees of freedom, whose construction represents a major achievement of Mathematical Physics \cite{hietarinta}--\cite{inozem}. The resulting bestiary comprises different classes of models of increasing complexity: separable linear systems, triangular systems, separable systems that are integrable by quadratures and some sporadic integrable systems. To begin with, we thus identified the nine families of potentials collected in Table \ref{tab:families}, whose solutions are explicit (in terms of $t$) up to the inversion of quadrature integrals, some of which are defined in the complex plane. In addition, the remarkable literature on two--dimensional dynamical systems led us to identify 26 sporadic models whose Liouville integrability is guaranteed by a second conserved charge accompanying their Hamiltonians. Although their explicit solutions would be far more complicated, the intuition gained from the simpler cases in Table \ref{tab:families} made their qualitative features rather transparent.

All the integrable potentials that we have identified involve combinations of exponential functions. As we anticipated in the preceding sections, these are ubiquitous in truncations to Cartan sectors of gauged supergravity models with scalar manifolds that are (necessarily non-compact) homogenous spaces $\mathrm{G/H}$. For $\mathcal{N}\ge 3$ this property is inevitable, while for $\mathcal{N}\le 2$ it identifies an important and widely investigated subset of the possible models. Furthermore, exponential potentials also emerge in orientifold models with ``brane supersymmetry breaking'' (BSB), where supersymmetry is broken at the string scale and is non--linearly realized in the low--energy Supergravity \cite{dmpr}, so that little is known to date about their low--energy manifestations in lower dimensions.
This work therefore revealed some interesting features, but it also raised some questions that remain unanswered and appear urgent and potentially very instructive. Generally speaking, the most urgent question is the following: \textit{can some of our integrable potentials emerge from Supergravity?} This is the case for the single exponential, albeit in an unconventional context where supersymmetry is non--linearly realized, but the general pattern is not known in detail and taking a closer look at it starting from the more conventional case of linear supersymmetry appears timely and potentially very interesting.

One approach to the problem that could be christened \textit{minimalist} posits that perhaps the integrable cosmological models cannot be deduced from Supergravity, and yet their solutions can provide good approximations to similar field equations that emerge from it under appropriate conditions. In a future publication \cite{prep} we shall elaborate on the effectiveness of this approach, drawing comparisons between cosmological solutions arising from integrable potentials and others arising from similar, non integrable, ones. One could provide manifold instances of this approach, which also drove mathematicians to discover some of the integrable two dimensional models that we considered here. For instance, it is well known that the study of sporadic integrable potentials of the H\'enon--Heiles type was motivated by some equations that describe the motion of stars in a plane near the galactic center, albeit with $a,b,c$ coefficients that are different from those of eq.~(\ref{GIapar}) that guarantee integrability. Historically the minimalist approach was fully justified, since after all the physical H\'enon--Heiles equations do not concern the fundamental equations of a theory (in this case Newton's theory) but emerge in applications to a particular physical system, so that finding approximate integrable models capturing some essential features is certainly very rewarding in this context.

A minimalist approach is however less satisfactory in attempts to address problems related to the origin and fate of our Universe, so that in this context the final aim should perhaps be framed within a \textit{maximalist approach}. In this respect, one should keep in mind that integrability reflects hidden symmetries bringing along additional conserved charges, and when touching upon Fundamental Physics one should not be indifferent to their role. Rather, one should try to connect hidden symmetries granting integrability and key symmetry principles of the complete theory, and in this respect clarifying whether and how integrable cosmological models might fit into candidate unified theories of all interactions appears a matter of utmost relevance, to which we plan to return elsewhere \cite{prep,FayetIlio}. In the first of these works we shall explore the possibility of deriving integrable cosmological models within $\mathcal{N}=1$ Supergravity by suitable choices of superpotential. Notwithstanding the difficulties that we met in this task, whose technical origin will be emphasized in \cite{prep}, we have already identified a small number of supersymmetric integrable models based on the coupling of $\mathcal{N}=1$ Supergravity to a single Wess-Zumino multiplet. Clearly there is a far wider hunting ground than what is provided by a single multiplet, and appropriate strategies should be developed to classify and explore one--field truncations of more general supergravity models. This sets up a program that overlaps to a considerable extent with efforts aimed at classifying supergravity gaugings with due attention to their vacua and to the key issue of moduli stabilization. The second planned paper \cite{FayetIlio} will be devoted to the classification of $\mathcal{N}=2$ gaugings with the method of the embedding tensor. While the identification of integrable one--field truncations will represent a side issue there, we shall provide nonetheless a proof of the non--integrability of Fayet--Iliopoulos abelian gaugings of $\mathcal{N}=2$ models based on $\mathrm{G_4}/\mathrm{H_4}$ scalar manifolds. Such a proof emerges from a close and challenging comparison between cosmological potentials induced by abelian gaugings and corresponding black--hole geodesic potentials of the same theory. In both cases the complete field equations can be reformulated in terms of a Poissonian system on the coadjoint orbits of the Borel subalgebra of the numerator group $\mathcal{B}(\mathrm{G_3}) \subset \mathrm{G_3}$, where the enlarged group $\mathrm{G_3} \supset \mathrm{G_4}$ emerges in both time--like and space--like dimensional reductions of Supergravity to three dimensions. The Poissonian structure is uniquely identified in both cases by the Lie--Poisson tensor defined by the structure constants of $\mathcal{B}(\mathrm{G})$, while the difference resides in the quadratic form that determines the Hamiltonian of the system. In the black--hole case this Hamiltonian is related to the solvable group manifold $\mathcal{B}(\mathrm{G_3})$, and thus to the invariant metric of a symmetric space $\mathrm{G_3}/\mathrm{H}^\star$, which implies the existence of a Lax--pair representation and complete integrability of the dynamical system, while on the contrary the quadratic form that enters the cosmological model corresponds to a metric on a solvable group manifold that is not even Einstein, and whose isometries cannot be extended to the full $\mathrm{G_3}$. This argument excludes the Lax--pair representation and the full integrability of the multi--field dynamical system. Strictly speaking, however, it does not exclude the integrability of special one--field truncations, which appears nonetheless somewhat unlikely, while in the black--hole case all consistent truncations inherit this property. This negative result \cite{FayetIlio} is consistent with the positive ones to be presented in \cite{prep}. For instance, an integrable gauging of the $\mathcal{S}^3$ model that we have found does not respect its $\mathcal{N}=2$ special K\"ahler structure but only its $\mathcal{N}=1$ K\"ahler structure, and indeed the specific superpotential leading to an integrable cosmological model is not a Fayet--Iliopoulos gauging.

All in all, the present indications are that integrable cosmological models have a better chance of fitting in Supergravity only after a breaking to $\mathcal{N}=1$
supersymmetry induced, for instance, by orbifold and/or orientifold projections within suitable flux compactifications. In this case, the  superpotential $W(z)$ for the residual moduli fields, for which explicit expressions are available in the literature in terms of fluxes and group structure constants, might possess special properties reflecting the hidden symmetries underlying integrability.
\par
 How about, then, if an integrable cosmological model were not to fit in Supergravity while observational data turned out, instead, to be consistent with it? Perhaps even this should not be read as circumstantial evidence for the minimalist approach, and indeed one might even conceive to turn the requirement of consistency with integrable cosmological models into an indication to discriminate between different candidate theories, together with several other constraints that draw their origin from Particle Physics or Astrophysics. After all, if integrability can reflect hidden symmetries, uncovering their threads and intertwining them into the tapestry of the fundamental laws of Physics can perhaps add to the well-established lessons that manifest symmetries have already provided over the years. The list of integrable cosmological models that we have been able to compile in the present paper may be regarded as a first concrete step in this direction. While we cannot claim that it be exhaustive, the integrable one--field potentials that we have investigated and the corresponding exact solutions constitute clearly a firm starting point for future investigations along these lines.

In this respect, we ought to stress that a new perspective emerged after the original version of this paper was sent to the ArXiv. Developing ideas put forward in \cite{minimalsergioKLP} (see also the related works \cite{Kallosh:2013hoa} -- \cite{Farakos:2013cqa}), two of us showed in \cite{copernicana} that positive--definite potentials, and among them all the integrable ones presented in this paper, can find a place in $\mathcal{N}=1$ supergravity as $D$-terms related to suitable one--dimensional K\"ahler manifolds. Such manifolds will be studied further in a forthcoming publication \cite{fromCopernicToKepler}, where they will be termed $D$--map images of the corresponding potentials. In particular, the $D$--maps of the potentials belonging to the second integrable series in the Table,
\begin{equation}\label{goodnewsgamma}
  V \, = \, c_1 \, e^{\, 2 \, \gamma \, \varphi} \, + \, c_2 \, e^{ \, (\gamma+ 1) \, \varphi} \ ,
\end{equation}
have a remarkable mathematical profile. However, the higher--dimensional origin of these K\"ahler manifolds and their links with String Theory remain a challenge for future work.

 Let us conclude by summarizing the main results of our study. Insofar as the exact solutions are concerned, they are the following:
 \begin{itemize}
   \item[1. ] the \emph{``climbing phenomenon''} is a generic property of ``critical'' or ``over--critical'' potentials that possess an asymptotically exponential behavior. This was first observed in \cite{dks} and here we have produced a number of illustrative examples involving more complicated potentials. As these systems emerge from the Big Bang, the scalar field cannot descend along such steep grades, or alternatively these potentials do not let the scalar field escape to infinity from their pull when the system proceeds toward a Big Crunch. The explicit analysis recently presented in \cite{cd} revealed this fact clearly for the ``critical'' $\cosh$--well;
   \item[2. ] \emph{a Big Crunch follows inevitably a Big Bang} whenever the scalar field tries to settle at a \emph{negative} extremum of the potential, consistently with our initial assumption \eqref{piatttosa} on the space--time metric and with the absence of flat AdS slices. On the other hand, an \emph{eventual de Sitter phase} is inevitably attained whenever the scalar field tries to settle at a \emph{positive} extremum of the potential. This was expected, since AdS has no spatially flat metrics, or alternatively negative extrema are non--admissible fixed points for the corresponding dynamical systems.
 \end{itemize}

Insofar as the links with String Theory are concerned, although our current grasp of the dynamics in the presence of non--linear supersymmetry is undeniably rather primitive, we have clarified how the available brane types reflect themselves, in lower dimensions, in corresponding exponential potentials. Our key result rests on the combined effects of the dilaton and of the breathing mode of the extra dimensions within a class of orientifold models \cite{orientifolds} which exhibit BSB \cite{bsb}, on the neglect of higher--derivative terms and on a simplifying assumption that is not more justified at present than any other reductions to one--field models of inflation. This posits that one combination of these fields, the one entering the steep exponential inherited from ten dimensions, dominates the early cosmological phase. Once this is done, however, an interesting prediction follows for the four--dimensional potentials
\be
V \ = \ V_0 \left( e^{\, 2\, \varphi} \  +  \ e^{\, 2\, \gamma \, \varphi} \right)
\ee
that were explored in \cite{dkps}. Here the first, steep exponential, is the signature of BSB and forces the scalar to climb up as it emerges from the initial singularity, while the other drives the subsequent inflationary phase. Under the assumptions discussed in Section 6, one arrives at the simple expression
\be
\gamma \ = \ \frac{1}{12} \ \big(p \ + \ 9 \ - \ 6\, \alpha \big)
\ee
for the values of $\gamma$ that are allowed in String Theory.
Here $p$ is the number of unwrapped dimensions of the brane and $\alpha$ is another integer, the inverse power of the string coupling that enters its world--volume action. All in all, the values of $\gamma$ associated to the couplings of various types of branes, a wide scan of which was recently presented in \cite{branescan}, to the field $\Phi_t$ of Section 6 are thus quantized in units of $\frac{1}{12}$.


\vskip 48pt
\section*{Acknowledgments}
We are grateful to E.~Dudas, S.~Ferrara, S.~P.~Patil, G.~Pradisi, F.~Riccioni and especially to N.~Kitazawa and M.~Trigiante for several stimulating discussions, and to CERN, the \'Ecole Polytechnique, Scuola Normale Superiore and the Lebedev Institute for the kind hospitality extended to one or more of us during the course of this work. We are also grateful to the Referee for his/her very useful comments. This work was supported in part by the ERC Advanced Grants n. 226455 (SUPERFIELDS) and n. 226371 (MassTeV), by Scuola Normale Superiore, by INFN (I.S. TV12), by the contract PITN-GA-2009-237920 and by the MIUR-PRIN contract 2009-KHZKRX. The work of A.S. was supported in part by the RFBR Grants No. 11-02-01335-a, No. 13-02-91330-NNIO-a and No. 13-02-90602-Arm-a.

\end{document}